\documentclass[a4paper,11pt]{article}
\usepackage{jheppub}

\usepackage[T1]{fontenc} 
\usepackage[utf8]{inputenc} 
\usepackage{amssymb, latexsym, amsmath, amsthm}
\usepackage{mathtools}
\usepackage{gensymb}
\usepackage{graphicx}
\usepackage{subcaption}
\usepackage[colorlinks=true]{hyperref}
\usepackage{array}
\usepackage{rotating,float,caption}
\usepackage{multirow}
\usepackage{braket}
\usepackage{bm}
\usepackage{tikz}
\usepackage{caption}
\usepackage{slashed}
\usepackage{makecell}
\usepackage{placeins}

\usepackage{color}

\usepackage{hyperref}
\hypersetup{
	colorlinks = true,
	linkcolor = {blue}
}

\renewcommand{\arraystretch}{1.3}

\DeclareOldFontCommand{\bf}{\normalfont\bfseries}{\textbf}
\DeclareOldFontCommand{\tt}{\normalfont\bfseries}{\texttt}
\DeclareMathAlphabet\mathbfcal{OMS}{cmsy}{b}{n}

\newcommand{\ampnum}[2]{\mathcal{A}_{#1}^{#2}}
\newcommand{\ampnumstraight}[2]{A_{#1}^{#2}}
\newcommand{\anum}[2]{a_{#1}^{#2}}
\newcommand{\me}[2]{\mathcal{M}_{#1}^{#2}}
\newcommand{\ketampdep}[3]{\ket{\ampnum{#1}{#2}(#3)}}

\newcommand{\ketamp}[2]{\ket{\ampnum{#1}{#2}}}
\newcommand{\braamp}[2]{\bra{\ampnum{#1}{#2}}}
\newcommand{\Iop}[2]{\bm{I}^{(#1)}\left(#2\right)}
\newcommand{\Iopd}[2]{\bm{I}^{(#1),\dagger}\left(#2\right)}
\newcommand{\Hop}[2]{\bm{H}^{(#1)}\left(#2\right)}
\newcommand{\Hopd}[2]{\bm{H}^{(#1),\dagger}\left(#2\right)}

\newcommand{\ourIop}[3]{\mathcal{I}^{(#1)}_{#2}\left(#3\right)}

\newcommand{\ourHopnodep}[2]{\mathcal{H}^{(#1)}_{#2}}
\newcommand{\ourHop}[3]{\mathcal{H}^{(#1)}_{#2}\left(#3\right)}
\newcommand{\ourHopd}[3]{\mathcal{H}^{(#1)}_{#2}\left(#3\right)}

\newcommand{\Itwoepsiloncoeff}{\text{e}^{-\epsilon\gamma_E}\dfrac{\Gamma(1-2\epsilon)}{\Gamma(1-\epsilon)}\left(\betaoe+K\right)}

\newcommand{\Jcol}[1]{\mathbfcal{J}^{(#1)}}
\newcommand{\Jcolb}[1]{\overline{\mathbfcal{J}}^{(#1)}}

\newcommand{\Jfull}[1]{\mathcal{J}_{2}^{(#1)}}
\newcommand{\Jfullb}[1]{\overline{\mathcal{J}}_{2}^{(#1)}}
\newcommand{\Jfullic}[2]{\mathcal{J}_{2,#2}^{(#1)}}
\newcommand{\Jfulltotic}[2]{\mathcal{J}_{#2}^{(#1)}}

\newcommand{\J}[1]{J_{2}^{(#1)}}
\newcommand{\Jh}[1]{\hat{J_{2}}^{(#1)}}
\newcommand{\Jt}[1]{\tilde{J_{2}}^{(#1)}}
\newcommand{\Jht}[1]{\hat{\tilde{J_{2}}}^{(#1)}}
\newcommand{\Jhh}[1]{\hat{\hat{J_{2}}}^{(#1)}}
\newcommand{\Jb}[1]{\bar{J_{2}}^{(#1)}}
\newcommand{\Jtb}[1]{\tilde{\bar{J_{2}}}^{(#1)}}

\newcommand{\Jic}[2]{J_{2,#2}^{(#1)}}
\newcommand{\Jhic}[2]{\hat{J}_{2,#2}^{(#1)}}
\newcommand{\Jtic}[2]{\tilde{J}_{2,#2}^{(#1)}}
\newcommand{\Jhtic}[2]{\hat{\tilde{J}}_{2,#2}^{(#1)}}

\newcommand{\XFFint}[3]{\mathcal{#1}_{#2}^{#3}}
\newcommand{\XIFint}[4]{\mathcal{#1}_{#2,#4}^{#3}}
\newcommand{\XIIint}[4]{\mathcal{#1}_{#2,#4}^{#3}}

\newcommand{\XhFFint}[3]{\wh{\mathcal{#1}}_{#2}^{#3}}
\newcommand{\XhIFint}[4]{\wh{\mathcal{#1}}_{#2,#4}^{#3}}
\newcommand{\XhIIint}[4]{\wh{\mathcal{#1}}_{#2,#4}^{#3}}

\newcommand{\XtFFint}[3]{\wt{\mathcal{#1}}_{#2}^{#3}}
\newcommand{\XtIFint}[4]{\wt{\mathcal{#1}}_{#2,#4}^{#3}}
\newcommand{\XtIIint}[4]{\wt{\mathcal{#1}}_{#2,#4}^{#3}}

\newcommand{\Gammaoneconv}[2]{\Gamma^{(1)}_{#1}(x_{#2})}
\newcommand{\Gammaone}[2]{\Gamma^{(1)}_{#1}\left(#2\right)}
\newcommand{\BGammaone}[2]{\mathbf{\Gamma}^{(1)}_{#1}\left(#2\right)}
\newcommand{\BGammaonenodep}[1]{\mathbf{\Gamma}^{(1)}_{#1}}

\newcommand{\Gammatwo}[2]{\overline{\Gamma}^{(2)}_{#1}\left({#2}\right)}
\newcommand{\BGammatwo}[2]{\overline{\mathbf{\Gamma}}^{(2)}_{#1}\left(#2\right)}

\newcommand{\GammaoneFconv}[2]{\wh{\Gamma}^{(1)}_{#1}(x_{#2})}
\newcommand{\GammaoneF}[2]{\wh{\Gamma}^{(1)}_{#1}\left({#2}\right)}
\newcommand{\GammatwoF}[2]{\wh{\overline{\Gamma}}^{(2)}_{#1}\left(x_{#2}\right)}
\newcommand{\GammatwoFt}[2]{\wh{\wt{\overline{\Gamma}}}^{(2)}_{#1}\left(x_{#2}\right)}
\newcommand{\GammatwoFF}[2]{\wh{\wh{\overline{\Gamma}}}^{(2)}_{#1}\left(x_{#2}\right)}
\newcommand{\Gammatwot}[2]{\wt{\overline{\Gamma}}^{(2)}_{#1}\left(x_{#2}\right)}
\newcommand{\Gammatwott}[2]{\wt{\wt{\overline{\Gamma}}}^{(2)}_{#1}\left(x_{#2}\right)}

\newcommand{\deltaone}{\delta_1}
\newcommand{\deltatwo}{\delta_2}

\newcommand{\betaoe}{\dfrac{\beta_0}{\epsilon}}

\newcommand{\boelite}{\frac{b_0}{\epsilon}}

\newcommand{\bFoelite}{\frac{b_{0,F}}{\epsilon}}

\newcommand{\QQslite}[1]{\left(\frac{|s_{#1}|}{\mu_r^2}\right)^{-\epsilon}}

\newcommand{\as}{\alpha_{s}}
\newcommand{\e}{\epsilon}
\newcommand{\lb}{\left\lbrace}
\newcommand{\rb}{\right\rbrace}
\newcommand{\pset}{\lb p \rb_n}
\newcommand{\Sgtoq}{S_{g\to q}}
\newcommand{\Sqtog}{S_{q\to g}}

\newcommand{\qb}{{\bar{q}}}

\newcommand{\nn}{\nonumber}
\newcommand{\dd}{\text{d}}

\newcommand{\poles}{\mathcal{P}oles}
\newcommand{\T}{\bm{T}}
\newcommand{\dr}[1]{{\dfrac{\dd #1}{#1}}}
\newcommand{\sigpart}[2]{{\hat{\sigma}^{#1}_{#2}}}
\newcommand{\ceps}{\overline{C}(\epsilon)}
\newcommand{\coeff}{\left(\dfrac{\as\ceps}{2\pi}\right)}
\newcommand{\coeffLO}{\mathcal{N}_{LO}}
\newcommand{\coeffVNLO}{\mathcal{N}^{V}_{NLO}}
\newcommand{\coeffRNLO}{\mathcal{N}^{R}_{NLO}}
\newcommand{\coeffVVNNLO}{\mathcal{N}^{VV}_{NNLO}}
\newcommand{\coeffRVNNLO}{\mathcal{N}^{RV}_{NNLO}}
\newcommand{\coeffRRNNLO}{\mathcal{N}^{RR}_{NNLO}}
\newcommand{\jet}[3]{J_{#1}^{(#2)}(#3)}
\newcommand{\dphi}[1]{\dd\Phi_{#1}}
\newcommand{\wt}[1]{\widetilde{#1}}
\newcommand{\wh}[1]{\widehat{#1}}

\newcommand{\dsigNNLO}[1]{\dd\sigpart{}{#1,\mathrm{NNLO}}}

\newcommand{\dsigV}[1]{\dd\sigpart{V}{#1,\mathrm{NLO}}}
\newcommand{\dsigR}[1]{\dd\sigpart{R}{#1,\mathrm{NLO}}}
\newcommand{\dsigTNLO}[1]{\dd\sigpart{T}{#1,\mathrm{NLO}}}
\newcommand{\dsigSNLO}[1]{\dd\sigpart{S}{#1,\mathrm{NLO}}}

\newcommand{\dsigVV}[1]{\dd\sigpart{VV}{#1,\mathrm{NNLO}}}
\newcommand{\dsigRV}[1]{\dd\sigpart{RV}{#1,\mathrm{NNLO}}}
\newcommand{\dsigRR}[1]{\dd\sigpart{RR}{#1,\mathrm{NNLO}}}
\newcommand{\dsigUNNLO}[1]{\dd\sigpart{U}{#1,\mathrm{NNLO}}}
\newcommand{\dsigTNNLO}[1]{\dd\sigpart{T}{#1,\mathrm{NNLO}}}
\newcommand{\dsigSNNLO}[1]{\dd\sigpart{S}{#1,\mathrm{NNLO}}}
\newcommand{\dsigSNNLOspe}[2]{\dd\sigpart{S,#2}{#1,\mathrm{NNLO}}}
\newcommand{\dsigVSNNLO}[1]{\dd\sigpart{VS}{#1,\mathrm{NNLO}}}
\newcommand{\dsigUNNLOshort}[1]{\dd\sigpart{U,#1}{ab,\mathrm{NNLO}}}
\newcommand{\dsigTNNLOshort}[1]{\dd\sigpart{T,#1}{ab,\mathrm{NNLO}}}
\newcommand{\dsigSNNLOshort}[1]{\dd\sigpart{S,#1}{ab,\mathrm{NNLO}}}

\newcommand{\dsigMFNLO}[1]{\dd\sigpart{MF}{#1,\mathrm{NLO}}}
\newcommand{\dsigMFNNLO}[2]{\dd\sigpart{MF,#2}{#1,\mathrm{NNLO}}}

\newcommand{\ins}[1]{\mathcal{I}ns\left[#1\right]}
\newcommand{\insdouble}[1]{\mathcal{I}ns_{2}\left[#1\right]}

\newcommand{\ugo}{{u_{g_1}}}
\newcommand{\ugt}{{u_{g_2}}}

\newcommand{\proj}[1]{\boldsymbol{\mathcal{P}}_{#1}}

\preprint{{\raggedleft%
		IPPP/23/63\\
		ZU-TH 70/23\\
}}

\title{The colourful antenna subtraction method}

\author{T.\ Gehrmann$^{a}$, E.W.N.\ Glover$^{b}$, M.\ Marcoli$^{a,b}$}

\affiliation{
	$^a$Physik-Institut, Universit\"at Z\"urich, Winterthurerstrasse 190, 8057 Z\"urich, Switzerland\\
	$^b$Institute for Particle Physics Phenomenology, Department of Physics, University of Durham, Durham, DH1 3LE, UK\\
}

\emailAdd{thomas.gehrmann@uzh.ch}
\emailAdd{e.w.n.glover@durham.ac.uk}
\emailAdd{matteo.marcoli@durham.ac.uk}

\abstract{
	We present a general subtraction scheme for NNLO calculations in massless QCD: the \textit{colourful antenna subtraction method}. It is a reformulation of the antenna subtraction approach designed to address some of the limitations of the traditional framework, especially aiming at high-multiplicity processes. In the context of the new formalism, structures needed to locally subtract the infrared-divergent behaviour of real emission corrections are systematically inferred from virtual subtraction terms, relying on the cancellation of infrared singularities and on the correspondence between integrated and unintegrated antenna functions. We illustrate in detail how the colourful antenna subtraction method works up to NNLO. The algorithm is particularly suited to be fully automated for the generation of NNLO subtraction terms for generic processes. We employ the new formalism to assemble the subtraction terms required for the calculation of the NNLO correction to hadronic three-jet production and describe their validation procedure.
}

\begin{document} 

	\maketitle
	
	\section{Introduction}\label{sec:intro}
	
	The successful interpretation of particle collider data demands high-precision theoretical predictions for Standard Model cross sections. Quantum Chromodynamics (QCD) effects dominate scattering events at particle colliders and need to be described accurately  to reach the desired precision. Computations in QCD for high-energy collisions are performed by power series expansion in the strong coupling constant $\alpha_s$. According to how many terms are computed in this expansion, one obtains increasingly more accurate predictions, denominated Leading-Order (LO),  Next-to-Leading-Order (NLO), Next-to-Next-to-Leading-Order (NNLO), and so forth. The complexity grows with the accuracy: the emergence of \textit{infrared divergences} in theoretical calculations poses an obstacle to the computation of higher-order terms in the $\alpha_s$ expansion. 
	
	LO predictions only provide a naive estimate and are not accurate enough for precision experiments. More precise calculations at NLO are required for any precision analysis. For more than two decades, fully general techniques to reach NLO accuracy have been available~\cite{Catani:1996jh,Frixione:1995ms}, as well as public multi-purpose computational frameworks for cross-section calculations~\cite{Alwall:2014hca,Sherpa:2019gpd,Bellm:2019zci,Alioli:2010xd}. However, for some observables NLO predictions are still not precise enough and NNLO calculations are needed to approach the percent-level accuracy goal. Several methods have been proposed to compute NNLO corrections~\cite{Gehrmann-DeRidder:2005btv,Currie:2013vh,DelDuca:2016ily,Catani:2007vq,Czakon:2010td,Czakon:2014oma,Gaunt:2015pea,Cacciari:2015jma,Caola:2017dug,Magnea:2018hab,Bertolotti:2022aih,Devoto:2023rpv} and NNLO-accurate predictions for essentially all $2\to 2$ processes at hadron colliders are nowadays available. From $2020$ onwards, NNLO calculations for $2\to 3$ processes have started to appear~\cite{Chawdhry:2019bji,Kallweit:2020gcp,Chawdhry:2021hkp,Czakon:2021mjy,Chen:2022ktf,Hartanto:2022qhh,Alvarez:2023fhi,Badger:2023mgf,Catani:2022mfv,Buonocore:2022pqq,Buonocore:2023ljm}, thanks to the calculation of two-loop five-point amplitudes~\cite{Chicherin:2017dob,Gehrmann:2018yef,Chicherin:2020oor,Abreu:2020cwb,Abreu:2021oya,Abreu:2021asb,Badger:2021ega,Badger:2021imn,Abreu:2023rco}. These results represent the current state-of-the-art for NNLO QCD corrections for LHC processes.
	
	The currently available results at NNLO have been obtained proceeding on a case-by-case basis, usually demanding a considerable amount of work to extend the existing formalisms to new classes of processes. At NLO, the development of process-independent and automatable techniques led to the industrialization of NLO-accurate predictions. Nowadays, NLO computations are systematically performed using public codes for any precision analysis at the LHC, thereby raising the standards of particle phenomenology. It is foreseeable that a similar revolution could occur at NNLO if general frameworks and tools for NNLO calculations become available. 
	
	In this paper we formulate a process-independent approach to perform NNLO calculations in QCD: the \textit{colourful antenna subtraction method}, which was originally presented in~\cite{Chen:2022ktf} for gluon scattering. The traditional antenna subtraction method~\cite{Gehrmann-DeRidder:2005btv,Currie:2013vh} makes use of \textit{antenna functions}~\cite{Gehrmann-DeRidder:2004ttg,Gehrmann-DeRidder:2005alt,Gehrmann-DeRidder:2005svg} to assemble suitable counterterms to deal with infrared divergences up to NNLO. These antenna functions 
	are based on colour-ordered matrix elements describing radiation from processes with two coloured particles: $\gamma^* \to q\bar{q}$, $\tilde{\chi} \to \tilde{g}g$ and $H \to gg$, covering the cases where the coloured particles are massless quarks and gluons.  It has been successfully applied to compute the NNLO correction to a series of phenomenologically relevant processes at electron-positron, electron-proton and hadron colliders~\cite{Gehrmann-DeRidder:2007foh,Currie:2017tpe,Currie:2017eqf,Chen:2022clm,Chen:2022ktf,Chen:2019zmr,Gehrmann:2020oec,Gehrmann-DeRidder:2015wbt,Gehrmann-DeRidder:2016cdi,Gehrmann-DeRidder:2017mvr,Gauld:2020deh,Gauld:2023zlv,Chen:2016zka,Cruz-Martinez:2018rod,Gauld:2019yng,Gauld:2021ule}. However, it has been gradually extended throughout the past fifteen years to deal with new processes, demanding a substantial amount of time and manpower for each new process. Despite being quite flexible, it presents some intrinsic limitations. 
	In particular, as many of the proposed NNLO subtraction schemes, it scales poorly with the number of coloured particles involved in the process and much of its simplicity is lost when dealing with contributions beyond the so-called \textit{leading-colour} approximation~\cite{Currie:2013vh} where one cannot directly rely on colour connections of squared amplitudes as the guiding principle for the construction of the subtraction terms, especially at high multiplicities.     
	
	The colourful antenna subtraction method derived here is a reformulation of the antenna subtraction approach designed to systematise the construction of the real-emission subtraction term. The main idea behind it is to exploit the predictability of the singularity structure of virtual amplitudes in colour space to straightforwardly construct the virtual subtraction terms in a completely general way. 
	This helps overcome the aforementioned limitations of the traditional antenna subtraction scheme in two ways.
	First, the complete structure of colour correlations among external QCD particles is retained by working in colour-space, making the construction of subtraction terms for {\em all} colour layers straightforward. This improves on the traditional approach using colour-decomposed antenna functions, which works best only for the leading-colour contribution. 
	Secondly, we exploit the correspondence between integrated and unintegrated antenna functions and the known infrared structure of the virtual subtraction terms to infer the subtraction terms for the real emission corrections. This strategy goes in the opposite direction with respect to the typical procedure implemented in infrared subtraction schemes, in which the real emission subtraction term is constructed first, and then analytically integrated to yield the virtual subtraction term. Nonetheless it is firmly supported by the same principle: the cancellation of infrared singularities between real and virtual corrections. The construction of local subtraction terms for real radiation starting from the infrared factorization properties of virtual amplitudes has also been investigated in~\cite{Magnea:2018ebr}.
	
	The colourful antenna approach is fully compatible with other (parallel) attempts to refine the antenna subtraction scheme by building the antenna functions directly from a list of desired infrared limits~\cite{Braun-White:2023sgd,Braun-White:2023zwd,Fox:2023bma}. This so-called \textit{designer antenna} approach avoids the need to decompose the antenna functions into sub-antennae, it streamlines (and reduces the size of) the subtraction terms by avoiding the introduction of spurious limits that are inevitably present in the matrix-element based antennae when applied to processes with multiple coloured particles. So far, designer antenna functions and their integrated counterparts have been derived for use in electron-positron annihilation, electron-hadron and hadron-hadron collisions at NLO~\cite{Braun-White:2023sgd,Fox:2023bma}, and at NNLO for electron-positron annihilation~\cite{Braun-White:2023sgd,Braun-White:2023zwd}. In the future, one can imagine an optimal approach  in which the colourful antenna subtraction method proposed here is supplemented by the use of designer antenna functions, and vice versa. 
	
	The outline of the paper is as follows. In Section~\ref{sec:antenna_general} we give a brief overview of the basic ingredients and structures of the traditional antenna subtraction method. Section~\ref{sec:col_space}, after reviewing the colour space formalism and the infrared singularity structure of QCD virtual amplitudes, introduces \textit{integrated dipoles} as operators in colour space, which play a fundamental role in the formalism we propose. In Sections~\ref{sec:subNLO} and~\ref{sec:subNNLO} we present the colourful antenna subtraction method, respectively at NLO and NNLO, supplemented by the appendices. The application of the new colourful formalism for the construction of NNLO subtraction terms for three-jet production at hadron colliders is then discussed in Section~\ref{sec:3jet}. Concluding remarks and an outlook are presented in Section~\ref{sec:conc}.

	\section{Basics of antenna subtraction}\label{sec:antenna_general}
	
	The antenna subtraction method has been developed in detail in~\cite{Gehrmann-DeRidder:2005btv,Gehrmann-DeRidder:2007foh,Currie:2013vh}. We summarize its essential features in the following. 
	
	The core ingredients of the antenna subtraction method are the \textit{antenna functions}. These objects capture the singular behaviour of matrix elements in the presence of unresolved emissions and are directly extracted from colour-ordered squared matrix elements for simple processes involving the radiation of QCD particles between a pair of hard radiators~\cite{Gehrmann-DeRidder:2004ttg,Gehrmann-DeRidder:2005alt,Gehrmann-DeRidder:2005svg}. According to the partonic species of the hard radiators, antenna functions are classified as quark-antiquark, quark-gluon and gluon-gluon antenna functions. 
	
	An unintegrated $\ell$-loop antenna function is a function of $n$ partonic momenta, a subset $I$ of which are in the initial state, generically indicated as:
	\begin{equation}
		X_{n,\lb i \rb_{i\in I}}^{\ell}(j_1,\dots,j_n).
	\end{equation}
	Configurations with zero, one or two partons in the initial state are respectively denoted by final-final (FF), initial-final (IF) and initial-initial (II) antenna functions. Different partonic arrangements are described by different antenna functions, characterized with a specific upper-case letter replacing the generic $X$. At one loop, $\ell=1$, 
	antenna functions containing a closed fermion loop are indicated by  
a \textit{hat} ($\wh{\phantom{X}}$), and  antenna functions corresponding to the subleading-colour virtual configurations are indicated 
with a \textit{tilde} ($\wt{\phantom{X}}$). Some of the antenna functions are further split into so-called \textit{sub-antennae}. This is done for several reasons such as the identification of two hard radiators, the separation of different reduced matrix elements or the separation of different momentum mappings. Typically sub-antennae have a more complicated structure than the full result. Sub-antennae are denoted with the same notation as full ones, but with lower-case letters or with a dedicated subscript.

	Exploiting the exact factorization of the phase space~\cite{Gehrmann-DeRidder:2005alt,Daleo:2006xa,Daleo:2009yj,Boughezal:2010mc,Gehrmann-DeRidder:2012too}, antenna functions can be analytically integrated in dimensional regularization to obtain their integrated counterparts. Integrated antenna functions are denoted as:
	\begin{equation}
		\mathcal{X}_{n,\lb i \rb_{i\in I}}^{\ell}(s_{j_1\dots j_n}),
	\end{equation}
	where $s_{j_q\dots j_n}$ is the invariant mass of the considered partonic system:
	\begin{equation}
		s_{j_1\dots j_n}=\left(\sum_{j=1}^n \lambda_j p_j\right)^2,
	\end{equation}
	with $\lambda_j=+1$ for final-state partons and $\lambda_j=-1$ for initial-state ones. 
	
	All the integrated and unintegrated antenna functions and sub-antennae used in this paper were derived in~\cite{Gehrmann-DeRidder:2005btv,Gehrmann-DeRidder:2007foh,Glover:2010kwr,Daleo:2006xa,Daleo:2009yj,Boughezal:2010mc,Gehrmann:2011wi,Gehrmann-DeRidder:2012too}.
	
	\subsection{Structure of antenna subtraction}\label{sec:struct_sub}
	
	In the following we review the infrastructure of the antenna subtraction method and of the subtraction counterterms which are constructed to remove the infrared singularities in different layers of a calculation. The detailed construction of the subtraction counterterms is described in Section~\ref{sec:subNLO} and~\ref{sec:subNNLO}. 
		
	\subsubsection{Subtraction at NLO}\label{sec:subNLO_pre}
	
	The NLO QCD correction to an $n$-jet partonic cross section with parton species $a$ and $b$ in the initial state is given by:
	\begin{equation}\label{NLOcs}
		\dd\sigpart{}{ab,\mathrm{NLO}}=\int_{n}\left(\dd\sigpart{V}{ab,\mathrm{NLO}}+\dd\sigpart{MF}{ab,\mathrm{NLO}}\right)+\int_{n+1}\dd\sigpart{R}{ab,\mathrm{NLO}},
	\end{equation}
	where $\dsigV{ab}$ and $\dsigR{ab}$ respectively represent the virtual and real corrections, while $\dsigMFNLO{ab}$ is the NLO mass factorization counterterm. The symbol $\int_n$ indicates an integration over a $n$-particle phase space. The NLO cross section in~\eqref{NLOcs}, despite being well defined and finite, is not suitable for numerical integration in this form, due to the emergence of infrared singularities. These cancel in the final result, but a proper subtraction procedure is needed to separately remove the singularities in the real and virtual corrections and make both integrals in~\eqref{NLOcs} computable with numerical methods. 
	
	In the context of antenna subtraction, this is achieved constructing a real subtraction term $\dsigSNLO{ab}$~\cite{Currie:2013vh}, which locally removes the singular behaviour of $\dsigR{ab}$ in the infrared (IR) limits and can be analytically integrated over the phase space of the unresolved radiation. For the regularization of the IR singularities, we rely on dimensional regularization with the customary choice $d=4-2\e$. After this integration, 
	$\dsigSNLO{ab}$ serves as ingredient to 
the virtual subtraction term $\dsigTNLO{ab}$, which cancels the explicit $\e$-poles of the virtual correction and contains the mass factorization contribution. The NLO cross section can then be reformulated as \cite{Currie:2013vh}:
	\begin{equation}\label{NLOcssub}
		\dd\sigpart{}{ab,\mathrm{NLO}}=\int_{n}\left[\dd\sigpart{V}{ab,\mathrm{NLO}}-\dsigTNLO{ab}\right]+\int_{n+1}\left[\dd\sigpart{R}{ab,\mathrm{NLO}}-\dsigSNLO{ab}\right],
	\end{equation}
	with 
	\begin{equation}\label{sigTNLO}
		\dsigTNLO{ab}=-\int_1\dsigSNLO{ab}-\dsigMFNLO{ab}.
	\end{equation}
	At NLO the mass factorization counterterm is given by:
	\begin{equation}\label{MF}
		\dd\sigpart{MF}{ab,\mathrm{NLO}}=-\coeff\sum_{c,d}\int\dr{x_1}\dr{x_2}\Gamma^{(1)}_{ab;cd}(x_1,x_2)\,\dd\sigpart{}{cd,\mathrm{LO}},
	\end{equation}
	where $x_1$ and $x_2$ represent the momentum fractions transferred to the hard process, while $\Gammaone{ab;cd}{x_1,x_2}$ denotes the NLO mass factorization kernel:
	\begin{equation}\label{split_ker_NLO}
		\Gammaone{ab;cd}{x_1,x_2}=\Gammaone{ca,\text{full}}{x_1}\delta_{db}\delta(1-x_2)+\Gammaone{db,\text{full}}{x_1}\delta_{ca}\delta(1-x_1).
	\end{equation}
	which can be organized into different terms corresponding to different colour factors:
	\begin{eqnarray}
		\Gammaone{qq,\text{full}}{x}&=&\left(\dfrac{N_c^2-1}{N_c}\right)\Gammaone{qq}{x}, \\
		\Gammaone{gq,\text{full}}{x}&=&\left(\dfrac{N_c^2-1}{N_c}\right)\Gammaone{gq}{x}, \\
		\Gammaone{qg,\text{full}}{x}&=&\Gammaone{qg}{x}, \\
		\Gammaone{gg,\text{full}}{x}&=&N_c\Gammaone{gg}{x}+N_f\GammaoneF{gg}{x}.
	\end{eqnarray}
	The NLO mass factorization terms are directly related to regularized LO Altarelli-Parisi splitting kernels~\cite{Altarelli:1977zs}:
	\begin{equation}
		\Gammaone{ab,\text{full}}{x}=-\dfrac{1}{\epsilon}P^{0}_{ab}(x),
	\end{equation}
	where
	\begin{eqnarray}
		P^{0}_{qq}(x)&=&\left(\dfrac{N_c^2-1}{N_c}\right)p^{0}_{qq}(x), \\
		P^{0}_{gq}(x)&=&\left(\dfrac{N_c^2-1}{N_c}\right)p^{0}_{gq}(x), \\
		P^{0}_{qg}(x)&=&p^{0}_{qg}(x), \\
		P^{0}_{gg}(x)&=&N_c p^{0}_{gg}(x)+N_f\hat{p}^{0}_{gg}(x).
	\end{eqnarray}
	
	\subsubsection{Subtraction at NNLO}\label{sec:subNNLO_pre}
	
	The NNLO QCD correction to an $n$-jet cross section is given by:
	\begin{eqnarray}\label{NNLOcs}
		\dsigNNLO{ab}&=&\int_n\left(\dsigVV{ab}+\dsigMFNNLO{ab}{2}\right)\nonumber\\
		&+&\int_{n+1}\left(\dsigRV{ab}+\dsigMFNNLO{ab}{1}\right) + \int_{n+2}\dsigRR{ab},
	\end{eqnarray}
	where $\dsigVV{ab}$ represents the double-virtual correction, $\dsigRV{ab}$ the real-virtual correction and $\dsigRR{ab}$ the double-real correction. The mass factorization counterterm at this perturbative order is split into two terms associated with $n$- and $(n+1)$-particle final states, respectively $\dsigMFNNLO{ab}{2}$ and $\dsigMFNNLO{ab}{1}$.
	
	As in the NLO case, the quantity in~\eqref{NNLOcs} cannot be computed directly with numerical methods. The singular behaviour of both the double-real and real-virtual corrections in infrared limits has to be subtracted and the explicit poles in the double-virtual and real-virtual matrix elements need to be removed properly. To achieve this, the NNLO cross section is rewritten in the context of antenna subtraction as~\cite{Currie:2013vh}:
	\begin{eqnarray}\label{NNLOcssub}
		\dsigNNLO{ab}&=&\int_n\left[\dsigVV{ab}-\dsigUNNLO{ab}\right]\nonumber\\
		&+&\int_{n+1}\left[\dsigRV{ab}-\dsigTNNLO{ab}\right]\nonumber\\
		&+&\int_{n+2}\left[\dsigRR{ab}-\dsigSNNLO{ab}\right],
	\end{eqnarray}
	where the subtracted quantities are the double-virtual, the real-virtual and the double-real subtraction term. These contributions have the following form~\cite{Currie:2013vh}:
	\begin{eqnarray}\label{subtermsNNLO}
		\dsigSNNLO{ab}&=&\dsigSNNLOspe{ab}{1}+\dsigSNNLOspe{ab}{2}\,,\nonumber\\
		\dsigTNNLO{ab}&=&\dsigVSNNLO{ab}-\int_1 \dsigSNNLOspe{ab}{1}-\dsigMFNNLO{ab}{1}\,,\nonumber\\
		\dsigUNNLO{ab}&=&-\int_1 \dsigVSNNLO{ab}-\int_2 \dsigSNNLOspe{ab}{2}-\dsigMFNNLO{ab}{2}\,.
	\end{eqnarray}
	The double-real subtraction term has been decomposed into two contributions which contain single and double unresolved IR limits: $\dsigSNNLOspe{ab}{1}$ and $\dsigSNNLOspe{ab}{2}$. In the real-virtual subtraction term, $\dsigVSNNLO{ab}$ cancels the implicit singular behaviour of the real-virtual correction in single unresolved limits. The remaining terms in $\dsigTNNLO{ab}$ and $\dsigUNNLO{ab}$, which are not mass factorization counterterms, are obtained via analytical integration of the aforementioned contributions, over the phase space of a single or double unresolved emission. These terms cancel explicit $\e$-poles in the virtual corrections. 
	
	Within the traditional antenna subtraction approach~\cite{Gehrmann-DeRidder:2005btv,Currie:2013dwa}, the infrastructure above is constructed starting from $\dsigSNNLOspe{ab}{1}$, $\dsigSNNLOspe{ab}{2}$ and $\dsigVSNNLO{ab}$, which are obtained by studying the infrared behaviour of the double-real and real-virtual matrix elements. Afterwards, analytical integrations are performed to complete the subtraction. Several adjustments are needed during this process to ensure the removal of spurious divergences and prevent the oversubtraction of singular behaviour. Examples of the typical structures appearing in each layer can be found in~\cite{Gehrmann-DeRidder:2005btv,Gehrmann-DeRidder:2007foh,Glover:2010kwr,Gehrmann-DeRidder:2011jwo,Gehrmann-DeRidder:2012dog,Currie:2013vh}.  
	
	At NNLO, we have two different contributions to the mass factorization counterterm: the double-virtual and the real-virtual mass factorization terms. The real-virtual mass factorization counterterm is given by~\cite{Currie:2013vh}:
	\begin{equation}\label{MFRV}
		\dd\sigpart{MF,1}{ab,\mathrm{NNLO}}=-\coeff\sum_{c,d}\int\dr{x_1}\dr{x_2}\Gammaone{ab;cd}{x_1,x_2}\left(\dd\sigpart{R}{cd,\mathrm{NLO}}-\dd\sigpart{S}{cd,\mathrm{NLO}}\right),
	\end{equation}
	where the required mass factorization kernels are the ones used at NLO. 
	
	The double-virtual mass factorization counterterm reads~\cite{Currie:2013vh}:
	\begin{eqnarray}\label{MFVV}
		\dd\sigpart{MF,2}{ab,\mathrm{NNLO}}&=&-\int\dr{x_1}\dr{x_2}\sum_{c,d}\Bigg\lbrace\Bigg.\coeff\left[\Gammaone{ab;cd}{x_1,x_2}\left(\dd\sigpart{V}{cd,\mathrm{NLO}}-\dd\sigpart{T}{cd,\mathrm{NLO}}\right)\right]\nonumber\\
		&&\hspace{0.5cm}+\coeff^2\Big[\Big.\Gammatwo{ab;cd}{x_1,x_2}-\dfrac{\beta_0}{\e}\Gammaone{ab;cd}{x_1,x_2}\nonumber\\
		&&\hspace{3cm}+\dfrac{1}{2}\sum_{\alpha,\beta}\left[\Gamma^{(1)}_{ab;\alpha\beta}\otimes\Gamma^{(1)}_{\alpha\beta;cd}\right](x_1,x_2)\Big.\Big]\dd\sigpart{}{cd,\mathrm{LO}}\Bigg.\Bigg\rbrace.
	\end{eqnarray}
	The reduced two-loop mass factorization kernel is defined as~\cite{Currie:2013vh}:
	\begin{equation}
		\Gammatwo{ab;cd}{x_1,x_2}=\Gammatwo{ca,\text{full}}{x_1}\delta_{db}\delta(1-x_2)+\Gammatwo{db,\text{full}}{x_2}\delta_{ca}\delta(1-x_1),
	\end{equation}
	where $\Gammatwo{ca,\text{full}}{x_i}$ are directly related to the LO and NLO Altarelli-Parisi spitting kernels~\cite{Altarelli:1977zs,Curci:1980uw,Furmanski:1980cm}:
	\begin{equation}\label{Gamma2}
		\Gammatwo{ab,\text{full}}{x}=-\dfrac{1}{2\e}\left(P^1_{ab}(x)+\dfrac{\beta_0}{\e}P^0_{ab}(x)\right),
	\end{equation}
	and can be decomposed into colour layers as~\cite{Currie:2013vh}:
	\begingroup
	\allowdisplaybreaks
	\begin{eqnarray}
		{\overline{\Gamma}}_{qq,\text{full}}^{(2)}(x)&=&\bigg(\frac{N_c^{2}-1}{N_c}\bigg)\bigg[N_c\overline{\Gamma}_{qq}^{(2)}(x)+\wt{\overline{\Gamma}}_{qq}^{(2)}(x)+\frac{1}{N_c}\wt{\wt{\overline{\Gamma}}}_{qq}^{(2)}(x)+N_f\wh{\overline{\Gamma}}_{qq}^{(2)}(x)\bigg],\\
		{\overline{\Gamma}}_{q\qb,\text{full}}^{(2)}(x)&=&\bigg(\frac{N_c^{2}-1}{N_c}\bigg)\bigg[\overline{\Gamma}_{q\qb}^{(2)}(x)+\frac{1}{N_c}\wt{\overline{\Gamma}}_{q\qb}^{(2)}(x)\bigg],\\
		{\overline{\Gamma}}_{qq',\text{full}}^{(2)}(x)&=&\bigg(\frac{N_c^{2}-1}{N_c}\bigg)\ \overline{\Gamma}_{qq'}^{(2)}(x),\\
		{\overline{\Gamma}}_{q\qb',\text{full}}^{(2)}(x)&=&\bigg(\frac{N_c^{2}-1}{N_c}\bigg)\ \overline{\Gamma}_{q\qb'}^{(2)}(x),\\
		{\overline{\Gamma}}_{gq,\text{full}}^{(2)}(x)&=&\bigg(\frac{N_c^{2}-1}{N_c}\bigg)\bigg[N_c\overline{\Gamma}_{gq}^{(2)}(x)+\frac{1}{N_c}\wt{\overline{\Gamma}}_{gq}^{(2)}(x)+N_f\wh{\overline{\Gamma}}_{gq}^{(2)}(x)\bigg],\\
		{\overline{\Gamma}}_{qg,\text{full}}^{(2)}(x)&=&N_c\overline{\Gamma}_{qg}^{(2)}(x)+\frac{1}{N_c}\wt{\overline{\Gamma}}_{qg}^{(2)}(x)+N_f\wh{\overline{\Gamma}}_{qg}^{(2)}(x),\\
		{\overline{\Gamma}}_{gg,\text{full}}^{(2)}(x)&=&N_c^{2}\overline{\Gamma}_{gg}^{(2)}(x)+N_cN_f\wh{\overline{\Gamma}}_{gg}^{(2)}(x)+\frac{N_f}{N_c}\wh{\wt{\overline{\Gamma}}}_{gg}^{(2)}(x)+N_f^{2}\wh{\wh{\overline{\Gamma}}}_{gg,}^{(2)}(x).
	\end{eqnarray}
	\endgroup
	The expressions for the two-loop mass factorization kernels are rather lengthy and therefore we do not report them here. Their expressions as well as detailed explanations about how they are extracted from the associated LO and NLO splitting kernels~\cite{Furmanski:1980cm}  can be found in Appendix A.2 of~\cite{Currie:2013vh}.
	
	\section{Integrated dipoles in colour space}\label{sec:col_space}
	
	In this section we introduce key ingredients for the colourful antenna subtraction approach: \textit{integrated dipoles}~\cite{Currie:2013vh,Currie:2013dwa,Chen:2022ktf,Chen:2022tpk}. Such objects conveniently collect integrated antenna functions and mass factorization kernels and are cast as insertion operators in colour space. Integrated dipoles allow to systematically describe the infrared singularity structure of one- and two-loop matrix elements in terms of integrated antenna functions. This is a pivotal prerequisite for the formulation of the colourful antenna subtraction method.
	
	Before illustrating the integrated dipoles, we provide a brief overview of the colour space formalism and of the infrared singularities of virtual amplitudes. 
	
	\subsection{Colour space formalism}
	
	The treatment of QCD amplitudes in colour space provides a universal description of the infrared singularity structure, in the presence of both virtual and real corrections~\cite{Bassetto:1983mvz,Catani:1996jh,Catani:1998bh,Catani:1999ss,Becher:2009qa,Gardi:2009qi,Gardi:2009zv}. 
	
	In colour space, an $\ell$-loop amplitude with $n$ external partons is represented by an abstract vector $\ket{\ampnum{n}{\ell}(\lb p\rb_n)}$, whose components carry the colour indices of the external partons in a suitable representation of SU$(N_c)$. If a set of generating vectors $\lb \ket{\mathcal{C}_{n,c}^{\ell}} \rb$ is defined, which span the $n$-parton colour space, any amplitude can be decomposed as:
	\begin{equation}\label{coldec}
		\ket{\ampnum{n}{\ell}(\lb p\rb_n)}=\sum_{c\in I^{\ell}} A^{\ell}_{n,c}(\lb p\rb_n)\,\ket{\mathcal{C}_{n,c}^{\ell}},
	\end{equation}
	where $I^{\ell}$ indicates a suitable subset of generating vectors. The scalar quantities $A^{\ell}_{n,c}(\lb p\rb_n)$ are \textit{colour-ordered partial amplitudes}. In equation~\eqref{coldec}, the dependence on the helicities of the external partons is implicit and in the following a sum over helicity configurations is always assumed when squared quantities are considered. For $\ell\geq1$, the dependence on the renormalization scale $\mu_r$ is understood. 
	
	An arbitrary number of colourless particles, such as photons, weak bosons or Higgs bosons, could be involved in the considered scattering process, however this does not modify the structure of the amplitude in colour space or the construction of the subtraction terms for infrared singularities in QCD. The presence of colourless emissions only affects the colour-ordered partial amplitudes, which clearly change for different processes. For this reason, in the following we restrict the discussion and the notation to $n$-parton amplitudes, modulo additional colourless particles.
	
	Our convention is to strip the partial amplitudes of overall coefficients such as couplings and incoming particles average factors, which are inserted later at the cross section level. In particular, we strip an $\ell$-loop amplitude of an overall factor $\left(\tfrac{\alpha_s \bar{C}(\e)}{2\pi}\right)^\ell$ with respect to the corresponding tree-level amplitude, where $\ceps=(4\pi)^{\e} e^{-\e\gamma_E}$.
	
	The description of infrared singularities of QCD amplitudes can be achieved through the evaluation of infrared insertion operators in colour space~\cite{Catani:1996jh,Catani:1998bh,Becher:2009qa,DelDuca:2016ily}. The action of an operator $\boldsymbol{\mathcal{O}}$ in colour space can be generically computed as
	\begin{equation}\label{sandwich}
		\braket{\ampnum{n}{\ell_1}(\lb p \rb_n)|\boldsymbol{\mathcal{O}}|\ampnum{n}{\ell_2}(\lb p\rb_n)}.
	\end{equation}
	In particular, the coherent emission of a gluon between a dipole formed by parton $i$ and parton $j$ is described in colour space by the colour charge dipole operator:
	\begin{equation}\label{TiTj}
		\T_i\cdot\T_j,
	\end{equation}
	where we use bold symbols to emphasize that $\T_i=(T_i)^a_{bc}$ is a vector of $a$ SU$(N_c)$ generators in the appropriate representation, with $a$ indicating the colour index of the emitted gluon. In general, in the remainder of the paper, bold symbols are used to denote operators in colour space, while Roman symbols indicate scalars in colour space. For different partonic species we have:
	\begin{equation}
		(T_i)^a_{bc} = 
		\begin{cases}
			i f_{bac}\quad&\text{for }i=\text{ gluon} \\
			t^a_{bc}\quad&\text{for }i=\text{ final-state quark or initial-state antiquark} \\
			(t^a_{bc})^{\dagger}=-t^a_{cb}\quad&\text{for }i=\text{ final-state antiquark or initial state quark} \\
		\end{cases}
	\end{equation}
	where $f_{abc}$ and $t^a_{bc}$ are the colour-charge matrices in the adjoint and fundamental representations of SU$(N_c)$. The following properties hold:
	\begin{eqnarray}
		\label{id_dip_1}\T_i\cdot\T_j&=&\T_j\cdot\T_i,\\
		\label{id_dip_2}\T_i\cdot\T_i&=&\T_i^2=C_i\,\mathbf{Id},
	\end{eqnarray}
	where $\mathbf{Id}$ represents the identity operator in colour space and $C_i$ indicates the Casimir coefficient for the SU$(N_c)$ representation associated to parton $i$: $C_g=C_A=N_c$ and $C_q=C_{\bar{q}}=C_F=(N_c^2-1)/(2 N_c)$.
	
	Each state $\ketampdep{n}{\ell}{\lb p \rb_n}$ is a colour singlet and colour conservation implies:
	\begin{equation}\label{colour_conservation}
		\sum_{i=1}^{n}\T_i\ketampdep{n}{\ell}{\lb p \rb_n}=0.
	\end{equation}
	Since in what follows we always consider colour singlet states, we can employ the previous identity as $\sum_{j\ne i}\T_j=-\T_i$.
	
	The explicit evaluation of squared matrix elements or colour operator insertions depends on the form of the generating vectors $\lb \ket{\mathcal{C}_{n,i}^{\ell}} \rb$, and, in general, produces numerous terms and non-trivial structures, especially for high-multiplicity processes. Nevertheless one can observe that the result of such calculations is given by real functions of the external momenta and possibly the renormalization scale. At tree-level, the general form of such a function is:
	\begin{equation}\label{func0}
		f_0\left(\set{p}_n\right)=\sum_{c,c'\in I^0} C^{0}_{n}(c,c')\,\anum{n}{0}(c,c';\pset), 
	\end{equation}
	where we generically assume that the indices $c$ and $c'$ here run over the set of possible colour orderings, along with the possible colour structures. The coefficients $C^{0}_{n}(c,c')$ are colour factors which depend on $N_c$ and the number of partons (possibly depending on flavour) and 
	\begin{equation}
		\anum{n}{0}(c,c';\pset)=
		\begin{cases}
			\left|\ampnumstraight{n,c}{0}(\pset)\right|^2\quad&\text{if } c=c',\\
			2\text{Re}\left[\ampnumstraight{n,c}{0}(\pset)^{\dagger}\ampnumstraight{n,c'}{0}(\pset)\right]\quad&\text{if }c\neq c'.
		\end{cases}
	\end{equation}
	This quantity represents the squared interference of two colour-ordered partial amplitudes, with generic colour structures dictated by the indices $c$ and $c'$. When $c=c'$, we have squared coherent partial amplitudes, while we refer to the case $c\neq c'$ as \textit{incoherent interference}. In general, $\ampnumstraight{n,c}{0}(\pset)^{\dagger}\ampnumstraight{n,c'}{0}(\pset)$ is not real. However, in the calculation of any quantity like~\eqref{sandwich}, its complex conjugate (obtained by swapping $c$ and $c'$) always appears with the same prefactor, hence only the real part matters. Analogously, at one loop we have:
	\begin{equation}\label{func1}
		f_1\left(\set{p}_n\right)=\sum_{c\in I^0,c'\in I^1} C^{1}_{n}(c,c')\,\anum{n}{1}(c,c';\pset),  
	\end{equation}
	where
	\begin{equation}
		\anum{n}{1}(c,c';\pset)=2\text{Re}\left[\ampnumstraight{n,c}{0}(\pset)^{\dagger}\ampnumstraight{n,c'}{1}(\pset)\right],
	\end{equation}
	which is manifestly a real quantity.
	
	\subsection{Infrared singularity structure of QCD loop amplitudes}\label{sec:IR_poles_virt}
	
	We review here how the infrared singularities of virtual corrections amplitudes in QCD can be predicted in a general way by means of infrared operators in colour space~\cite{Catani:1998bh,Becher:2009qa,Gardi:2009qi,Gardi:2009zv}. 
	
	\subsubsection{Infrared singularity structure at one loop}\label{sec:IR_1loop}
	
	The singularity structure of renormalized $(n+2)$-parton one-loop amplitudes in QCD can be described in colour space with~\cite{Catani:1998bh}:
	\begin{equation}\label{1l-sing}
		\ketamp{n+2}{1}=\Iop{1}{\epsilon,\mu_r^2}\ketamp{n+2}{0}+\ketampdep{n+2}{1,\text{fin}}{\mu_r^2},
	\end{equation}
	where $\mu_r$ is the renormalization scale, $\ketampdep{n+2}{1,\text{fin}}{\mu_r^2}$ is a finite remainder and $\Iop{1}{\e,\mu_r^2}$ is Catani's infrared insertion operator given by~\cite{Catani:1998bh}:
	\begin{equation}\label{I1}
		\Iop{1}{\epsilon,\mu_r^2}=\dfrac{1}{2}\dfrac{e^{\e\gamma_E}}{\Gamma(1-\e)}\sum_{i=1}^{n+2}\dfrac{1}{\T_i^2}\mathcal{V}_{i}(\e)\sum_{j\ne i}(\T_i\cdot \T_j)\left(\dfrac{-s_{ij}}{\mu_r^2}\right)^{-\e}\,.
	\end{equation}
	The singular functions $\mathcal{V}_{i}(\e)$ contain double and single $\e$-poles:
	\begin{equation}
		\mathcal{V}_{i}(\e)=\T^2_i\dfrac{1}{\e^2}+\gamma_i\dfrac{1}{\e},
	\end{equation}
	with
	\begin{eqnarray}
		\T^2_q=\T^2_{\bar{q}}=C_F,\quad\T^2_g=C_A,\quad\gamma_q=\gamma_{\bar{q}}=\dfrac{3}{2}C_F,\quad\gamma_g=\dfrac{11}{6}C_A-\dfrac{1}{3}N_f.
	\end{eqnarray}
	Equation \eqref{I1} can be rewritten as
	\begin{eqnarray}\label{I1v2}
		\Iop{1}{\epsilon,\mu_r^2}&=&\dfrac{1}{2}\sum_{i=1}^{n+2}\sum_{j\ne i}\left(\T_i\cdot\T_j\right)\ourIop{1}{ij}{\e,\mu_r^2}\nonumber\\
		&=&\sum_{(i,j)}\left(\T_i\cdot\T_j\right)\ourIop{1}{ij}{\e,\mu_r^2},
	\end{eqnarray}
	where in the last line the sum runs over unordered pairs of partons and:
	\begin{eqnarray}\label{Iij}
		\ourIop{1}{i_gj_g}{\e,\mu_r^2}&=&\dfrac{e^{\e\gamma_E}}{\Gamma(1-\e)}\left[\dfrac{1}{\e^2}+\dfrac{1}{\e}\left(\dfrac{11}{6}-\dfrac{1}{3}\dfrac{N_f}{N_c}\right)\right]\left(\dfrac{-s_{ij}}{\mu_r^2}\right)^{-\e},\\
		\ourIop{1}{i_gj_q}{\e,\mu_r^2}=\ourIop{1}{i_qj_g}{\e,\mu_r^2}&=&\dfrac{e^{\e\gamma_E}}{\Gamma(1-\e)}\left[\dfrac{1}{\e^2}+\dfrac{1}{\e}\left(\dfrac{5}{3}-\dfrac{1}{6}\dfrac{N_f}{N_c}\right)\right]\left(\dfrac{-s_{ij}}{\mu_r^2}\right)^{-\e},\\
		\ourIop{1}{i_qj_{q'}}{\e,\mu_r^2}&=&\dfrac{e^{\e\gamma_E}}{\Gamma(1-\e)}\left[\dfrac{1}{\e^2}+\dfrac{3}{2\e}\right]\left(\dfrac{-s_{ij}}{\mu_r^2}\right)^{-\e},
	\end{eqnarray}
	independently from the flavour of $q$ and $q'$. $N_f$ indicates the number of light quark flavours.
	
	Using~\eqref{1l-sing} it is possible to extract the poles of one-loop matrix elements in the following way:
	\begin{eqnarray}
		\poles\left(\me{n+2}{1}\right)&=&\poles\left(\braket{\ampnum{n+2}{0}|\ampnum{n+2}{1}}+\braket{\ampnum{n+2}{1}|\ampnum{n+2}{0}}\right)\nonumber\\
		&=&\poles(\braket{\ampnum{n+2}{0}|\Iop{1}{\e}+\Iopd{1}{\e}|\ampnum{n+2}{0}})\,.
	\end{eqnarray}
	The appearance of the sum $\Iop{1}{\e}+\Iopd{1}{\e}$ indicates that only the real part of the insertion operator affects the description of the poles at the matrix element level, as expected. We can then write
	\begin{equation}
		\poles\left(\me{n+2}{1}\right)=\poles\left[\sum_{(i,j)}\braket{\ampnum{n+2}{0}|\T_{i}\cdot \T_{j}|\ampnum{n+2}{0}}\,2\text{Re}\left(\ourIop{1}{i j}{\e,\mu_r^2}\right)\right]\,,
	\end{equation}
	where we considered that the colour charge evaluation on the tree-level amplitude gives a real quantity, as argued in Section \ref{sec:col_space}. At the cross section level we have
	\begin{eqnarray}\label{Vpoles}
		\poles\left(\sigpart{V}{NLO}\right)&=&\coeffVNLO\int\dphi{n}(p_3,\dots,p_{n+2};p_1,p_2)\,\jet{n}{n}{\lb p \rb_n}\nonumber\\
		&&\times\poles\left[\sum_{(i,j)}\braket{\ampnum{n+2}{0}|\T_{i}\cdot \T_{j}|\ampnum{n+2}{0}}2\,\text{Re}\left(\ourIop{1}{i j}{\e,\mu_r^2}\right)\right],
	\end{eqnarray}
	where $\jet{n}{m}{\lb p \rb_m}$ represents the jet algorithm, which reconstructs $n$ resolved jets from $m$ final-state partons with momenta $\lb p \rb_m$ (here $m=n$). The factor $\coeffVNLO$ is given by
	\begin{equation}
		\coeffVNLO=\coeff\coeffLO,
	\end{equation}
	where $\coeffLO$ contains the overall factors appropriate for the LO process, such as the strong coupling, symmetry factors and the spin- and colour-average over the initial-state partons.

	\subsubsection{Infrared singularity structure at two loops}\label{sec:IR_2loop}
	
	The singularity structure of renormalized two-loop amplitudes in QCD can be described in colour space by~\cite{Catani:1998bh,Becher:2009qa,Gardi:2009qi,Gardi:2009zv}:
	\begin{equation}\label{2l-sing}
		\ketamp{n+2}{2}=\Iop{1}{\epsilon,\mu_r^2}\ketamp{n+2}{1}+\Iop{2}{\epsilon,\mu_r^2}\ketamp{n+2}{0}+\ketampdep{n+2}{2,\text{fin}}{\mu_r^2},
	\end{equation}
	where, as before, $\ketampdep{n+2}{2,\text{fin}}{\mu_r^2}$ is a finite remainder. The two-loop infrared insertion operator has the following expression~\cite{Catani:1998bh}:
	\begin{eqnarray}\label{I2}
		\Iop{2}{\epsilon,\mu_r^2}&=&-\dfrac{\beta_0}{\e}\Iop{1}{\epsilon,\mu_r^2;\pset}-\dfrac{1}{2}\Iop{1}{\epsilon,\mu_r^2}\Iop{1}{\epsilon,\mu_r^2}\nonumber\\
		&&+e^{-\e\gamma_E}\dfrac{\Gamma(1-2\e)}{\Gamma(1-\e)}\left(\dfrac{\beta_0}{\e}+K\right)\Iop{1}{2\epsilon,\mu_r^2}\nonumber\\
		&&+\Hop{2}{\epsilon,\mu_r^2}\,.
	\end{eqnarray}
	where
	\begin{eqnarray}
		\beta_0&=&\dfrac{11}{6}N_c-\dfrac{1}{3}N_f,\\
		K&=&\left(\dfrac{67}{18}-\dfrac{\pi^2}{6}\right)N_c-\dfrac{5}{9}N_f.
	\end{eqnarray}
	
	The colour structure of~\eqref{I2} is more involved than the colour charge dipole structure of~\eqref{I1}, with products of two colour charge dipoles appearing. The last line of~\eqref{I2} contains the hard radiation function $\Hop{2}{\epsilon,\mu_r^2}$~\cite{Catani:1998bh,Bern:2003ck,Becher:2009qa}, which can be decomposed in the following manner:
	\begin{equation}\label{H2}
		\Hop{2}{\epsilon,\mu_r^2}=\sum_{i}C_i\ourHop{2}{i}{\e}\mathbf{Id}+\check{\bm{H}}^{(2)}(\epsilon,\mu_r^2),
	\end{equation}
	where the sum runs over the $n+2$ external partons and $C_i$ are Casimir coefficients. The first term in~\eqref{H2} is proportional to the identity in colour space, while the second term has a non-trivial colour structure, which cannot be in general expressed in terms of colour charge dipoles. However, the second term vanishes when evaluated between tree-level states after summing over helicities~\cite{Bern:2003ck,Seymour:2008xr,Becher:2009qa,Czakon:2013hxa}:
	\begin{equation}\label{H2vanish}
		\braamp{n+2}{0}\check{\bm{H}}^{(2)}(\epsilon,\mu_r^2)\ketamp{n+2}{0}=0.
	\end{equation}
	For the purpose of describing the infrared singularity structure of two-loop squared matrix elements, the hard radiation function $\Hop{2}{\epsilon,\mu_r^2}$ needs to be evaluated on tree-level states and therefore it is possible to neglect $\check{\bm{H}}^{(2)}(\epsilon,\mu_r^2)$ in its decomposition. We can express a colour operator proportional to the identity as a sum of colour charge dipoles using colour conservation~\eqref{colour_conservation}. We can then rewrite:
	\begin{eqnarray}\label{H2todipole}
		\sum_{i}C_i\ourHop{2}{i}{\e}\mathbf{Id}&=&-\sum_{i}\ourHop{2}{i}{\e}\sum_{j\ne i}\T_i\cdot\T_j\nonumber\\
		&=&-\sum_{(i,j)}\ourHop{2}{ij}{\e}\T_i\cdot\T_j\,,
	\end{eqnarray}
	where, as usual, the sum runs over pairs of partons and $\ourHopnodep{2}{ij}=\ourHopnodep{2}{i}+\ourHopnodep{2}{j}$. The hard radiation functions are given by~\cite{Catani:1998bh,Becher:2009qa}: 
	\begin{eqnarray}
		\label{H2g}\ourHop{2}{g}{\epsilon}= \dfrac{e^{\e\gamma_E}}{4\Gamma(1-\e)\e}N_c\Biggl\{  &&\left[\dfrac{5}{12}+\dfrac{11}{144}\pi^2+\dfrac{\zeta_3}{2}\right]+\dfrac{N_f}{N_c}\left[-\dfrac{89}{108}-\dfrac{\pi^2}{72}\right]\nonumber \\
		&&+\dfrac{N_f}{N_c^3}\left[\dfrac{-1}{4}\right]+\dfrac{N_f^2}{N_c^2}\left[\dfrac{5}{27}\right]\Biggr\}\, ,\\
		\label{H2q}\ourHop{2}{q}{\epsilon}= \dfrac{e^{\e\gamma_E}}{4\Gamma(1-\e)\e}N_c\Biggl\{  &&\left[\dfrac{409}{432}-\dfrac{11}{48}\pi^2+\dfrac{7}{2}\zeta_3\right]+\dfrac{N_f}{N_c}\left[-\dfrac{25}{108}+\dfrac{\pi^2}{24}\right]\nonumber \\
		&&+\dfrac{1}{N_c^2}\left[\dfrac{3}{16}-\dfrac{\pi^2}{4}+3\zeta_3\right]\Biggr\}\, .
	\end{eqnarray}
	Using~\eqref{I1v2} and~\eqref{H2todipole} and neglecting $\check{\bm{H}}^{(2)}(\epsilon,\mu_r^2)$, we can rearrange equation~\eqref{I2} as:
	\begin{eqnarray}\label{I2v2}
		\Iop{2}{\epsilon,\mu_r^2}&=&-\dfrac{\beta_0}{\e}\sum_{(i,j)}\ourIop{1}{ij}{\e,\mu_r^2}\T_i\cdot\T_j\nonumber\\
		&&-\dfrac{1}{2}\sum_{(i,j)}\sum_{(k,l)}\ourIop{1}{ij}{\e,\mu_r^2}\ourIop{1}{kl}{\e,\mu_r^2}(\T_i\cdot\T_j)(\T_k\cdot\T_l)\nonumber\\
		&&+\sum_{(i,j)}\ourIop{2}{ij}{\e,\mu_r^2}\T_i\cdot\T_j,
	\end{eqnarray}
	where 
	\begin{equation}
		\ourIop{2}{ij}{\e,\mu_r^2}=e^{-\e\gamma_E}\dfrac{\Gamma(1-2\e)}{\Gamma(1-\e)}\left(\dfrac{\beta_0}{\e}+K\right)\ourIop{1}{ij}{2\e,\mu_r^2}-\ourHop{2}{ij}{\epsilon}\,.
	\end{equation}
	
	We can now use~\eqref{1l-sing},~\eqref{2l-sing} and~\eqref{I2} to express the singularity structure of a two-loop matrix element:
	\begin{eqnarray}\label{M2poles}
		\poles\left(\me{n+2}{2}\right)&=&\poles\left(\braket{\ampnum{n+2}{2}|\ampnum{n+2}{0}}+\braket{\ampnum{n+2}{0}|\ampnum{n+2}{2}}+\braket{\ampnum{n+2}{1}|\ampnum{n+2}{1}}\right)\nonumber\\
		&=&\poles\Big\lbrace\Big.\braket{\ampnum{n+2}{1}|\Iop{1}{\e}+\Iopd{1}{\e}|\ampnum{n+2}{0}}+\braket{\ampnum{n+2}{0}|\Iop{1}{\e}+\Iopd{1}{\e}|\ampnum{n+2}{1}}\nonumber\\
		&&-\dfrac{1}{2}\braket{\ampnum{n+2}{0}|\left(\Iop{1}{\e}+\Iopd{1}{\e}\right)\left(\Iop{1}{\e}+\Iopd{1}{\e}\right)|\ampnum{n+2}{0}}\nonumber\\
		&&-\betaoe\braket{\ampnum{n+2}{0}|\Iop{1}{\e}+\Iopd{1}{\e}|\ampnum{n+2}{0}}\nonumber\\
		&&+\Itwoepsiloncoeff\braket{\ampnum{n+2}{0}|\Iop{1}{2\e}+\Iopd{1}{2\e}|\ampnum{n+2}{0}}\nonumber\\
		&&+\braket{\ampnum{n+2}{0}|\Hop{2}{\e}+\Hopd{2}{\e}|\ampnum{n+2}{0}}\Big.\Big\rbrace\,.
	\end{eqnarray}
	We see again that only the real parts of the insertion operators are needed to describe the singularity structure. Using~\eqref{I1v2} and~\eqref{I2v2} it is possible to recast equation~\eqref{M2poles} as:
	\begingroup
	\allowdisplaybreaks
	\begin{eqnarray}\label{M2poles2}
		\poles\left(\me{n+2}{2}\right)&=&\poles\Big\lbrace\Big.\sum_{(i,j)}2\text{Re}\left[\ourIop{1}{ij}{\e,\mu_r^2}\right]\left[\braket{\ampnum{n+2}{1}|\T_i\cdot\T_j|\ampnum{n+2}{0}}+\braket{\ampnum{n+2}{0}|\T_i\cdot\T_j|\ampnum{n+2}{1}}\right]\nonumber\\
		&&-\dfrac{1}{2}\sum_{(i,j)}\sum_{(k,l)}2\text{Re}\left[\ourIop{1}{ij}{\e,\mu_r^2}\right]2\text{Re}\left[\ourIop{1}{lk}{\e,\mu_r^2}\right]\braket{\ampnum{n+2}{0}|(\T_i\cdot\T_j)(\T_k\cdot\T_l)|\ampnum{n+2}{0}}\nonumber\\
		&&-\betaoe\sum_{(i,j)}2\text{Re}\left[\ourIop{1}{ij}{\e,\mu_r^2}\right]\braket{\ampnum{n+2}{0}|\T_i\cdot\T_j|\ampnum{n+2}{0}}\nonumber\\
		&&+\sum_{(i,j)}2\text{Re}\left[\ourIop{2}{ij}{\e,\mu_r^2}\right]\braket{\ampnum{n+2}{0}|\T_i\cdot\T_j|\ampnum{n+2}{0}}\Big.\Big\rbrace.
	\end{eqnarray}
	\endgroup
	Therefore, the poles of the double-virtual contribution to the cross section are given by:
	\begin{eqnarray}\label{VVpoles}
		\lefteqn{\poles\left(\sigpart{VV}{NNLO}\right)=\coeffVVNNLO\int\dphi{n+2}(p_3,\dots,p_{n+2};p_1,p_2)\,\jet{n}{n}{\lb p \rb_n}}\nonumber\\
		&\times&\poles\Big\lbrace\Big.\sum_{(i,j)}2\text{Re}\left[\ourIop{1}{ij}{\e,\mu_r^2}\right]\left[\braket{\ampnum{n+2}{1}|\T_{i}\cdot\T_{j}|\ampnum{n+2}{0}}+\braket{\ampnum{n+2}{0}|\T_{i}\cdot\T_{j}|\ampnum{n+2}{1}}\right]\nonumber\\
		&&-\dfrac{1}{2}\sum_{(i,j)}\sum_{(k,l)}2\text{Re}\left[\ourIop{1}{ij}{\e,\mu_r^2}\right]2\text{Re}\left[\ourIop{1}{lk}{\e,\mu_r^2}\right]\braket{\ampnum{n+2}{0}|(\T_{i}\cdot\T_{j})(\T_{k}\cdot\T_{l})|\ampnum{n+2}{0}}\nonumber\\
		&&-\betaoe\sum_{(i,j)}2\text{Re}\left[\ourIop{1}{ij}{\e,\mu_r^2}\right]\braket{\ampnum{n+2}{0}|\T_{i}\cdot\T_{j}|\ampnum{n+2}{0}}\nonumber\\
		&&+\sum_{(i,j)}2\text{Re}\left[\ourIop{2}{ij}{\e,\mu_r^2}\right]\braket{\ampnum{n+2}{0}|\T_{i_g}\cdot\T_{j}|\ampnum{n+2}{0}}\Big.\Big\rbrace,
	\end{eqnarray}
	where 
	\begin{equation}
		\coeffVVNNLO=\coeff^2\coeffLO\,.
	\end{equation}
	

	\subsection{Integrated dipoles from antenna functions}\label{sec:int_dip}
	
	We are now ready to discuss the construction of integrated dipoles in colour space. We distinguish two types of integrated dipoles: identity-preserving (IP) and identity-changing (IC). The former reproduce the infrared singularity structure of virtual corrections and are naturally cast as operators in colour space. The latter address identity-changing initial-state collinear singularities. 
	
	\subsubsection{One-loop integrated dipoles}\label{sec:int_dip_1}
	
	We begin by discussing one-loop integrated dipoles, originally presented in~\cite{Currie:2013vh}. They are obtained combining integrated three-parton tree-level antenna functions $\mathcal{X}_3^0$ and NLO mass factorization kernels $\Gamma_{ab}^{(1)}$. 
	
	\paragraph{Identity-preserving dipoles}
	
	We define the following one-loop singularity dipole operator in colour space for an $(n+2)$-parton process:
	\begin{eqnarray}\label{J21}
		\Jcol{1}(\e)&&=\sum_{(i,j)\geq 3}(\T_i\cdot\T_j)\,\Jfull{1}(i,j)+\sum_{i\ne 1,2}(\T_1\cdot\T_i)\,\Jfull{1}(1,i)\nonumber\\
		&&\,+\sum_{i\ne 1,2}(\T_2\cdot\T_i)\,\Jfull{1}(2,i)+(\T_1\cdot\T_2)\,\Jfull{1}(1,2)\,.
	\end{eqnarray}
	The first sum runs over all pairs of partons in the final state, the second and the third sums include all pairs with an initial-state parton (respectively $1$ or $2$) and a final-state one and the last term addresses the configuration where both partons are in the initial state. The scalar functions $\Jfull{1}(i,j)$ are identity-preserving \textit{colour stripped one-loop integrated dipoles}. These can be classified according to the flavour of the partons $(i,j)$ and their kinematical configuration: final-final (FF), initial-final (IF) or initial-initial (II). They can be further decomposed as:
	\begin{eqnarray}
		\Jfull{1}(q,\bar{q})&=&\J{1}(q,\bar{q}), \\
		\Jfull{1}(i,g)&=&\J{1}(i,g)+\dfrac{N_f}{N_c}\Jh{1}(i,g),\quad i=q,g\,.
	\end{eqnarray}
	We list the IP quark-antiquark, quark-gluon and gluon-gluon one-loop colour-stripped integrated dipoles in Tables~\ref{tab:J21IPqqb}-\ref{tab:J21IPgg}.  We note that IF and II integrated antenna functions exhibit a dependence on the momentum fractions $x_1$ and $x_2$, which is omitted in the previous expressions and in the remainder of the paper for simplicity. We use the following shorthand notation: $\delta_{i}=\delta(1-x_i)$. We introduced, where needed, dedicated subscripts which accompany the kinematical configuration labels IF and II, to distinguish between different choices of initial-state partons.
	
	\begingroup
	\renewcommand{\arraystretch}{1.8} 
	\begin{table}[t]
		\centering\small
		\begin{tabular}{c|l}
			& Integrated dipoles
			\\ \hline
			
			\multirow{1}{*}{FF}  
			& $\J{1}(i_q,j_{\bar{q}})=\XFFint{A}{3}{0}(s_{ij})$  \\\cline{2-2}
			\hline
			
			\multirow{1}{*}{IF}  
			& $\J{1}(1_q,i_{\bar{q}})=\XIFint{A}{3}{0}{q}(s_{1i})-\Gammaone{qq}{x_1}\delta_2$  \\\cline{2-2}
			\hline
			
			\multirow{1}{*}{II}  
			& $\J{1}(1_q,2_{\bar{q}})=\XIIint{A}{3}{0}{q\bar{q}}(s_{12})-\Gammaone{qq}{x_1}\delta_2-\Gammaone{qq}{x_2}\delta_1$  \\\cline{2-2}
			\hline
		\end{tabular}
		\caption{Identity-preserving quark-antiquark one-loop colour-stripped integrated dipoles.}
		\label{tab:J21IPqqb}
	\end{table}
	\endgroup
	
	\begingroup
	\renewcommand{\arraystretch}{1.8} 
	\begin{table}[t]
		\centering\small
		\begin{tabular}{c|l}
			& Integrated dipoles
			\\ \hline
			
			\multirow{2}{*}{FF} 
			& $\J{1}(i_q,j_g)=\frac{1}{2}\XFFint{D}{3}{0}(s_{ij})$  \\\cline{2-2}
			& $\Jh{1}(i_q,j_g)=\frac{1}{2}\XFFint{E}{3}{0}(s_{ij})$  \\\cline{2-2}
			\hline
			
			\multirow{2}{*}{IF$_q$} 
			& $\J{1}(1_q,i_g)=\frac{1}{2}\XIFint{D}{3}{0}{q}(s_{1i})-\Gammaone{qq}{x_1}\delta_2$  \\\cline{2-2}
			& $\Jh{1}(1_q,i_g)=\frac{1}{2}\XIFint{E}{3}{0}{q}(s_{1i})$  \\\cline{2-2}
			\hline
			
			\multirow{2}{*}{IF$_g$}
			& $\J{1}(1_g,i_q)=\XIFint{D}{3}{0}{g,gq}(s_{1i})-\frac{1}{2}\Gammaone{gg}{x_1}\delta_2$  \\\cline{2-2}
			& $\Jh{1}(1_g,i_q)=-\frac{1}{2}\GammaoneF{gg}{x_1}\delta_2$  \\\cline{2-2}
			\hline
			
			\multirow{2}{*}{II}
			& $\J{1}(1_q,2_g)=\XIIint{D}{3}{0}{qg}(s_{12})-\Gammaone{qq}{x_1}\delta_2-\frac{1}{2}\Gammaone{gg}{x_2}\delta_1$  \\\cline{2-2}
			& $\Jh{1}(1_q,2_g)=-\frac{1}{2}\GammaoneF{gg}{x_2}\delta_1$  \\\cline{2-2}
			\hline
		\end{tabular}
		\caption{Identity-preserving quark-gluon one-loop colour-stripped integrated dipoles. The subscripts indicate different choices of initial-state partons.}
		\label{tab:J21IPqg}
	\end{table}
	\endgroup
	
	\begingroup
	\renewcommand{\arraystretch}{1.8} 
	\begin{table}[t]
		\centering\small
		\begin{tabular}{c|l}
			& Integrated dipoles
			\\ \hline
			
			\multirow{2}{*}{FF}  
			& $\J{1}(i_g,j_g)=\frac{1}{3}\XFFint{F}{3}{0}(s_{ij})$  \\\cline{2-2}
			& $\Jh{1}(i_g,j_g)=\XFFint{G}{3}{0}(s_{ij})$  \\\cline{2-2}
			\hline
			
			\multirow{2}{*}{IF}  
			& $\J{1}(1_g,i_g)=\frac{1}{2}\XIFint{F}{3}{0}{g}(s_{1i})-\frac{1}{2}\Gammaone{gg}{x_1}\delta_2$  \\\cline{2-2}
			& $\Jh{1}(1_g,i_g)=\frac{1}{2}\XIFint{G}{3}{0}{g}(s_{1i})-\frac{1}{2}\GammaoneF{gg}{x_1}\delta_2$  \\\cline{2-2}
			\hline
			
			\multirow{2}{*}{II} 
			& $\J{1}(1_g,2_g)=\XIIint{F}{3}{0}{gg}(s_{12})-\frac{1}{2}\Gammaone{gg}{x_1}\deltatwo-\frac{1}{2}\Gammaone{gg}{x_2}\deltaone$  \\\cline{2-2}
			& $\Jh{1}(1_g,2_g)=-\frac{1}{2}\GammaoneF{gg}{x_1}\deltatwo-\frac{1}{2}\GammaoneF{gg}{x_2}\deltaone$  \\\cline{2-2}
			\hline
		\end{tabular}
		\caption{Identity-preserving gluon-gluon one-loop colour-stripped integrated dipoles.}
		\label{tab:J21IPgg}
	\end{table}
	\endgroup
	\FloatBarrier
	
	As detailed in the next paragraph, the presence of splitting kernels in association with integrated IF and II antenna functions is related to the mass 
	factorization of initial-state collinear singularities, which is part of the definition of the integrated dipoles. 
	 One can therefore understand the appearance of certain splitting kernels in a given integrated dipole by considering the possible emissions occurring from the initial-state partons. For example, within the IP dipoles considered above, only $\Gamma^{(1)}_{qq}$, $\Gamma^{(1)}_{gg}$, and $\wh{\Gamma}^{(1)}_{gg}$ can appear, since the emission of a gluon does not change the species of the parton entering the hard process.

	\paragraph{Cancellation of infrared singularities at one-loop}
	
	The structure of the IP integrated dipoles is chosen in such a way that the singularities carried by the mass factorization kernels cancel with poles in the integrated IF and II antenna functions associated with initial-state collinear divergences. The remaining $\e$-poles exactly match the ones of the virtual matrix element, once the operator in~\eqref{J21} is evaluated on the corresponding LO amplitude in colour space. In particular, at one loop the following relation holds:
	\begin{equation}\label{J21relation}
		\poles\left[\Jfull{1}(i,j)\right]=\poles\left[\text{Re}\left(\ourIop{1}{i j}{\e,\mu_r^2}\right)\right],
	\end{equation}
	where the integrated dipole on the left-hand-side can be in the FF, IF or II configuration. We have explicitly verified that the relation above is satisfied for all the partonic and kinematical configurations. Given that unintegrated antenna functions capture the singular behaviour associated to unresolved emissions between a pair of hard radiators, it is not surprising that we can systematically match the singularity structure of dipole insertion operators onto integrated antenna functions~\cite{Gehrmann-DeRidder:2005btv}.
	
	Equation~\eqref{J21relation} establishes a correspondence between the operators defined in~\eqref{J21} and~\eqref{I1}, which will be crucial for the formulation of the colourful antenna subtraction method. 
	
	\paragraph{Identity-changing dipoles}
	
	For the IF and II configurations, we also introduce identity-changing integrated dipoles, obtained from IC antenna functions and splitting kernels. Contrary to the IP ones, they are not defined as operators in colour space, for reasons that will be explained in Section~\ref{sec:subNLO}. Their structure can however be inferred from the one of splitting kernels at NLO in~\eqref{split_ker_NLO}:
	\begin{equation}\label{J21ic}
		\Jfulltotic{1}{ab;cd}(x_1,x_2)=\Jfullic{1}{a\to c}(x_1)\delta_{db}\delta(1-x_2)+\Jfullic{1}{b\to d}(x_2)\delta_{ca}\delta(1-x_1).
	\end{equation}	
	We can distinguish between the \textit{quark-to-gluon} and \textit{gluon-to-quark} dipoles, which reflect the decomposition of the splitting kernels associated to the corresponding IC splitting:
	\begin{eqnarray}
		\Jfullic{1}{q\to g}(g,i)&=&\left(\dfrac{N_c^2-1}{N_c}\right)\Jic{1}{q\to g}(g,i), \\
		\Jfullic{1}{g\to q}(q,i)&=&\Jic{1}{g\to q}(q,i)\,,
	\end{eqnarray}
	where the first argument denotes the parton species which enters the hard scattering, while the second argument $i=q,g$ is a spectator parton which is not directly involved in the initial-state collinear emission and can be either in the final (IF) or initial (II) state. For identity-changing configurations we need to consider that the spin-averaging factor for a gluon and a quark differ in $d$-dimensions. To properly take this into account and compensate for the mismatch within antenna functions and splitting kernels, we introduce the following factors:
	\begin{equation}
		\Sgtoq = \dfrac{2-2\e}{2}=1-\e,\quad\Sqtog=\dfrac{2}{2-2\e}=\dfrac{1}{1-\e}.
	\end{equation}
	The expressions for the IC quark-antiquark, quark-gluon and gluon-gluon one-loop integrated dipoles are given in Tables~\ref{tab:J21ICqqb}--\ref{tab:J21ICgg}. 
	
	\begingroup
	\renewcommand{\arraystretch}{1.8} 
	\begin{table}[t]
		\centering\small
		\begin{tabular}{c|l}
			& Integrated dipoles
			\\ \hline
			
			\multirow{1}{*}{IF}  
			& $\Jic{1}{g\to q}(1_q,i_{\bar{q}})=-\frac{1}{2}\XIFint{A}{3}{0}{g}(s_{1i})-\Sgtoq\Gammaone{qg}{x_1}\delta_2$  \\\cline{2-2}
			\hline
			
			\multirow{1}{*}{II}  
			& $\Jic{1}{g\to q}(1_q,2_{\bar{q}})=-\XIIint{A}{3}{0}{gq}(s_{12})-\Sgtoq\Gammaone{qg}{x_1}\delta_2$  \\\cline{2-2}
			\hline
		\end{tabular}
		\caption{Identity-changing quark-antiquark one-loop colour-stripped integrated dipoles. The subscripts indicate different choices of initial-state partons.}
		\label{tab:J21ICqqb}
	\end{table}
	\endgroup
	
	\begingroup
	\renewcommand{\arraystretch}{1.8} 
	\begin{table}[t]
		\centering\small
		\begin{tabular}{c|l}
			& Integrated dipoles
			\\ \hline
			
			\multirow{1}{*}{IF$_{g\to q}$} 
			& $\Jic{1}{g\to q}(1_q,i_g)=-\XIFint{D}{3}{0}{g\to q}(s_{1i})-\Sgtoq\Gammaone{qg}{x_1}\delta_2$  \\\cline{2-2}
			\hline
			
			\multirow{1}{*}{IF$_{q\to g}$}
			& $\Jic{1}{q\to g}(1_g,i_q)=-\XIFint{E}{3}{0}{q'}(s_{1i})-\Sqtog\Gammaone{gq}{x_1}\delta_2$  \\\cline{2-2}
			\hline
			
			\multirow{1}{*}{II$_{g\to q}$}
			& $\Jic{1}{g\to q}(1_q,2_g)=-\XIIint{D}{3}{0}{gg}(s_{12})-\Sgtoq\Gammaone{qg}{x_1}\delta_2$  \\\cline{2-2}
			\hline
			
			\multirow{1}{*}{II$_{q\to g}$}
			& $\Jic{1}{q\to g}(1_g,2_q)=-\XIIint{E}{3}{0}{q'q}(s_{12})-\Sqtog\Gammaone{gq}{x_1}\delta_2$  \\\cline{2-2}
			\hline
		\end{tabular}
		\caption{Identity-changing quark-gluon one-loop colour-stripped integrated dipoles.}
		\label{tab:J21ICqg}
	\end{table}
	\endgroup
	
	\begingroup
	\renewcommand{\arraystretch}{1.8} 
	\begin{table}[t]
		\centering\small
		\begin{tabular}{c|l}
			& Integrated dipoles
			\\ \hline
			
			\multirow{1}{*}{IF}  
			& $\Jic{1}{q\to g}(1_g,i_g)=-\XIFint{G}{3}{0}{q}(s_{1i})-\Sqtog\Gammaone{gq}{x_1}\delta_2$  \\\cline{2-2}
			\hline
			
			\multirow{1}{*}{II} 
			& $\Jic{1}{q\to g}(1_g,2_g)=-\XIFint{G}{3}{0}{qg}(s_{12})-\Sqtog\Gammaone{gq}{x_1}\delta_2$  \\\cline{2-2}
			\hline
		\end{tabular}
		\caption{Identity-changing gluon-gluon one-loop colour-stripped integrated dipoles.}
		\label{tab:J21ICgg}
	\end{table}
	\endgroup
	\FloatBarrier
	As expected, we see the appearance of IC splitting kernels $\Gamma^{(1)}_{gq}$ and $\Gamma^{(1)}_{qg}$, associated to a final-quark becoming collinear to an initial-state parton.
	
	The IC integrated dipoles are free of explicit infrared singularities: 
	\begin{equation}\label{J21icrelation}
		\poles\left[\Jfullic{1}{a\to b}(i,j)\right]=0,
	\end{equation}
	because the only $\e$-poles which are present in integrated IC antenna functions have an initial-state collinear origin and completely cancel against the analogous ones in the IC splitting kernels.
	
	\subsubsection{Two-loop integrated dipoles}\label{sec:int_dip_2}
	
	The characteristic new ingredients of two-loop integrated dipoles are integrated four-parton tree-level antenna functions $\mathcal{X}_4^0$, three-parton one-loop antenna functions $\mathcal{X}_3^1$, convolutions of two three-parton tree-level antenna functions $\mathcal{X}_3^0\otimes \mathcal{X}_3^0$ and two-loop mass factorization kernels $\overline{\Gamma}_{ab}^{(2)}$. Analogously to the one-loop case, the presence of certain two-loop splitting kernels in association with integrated antenna functions can be understood considering the possible emissions from initial-state partons. 
	
	Also at two-loop we distinguish between identity-preserving and identity-changing integrated dipoles, however a significant difference with respect to the one-loop case has to be highlighted. Explicit integrated results are available at NLO, not only for all the three-parton tree-level antenna functions, but also for any sub-antenna, given the simplicity of the analytical integration. This means that both at the unintegrated and integrated level one has complete control over which infrared-divergent configurations are considered. This is not true at NNLO, where the unavailability of individual integrated sub-antennae forces one to include additional infrared singularities within the identity-preserving two-loop integrated dipoles. These singularities have to be properly identified and subtracted. This is achieved by introducing spurious terms which target the undesired structures. We identify three classes of such terms:
	\begin{itemize}
		\item \textbf{flip-flopping contributions:} address IC initial-state collinear limits like $g\to q\to g$ or $q\to g \to q$, where the identity of the initial-state emitter formally coincides with the original one. These contributions are denoted in the following by the subscript $\mathrm{f/f}$; 
		\item \textbf{IC corrective contributions:} target genuine IC limits which need to be removed from IP structures, denoted with the subscript $\mathrm{IC}\,\mathrm{corr}.$;
		\item \textbf{unphysical triple-collinear limits:} limits existing in the supersymmetric matrix elements used to extract antenna functions, which are not present in QCD. They are treated introducing suitable insertion operators.
	\end{itemize}
	We will comment further about each class in the paragraphs below. Expressions for the two-loop colour-stripped integrated dipoles used for the construction of NNLO subtraction terms for di-jet production can be found in~\cite{Chen:2022tpk}. In the following we review them and discuss in more detail how to generalize their construction for any process. To simplify the expressions, we omit the scale $s_{ij}$ dependence of the integrated antenna functions.
	
	We mention that the complete development of designer antenna functions~\cite{Braun-White:2023sgd,Braun-White:2023zwd} at NNLO will bring significant simplifications to some of the structures presented below. Indeed, the assemblage of antenna functions from specific infrared limits, rather than from physical matrix elements, naturally prevents the introduction of spurious divergences and therefore the necessity of the corrective terms introduced above.
	
	\paragraph{Identity-preserving dipoles}
	
	In analogy with~\eqref{J21}, we can define a two-loop insertion operator in colour space:
	\begin{eqnarray}\label{J22}
		\Jcol{2}(\e)&&=N_c\sum_{(i,j)\geq 3}(\T_i\cdot\T_j)\,\Jfull{2}(i_g,j_g)+N_c\sum_{i\ne 1,2}(\T_1\cdot\T_i)\,\Jfull{2}(\hat{1}_g,i_g)\nonumber\\
		&&\,+N_c\sum_{i\ne 1,2}(\T_2\cdot\T_i)\,\Jfull{2}(\hat{2}_g,i_g)+N_c\,(\T_1\cdot\T_2)\,\Jfull{2}(\hat{1}_g,\hat{2}_g)\,.
	\end{eqnarray}
	where the functions accompanying the charge dipoles are identity-preserving \textit{two-loop colour-stripped integrated dipoles}. They can be decomposed according to:
	\begin{eqnarray}
		\Jfull{2}\left(q,\bar{q}\right)&=&\J{2}\left(q,\bar{q}\right)-\dfrac{1}{N_c^2}\Jt{2}\left(q,\bar{q}\right)+\dfrac{N_f}{N_c}\Jh{2}\left(q,\bar{q}\right)\\
		\Jfull{2}\left(g,i\right)&=&\J{2}\left(g,i\right)+\dfrac{N_f}{N_c}\,\Jh{2}\left(g,i\right)\nonumber \\
		&-&\dfrac{N_f}{N_c^3}\Jht{2}\left(g,i\right)+\dfrac{N_f^2}{N_c^2}\,\Jhh{2}\left(g,i\right),\quad i=g,q\, .
	\end{eqnarray}
	
	\paragraph{Quark-antiquark integrated dipoles}
	
The	IP two-loop quark-antiquark colour-stripped integrated dipoles can be  constructed in a straightforward manner, since the quark-antiquark NNLO antenna functions specifically contain only the desired physical limits in any kinematical configuration and allow for a unique identification of the 
	hard radiator partons. Even if sub-antennae need to be employed, for example for $\wt{A}_4^0$~\cite{Gehrmann-DeRidder:2007foh}, this is only required to properly define a colour-ordered momentum mapping. Therefore, for a given hard quark-antiquark dipole, no spurious structure is present in the associated integrated antenna functions. The IP quark-antiquark two-loop integrated dipoles are listed in Table~\ref{tab:J22IPqqb}.
	\begingroup
	\renewcommand{\arraystretch}{1.8} 
	\begin{table}[t]
		\centering\small
		\begin{tabular}{c|l}
			& Integrated dipoles
			\\ \hline
			
			\multirow{3}{*}{FF}  
			& $\J{2}\left(i_{q},j_{\bar{q}}\right)=\XFFint{A}{4}{0}+\XFFint{A}{3}{1}+\boelite\QQslite{ij}\XFFint{A}{3}{0}-\frac{1}{2}\left[\XFFint{A}{3}{0}\otimes\XFFint{A}{3}{0}\right]$\\\cline{2-2}
			& $\Jt{2}\left(1_{q},2_{\bar{q}}\right)=\frac{1}{2}\XtFFint{A}{4}{0}+2\XFFint{C}{4}{0}+\XtFFint{A}{3}{1}-\frac{1}{2}\left[\XFFint{A}{3}{0}\otimes\XFFint{A}{3}{0}\right]$\\\cline{2-2}
			& $\Jh{2}\left(1_{q},2_{\bar{q}}\right)=\XFFint{B}{4}{0}+\XhFFint{A}{3}{1}+\bFoelite\QQslite{ij}\XFFint{A}{3}{0}$  \\\cline{2-2}
			\hline
			
			\multirow{5}{*}{IF}  
			& $\J{2}\left(1_{q},i_{\bar{q}}\right)=\XIFint{A}{4}{0}{q}+\XIFint{A}{3}{1}{q}+\boelite\QQslite{1i}\XIFint{A}{3}{0}{q}-\frac{1}{2}\left[\XIFint{A}{3}{0}{q}\otimes\XIFint{A}{3}{0}{q}\right]$ \\
			& $\phantom{\J{2}\left(1_{q},i_{\bar{q}}\right)}-\Gammatwo{qq}{1}\deltatwo$ \\\cline{2-2}
			& $\Jt{2}\left(1_q,i_{\bar{q}}\right)=\frac{1}{2}\XtIFint{A}{4}{0}{q}+2\XIFint{C}{4}{0}{q}+\XIFint{C}{4}{0}{\bar{q}}+\XtIFint{A}{3}{1}{q}-\frac{1}{2}\left[\XIFint{A}{3}{0}{q}\otimes\XIFint{A}{3}{0}{q}\right]$ \\
			& $\phantom{\Jt{2}\left(1_q,i_{\bar{q}}\right)}+\Gammatwott{qq}{1}\deltatwo$ \\\cline{2-2}
			& $\Jh{2}\left(1_q,i_{\bar{q}}\right)=\XIFint{B}{4}{0}{q}+\XhIFint{A}{3}{1}{q}+\bFoelite\QQslite{1i}\XIFint{A}{3}{0}{q}-\GammatwoF{qq}{1}\deltatwo$  \\\cline{2-2}
			\hline
			
			\multirow{6}{*}{II}  
			& $\J{2}\left(1_{q},2_{\bar{q}}\right)=\XIIint{A}{4}{0}{q\bar{q}}+\XIIint{A}{3}{1}{q\bar{q}}+\boelite\QQslite{12}\XIIint{A}{3}{0}{q\bar{q}}-\frac{1}{2}\left[\XIIint{A}{3}{0}{q\bar{q}}\otimes\XIIint{A}{3}{0}{q\bar{q}}\right]$ \\ 
			&$\phantom{\J{2}\left(1_{q},2_{\bar{q}}\right)}-\Gammatwo{qq}{1}\deltatwo-\Gammatwo{qq}{2}\deltaone$ \\\cline{2-2}
			& $\Jt{2}\left(1_q,2_{\bar{q}}\right)=\frac{1}{2}\XtIIint{A}{4}{0}{q\bar{q}}+2\XIIint{C}{4}{0}{q\bar{q}}+2\XIIint{C}{4}{0}{\bar{q}q}+\XtIIint{A}{3}{1}{q\bar{q}}-\frac{1}{2}\left[\XIIint{A}{3}{0}{q\bar{q}}\otimes\XIIint{A}{3}{0}{q\bar{q}}\right]$ \\
			& $\phantom{\Jt{2}\left(1_q,2_{\bar{q}}\right)}+\Gammatwott{qq}{1}\deltatwo+\Gammatwott{qq}{2}\deltaone$ \\\cline{2-2}
			& $\Jh{2}\left(1_q,2_{\bar{q}}\right)=\XIIint{B}{4}{0}{q\qb}+\XhIIint{A}{3}{1}{qq}+\bFoelite\QQslite{12}\XIIint{A}{3}{0}{qq}$ \\
			& $\phantom{\Jh{2}\left(1_q,2_{\bar{q}}\right)}-\GammatwoF{qq}{1}\deltatwo-\GammatwoF{qq}{2}\deltaone$  \\\cline{2-2}
			\hline
		\end{tabular}
		\caption{Identity-preserving quark-antiquark two-loop colour-stripped integrated dipoles.}
		\label{tab:J22IPqqb}
	\end{table}
	\endgroup
	
	\paragraph{Quark-gluon integrated dipoles}
	
The	IP quark-gluon dipoles are particularly affected by the presence of undesired singularities. This is mostly due to quark-gluon NNLO antenna functions containing any allowed unresolved limit among the involved partons, with no possibility to clearly disentangle the hard radiators from the unresolved partons 
 at the integrated level. The IP quark-gluon two-loop colour stripped integrated dipoles are given in Table~\ref{tab:J22IPqg}.
 
	\begingroup
	\renewcommand{\arraystretch}{1.8} 
	\begin{table}[t]
		\centering\small
		\begin{tabular}{c|l}
			& Integrated dipoles
			\\ \hline
			
			\multirow{5}{*}{FF} 
			& $\J{2}\left(i_q,j_g\right)=\frac{1}{2}\XFFint{D}{4}{0}+\frac{1}{2}\XFFint{D}{3}{1}+\frac{1}{2}\boelite\QQslite{ij}\XFFint{D}{3}{0}-\frac{1}{4}\left[\XFFint{D}{3}{0}\otimes\XFFint{D}{3}{0}\right]$  \\\cline{2-2}
			& $\Jh{2}\left(i_q,j_g\right)=\XFFint{E}{4}{0}+\frac{1}{2}\XhFFint{D}{3}{1}+\frac{1}{2}\XFFint{E}{3}{1}+\frac{1}{2}\bFoelite\QQslite{ij}\XFFint{D}{3}{0} +\frac{1}{2}\boelite\QQslite{ij}\XFFint{E}{3}{0}$ \\
			& $\phantom{\Jh{2}\left(i_q,j_g\right)}-\frac{1}{2}\left[\XFFint{E}{3}{0}\otimes\XFFint{D}{3}{0}\right]$ \\\cline{2-2}
			& $\Jht{2}\left(i_q,j_g\right)=\frac{1}{2}\XtFFint{E}{4}{0}+\frac{1}{2}\XtFFint{E}{3}{1}$  \\\cline{2-2}
			& $\Jhh{2}\left(i_q,j_g\right)=\frac{1}{2}\XhFFint{E}{3}{1}+\frac{1}{2}\bFoelite\QQslite{ij}\XFFint{E}{3}{0}-\frac{1}{4}\left[\XFFint{E}{3}{0}\otimes\XFFint{E}{3}{0}\right]$  \\\cline{2-2}
			\hline
			
			\multirow{6}{*}{IF$_q$} 
			& 
			$\J{2}\left(1_q,i_g\right)=\frac{1}{2}\XIFint{D}{4}{0}{q}+\frac{1}{2}\XIFint{D}{3}{1}{q}+\frac{1}{2}\boelite\QQslite{1i}\XIFint{D}{3}{0}{q}-\frac{1}{4}\left[\XIFint{D}{3}{0}{q}\otimes\XIFint{D}{3}{0}{q}\right]$ \\
			& $\phantom{\J{2}\left(1_q,i_g\right)}-\Gammatwo{qq}{1}\deltatwo$  \\\cline{2-2}
			& $\Jh{2}\left(1_q,i_g\right)=\XIFint{E}{4}{0}{q}+\frac{1}{2}\XhIFint{D}{3}{1}{q}+\frac{1}{2}\XIFint{E}{3}{1}{q}+\frac{1}{2}\bFoelite\QQslite{1i}\XIFint{D}{3}{0}{q}$ \\ 
			& $\phantom{\Jh{2}\left(1_q,i_g\right)}+\frac{1}{2}\boelite\QQslite{1i}\XIFint{E}{3}{0}{q}-\frac{1}{2}\left[\XIFint{E}{3}{0}{q}\otimes\XIFint{D}{3}{0}{q}\right]-\GammatwoF{qq}{1}\deltatwo$  \\\cline{2-2}
			& $\Jht{2}\left(1_q,i_g\right)=\frac{1}{2}\XtIFint{E}{4}{0}{q}+\frac{1}{2}\XtIFint{E}{3}{1}{q}$ \\\cline{2-2}
			& $\Jhh{2}\left(1_q,i_g\right)=\frac{1}{2}\XhIFint{E}{3}{1}{q}+\frac{1}{2}\bFoelite\QQslite{1i}\XIFint{E}{3}{0}{q}-\frac{1}{4}\left[\XIFint{E}{3}{0}{q}\otimes\XIFint{E}{3}{0}{q}\right]$  \\\cline{2-2}
			\hline
			
			\multirow{6}{*}{IF$_g$}
			& $\J{2}\left(1_g,i_q\right)=\XIFint{D}{4}{0}{g}+\frac{1}{2}\XIFint{D}{4}{0}{g'}+\XIFint{D}{3}{1}{g}+\boelite\QQslite{1i}\XIFint{D}{3}{0}{g}$ \\ 
			& $\phantom{\J{2}\left(1_g,i_q\right)}-\left[\XIFint{D}{3}{0}{g\to g}\otimes\XIFint{D}{3}{0}{g\to g}\right]-\frac{1}{2}\Gammatwo{gg}{1}\deltatwo
			+\Jic{2}{\mathrm{IC}\,\mathrm{corr}.}(q,\qb)$ \\\cline{2-2}
			& $\Jh{2}\left(1_g,i_q\right)=\XIFint{E}{4}{0}{g}+\XhIFint{D}{3}{1}{g}+\bFoelite\QQslite{1i}\XIFint{D}{3}{0}{g}$ \\ 
			& $\phantom{\Jh{2}\left(1_g,i_q\right)}-\frac{1}{2}\GammatwoF{gg}{1}\deltatwo
			+\Jhic{2}{\mathrm{f/f}}\left(1_g,i_q\right)
			+\Jhic{2}{\mathrm{IC}\,\mathrm{corr}.}(q,\qb)$ \\\cline{2-2}
			& $\Jht{2}\left(1_g,i_q\right)=\frac{1}{2}\XtIFint{E}{4}{0}{g}+\frac{1}{2}\GammatwoFt{gg}{1}\deltatwo
			+\Jhtic{2}{\mathrm{f/f}}\left(1_g,i_q\right)$  \\\cline{2-2}
			& $\Jhh{2}\left(1_g,i_q\right)=-\frac{1}{2}\GammatwoFF{gg}{1}\deltatwo$  \\\cline{2-2}
			\hline
			
			\multirow{5}{*}{II}
			& $\J{2}\left(1_q,2_g\right)=\XIIint{D}{4}{0}{qg}+\frac{1}{2}\XIIint{D}{4}{0}{qg'}+\XIIint{D}{3}{1}{qg}+\boelite\QQslite{12}\XIIint{D}{3}{0}{qg}$ \\ 
			& $\phantom{\J{2}\left(1_q,2_g\right)}-\left[\XIIint{D}{3}{0}{qg}\otimes\XIIint{D}{3}{0}{qg}\right]-\GammatwoF{qq}{1}\deltatwo-\frac{1}{2}\GammatwoF{gg}{2}\deltaone$ \\\cline{2-2}
			& $\Jh{2}\left(1_q,2_g\right)=\XIIint{E}{4}{0}{qg}+\frac{1}{2}\XhIIint{D}{3}{1}{qg}+\bFoelite\QQslite{12}\XIIint{D}{3}{0}{qg}-\GammatwoF{qq}{1}\deltatwo$ \\ 
			& $\phantom{\Jh{2}\left(1_q,2_g\right)}-\frac{1}{2}\GammatwoF{gg}{2}\deltaone
			+\Jhic{2}{\mathrm{f/f}}\left(1_q2_g\right)$ \\\cline{2-2}
			& $\Jht{2}\left(1_q,2_g\right)=\frac{1}{2}\XtIIint{E}{4}{0}{qg}+\frac{1}{2}\GammatwoFt{gg}{2}\deltaone+\Jhtic{2}{\mathrm{f/f}}\left(1_q2_g\right)$  \\\cline{2-2}
			\hline
		\end{tabular}
		\caption{Identity-preserving quark-gluon two-loop colour-stripped integrated dipoles. The subscripts indicate different choices of initial-state partons.}
		\label{tab:J22IPqg}
	\end{table}
	\endgroup

	As anticipated, we see the appearance of corrective contributions. The flip-flopping terms remove triple-collinear limits $g\parallel q' \parallel \qb'$ with the gluon in the initial state, which are present in the $\XIFint{E}{4}{0}{g}$, $\XtIFint{E}{4}{0}{g}$, $\XIIint{E}{4}{0}{qg}$ and $\XtIIint{E}{4}{0}{qg}$ integrated antenna functions. Technically, the identity of the hard initial-state parton after integration is preserved, but the infrared divergences associated to such an unresolved configuration are absorbed by IC spitting kernels, rather than by virtual corrections. Therefore, in the definition of genuinely IP integrated dipoles, these singularities need to be removed. The flip-flopping terms are given in Table~\ref{tab:J22qgff}.
	
	\begingroup
	\renewcommand{\arraystretch}{1.8} 
	\begin{table}[t]
		\centering\small
		\begin{tabular}{c|l}
			& Integrated dipoles
			\\ \hline
			
			\multirow{2}{*}{IF}  
			& $\Jhic{2}{\mathrm{f/f}}\left(1_g,i_q\right)=\Sgtoq\left[\Gammaoneconv{qg}{1}\otimes\XIFint{E}{3}{0}{q'}\right]+\frac{1}{2}\left[\Gammaoneconv{qg}{1}\otimes\Gammaoneconv{gq}{1}\right]$ \\\cline{2-2}
			& $\Jhtic{2}{\mathrm{f/f}}\left(1_g,i_q\right)=\Sgtoq\left[\Gammaoneconv{qg}{1}\otimes\XIFint{E}{3}{0}{q'}\right]+\frac{1}{2}\left[\Gammaoneconv{qg}{1}\otimes\Gammaoneconv{gq}{1}\right]$  \\\cline{2-2}
			\hline
			
			\multirow{2}{*}{II} 
			& $\Jhic{2}{\mathrm{f/f}}\left(1_g,2_q\right)=\Sgtoq\left[\Gammaoneconv{qg}{2}\otimes\XIFint{E}{3}{0}{qq'}\right]+\frac{1}{2}\left[\Gammaoneconv{qg}{2}\otimes\Gammaoneconv{gq}{2}\right]$  \\\cline{2-2}
			& $\Jhtic{2}{\mathrm{f/f}}\left(1_g,2_q\right)=\Sgtoq\left[\Gammaoneconv{qg}{2}\otimes\XIFint{E}{3}{0}{qq'}\right]+\frac{1}{2}\left[\Gammaoneconv{qg}{2}\otimes\Gammaoneconv{gq}{2}\right]$  \\\cline{2-2}
			\hline
		\end{tabular}
		\caption{Flip-flopping contributions to identity-preserving quark-gluon two-loop integrated dipoles.}
		\label{tab:J22qgff}
	\end{table}
	\endgroup
	Analogously, spurious IC singularities are present in the $\XIFint{D}{3}{0}{g}$, $\XIFint{D}{3}{1}{g}$, $\XhIFint{D}{3}{1}{g}$ and $\XIFint{D}{4}{0}{g}$ antenna functions. In particular, these gluon-initiated antenna functions contain $\e$-poles due to the final-state quark becoming collinear to the initial-state gluon. This is an IC unresolved limit, which cannot be isolated from the remaining IP ones due to the unavailability of dedicated integrated sub-antennae. It is possible to remove such singularities by a suitable combination of gluon-initiated IC quark-gluon and quark-antiquark antenna functions. The former have the same gluon and quark involved in the dipole as hard radiators. The latter have one quark (antiquark) coinciding with the one in the quark-gluon dipole, and so colour-connected to the initial-state gluon, while the other antiquark (quark) is its colour-connected partner. We note that in such quark-antiquark antennae, no divergent behaviour is associated with the antiquark (quark) becoming collinear to the initial-state gluon. If the antiquark (quark) is in the final state, the introduced antenna functions will be in the initial-final configurations, otherwise, if it is in the initial state, the antenna functions will be in the initial-initial configuration. The explicit expressions according to the kinematical configurations of the quark-antiquark pair are given in Table~\ref{tab:J22corr}.
	\begingroup
	\renewcommand{\arraystretch}{1.8} 
	\begin{table}[t]
		\centering\small
		\begin{tabular}{c|l}
			& Integrated dipoles
			\\ \hline
			
			\multirow{8}{*}{IF}  
			& $\Jic{2}{\mathrm{IC}\,\mathrm{corr}}\left(1_q,i_{\qb}\right)=-\left[\XIFint{D}{3}{0}{q}\otimes\XIFint{D}{3}{0}{g\to q}\right]-\left[\Gammaoneconv{gg}{1}\otimes\XIFint{D}{3}{0}{g\to q}\right]$ \\
			& $\phantom{\Jic{2}{\mathrm{IC}\,\mathrm{corr}}\left(1_q,i_{\qb}\right)}+2\left[\Gammaoneconv{qq}{1}\otimes\XIFint{D}{3}{0}{g\to q}\right]-\XIFint{A}{4}{0}{g}-\frac{1}{2}\XtIFint{A}{4}{0}{g}-\frac{1}{2}\XIFint{A}{3}{1}{g}$ \\
			& $\phantom{\Jic{2}{\mathrm{IC}\,\mathrm{corr}}\left(1_q,i_{\qb}\right)}-\frac{1}{2}\XtIFint{A}{3}{1}{g}-\frac{1}{2}\boelite\left(\QQslite{1i}-1\right)\XIFint{A}{3}{0}{g}+\left[\XIFint{A}{3}{0}{q}\otimes\XIFint{A}{3}{0}{g}\right]$ \\
			& $\phantom{\Jic{2}{\mathrm{IC}\,\mathrm{corr}}\left(1_q,i_{\qb}\right)}+\frac{1}{2}\left[\Gammaoneconv{gg}{1}\otimes\XIFint{A}{3}{0}{g}\right]-\left[\Gammaoneconv{qq}{1}\otimes\XIFint{A}{3}{0}{g}\right]$ \\
			& $\phantom{\Jic{2}{\mathrm{IC}\,\mathrm{corr}}\left(1_q,i_{\qb}\right)}-\boelite\left(\XIFint{D}{3}{0}{g}-\XIFint{D}{3}{0}{g\to g}\right)$  \\\cline{2-2}
			& $\Jhic{2}{\mathrm{IC}\,\mathrm{corr}}\left(1_q,i_{\qb}\right)=-\left[\XIFint{E}{3}{0}{q}\otimes\XIFint{D}{3}{0}{g\to q}\right]-\left[\GammaoneFconv{gg}{1}\otimes\XIFint{D}{3}{0}{g\to q}\right]$ \\
			& $\phantom{\Jhic{2}{\mathrm{IC}\,\mathrm{corr}}\left(1_g,i_{\qb}\right)}-\frac{1}{2}\XhIFint{A}{3}{1}{g}-\frac{1}{2}\bFoelite\left(\QQslite{1i}-1\right)\XIFint{A}{3}{0}{g}$ \\
			& $\phantom{\Jhic{2}{\mathrm{IC}\,\mathrm{corr}}\left(1_g,i_{\qb}\right)}+\frac{1}{2}\left[\GammaoneFconv{gg}{1}\otimes\XIFint{A}{3}{0}{g}\right]-\bFoelite\left(\XIFint{D}{3}{0}{g}-\XIFint{D}{3}{0}{g\to g}\right)$  \\\cline{2-2}
			\hline
			
			\multirow{9}{*}{II}  
			& $\Jic{2}{\mathrm{IC}\,\mathrm{corr}}\left(1_q,2_{\qb}\right)=-\left[\XIFint{D}{3}{0}{q}\otimes\XIFint{D}{3}{0}{g\to q}\right]-\left[\Gammaoneconv{gg}{1}\otimes\XIFint{D}{3}{0}{g\to q}\right]$ \\
			& $\phantom{\Jic{2}{\mathrm{IC}\,\mathrm{corr}}\left(1_q,2_{\qb}\right)}+2\left[\Gammaoneconv{qq}{1}\otimes\XIFint{D}{3}{0}{g\to q}\right]-\XIIint{A}{4}{0}{qg}-\XIIint{A}{4}{0}{qg'}$ \\
			& $\phantom{\Jic{2}{\mathrm{IC}\,\mathrm{corr}}\left(1_q,2_{\qb}\right)}-\frac{1}{2}\XtIIint{A}{4}{0}{qg}-\frac{1}{2}\XIIint{A}{3}{1}{qg}-\frac{1}{2}\XtIIint{A}{3}{1}{qg}$ \\
			& $\phantom{\Jic{2}{\mathrm{IC}\,\mathrm{corr}}\left(1_q,2_{\qb}\right)}-\boelite\left(\QQslite{12}-1\right)\XIIint{A}{3}{0}{gq}+2\left[\XIIint{A}{3}{0}{gq}\otimes\XIIint{A}{3}{0}{gq}\right]$ \\
			& $\phantom{\Jic{2}{\mathrm{IC}\,\mathrm{corr}}\left(1_q,2_{\qb}\right)}+\left[\Gammaoneconv{gg}{1}\otimes\XIIint{A}{3}{0}{gq}\right]-2\left[\Gammaoneconv{qq}{1}\otimes\XIIint{A}{3}{0}{gq}\right]$ \\
			& $\phantom{\Jic{2}{\mathrm{IC}\,\mathrm{corr}}\left(1_q,2_{\qb}\right)}-\boelite\left(\XIFint{D}{3}{0}{g}-\XIFint{D}{3}{0}{g\to g}\right)$  \\\cline{2-2}
			& $\Jhic{2}{\mathrm{IC}\,\mathrm{corr}}\left(1_q,2_{\qb}\right)=-\left[\XIFint{E}{3}{0}{q}\otimes\XIFint{D}{3}{0}{g\to q}\right]-\left[\GammaoneFconv{gg}{1}\otimes\XIFint{D}{3}{0}{g\to q}\right]$ \\
			& $\phantom{\Jhic{2}{\mathrm{IC}\,\mathrm{corr}}\left(1_q,2_{\qb}\right)}-\XhIIint{A}{3}{1}{gq}-\bFoelite\left(\QQslite{12}-1\right)\XIIint{A}{3}{0}{gq}$ \\
			& $\phantom{\Jhic{2}{\mathrm{IC}\,\mathrm{corr}}\left(1_q,2_{\qb}\right)}+\left[\GammaoneFconv{gg}{1}\otimes\XIIint{A}{3}{0}{gq}\right]-\bFoelite\left(\XIFint{D}{3}{0}{g}-\XIFint{D}{3}{0}{g\to g}\right)$  \\\cline{2-2}
			\hline
		\end{tabular}
		\caption{Corrective terms required to remove spurious identity-changing singularities from integrated identity-preserving gluon-initiated quark-gluon antenna functions.}
		\label{tab:J22corr}
	\end{table}
	\endgroup
	
	These corrective terms display a quite involved structure. Indeed, even if the presence of spurious singularities within integrated antenna functions is well understood and formally under control, they inevitably complicate the structure of IP integrated dipoles as well as of the subtraction terms. In particular, such corrections propagate into all the layers of the subtraction infrastructure and require careful treatment in any practical implementation.
	
	Finally, the $D$-type tree-level four-parton antenna functions $D_4^0(q,g_1,g_2,g_3)$ (FF, IF or II) present a triple-collinear $g_3\parallel q \parallel g_1$ limit which is an unphysical remnant due to the cyclicity of the super-symmetric matrix element used to extract the antenna functions~\cite{Gehrmann-DeRidder:2005svg}. This limit shows up at the integrated level as spurious $\e$-poles which need to be removed from the integrated dipoles. The collinear poles in a quark-gluon dipole are described by the hard radiation function $\ourHopd{2}{gq}{\e}=\ourHopd{2}{g}{\e}+\ourHopd{2}{q}{\e}$, with $\ourHopd{2}{g}{\e}$ and $\ourHopd{2}{q}{\e}$ given in~\eqref{H2g} and~\eqref{H2q}. On the contrary, for the decay of a neutralino to a gluon and a gluino $\chi\to g\tilde{g}$, the process used to extract the antenna functions, the collinear poles are described by  $\ourHopd{2}{g\tilde{g}}{\e}=\ourHopd{2}{g}{\e}+\ourHopd{2}{\tilde{g}}{\e}$, with the quark hard radiation function replace by the gluino one~\cite{Gehrmann-DeRidder:2005svg,Chen:2023egx}:
	\begin{equation}\label{H2gluino}
		\ourHop{2}{\tilde{g}}{\epsilon}= \dfrac{e^{\e\gamma_E}}{4\Gamma(1-\e)\e}N_c\left\lbrace  \left[-\dfrac{187}{216}+\dfrac{13}{48}\pi^2-\dfrac{1}{2}\zeta_3\right]+\dfrac{N_f}{N_c}\left[-\dfrac{25}{108}+\dfrac{\pi^2}{24}\right]\right\rbrace.
	\end{equation}
    The unphysical poles which are not present in QCD matrix elements are identified by the mismatch between the $N_f=0$ components of the quark and the gluino hard radiation functions:
	\begin{eqnarray}
		\ourHop{2}{q}{\epsilon}-\ourHop{2}{\tilde{g}}{\epsilon}&=&\dfrac{e^{\e\gamma_E}N_c}{4\Gamma(1-\e)\e}\Biggl\{
		\left[\dfrac{29}{16}-\dfrac{\pi^2}{2}+4\zeta_3\right]+\dfrac{1}{N_c^2}\left[\dfrac{3}{16}-\dfrac{\pi^2}{4}+3\zeta_3\right]\Biggr\}.
	\end{eqnarray}
	To compensate for the discrepancy above we introduce an auxiliary quark-antiquark two-loop integrated dipole:
	\begin{equation}\label{J2bar}
		\Jfullb{2}(q,\qb)=\dfrac{1}{2}\Jb{2}(q,\qb)+\dfrac{1}{2N_c^2}\Jtb{2}(q,\qb), 
	\end{equation}
	which satisfies:
	\begin{equation}
		\Jfullb{2}(q,\qb)=\ourHop{2}{q}{\epsilon}\Bigg|_{N_f=0}-\ourHop{2}{\tilde{g}}{\epsilon}\Bigg|_{N_f=0}+\mathcal{O}(\e^0).
	\end{equation}
	The integrated dipoles in~\eqref{J2bar} can be written in terms of integrated antenna functions and are listed in Table~\ref{tab:J22bar}.
	\begingroup
	\renewcommand{\arraystretch}{1.8} 
	\begin{table}[t]
		\centering\small
		\begin{tabular}{c|l}
			& Integrated dipoles
			\\ \hline
			
			\multirow{2}{*}{FF}  
			& $\Jb{2}(i_q,j_\qb)=\frac{1}{2}\XtFFint{A}{4}{0}+\XtFFint{A}{3}{1}-\frac{1}{2}\left[\XFFint{A}{3}{0}\otimes\XFFint{A}{3}{0}\right]$  \\\cline{2-2}
			& $\Jtb{2}(i_q,j_\qb)=\Jt{2}(i_q,j_\qb)$  \\\cline{2-2}
			\hline
			
			\multirow{2}{*}{IF}  
			& $\Jb{2}(1_q,i_\qb)=\frac{1}{2}\XtIFint{A}{4}{0}{q}+\XtIFint{A}{3}{1}{q}-\frac{1}{2}\left[\XIFint{A}{3}{0}{q}\otimes\XIFint{A}{3}{0}{q}\right]$  \\\cline{2-2}
			& $\Jtb{2}(1_q,i_\qb)=\Jt{2}(i_q,j_\qb)$  \\\cline{2-2}
			\hline
			
			\multirow{2}{*}{II}  
			& $\Jb{2}(1_q,2_\qb)=\frac{1}{2}\XtIIint{A}{4}{0}{q\qb}+\XtIIint{A}{3}{1}{q\qb}-\frac{1}{2}\left[\XIIint{A}{3}{0}{q\qb}\otimes\XIIint{A}{3}{0}{q\qb}\right]$  \\\cline{2-2}
			& $\Jtb{2}(1_q,2_\qb)=\Jt{2}(i_q,j_\qb)$  \\\cline{2-2}
			\hline
		\end{tabular}
		\caption{Auxiliary quark-antiquark two-loop integrated dipoles needed to remove spurious poles present in integrated quark-gluon antenna functions, which are not present in physical matrix elements.}
		\label{tab:J22bar}
	\end{table}
	\endgroup

	The combination of antenna functions which gives the auxiliary dipoles was found previously in~\cite{Gehrmann-DeRidder:2007foh} and in~\cite{Currie:2013vh} during the construction of double-real subtraction terms for $e^{+}e^{-}\to 3j$ and $pp\to jj$, as a correction to remove the unphysical triple-collinear limits of the $D$-type antenna functions. The presence of such a combination can be easily tracked down due to the anomalous appearance of subleading-colour quark-antiquark antenna functions within the leading-colour contribution to the subtraction terms. 
	
	The newly introduced $\Jfullb{2}(q,\qb)$ can be promoted to an operator in colour space, however its structure depends on the number of fermionic lines present in the process. For a process involving a single quark-antiquark pair $(q,\qb)$ and an arbitrary number of gluons, we have:
	\begin{equation}\label{J2barcol}
		\Jcolb{2}=N_c\sum_{g}\left(\T_q+\T_\qb\right)\T_g\,\Jfullb{2}(q,\qb),
	\end{equation}
	which has the peculiarity of combining quark-gluon dipole charge operators with quark-antiquark integrated dipoles. This is once again due to quark-gluon integrated antenna functions not exactly reproducing the QCD radiation among the physical quark-gluon dipole. 
	
	For processes involving more than one quark-antiquark pair, the appropriate structure in colour space is more complicated. This is explained by the non-trivial colour connections between quarks belonging to different quark-antiquark pairs. The extension of~\eqref{J2bar} to multiple fermionic lines is first required addressing NNLO calculations with $5$ partons at Born-level, such as $pp\to 3j$, for the sub-processes with two fermionic lines and one gluon: $q\qb q'\qb' g$. Configurations with two fermionic lines and more than one gluon or three fermionic lines with at least one gluon would appear, for example, in the computation of the NNLO correction to $pp\to 4j$ and $pp\to 5j$, respectively. Since at present such calculations are well beyond the reach of available two-loop matrix elements and, arguably, computational techniques, we focus on the five-parton configuration. The appropriate colour space structure  for two quark-antiquark pairs (possibly of the same flavour), denoted with $(q,\qb)$ and $(q',\qb')$ and a gluon $g$ is then:
	\begin{eqnarray}\label{J2barcol2}
		\Jcolb{2}&=&N_c\left(\T_q+\T_\qb\right)\T_g\,\Jfullb{2}(q,\qb)+N_c\left(\T_{q'}+\T_{\qb'}\right)\T_g\,\Jfullb{2}(q',\qb')\nn\\
		&+&N_c\left(\T_q-\T_\qb\right)\T_g\,\left(\Jfullb{2}(q,\qb')-\Jfullb{2}(q',\qb)\right)(\delta_{q,1}+\delta_{q,2})\nn\\
		&+&N_c\left(\T_{q'}-\T_{\qb'}\right)\T_g\,\left(\Jfullb{2}(q,\qb')-\Jfullb{2}(q',\qb)\right)(\delta_{q',1}+\delta_{q',2}),
	\end{eqnarray}
	where the two last lines are only present if $q$ or $q'$ are in the initial state. We note that the structure above is in principle not unique and other arrangements can correctly reproduce the same singularity structure, thanks to relations among the $\Jfullb{2}$ integrated dipoles, such as:
	\begin{equation}
		\Jfullb{2}(1,2)=\Jfullb{2}(1,i)+\Jfullb{2}(2,j)-\Jfullb{2}(i,j),\quad i,j\geq 3,
	\end{equation}
	and colour conservation
	\begin{equation}
		\T_i=-\sum_{j\ne i}\T_j.
	\end{equation}
	The configuration of antenna functions in~\eqref{J2barcol2}, in their unintegrated form, correctly subtracts the singular behaviour also at the real-virtual and double-real levels, in infrared limits involving a $q\parallel g$ collinear pair. 
	
	\paragraph{Gluon-gluon integrated dipoles}
	
	The IP gluon-gluon dipoles only require minimal adjustments to the natural arrangement of integrated antenna functions. They are given in Table~\ref{tab:J22IPgg}.
	\begingroup
	\renewcommand{\arraystretch}{1.8} 
	\begin{table}[t]
		\centering\small
		\begin{tabular}{c|l}
			& Integrated dipoles
			\\ \hline
			
			\multirow{5}{*}{FF}  
			& $\J{2}\left(i_g,j_g\right)=\frac{1}{4}\XFFint{F}{4}{0}+\frac{1}{3}\XFFint{F}{3}{1}+\frac{1}{3}\boelite\QQslite{ij}\XFFint{F}{3}{0}-\frac{1}{9}\left[\XFFint{F}{3}{0}\otimes\XFFint{F}{3}{0}\right]$  \\\cline{2-2}
			& $\Jh{2}\left(i_g,j_g\right)=\XFFint{G}{4}{0}+\frac{1}{3}\XhFFint{F}{3}{1}+\XFFint{G}{3}{1}+\frac{1}{3}\bFoelite\QQslite{ij}\XFFint{F}{3}{0}
			+\boelite\QQslite{ij}\XFFint{G}{3}{0}$ \\ 
			& $\phantom{\Jh{2}\left(i_g,j_g\right)}-\frac{2}{3}\left[\XFFint{G}{3}{0}\otimes\XFFint{F}{3}{0}\right]$  \\\cline{2-2}
			& $\Jht{2}\left(i_g,j_g\right)=\frac{1}{2}\XtFFint{G}{4}{0}+\XtFFint{G}{3}{1}$  \\\cline{2-2}
			& $\Jhh{2}\left(i_g,j_g\right)=\frac{1}{2}\XFFint{H}{4}{0}+\XhFFint{G}{3}{1}+\bFoelite\QQslite{ij}\XFFint{G}{3}{0}-\left[\XFFint{G}{3}{0}\otimes\XFFint{G}{3}{0}\right]$  \\\cline{2-2}
			\hline
			
			\multirow{8}{*}{IF}  
			& 
			$\J{2}\left(1_g,i_g\right)=\frac{1}{2}\XIFint{F}{4}{0}{g}+\frac{1}{2}\XIFint{F}{3}{1}{g}+\frac{1}{2}\boelite\QQslite{1i}\XIFint{F}{3}{0}{g}
			-\frac{1}{4}\left[\XIFint{F}{3}{0}{g}\otimes\XIFint{F}{3}{0}{g}\right]$ \\ 
			& $\phantom{\J{2}\left(1_g,i_g\right)}-\frac{1}{2}\Gammatwo{gg}{1}\deltatwo$  \\\cline{2-2}
			& $\Jh{2}\left(1_g,i_g\right)=\XIFint{G}{4}{0}{g}+\frac{1}{2}\XhIFint{F}{3}{1}{g}+\frac{1}{2}\XIFint{G}{3}{1}{g}+\frac{1}{2}\bFoelite\QQslite{1i}\XIFint{F}{3}{0}{g}
			$ \\ 
			& $\phantom{\Jh{2}\left(1_g,i_g\right)}+\frac{1}{2}\boelite\QQslite{1i}\XIFint{G}{3}{0}{g}-\frac{1}{2}\left[\XIFint{G}{3}{0}{g}\otimes\XIFint{F}{3}{0}{g}\right]-\frac{1}{2}\GammatwoF{gg}{1}\deltatwo$ \\ 
			& $\phantom{\Jh{2}\left(1_g,i_g\right)}+\Jhic{2}{\mathrm{f/f}}\left(1_g,i_g\right)$  \\\cline{2-2}
			& $\Jht{2}\left(1_g,i_g\right)=\frac{1}{2}\XtIFint{G}{4}{0}{g}+\frac{1}{2}\XtIFint{G}{3}{1}{g}+\frac{1}{2}\GammatwoFt{gg}{1}\deltatwo
			+\Jhtic{2}{\mathrm{f/f}}\left(1_g,i_g\right)$  \\\cline{2-2}
			& $\Jhh{2}\left(1_g,i_g\right)=\frac{1}{2}\XhIFint{G}{3}{1}{g}+\frac{1}{2}\bFoelite\QQslite{1i}\XIFint{G}{3}{0}{g}
			-\frac{1}{4}\left[\XIFint{G}{3}{0}{g}\otimes\XIFint{G}{3}{0}{g}\right]$ \\ 
			& $\phantom{\Jhh{2}\left(1_g,i_g\right)}-\frac{1}{2}\GammatwoFF{gg}{1}\deltatwo$  \\\cline{2-2}
			\hline
			
			\multirow{6}{*}{II} 
			& $\J{2}\left(1_g,2_g\right)=\XIIint{F}{4}{0}{gg}+\frac{1}{2}\XIIint{F}{4}{0}{gg'}+\XIIint{F}{3}{1}{gg}+\boelite\QQslite{12}\XIIint{F}{3}{0}{gg}$ \\ 
			& $\phantom{\J{2}\left(1_g,2_g\right)}-\left[\XIIint{F}{3}{0}{gg}\otimes\XIIint{F}{3}{0}{gg}\right]-\frac{1}{2}\Gammatwo{gg}{1}\deltatwo-\frac{1}{2}\Gammatwo{gg}{2}\deltaone$  \\\cline{2-2}
			& $\Jh{2}\left(1_g,2_g\right)=\XIIint{G}{4}{0}{gg}+\XhIIint{F}{3}{1}{gg}+\bFoelite\QQslite{12}\XIIint{F}{3}{0}{gg}
			-\frac{1}{2}\GammatwoF{gg}{1}\deltatwo$ \\ 
			& $\phantom{\Jh{2}\left(1_g,2_g\right)}-\frac{1}{2}\GammatwoF{gg}{2}\deltaone
			+\Jhic{2}{\mathrm{f/f}}\left(1_g,2_g\right)$  \\\cline{2-2}
			& $\Jht{2}\left(1_g,2_g\right)=\frac{1}{2}\XtIIint{G}{4}{0}{gg}+\frac{1}{2}\GammatwoFt{gg}{1}\deltatwo+\frac{1}{2}\GammatwoFt{gg}{2}\deltaone
			+\Jhtic{2}{\mathrm{f/f}}\left(1_g,2_g\right)$  \\\cline{2-2}
			& $\Jhh{2}\left(1_g,2_g\right)=-\frac{1}{2}\GammatwoFF{gg}{1}\deltatwo-\frac{1}{2}\GammatwoFF{gg}{2}\deltaone$  \\\cline{2-2}
			\hline
		\end{tabular}
		\caption{Identity-preserving gluon-gluon two-loop colour-stripped integrated dipoles.}
		\label{tab:J22IPgg}
	\end{table}
	\endgroup

	As for the $E$-type IF and II antenna functions, the $G$-type ones $\XIFint{G}{4}{0}{g}$, $\XtIFint{G}{4}{0}{g}$, $\XIIint{G}{4}{0}{gg}$ and $\XtIIint{G}{4}{0}{gg}$ need to have the flip-flopping limits removed. When the final-state quark-antiquark pair becomes collinear to an initial-state gluon, an IC $g \to q \to g$ configuration is generated. The required flip-flopping terms are given in Table~\ref{tab:J22ggff}.

	\begingroup
	\renewcommand{\arraystretch}{1.8} 
	\begin{table}[t]
		\centering\small
		\begin{tabular}{c|l}
			& Integrated dipoles
			\\ \hline
			
			\multirow{2}{*}{IF}  
			& $\Jhic{2}{\mathrm{f/f}}\left(1_g,i_g\right)=\Sgtoq\left[\Gammaoneconv{qg}{1}\otimes\XIFint{G}{3}{0}{q'}\right]+\frac{1}{2}\left[\Gammaoneconv{qg}{1}\otimes\Gammaoneconv{gq}{1}\right]$ \\\cline{2-2}
			& $\Jhtic{2}{\mathrm{f/f}}\left(1_g,i_g\right)=\Sgtoq\left[\Gammaoneconv{qg}{1}\otimes\XIFint{G}{3}{0}{q'}\right]+\frac{1}{2}\left[\Gammaoneconv{qg}{1}\otimes\Gammaoneconv{gq}{1}\right]$  \\\cline{2-2}
			\hline
			
			\multirow{4}{*}{II} 
			& $\Jhic{2}{\mathrm{f/f}}\left(1_g,2_g\right)=\Sgtoq\left[\Gammaoneconv{qg}{1}\otimes\XIIint{G}{3}{0}{qg}\right]+\frac{1}{2}\left[\Gammaoneconv{qg}{1}\otimes\Gammaoneconv{gq}{1}\right]$ \\
			& $\phantom{\Jhic{2}{\mathrm{f/f}}\left(1_g,2_g\right)}+\Sgtoq\left[\Gammaoneconv{qg}{2}\otimes\XIIint{G}{3}{0}{gq}\right]+\frac{1}{2}\left[\Gammaoneconv{qg}{2}\otimes\Gammaoneconv{gq}{2}\right]$  \\\cline{2-2}
			& $\Jhtic{2}{\mathrm{f/f}}\left(1_g,2_g\right)=\Sgtoq\left[\Gammaoneconv{qg}{1}\otimes\XIIint{G}{3}{0}{qg}\right]+\frac{1}{2}\left[\Gammaoneconv{qg}{1}\otimes\Gammaoneconv{gq}{1}\right]$ \\
			& $\phantom{\Jhtic{2}{\mathrm{f/f}}\left(1_g,2_g\right)}+\Sgtoq\left[\Gammaoneconv{qg}{2}\otimes\XIIint{G}{3}{0}{gq}\right]+\frac{1}{2}\left[\Gammaoneconv{qg}{2}\otimes\Gammaoneconv{gq}{2}\right]$  \\\cline{2-2}
			\hline
		\end{tabular}
		\caption{Flip-flopping contributions to identity-preserving gluon-gluon two-loop integrated dipoles.}
		\label{tab:J22ggff}
	\end{table}
	\endgroup
	
	\paragraph{Cancellation of infrared singularities at two-loop}
	 
	 In analogy with~\eqref{J21relation}, the $\e$-poles of IP two-loop colour-stripped integrated dipoles can be related to the to the infrared singularities of two-loop matrix elements: any pole of initial-state collinear origin cancel against two-loop mass factorization kernels, and the remaining singularities satisfy:
	 \begingroup
	 \allowdisplaybreaks
	\begin{eqnarray}
		\label{dipids1}&&\poles\left[N_c\,\Jfull{2}(q,\qb)-\dfrac{\beta_0}{\e}\Jfull{1}(q,\qb)\right]=\nn\\
		&&\hspace{5cm}\poles\left[\text{Re}\left(\ourIop{2}{q\qb}{\e,\mu_r^2}-\dfrac{\beta_0}{\e}\ourIop{1}{q\qb}{\e,\mu_r^2}\right)\right],\\
		\label{dipids2}&&\poles\left[N_c\,\Jfull{2}(g,g)-\dfrac{\beta_0}{\e}\Jfull{1}(g,g)\right]=\nn\\
		&&\hspace{5cm}\poles\left[\text{Re}\left(\ourIop{2}{gg}{\e,\mu_r^2}-\dfrac{\beta_0}{\e}\ourIop{1}{gg}{\e,\mu_r^2}\right)\right],\\
		\label{dipids3}&&\poles\Bigg[\Bigg.N_c\left(\Jfull{2}(q,g)+\Jfull{2}(g,\qb)-2\Jfullb{2}(q,\qb)\right)\nn\\&&\hspace{6cm}-\dfrac{\beta_0}{\e}\left(\Jfull{1}(q,g)+\Jfull{1}(g,\qb)\right)\Bigg.\Bigg]=\nn\\
		&&\poles\left[\text{Re}\left(\ourIop{2}{qg}{\e,\mu_r^2}+\ourIop{2}{g\qb}{\e,\mu_r^2}-\dfrac{\beta_0}{\e}\left(\ourIop{1}{gg}{\e,\mu_r^2}+\ourIop{1}{g\qb}{\e,\mu_r^2}\right)\right)\right].
	\end{eqnarray}
	\endgroup
	For quark-gluon dipoles we observe in the last equation the presence of  $\Jfullb{2}$ to compensate for unphysical poles. We have explicitly verified that the relations above are satisfied for all the partonic and kinematical configurations.

	\begingroup
	\renewcommand{\arraystretch}{1.8} 
	\begin{table}[t]
		\centering\small
		\begin{tabular}{c|l}
			& Integrated dipoles
			\\ \hline
			
			\multirow{8}{*}{IF$_{g\to q}$}
			& $\Jic{2}{g\to q}\left(1_q,i_\qb\right)=-\XIFint{A}{4}{0}{g}-\frac{1}{2}\XIFint{A}{3}{1}{g}-\frac{1}{2}\boelite\QQslite{1i}\XIFint{A}{3}{0}{g}$\\
			& $\hspace{-1cm}\phantom{\Jic{2}{g\to q}\left(1_q,i_\qb\right)}+\frac{1}{2}\left[\XIFint{A}{3}{0}{g}\otimes\XIFint{A}{3}{0}{q}\right]-\Sgtoq\Gammatwo{qg}{1}\deltatwo$\\
			& $\hspace{-1cm}\phantom{\Jic{2}{g\to q}\left(1_q,i_\qb\right)}+\frac{1}{2}\left[\XIFint{A}{3}{0}{g}\otimes\Gammaoneconv{gg}{1}\right]+\frac{1}{2}\Sgtoq\left[\Gammaoneconv{qg}{1}\otimes\Gammaoneconv{gg}{1}\right]$\\
			& $\hspace{-1cm}\phantom{\Jic{2}{g\to q}\left(1_q,i_\qb\right)}-\frac{1}{2}\left[\XIFint{A}{3}{0}{g}\otimes\Gammaoneconv{qq}{1}\right]-\frac{1}{2}\Sgtoq\left[\Gammaoneconv{qg}{1}\otimes\Gammaoneconv{qq}{1}\right]$\\
			& $\Jhic{2}{g\to q}\left(1_q,i_\qb\right)=-\frac{1}{2}\XhIFint{A}{3}{1}{g}-\frac{1}{2}\bFoelite\QQslite{1i}\XIFint{A}{3}{0}{g}-\Sgtoq\GammatwoF{qg}{1}\deltatwo$\\
			& $\hspace{-1cm}\phantom{\Jhic{2}{g\to q}\left(1_q,i_\qb\right)}+\frac{1}{2}\left[\XIFint{A}{3}{0}{g}\otimes\GammaoneFconv{gg}{1}\right]+\frac{1}{2}\Sgtoq\left[\Gammaoneconv{qg}{1}\otimes\GammaoneFconv{gg}{1}\right]$\\
			& $\Jtic{2}{g\to q}\left(1_q,i_\qb\right)=-\XtIFint{A}{4}{0}{g}-\XtIFint{A}{3}{1}{g}+\left[\XIFint{A}{3}{0}{q}\otimes \XIFint{A}{3}{0}{g}\right]-\left[\Gammaoneconv{qq}{1}\otimes\XIFint{A}{3}{0}{g}\right]$\\
			& $\hspace{-1cm}\phantom{\Jtic{2}{g\to q}\left(1_q,i_\qb\right)}-\Sgtoq\left[\Gammaoneconv{qg}{1}\otimes\Gammaoneconv{qq}{1}\right]+2\Sgtoq\Gammatwot{qg}{1}\delta_2$  \\\cline{2-2}
			\hline
			
			\multirow{1}{*}{IF$_{q\to \qb}$}
			& $\Jtic{2}{q\to \qb}\left(1_\qb,i_q\right)=\XIIint{C}{4}{0}{\qb'}+\Gammatwot{q\qb}{1}\deltatwo$  \\\cline{2-2}
			\hline
			
			\multirow{2}{*}{IF$_{q\to g\to q}$}
			& $\Jic{2}{q\to g\to q}\left(1_q,i_\qb\right)=\XIFint{B}{4}{0}{q'}+\Sqtog\left[\Gammaoneconv{gq}{1}\otimes\XIFint{A}{3}{0}{g}\right]$\\
			& $\hspace{-1cm}\phantom{\Jic{2}{q\to g\to q}\left(1_q,i_g\right)}+\left[\Gammaoneconv{gq}{1}\otimes\Gammaoneconv{qg}{1}\right]-2\Gammatwot{qq}{1}\delta_2$  \\\cline{2-2}
			\hline
			\end{tabular}
			\caption{Initial-final identity-changing quark-antiquark two-loop colour-stripped integrated dipoles.}
			\label{tab:J22ICqqbIF}
		\end{table}
	\endgroup

	\begingroup
	\renewcommand{\arraystretch}{1.8} 
	\begin{table}[t]
		\centering\small
		\begin{tabular}{c|l}
			& Integrated dipoles
			\\ \hline
			
			\multirow{9}{*}{II$_{g\to q}$}
			& $\Jic{2}{g\to q}\left(1_q,2_\qb\right)=-\XIIint{A}{4}{0}{gq}-\XIIint{A}{4}{0}{g'q}-\XIIint{A}{3}{1}{gq}-\boelite\QQslite{12}\XIIint{A}{3}{0}{gq}$\\
			& $\hspace{-1cm}\phantom{\Jic{2}{g\to q}\left(1_q,2_\qb\right)}+\left[\XIIint{A}{3}{0}{gq}\otimes\XIIint{A}{3}{0}{q\qb}\right]-\Sgtoq\Gammatwo{qg}{1}\deltatwo$\\
			& $\hspace{-1cm}\phantom{\Jic{2}{g\to q}\left(1_q,2_\qb\right)}+\left[\XIIint{A}{3}{0}{gq}\otimes\Gammaoneconv{gg}{1}\right]+\frac{1}{2}\Sgtoq\left[\Gammaoneconv{qg}{1}\otimes\Gammaoneconv{gg}{1}\right]$\\
			& $\hspace{-1cm}\phantom{\Jic{2}{g\to q}\left(1_q,2_\qb\right)}-\left[\XIIint{A}{3}{0}{gq}\otimes\Gammaoneconv{qq}{1}\right]-\frac{1}{2}\Sgtoq\left[\Gammaoneconv{qg}{1}\otimes\Gammaoneconv{qq}{1}\right]$\\
			& $\Jhic{2}{g\to q}\left(1_q,2_\qb\right)=-\frac{1}{2}\XhIIint{A}{3}{1}{gq}-\bFoelite\QQslite{12}\XIIint{A}{3}{0}{gq}-\Sgtoq\GammatwoF{qg}{1}\deltatwo$\\
			& $\hspace{-1cm}\phantom{\Jic{2}{g\to q}\left(1_q,2_\qb\right)}+\left[\XIIint{A}{3}{0}{gq}\otimes\GammaoneFconv{gg}{1}\right]+\frac{1}{2}\Sgtoq\left[\Gammaoneconv{qg}{1}\otimes\GammaoneFconv{gg}{1}\right]$\\
			& $\Jtic{2}{g\to q}\left(1_q,2_\qb\right)=-\XtIIint{A}{4}{0}{gq}-\XtIIint{A}{3}{1}{gq}$\\
			& $\hspace{-1cm}\phantom{\Jtic{2}{g\to q}\left(1_q,2_\qb\right)}+\left[\XIIint{A}{3}{0}{qq}\otimes \XIIint{A}{3}{0}{gq}\right]-\left[\Gammaoneconv{qq}{1}\otimes\XIIint{A}{3}{0}{gq}\right]$\\
			& $\hspace{-1cm}\phantom{\Jtic{2}{g\to q}\left(1_q,2_\qb\right)}-\frac{1}{2}\Sgtoq\left[\Gammaoneconv{qg}{1}\otimes\GammaoneFconv{qq}{1}\right]+\Sgtoq\Gammatwot{qg}{1}\delta_2$  \\\cline{2-2}
			\hline
			
			\multirow{2}{*}{II$_{q\to q'}$}
			& $\Jic{2}{q\to q'}\left(1_{q'},2_\qb\right)=\XIIint{B}{4}{0}{q'q}+\Sqtog\left[\Gammaoneconv{gq}{1}\otimes \XIIint{A}{3}{0}{gq}\right]$\\
			& $\hspace{-1cm}\phantom{\Jic{2}{q\to q'}\left(1_{q'},2_\qb\right)}+\frac{1}{2}\left[\Gammaoneconv{gq}{1}\otimes\Gammaoneconv{qg}{1}\right]-\Gammatwo{qq'}{x_1}\deltatwo$  \\\cline{2-2}
			\hline
			
			\multirow{1}{*}{II$_{q\to \qb}$}
			& $	\Jtic{2}{q\to \qb}\left(1_\qb,2_q\right)=\XIIint{C}{4}{0}{\qb'\qb}+\Gammatwot{q\qb}{1}\deltatwo$  \\\cline{2-2}
			\hline
			
			\multirow{2}{*}{II$_{q\to g\to q}$}
			& $\Jic{2}{q\to g\to q}\left(1_q,2_\qb\right)=\XIIint{B}{4}{0}{q'q}+\Sqtog\left[\Gammaoneconv{gq}{1}\otimes\XIIint{A}{3}{0}{gq}\right]$\\
			& $\hspace{-1cm}\phantom{\Jic{2}{q\to g\to q}\left(1_q,2_\qb\right)}+\frac{1}{2}\left[\Gammaoneconv{gq}{1}\otimes\Gammaoneconv{qg}{1}\right]-\Gammatwo{qq'}{x_1}\delta_2$  \\\cline{2-2}
			\hline
			
			\multirow{4}{*}{II$_{gg\to q\qb}$}
			& $\Jic{2}{gg\to q\qb}\left(1_q,2_\qb\right)=\XIIint{A}{4}{0}{gg}+\Sgtoq\left[\Gammaoneconv{qg}{1}\otimes \XIIint{A}{3}{0}{qg}\right]$\\
			& $\hspace{-1cm}\phantom{\Jic{2}{gg\to q\qb}\left(1_q,2_\qb\right)}+\Sgtoq\left[\Gammaoneconv{qg}{2}\otimes \XIIint{A}{3}{0}{gq}\right]+\Sgtoq^2\Gammaone{qg}{x_1}\Gammaone{qg}{x_2}$\\
			& $\Jtic{2}{gg\to q\qb}\left(1_q,2_\qb\right)=\XtIIint{A}{4}{0}{gg}+2\Sgtoq\left[\Gammaoneconv{gq}{1}\otimes \XIIint{A}{3}{0}{qg}\right]$\\
			& $\hspace{-1cm}\phantom{\Jtic{2}{gg\to q\qb}\left(1_q,2_\qb\right)}+2\Sgtoq\left[\Gammaoneconv{qg}{2}\otimes \XIIint{A}{3}{0}{gq}\right]+2\Sgtoq^2\Gammaone{qg}{x_1}\Gammaone{qg}{x_2}$  \\\cline{2-2}
			\hline
		\end{tabular}
		\caption{Initial-initial identity-changing quark-antiquark two-loop colour-stripped integrated dipoles.}
		\label{tab:J22ICqqbII}
	\end{table}
	\endgroup

	\paragraph{Identity-changing dipoles}
	
	Identity-changing integrated dipoles, as for the one-loop case, are not treated in colour space. We have two possible structures, one is associated to configurations where at least one initial state parton actually changes identity, while the other addresses flip-flopping contributions for which, after integration, the species of the initial-state partons is formally the same as before. The first type reads:
	\begin{eqnarray}\label{J22ic}
		\Jfulltotic{2}{ab;cd}(x_1,x_2)&=&\Jfullic{2}{a\to c}(x_1)\delta_{db}\delta(1-x_2)+\Jfullic{2}{b\to d}(x_2)\delta_{ca}\delta(1-x_1)\nn \\ 
		&+&\Jfullic{2}{ab\to cd}(x_1,x_2),
	\end{eqnarray}
	where either a single initial-state leg changes identity (first two terms) or both legs change identity (last term). The flip-flopping integrated dipoles are given by:
	\begin{eqnarray}\label{J22icff}
		\Jfulltotic{2}{ab;cd;ab}(x_1,x_2)&=&\Jfullic{2}{a\to c\to a}(x_1)\delta_{db}\delta(1-x_2)+\Jfullic{2}{b\to d\to b}(x_2)\delta_{ca}\delta(1-x_1).
	\end{eqnarray}
	The integrated dipoles in~\eqref{J22ic} and~\eqref{J22icff} have a colour decomposition which mirrors the one of two-loop IC splitting kernels:
	\begingroup
	\allowdisplaybreaks
	\begin{eqnarray}
		\Jfullic{2}{q\to g}(g,i)&=&\left(\dfrac{N_c^2-1}{N_c}\right)\Big[\Big.N_c\Jic{2}{q\to g}(g,i)+\dfrac{1}{N_c}\Jtic{2}{q\to g}(g,i) \nn\\
		&&\hspace{5.5cm}+N_f\Jhic{2}{q\to g}(g,i)\Big.\Big]\,, \\
		\Jfullic{2}{g\to q}(q,i)&=&N_c\Jic{2}{g\to q}(q,i)+\dfrac{1}{N_c}\Jtic{2}{g\to q}(q,i)+N_f\Jhic{2}{g\to q}(q,i)\,, \\
		\Jfullic{2}{q\to \qb}(\qb,i)&=&\left(\dfrac{N_c^2-1}{N_c}\right)\left[\Jic{2}{q\to \qb}(\qb,i)+\dfrac{1}{N_c}\Jtic{2}{q\to \qb}(\qb,i)\right]\,, \\
		\Jfullic{2}{q\to q'}(q',i)&=&\left(\dfrac{N_c^2-1}{N_c}\right)\Jic{2}{q\to q'}(q',i)\,, \\
		\Jfullic{2}{gg\to q\qb}(q,\qb)&=&N_c\,\Jic{2}{gg\to q\qb}(q,\qb)+\dfrac{1}{N_c}\Jtic{2}{gg\to q\qb}(q,\qb)\,, \\
		\Jfullic{2}{qq'\to gg}(g,g)&=&\left(\dfrac{N_c^2-1}{N_c}\right)\Jic{2}{q\qb\to gg}(g,g)\,, \\
		\Jfullic{2}{q'g\to gq}(g,q)&=&N_c\,\Jic{2}{q'g\to gq}(g,q)+\dfrac{1}{N_c}\Jtic{2}{q'g\to gq}(g,q)\,, \\
		\Jfullic{2}{q\to g\to q}(q,i)&=&\left(\dfrac{N_c^2-1}{N_c}\right)\Jic{2}{q\to g\to q}(q,i)\,.
	\end{eqnarray}
	\endgroup
	where $i=q,g$. For configurations where only one initial-state parton changes identity, the second hard parton in the dipole acts as a mere spectator and no unresolved limits are associated to it. This means that in any practical implementation one has the freedom to choose different spectators for the same IC limits. In particular, in any IC unresolved configuration resulting in a hard quark, such as $g\to q$, $q\to q'$, $q\to \qb$ and $q\to g \to q$, it is always possible to choose the fermionic partner of the hard quark as a spectator. For this reason, there is no need to define IC integrated dipoles for these configurations with a hard gluon as spectator. In the following we present the expressions for the two-loop IC integrated dipoles. We note that typically for a specific process only some of the listed dipoles are actually needed. However, we provide expressions for all of them for the sake of generality. The IC quark-antiquark, quark-gluon and gluon-gluon two-loop integrated dipoles are listed in Tables~\ref{tab:J22ICqqbIF}--\ref{tab:J22ICgg}. 

	In the $\Jic{2}{q\to g}(1_g,i_q)$ and in the $\Jic{2}{q'g\to gq}(1_g,2_q)$ integrated dipoles of Table~\ref{tab:J22ICqg}, corrective terms had to be included, as one can notice from the appearance of quark-antiquark integrated antenna functions. The origin of such terms is completely analogous to the ones of Table~\ref{tab:J22corr}: they are needed to remove the triple collinear $q'\parallel \qb' \parallel q$ limit present in the integrated E-type four-parton antenna functions. Indeed, such contributions yields a $q\to g\to q$ identity-changing limit which is not compatible with the considered configurations. 

\begingroup
\renewcommand{\arraystretch}{1.8} 
\begin{table}[t]
	\centering\small
	\begin{tabular}{c|l}
		& Integrated dipoles
		\\ \hline
		
		\multirow{8}{*}{IF$_{q\to g}$}
		& $\Jic{2}{q\to g}(1_g,i_{q})=-\XIFint{E}{4}{0}{q'}-\XIFint{E}{4}{0}{\qb'}-\XIFint{E}{3}{1}{q'}-\boelite\QQslite{1i}\XIFint{E}{3}{0}{q'}$\\
		& $\hspace{-1cm}\phantom{\Jic{2}{q\to g}(1_g,i_{q})}+2\left[\XIFint{D}{3}{0}{g}\otimes\XIFint{E}{3}{0}{q'}\right]+\left[\Gammaoneconv{qq}{1}\otimes\XIFint{E}{3}{0}{q'}\right]$\\
		& $\hspace{-1cm}\phantom{\Jic{2}{q\to g}(1_g,i_{q})}-\left[\Gammaoneconv{gg}{1}\otimes\XIFint{E}{3}{0}{q'}\right]+\frac{1}{2}\Sqtog\left[\Gammaoneconv{qq}{1}\otimes\Gammaoneconv{gq}{1}\right]$\\
		& $\hspace{-1cm}\phantom{\Jic{2}{q\to g}(1_g,i_{q})}-\frac{1}{2}\Sqtog\left[\Gammaoneconv{gg}{1}\otimes\Gammaoneconv{gq}{1}\right]-\Sqtog\Gammatwo{gq}{1}\deltatwo$\\
		& $\hspace{-1cm}\phantom{\Jtic{2}{q\to g}(1_g,i_{q})}+\XIFint{B}{4}{0}{q'}+2\Sgtoq\left[\Gammaoneconv{qg}{1}\otimes\XIFint{E}{3}{0}{q'}\right]$\\
		& $\hspace{-1cm}\phantom{\Jtic{2}{q\to g}(1_g,i_{q})}+\Sqtog\left[\Gammaoneconv{gq}{1}\otimes\XIFint{A}{3}{0}{g}\right]+2\left[\Gammaoneconv{gq}{1}\otimes\Gammaoneconv{qg}{1}\right]$ \\
		& $\Jhic{2}{q\to g}(1_g,i_{q})=-\XhIFint{E}{3}{1}{q'}-\bFoelite\QQslite{1i}\XIFint{E}{3}{0}{q'}-\left[\GammaoneFconv{gg}{1}\otimes\XIFint{E}{3}{0}{q'}\right]$\\
		& $\hspace{-1cm}\phantom{\Jhic{2}{q\to g}(1_g,i_{q})}-\frac{1}{2}\Sqtog\left[\GammaoneFconv{gg}{1}\otimes\Gammaoneconv{gq}{1}\right]-\Sqtog\GammatwoF{gq}{1}\deltatwo$\\
		& $\Jtic{2}{q\to g}(1_g,i_{q})=-\XtIFint{E}{4}{0}{q'}-\XtIFint{E}{3}{1}{q'}+\left[\Gammaoneconv{qq}{1}\otimes\XIFint{E}{3}{0}{q'}\right]$\\
		& $\hspace{-1cm}\phantom{\Jtic{2}{q\to g}(1_g,i_{q})}+\frac{1}{2}\Sqtog\left[\Gammaoneconv{qq}{1}\otimes\Gammaoneconv{gq}{1}\right]+\Sqtog\Gammatwot{gq}{1}\deltatwo$  \\\cline{2-2}
		\hline
		
		\multirow{10}{*}{II$_{q\to g}$}
		& $\Jic{2}{q\to g}(1_g,2_{q})=-\XIIint{E}{4}{0}{q'q}-\XIIint{E}{4}{0}{\qb'q}-\XIIint{E}{3}{1}{q'q}-\boelite\QQslite{12}\XIIint{E}{3}{0}{q'q}$\\
		& $\hspace{-1cm}\phantom{\Jic{2}{q\to g}(1_g,2_{q})}+2\left[\XIIint{D}{3}{0}{gq}\otimes\XIIint{E}{3}{0}{q'q}\right]+\left[\Gammaoneconv{qq}{1}\otimes\XIIint{E}{3}{0}{q'q}\right]$\\
		& $\hspace{-1cm}\phantom{\Jic{2}{q\to g}(1_g,i_{q})}-\left[\Gammaoneconv{gg}{1}\otimes\XIIint{E}{3}{0}{q'q}\right]+\frac{1}{2}\Sqtog\left[\Gammaoneconv{qq}{1}\otimes\Gammaoneconv{gq}{1}\right]$\\
		& $\hspace{-1cm}\phantom{\Jic{2}{q\to g}(1_g,i_{q})}-\frac{1}{2}\Sqtog\left[\Gammaoneconv{gg}{1}\otimes\Gammaoneconv{gq}{1}\right]$\\
		& $\hspace{-1cm}\phantom{\Jic{2}{q\to g}(1_g,2_{q})}-\Sqtog\Gammatwo{gq}{1}\deltatwo$\\
		& $\Jhic{2}{q\to g}(1_g,2_{q})=-\XhIIint{E}{3}{1}{q'q}-\bFoelite\QQslite{12}\XIIint{E}{3}{0}{q'q}$\\
		& $\hspace{-1cm}\phantom{\Jic{2}{q\to g}(1_g,i_{q})}-\left[\GammaoneFconv{gg}{1}\otimes\XIIint{E}{3}{0}{q'q}\right]$\\
		& $\hspace{-1cm}\phantom{\Jic{2}{q\to g}(1_g,i_{q})}-\frac{1}{2}\Sqtog\left[\GammaoneFconv{gg}{1}\otimes\Gammaoneconv{gq}{1}\right]-\Sqtog\GammatwoF{gq}{1}\deltatwo$\\
		& $\Jtic{2}{q\to g}(1_g,2_{q})=-\XtIIint{E}{4}{0}{q'q}-\XtIIint{E}{3}{1}{q'q}+\left[\Gammaoneconv{qq}{1}\otimes\XIIint{E}{3}{0}{q'q}\right]$\\
		& $\hspace{-1cm}\phantom{\Jtic{2}{q\to g}(1_g,2_{q})}+\frac{1}{2}\Sqtog\left[\Gammaoneconv{qq}{1}\otimes\Gammaoneconv{gq}{1}\right]+\Sqtog\Gammatwot{gq}{1}\deltatwo$  \\\cline{2-2}
		\hline
		
		\multirow{2}{*}{II$_{q'g\to gq}$}
		& $\Jic{2}{q'g\to gq}(1_g,2_q)=2\XIIint{E}{4}{0}{\qb'g}+2\Sqtog\left[\Gammaoneconv{gq}{1}\otimes \XIIint{D}{3}{0}{gg}\right]$\\
		& $\hspace{-1cm}\phantom{\Jic{2}{q'g\to gq}(1_g,2_q)}+2\Sgtoq\left[\Gammaoneconv{gq}{2}\otimes \XIIint{E}{3}{0}{q'q}\right]+\Gammaone{gq}{x_1}\Gammaone{qg}{x_2}$\\
		& $\hspace{-1cm}\phantom{\Jic{2}{q'g\to gq}(1_g,2_q)}-\XIFint{B}{4}{0}{q'}+\left[\XIFint{E}{3}{0}{q'}\otimes \XIFint{A}{3}{0}{g}\right]+2\Sgtoq\left[\Gammaoneconv{qg}{1}\otimes \XIFint{E}{3}{0}{q'}\right]$\\
		& $\hspace{-1cm}\phantom{\Jic{2}{q'g\to gq}(1_g,2_q)}+2\left[\Gammaoneconv{qg}{1}\otimes\Gammaoneconv{gq}{1}\right]$\\\cline{2-2}
		\hline
	\end{tabular}
	\caption{Identity-changing quark-gluon two-loop colour-stripped integrated dipoles.}
	\label{tab:J22ICqg}
\end{table}
\endgroup

\begingroup
\renewcommand{\arraystretch}{1.8} 
\begin{table}[t]
	\centering\small
	\begin{tabular}{c|l}
		& Integrated dipoles
		\\ \hline
		
		\multirow{10}{*}{IF$_{q\to g}$}
		& $\Jic{2}{q\to g}\left(1_g,i_g\right)=-\XIFint{G}{4}{0}{q}-\XIFint{G}{3}{1}{q}-\boelite\QQslite{1i}\XIFint{G}{3}{0}{q}$\\
		& $\hspace{-1cm}\phantom{\Jic{2}{q\to g}\left(1_g,i_g\right)}+\left[\XIFint{G}{3}{0}{q}\otimes\XIFint{F}{3}{0}{g}\right]-\Sqtog\Gammatwo{gq}{1}\deltatwo$\\
		& $\hspace{-1cm}\phantom{\Jic{2}{q\to g}\left(1_g,i_g\right)}+\left[\Gammaoneconv{qq}{1}\otimes\XIFint{G}{3}{0}{q}\right]-\left[\Gammaoneconv{gg}{1}\otimes\XIFint{G}{3}{0}{q}\right]$\\
		& $\hspace{-1cm}\phantom{\Jic{2}{q\to g}\left(1_g,i_g\right)}+\frac{1}{2}\Sqtog\left[\Gammaoneconv{qq}{1}\otimes\Gammaoneconv{gq}{1}\right]$\\
		& $\hspace{-1cm}\phantom{\Jic{2}{q\to g}\left(1_g,i_g\right)}-\frac{1}{2}\Sqtog\left[\Gammaoneconv{gg}{1}\otimes\Gammaoneconv{gq}{1}\right]$\\
		& $\Jhic{2}{q\to g}(1_g,i_g)=\XIFint{H}{4}{0}{q}-\XhIFint{G}{3}{1}{q}-\bFoelite\QQslite{1i}\XIFint{G}{3}{0}{q}$\\
		& $\hspace{-1cm}\phantom{\Jhic{2}{q\to g}(1_g,i_g)}+\Sqtog\left[\Gammaoneconv{gq}{1}\otimes\XIFint{G}{3}{0}{g}\right]+\frac{1}{2}\Sqtog\left[\Gammaoneconv{gq}{1}\otimes\GammaoneF{gg}{x_1}\right]$\\
		& $\hspace{-1cm}\phantom{\Jhic{2}{q\to g}(1_g,i_g)}-\Sqtog\GammatwoF{gq}{1}\deltatwo$\\
		& $\Jtic{2}{q\to g}\left(1_g,i_g\right)=-\frac{1}{2}\XtIFint{G}{4}{0}{q}-\XtIFint{G}{3}{1}{q}+\left[\XIFint{G}{3}{0}{q}\otimes\Gammaoneconv{qq}{1}\right]$\\
		& $\hspace{-1cm}\phantom{\Jtic{2}{q\to g}\left(1_g,i_g\right)}+\frac{1}{2}\Sqtog\left[\Gammaoneconv{qq}{1}\otimes\Gammaoneconv{gq}{1}\right]+\Sqtog\Gammatwot{gq}{1}\deltatwo$  \\\cline{2-2}
		\hline
		
		\multirow{10}{*}{II$_{q\to g}$}
		& $\Jic{2}{q\to g}\left(1_g,2_g\right)=-\XIIint{G}{4}{0}{qg}-\XIIint{G}{4}{0}{qg'}-\XIIint{G}{3}{1}{qg}-\boelite\QQslite{12}\XIIint{G}{3}{0}{qg}$\\
		& $\hspace{-1cm}\phantom{\Jic{2}{q\to g}\left(1_g,2_g\right)}+2\left[\XIIint{G}{3}{0}{qg}\otimes\XIIint{F}{3}{0}{gg}\right]-\Sqtog\Gammatwo{gq}{1}\deltatwo$\\
		& $\hspace{-1cm}\phantom{\Jic{2}{q\to g}\left(1_g,2_g\right)}-\Sgtoq\left[\Gammaoneconv{qg}{2}\otimes\XIIint{G}{3}{0}{qq}\right]+\left[\Gammaoneconv{qq}{1}\otimes\XIIint{G}{3}{0}{qg}\right]$\\
		& $\hspace{-1cm}\phantom{\Jic{2}{q\to g}\left(1_g,2_g\right)}-\left[\Gammaoneconv{gg}{1}\otimes\XIIint{G}{3}{0}{qg}\right]+\frac{1}{2}\Sqtog\left[\Gammaoneconv{qq}{1}\otimes\Gammaoneconv{gq}{1}\right]$\\
		& $\hspace{-1cm}\phantom{\Jic{2}{q\to g}\left(1_g,2_g\right)}-\frac{1}{2}\Sqtog\left[\Gammaoneconv{gg}{1}\otimes\Gammaoneconv{gq}{1}\right]$\\
		& $\Jhic{2}{q\to g}\left(1_g,2_g\right)=-\XhIIint{G}{3}{1}{qg}-\bFoelite\QQslite{12}\XIIint{G}{3}{0}{qg}-\Sqtog\GammatwoF{gq}{1}\deltatwo$\\
		& $\hspace{-1cm}\phantom{\Jhic{2}{q\to g}\left(1_g,2_g\right)}-\left[\XIIint{G}{3}{0}{qg}\otimes\GammaoneFconv{gg}{1}\right]-\frac{1}{2}\Sqtog\left[\Gammaoneconv{gq}{1}\otimes\GammaoneFconv{gg}{1}\right]$\\
		& $\Jtic{2}{q\to g}\left(1_g,2_g\right)=-\XtIIint{G}{4}{0}{qg}-\XtIIint{G}{3}{1}{qg}+\left[\XIIint{G}{3}{0}{qg}\otimes\Gammaoneconv{qq}{1}\right]$\\
		& $\hspace{-1cm}\phantom{\Jtic{2}{q\to g}\left(1_g,2_g\right)}-\Sgtoq\left[\Gammaoneconv{qg}{2}\otimes\XIIint{G}{3}{0}{qq}\right]$  \\
		& $\hspace{-1cm}\phantom{\Jtic{2}{q\to g}\left(1_g,2_g\right)}+\frac{1}{2}\Sqtog\left[\Gammaoneconv{qq}{1}\otimes\Gammaoneconv{gq}{1}\right]+\Sqtog\Gammatwot{gq}{1}\deltatwo$  \\\cline{2-2}
		\hline
		
		\multirow{2}{*}{II$_{q\qb \to gg}$}
		& $\Jic{2}{q\qb\to gg}\left(1_g,2_g\right)=\XIIint{H}{4}{0}{qq'}+\Sqtog\left[\Gammaoneconv{gq}{2}\otimes \XIIint{G}{3}{0}{qg}\right]$\\
		& $\hspace{-1cm}\phantom{\Jic{2}{q\to g}(1_g,i_{q})}+\Sqtog\left[\Gammaoneconv{gq}{1}\otimes \XIIint{G}{3}{0}{gq}\right]+\Sqtog^2\Gammaone{gq}{x_1}\Gammaone{gq}{x_2}$  \\\cline{2-2}
		\hline
	\end{tabular}
	\caption{Identity-changing gluon-gluon two-loop colour-stripped integrated dipoles.}
	\label{tab:J22ICgg}
\end{table}
\endgroup
\FloatBarrier

Contrary to the one-loop IC dipoles, two-loop ones are not free from singularities. This is understood considering that, with two unresolved partons, along with initial-state collinear singularities which cancel against the mass factorization kernels, additional soft or final-state collinear poles can be present. As we will see in more detail in Section~\ref{sec:subNNLO}, the singularities of the IC two-loop integrated dipoles cancel against explicit poles in other structures present in the two-loop mass factorization counterterm for identity-changing splittings. This cancellation yields an IC double-virtual subtraction term which is correctly free from explicit singularities.
		
	
	\section{Colourful antenna subtraction at NLO}\label{sec:subNLO}
	
	In the following, we describe the colourful antenna subtraction approach at NLO, aiming to introduce the underlying concepts and preparing for the 
	full NNLO formulation.

    The NLO correction to a partonic sub-process initiated by partons $a$ and $b$ reads as follow. 
	\begin{equation}\label{NLOcssub2}
		\dd\sigpart{}{ab,\mathrm{NLO}}=\int_{n}\left[\dd\sigpart{V}{ab,\mathrm{NLO}}-\dsigTNLO{ab}\right]+\int_{n+1}\left[\dd\sigpart{R}{ab,\mathrm{NLO}}-\dsigSNLO{ab}\right],
	\end{equation}
	with 
	\begin{equation}\label{sigTNLO2}
		\dsigTNLO{ab}=-\int_1\dsigSNLO{ab}-\dsigMFNLO{ab}.
 	\end{equation}
	Typically, the construction of the subtraction terms would begin by analysing the divergent behaviour of the real emission correction and by individuating the proper combination of antenna functions and reduced LO squared amplitudes which reproduces the implicit infrared singularities. Hence, the real subtraction term $\dsigSNLO{ab}$, up to an overall process-dependent normalisation, has the following form~\cite{Currie:2013vh}:
    \begin{eqnarray}\label{sigSexample}
    	\dsigSNLO{ab}&\sim&\sum_{c,c'}\sum_{i,j,k}\dphi{n+1}(p_3,\dots,p_{n+3};p_1,p_2)\,\nonumber\\
    	&&\hspace{1.5cm}\times\,X_3^0(i,k,j)\,a_{n+2}^0(c,c';\{.,\wt{ik},.,\wt{kj},.\})\jet{n}{n}{.,p_{\wt{ik}},.,p_{\wt{kj}},.}.
    \end{eqnarray}
    In the notation of Section~\ref{sec:col_space}, $a_{n+2}^0(c,c',\{\dots\})$ indicates a generic $(n+2)$-parton interference between colour ordering $c$ and $c'$. The first sum runs over the possible colour structures and orderings, while the inner sum covers permutations of external momenta. In~\eqref{sigSexample}, as well as analogous expressions in the remainder of the paper, each term in the sum can come with a different numerical factor in front of it. Such factors are not generalizable and not particularly relevant for the description of the structure of the subtraction terms. The symbol $\sim$ indicates their omissions.
    
    An analytical integration of~\eqref{sigSexample} over the phase space of the unresolved radiation (antenna phase space) yields:
    \begin{eqnarray}
    	\int_1\dsigSNLO{ab}&\sim&\sum_{c,c'}\sum_{i,j,k}\dphi{n}(.,p_{\wt{ik}},.,p_{\wt{kj}},.;p_1,p_2)\,\nonumber\\
    	&&\hspace{1.1cm}\times\,\mathcal{X}_3^0(s_{(\wt{ik})(\wt{kj})})\,a_{n+2}^0(c,c';\{.,\wt{ik},.,\wt{kj},.\})\jet{n}{n}{.,p_{\wt{ik}},.,p_{\wt{kj}},.}\,,
    \end{eqnarray}
    where the indices $\wt{ik}$ and $\wt{jk}$ are the mapped momenta that also feature in the unintegrated subtraction term~\eqref{sigSexample}. 
    Finally, the full virtual subtraction term \eqref{sigTNLO} is obtained by adding all mass factorization counterterms appropriate for the incoming partons.
    
   In the context of the antenna subtraction method, the construction of real-emission subtraction terms for a generic process may become cumbersome and require a considerable amount of process-specific work. In particular, beyond leading-colour, one cannot directly rely on colour connections of squared amplitudes as the guiding principle for the construction of the subtraction terms, especially at high multiplicities.     
    The colourful antenna subtraction approach is designed to systematise the construction of the real-emission subtraction term. The main idea behind it consists of exploiting the predictability of the singularity structure of virtual amplitudes in colour space to straightforwardly construct the virtual subtraction term in a completely general way. Subsequently, the real subtraction term is derived with a systematic procedure exploiting the correspondence between integrated and unintegrated antenna functions. To properly introduce how this is done we first need to recast the NLO mass factorization counterterm in colour space.

	\subsection{NLO mass factorization in colour space}\label{sec:MFNLOcol}
	
	We notice that equation~\eqref{MF} has both identity-preserving and identity-changing contributions, respectively when $(c,d)=(a,b)$ and $(c,d)\ne(a,b)$. For a given LO partonic matrix element, initiated by partons $a$ and $b$, the virtual infrared singularities in \eqref{Vpoles} and the identity-preserving contributions in \eqref{MF} factorize upon the same set of LO partonic colour correlations. Nevertheless, it is possible to express the identity-preserving mass factorization kernels in colour space, in analogy to  the singularity structure of one-loop amplitudes. We define:
	\begin{equation}
		\BGammaone{ab;ab}{x_1,x_2}=\BGammaone{aa,\text{full}}{x_1}\delta(1-x_2)+\BGammaone{bb,\text{full}}{x_2}\delta(1-x_1),
	\end{equation}
	where, as usual, bold symbols are used to indicate that the splitting kernels have been promoted to be operators in colour space:
	\begin{equation}
		\BGammaone{cc,\text{full}}{x_i}=-\Gammaone{cc,\text{full}}{x_i}\dfrac{1}{C_c}\sum_{j\ne i}\T_i\cdot\T_j,\quad c=a(b),\quad i=1(2).
	\end{equation}
	This colour operator is proportional to the identity in colour space when it acts on a colour-singlet vector, due to colour conservation:
		\begin{equation}
		\BGammaone{ab;ab}{x_1,x_2}\ket{\ampnum{n+2}{\ell}}=\Gammaone{ab;ab}{x_1,x_2}\ket{\ampnum{n+2}{\ell}},
	\end{equation}
	which restores the original result. We can then rewrite the identity-preserving (IP) sector of the mass factorization counterterms in~\eqref{MF} as:
	\begin{eqnarray}\label{MFIP}
		\dd\sigpart{MF,\mathrm{IP}}{ab,\mathrm{NLO}}&=&-\coeffVNLO\int\dr{x_1}\dr{x_2}\int\dphi{n}(p_3,\dots,p_{n+2};x_1 p_1,x_2 p_2)\,\jet{n}{n}{\lb p \rb_n}\nonumber\\
		&&\times\braket{\ampnum{n+2}{0}|\BGammaone{ab;ab}{x_1,x_2}|\ampnum{n+2}{0}}\,,
	\end{eqnarray}
 which is analogous to \eqref{Vpoles}.
	
	While it is possible to apply a similar treatment for the identity-changing mass factorization kernels in principle, it is not particularly convenient to rewrite them as operators in colour space. Hence, we choose to maintain the identity-changing mass factorization counterterm as colour scalars:
	\begin{equation}\label{MFIC}
		\dd\sigpart{MF,\mathrm{IC}}{ab,\mathrm{NLO}}=-\coeff\sum_{(c,d)\neq(a,b)}\int\dr{x_1}\dr{x_2}\Gamma^{(1)}_{ab;cd}(x_1,x_2)\,\dd\sigpart{}{cd,\mathrm{LO}}\,.
	\end{equation}
	
	\subsection{NLO virtual subtraction term}\label{sec:NLOV} 
	
	The virtual subtraction term at NLO, $\sigpart{T}{ab,\mathrm{NLO}}$ can be separated into IP and IC components:
	\begin{equation}
		\sigpart{T}{ab,\mathrm{NLO}}=\sigpart{T,\mathrm{IP}}{ab,\mathrm{NLO}}+\sigpart{T,\mathrm{IC}}{ab,\mathrm{NLO}}\,.
	\end{equation} 
	The IP component has to:
	\begin{itemize}
		\item remove the explicit poles of the virtual matrix element;
		\item contain the IP mass factorization counterterm $\dd\sigpart{MF,\mathrm{IP}}{ab,\mathrm{NLO}}$.
	\end{itemize}
	We know that the first requirement can be addressed with~\eqref{Vpoles}, while $\dd\sigpart{MF,\mathrm{IP}}{ab,\mathrm{NLO}}$ is given above in~\eqref{MFIP}. By~\eqref{J21}, \eqref{J21relation} and the explicit formulae given in Section~\ref{sec:int_dip_1}.
	One can see that the one-loop integrated dipoles that were introduced in  Section~\ref{sec:int_dip_1} address both tasks, yielding the 
	construction of the  NLO IP virtual subtraction term as
	\begin{eqnarray}\label{sigTIP}
		\dd\sigpart{T,\mathrm{IP}}{ab,\mathrm{NLO}}&=&\coeffVNLO\int\dr{x_1}\dr{x_2}\dphi{n}(p_3,\dots,p_{n+2};x_1 p_1,x_2 p_2)\jet{n}{n}{\lb p \rb_n}\nonumber\\
		&&\times 2\braket{\ampnum{n+2}{0}|\Jcol{1}(\e)|\ampnum{n+2}{0}}\,.
	\end{eqnarray}
	The equation above explains the important role of the integrated dipoles and justifies their definition. First of all, we can describe in a general way the singularity structure of one-loop matrix elements, naturally including also the IP mass factorization contribution. Moreover, we achieved this by means of integrated antenna functions, which will be crucial for the next step. The explicit form of~\eqref{sigTIP} is obtained computing $\braket{\ampnum{n+2}{0}|\T_i\cdot\T_j|\ampnum{n+2}{0}}$ and dressing them with the associated colour stripped integrated dipoles. We remark that~\eqref{sigTIP} is a completely process-independent result, which is valid for any multiplicity and retains the full $N_c$ dependence.
	
	The IC component of the virtual subtraction term is analogously obtained exploiting IC one-loop integrated dipoles. These were defined as scalars in colour space in~\eqref{J21ic}, since IC infrared singularities factorize onto full LO squared matrix elements. Hence, $\dd\sigpart{T,\mathrm{IC}}{ab,\mathrm{NLO}}$ is given by:
	\begin{equation}\label{sigTIC}
		\dd\sigpart{T,\mathrm{IC}}{ab,\mathrm{NLO}}=\coeff\sum_{(c,d)\neq(a,b)}\int\dr{x_1}\dr{x_2}\Jfulltotic{1}{ab;cd}(x_1,x_2)\,\dd\sigpart{}{cd,\mathrm{LO}}\,.
	\end{equation}
	The contribution above contains integrated initial-state collinear limits and the associated mass factorization kernels.

	\subsection{NLO real subtraction term}\label{sec:NLOR}
	
	In the colourful antenna approach, we aim for a systematic generation of $\dd\sigpart{S}{ab,\mathrm{NLO}}$, starting from the 
	$\dd\sigpart{T}{ab,\mathrm{NLO}}$ derived above. 
	 Each term in $\dd\sigpart{T}{ab,\mathrm{NLO}}$ must have an unintegrated counterpart in $\dd\sigpart{S}{ab,\mathrm{NLO}}$. Since $\dd\sigpart{T}{ab,\mathrm{NLO}}$ is completely written in terms of integrated antenna functions, there is a one-to-one relation with their unintegrated counterparts:
	\begin{equation}\label{insertion}
		\mathcal{X}_{3}^{0}(s_{ij})\,\anum{n+2}{0}(c,c',\{.,i,.,j,.\})\,\leftrightarrow\,X_{3}^{0}(i,u,j)\anum{n+2}{0}(c,c',\{.,\wt{iu},.,\wt{uj},.\}),
	\end{equation}
	where $\mathcal{X}_{3}^{0}(s_{ij})$ is the integrated antenna function obtained by integrating the tree-level three-parton antenna function $X_{3}^{0}(i,u,j)$ over the phase space of the unresolved parton $u$. Due to this correspondence, once the virtual subtraction term is obtained, the structure of the real subtraction term can be determined by inserting an unresolved emission between each pair of hard radiators appearing in the integrated dipoles. This involves a transition from an integrated NLO antenna function to an unintegrated one and from a genuine LO colour interference to a \textit{reduced} one where the ${(n+2)}$-particle momenta are meant to be obtained from a $(n+3)$-particle phase space through a suitable mapping, dictated by the accompanying antenna function. The right-hand-side of~\eqref{insertion} reproduces the divergent behaviour of the real interference $\anum{n+3}{0}(c,c';\{.,i,.,u,.,j,.\})$ when parton $u$ is unresolved between the hard pair $(i,j)$. 
	
	We will illustrate how the transition between integrated and unintegrated quantities is performed. The procedure can be summarized as follows:
	\begin{enumerate}
		\item[\textbf{1)}] remove the mass-factorization kernels from the colour-stripped integrated dipoles $\J{1}$ in $\dd\sigpart{T}{ab,\mathrm{NLO}}$;
		
		\item[\textbf{2)}]  replace each integrated antenna function $\mathcal{X}_{3}^{0}(s_{ij})$ with its unintegrated counterpart $X_{3}^{0}(i,u,j)$ (see below);
		
		\item[\textbf{3)}]  suitably replace the momenta in the colour interferences, according to the accompanying integrated antenna, following (\ref{insertion});
		
		\item[\textbf{4)}]  apply the same momenta relabelling to the jet function;

		\item[\textbf{5)}]  promote the phase-space measure to the appropriate one for $(n+1)$ final-state momenta;
		
		\item[\textbf{6)}]  replace the overall factor with an appropriate one for the real correction:
		\begin{equation}
			\coeffVNLO\to\coeffRNLO=s_{R}\left(4\pi\alpha_s\right)\coeffLO,
		\end{equation}
		where $s_{R}$ compensates the potentially different final-state symmetry factors in the presence of an extra emission;
		 
	\end{enumerate}
	All the steps above consist of rather simple manipulations, except for step $2$. The replacement of an integrated antenna function with its unintegrated counterpart stands at the core of the whole algorithm and dictates how to perform the subsequent steps. As indicated in~\eqref{insertion}, the transition occurs by inserting an unresolved parton within the hard dipole described by the integrated antenna function. We have to distinguish three different types of insertions:
	\begin{itemize}
		\item insertion of an unresolved gluon;
		\item insertion of an unresolved quark-antiquark pair, coming from a hard gluon splitting;
		\item insertion of unresolved partons in IC limits.
	\end{itemize}
	These three cases are described in detail in the following.
	
	\paragraph{Insertion of an unresolved gluon}
	
	The insertion of an unresolved gluon was already described in our  earlier work~\cite{Chen:2022ktf}. From a practical standpoint, it is the simplest type of insertion, since it affects all the external partons in the same way. It occurs within antenna functions coming from the integration over a soft or collinear gluon, namely $A$-, $D$-, and $F$-type three-parton tree-level antenna functions. 
	In Table~\ref{tab:insg} we indicate the transition rules to convert integrated dipoles and corresponding integrated antenna functions to the unintegrated antenna functions. The inserted unresolved gluon is denoted with the label $u_g$.
	\renewcommand{\arraystretch}{1.5} 
	\begin{table}
		\centering
		\begin{tabular}{c|ccc}
			& $\J{1}$ & $\mathcal{X}_3^0$ & $X_3^0$ 
			\\ \hline
			
			\multirow{3}{*}{$q-\bar{q}$} & 
			$\J{1}(i_q,j_{\qb})$	& $\XFFint{A}{3}{0}(s_{ij})$ & $A_3^0(i_q,u_g,j_{\qb})$ \\\cline{2-4}
			& $\J{1}(1_q,i_{\qb})$	& $\XIFint{A}{3}{0}{q}(s_{1i})$ & $A_{3,q}^0(1_q,u_g,i_{\qb})$ \\\cline{2-4}
			& $\J{1}(1_q,2_{\qb})$	& $\XIIint{A}{3}{0}{q\qb}(s_{12})$ & $A_{3,q\qb}^0(1_q,u_g,2_{\qb})$ \\\cline{2-4}
			\hline
			
			\multirow{4}{*}{$q-g$} & 
			$\J{1}(i_q,j_g)$	& $\frac{1}{2}\XFFint{D}{3}{0}(s_{ij})$ & $d_3^0(i_q,u_g,j_g)$ \\\cline{2-4}
			& $\J{1}(1_q,i_g)$	& $\frac{1}{2}\XIFint{D}{3}{0}{q}(s_{1i})$ & $d_{3,q}^0(1_q,u_g,i_g)$ \\\cline{2-4}
			& $\J{1}(1_g,i_q)$	& $\XIFint{D}{3}{0}{g\to g}(s_{1i})$ & $d_{3,g}^0(i_q,u_g,1_g)$ \\\cline{2-4}
			& $\J{1}(1_q,2_g)$	& $\XIIint{D}{3}{0}{qg}(s_{12})$ & $D_{3,qg}^0(1_q,u_g,2_g)$ \\\cline{2-4}
			\hline
			
			\multirow{3}{*}{$g-g$} & 
			$\J{1}(i_g,j_g)$	& $\frac{1}{3}\XFFint{F}{3}{0}(s_{ij})$ & $f_3^0(i_g,u_g,j_g)$ \\\cline{2-4}
			& $\J{1}(1_g,i_g)$	& $\frac{1}{2}\XIFint{F}{3}{0}{g}(s_{1i})$ & $f_{3,g}^0(1_g,u_g,i_g)$ \\\cline{2-4}
			& $\J{1}(1_g,2_g)$	& $\XIIint{F}{3}{0}{gg}(s_{12})$ & $F_{3,gg}^0(1_g,u_g,2_g)$ \\\cline{2-4}
			\hline
		\end{tabular}
		\caption{Replacement rules to convert integrated antenna functions to their unintegrated counterparts for the insertion of an unresolved gluon (denoted with $u_g$) between the pair of hard radiators.}
		\label{tab:insg}
	\end{table}
	
	 In a practical implementation, one typically requires a numerical indexing of the partons: for an $(n+2)$-parton LO process, one can chose $u_g=n+3$. We clarify how the insertion is performed with an example. We consider the following term which is part of the leading-colour NLO virtual subtraction term for $pp\to 4j$ production, specifically for $q\qb\to gggg$:
	\begin{eqnarray}\label{exg1}
		&&-\left(\J{1}(1_q,3_g)+\J{1}(3_g,4_g)+\J{1}(4_g,5_g)+\J{1}(5_g,6_g)+\J{1}(2_\qb,6_g)\right)\nn\\
		&&\hspace{5cm}\times A^{2q,0}_{6}(1_q,3_g,4_g,5_g,6_g,2_\qb)\jet{4}{4}{p_3,p_4,p_5,p_6}.
	\end{eqnarray}
	To insert the unresolved gluon $7_g$, we can rely on Table~\ref{tab:insg} to first convert the integrated dipoles to integrated antenna functions, and then replace them with their unintegrated counterparts. We obtain:
	\begin{eqnarray}\label{exg3}
		&-&d^0_{3,q}(1_q,7_g,3_g)  A^{2q,0}_{6}(\wt{(17)}_q,\wt{(37)}_g,4_g,5_g,6_g,2_\qb)\jet{4}{4}{\wt{p_{37}},p_4,p_5,p_6}\nn\\
		&-&f^0_{3}(3_g,7_g,4_g)    A^{2q,0}_{6}(1_q,\wt{(37)}_g,\wt{(47)}_g,5_g,6_g,2_\qb)\jet{4}{4}{\wt{p_{37}},\wt{p_{47}},p_5,p_6}\nn\\
		&-&f^0_{3}(4_g,7_g,5_g)    A^{2q,0}_{6}(1_q,3_g,\wt{(47)}_g,\wt{(57)}_g,6_g,2_\qb)\jet{4}{4}{p_3,\wt{p_{47}},\wt{p_{57}},p_6}\nn\\
		&-&f^0_{3}(5_g,7_g,6_g)    A^{2q,0}_{6}(1_q,3_g,4_g,\wt{(57)}_g,\wt{(67)}_g,2_\qb)\jet{4}{4}{p_3,p_4,\wt{p_{57}},\wt{p_{67}}}\nn\\
		&-&d^0_{3,q}(2_\qb,7_g,6_g)A^{2q,0}_{6}(1_q,3_g,4_g,5_g,\wt{(67)}_g,\wt{(27)}_\qb)\jet{4}{4}{p_3,p_4,p_5,\wt{p_{67}}}.
	\end{eqnarray}
 As another example, we consider a specific contribution to the subleading-colour part of the virtual subtraction terms: 
	\begin{eqnarray}\label{exg4}
		&&-2\left(\J{1}(1_q,2_{\qb})-\J{1}(1_q,6_g)-\J{1}(2_\qb,4_g)+\J{1}(4_g,6_g)\right)\nn\\
		&&\hspace{2cm}\times A^{2q,0}_{6}(1_q,3_g,4_g,5_g,6_g,2_\qb;1_q,4_g,5_g,6_g,3_g,2_\qb)\jet{4}{4}{p_3,p_4,p_5,p_6},
	\end{eqnarray}
	where we have an interference term between the two colour orderings $(1_q,3_g,4_g,5_g,6_g,2_\qb)$ and $(1_q,4_g,5_g,6_g,3_g,2_\qb)$, accompanied by a non-trivial dipole structure. Repeating the insertion steps yields:
	\begin{eqnarray}\label{exg5}
		&-&2A^0_{3,q\qb}(1_q,7_g,2_\qb)  A^{2q,0}_{6}(\wt{(17)}_q,3_g,4_g,5_g,6_g,\wt{(27)}_\qb;\wt{(17)}_q,4_g,5_g,6_g,3_g,\wt{(27)}_\qb)\nn\\
		&&\hspace{8.5cm}\times\jet{4}{4}{p_3,p_4,p_5,p_6}\nn\\
		&+&2d^0_{3,q}(1_q,7_g,6_g)       A^{2q,0}_{6}(\wt{(17)}_q,3_g,4_g,5_g,\wt{(67)}_g,2_\qb;\wt{(17)}_q,4_g,5_g,\wt{(67)}_g,3_g,2_\qb)\nn\\
		&&\hspace{8.5cm}\times\jet{4}{4}{p_3,p_4,p_5,\wt{p_{67}}}\nn\\
		&+&2d^0_{3,q}(2_\qb,7_g,4_g)     A^{2q,0}_{6}(1_q,3_g,\wt{(47)}_g,5_g,6_g,\wt{(27)}_\qb;1_q,\wt{(47)}_g,5_g,6_g,3_g,2_\qb)\nn\\
		&&\hspace{8.5cm}\times\jet{4}{4}{p_3,\wt{p_{47}},p_5,p_6}\nn\\
		&-&2f^0_{3}(4_g,7_g,6_g)         A^{2q,0}_{6}(1_q,3_g,\wt{(47)}_g,5_g,\wt{(67)}_g,2_\qb;1_q,\wt{(47)}_g,5_g,\wt{(67)}_g,3_g,2_\qb)\nn\\
		&&\hspace{8.5cm}\times\jet{4}{4}{p_3,\wt{p_{47}},p_5,\wt{p_{67}}}\,.
	\end{eqnarray}
	
	The resulting expressions are part of the NLO real subtraction term for $q\qb\to gggg$. Let's now assume that, instead of the particular terms considered in~\eqref{exg1} and~\eqref{exg4} one performs the insertion of the unresolved gluon $7_g$ within the full virtual subtraction term for $q\qb\to gggg$, namely summing over all possible colour orderings. The result obtained this way will still not contain the entire unresolved behaviour of the real matrix element associated to gluon $7_g$ becoming soft or collinear to hard partons. In particular, it will contain the entirety of the soft limits and the quark-gluon collinear limits, but for each possible gluon-gluon collinear pair $(i_g,7_g)$ it will only have part of the collinear limit, due to the construction of the three-parton tree-level sub-antenna functions. The full collinear behaviour is restored when an analogous term with unresolved gluon $i_g$ is added.

	To be more precise, the procedure described before for the insertion of the unresolved gluon $7_g$ generates a function of the external real-kinematics momenta:
	\begin{equation}
		f(p_1,p_2,p_3,p_4,p_5,p_6;p_7)\,,
	\end{equation}
	 where we isolated the last argument, since it is the one corresponding to the newly inserted gluon, which, for the moment, is the only one with infrared divergences associated to it. The full singular behaviour is obtained, in this specific example with five final-state gluons, taking:
	 \begin{eqnarray}\label{sumunresg}
	 	f(p_1,p_2,p_3,p_4,p_5,p_6;p_7)&+&f(p_1,p_2,p_4,p_5,p_6,p_7;p_3)\nn\\
	 	f(p_1,p_2,p_5,p_6,p_7,p_3;p_4)&+&f(p_1,p_2,p_6,p_7,p_3,p_4;p_5)\nn\\
	 	f(p_1,p_2,p_7,p_3,p_4,p_5;p_6)&.&
	 \end{eqnarray} 
	 The combination above allows each gluon to play the role of the unresolved parton, effectively reproducing all the soft and collinear limits in the considered process. We note that the actual ordering of the hard gluons in each instance of the function $f$ is irrelevant, since it already contains a sum over all possible colour orderings. Indeed, in~\eqref{sumunresg}, the only relevant position for a gluon momentum is the last one. 
	
	\paragraph{Insertion of an unresolved quark-antiquark pair}
	
	\newcommand{\nfsg}{{n_{g}}}
	
	The insertion of quark-antiquark pair is necessary when the LO process has gluons in the final state. $E$- and $G$-type antenna functions address such configuration. While the insertion of an unresolved gluon proceeds just by adding a parton on top of the LO partonic content, in this case we have a final-state gluon which is converted into a quark-antiquark pair. 
	
	We consider a $(n+2)$-parton process with $\nfsg$ final-state gluons, labelled as $(i,\dots,i+\nfsg-1)$, while we temporarily label the new quark and antiquark with $u_q$ and $u_\qb$. Any final-state gluon is allowed to split into $(u_q,u_\qb)$. To take this into account, we apply to the integrated expression the insertions described in Table~\ref{tab:insqqb}. In particular, within the FF $G_3^0$ antenna function, either gluon $i_g$ or $j_g$ can split into $(u_q,u_\qb)$ and both contributions are taken into account. Since we are inserting a quark-antiquark pair which is allowed to become collinear, it is convenient to directly perform a symmetrization over the quark and the antiquark indices already at this level. 
	\renewcommand{\arraystretch}{1.5} 
	\addtolength{\tabcolsep}{-8pt} 
	\begin{table}
		\centering
		\begin{tabular}{c|ccc}
			& $\J{1}$ & $\mathcal{X}_3^0$ & $X_3^0$ 
			\\ \hline
			
			\multirow{2}{*}{$q-g$} & 
			$\Jh{1}(i_q,j_g)$	& $\frac{1}{2}\XFFint{E}{3}{0}(s_{ij})$ & $\frac{1}{4}\left[E_3^0(i_q,u_{q},u_{\qb})+E_3^0(i_q,u_{\qb},u_{q})\right]$ \\\cline{2-4}
			& $\Jh{1}(1_q,i_g)$	& $\frac{1}{2}\XIFint{E}{3}{0}{q}(s_{1i})$ & $\frac{1}{4}\left[E_{3,q}^0(1_q,u_{q},u_{\qb})+E_{3,q}^0(1_q,u_{\qb},u_{q})\right]$ \\\cline{2-4}
			\hline
			
			\multirow{2}{*}{$g-g$} & 
			$\Jh{1}(i_g,j_g)$	& $\XFFint{G}{3}{0}(s_{ij})$ & \thead{$\frac{1}{2}\big[\Big.G_3^0(i_g,u_{q},u_{\qb})+G_3^0(i_g,u_{\qb},u_{q})$ \\ $\hspace{0.5cm}+G_3^0(u_{q},u_{\qb},j_q)+G_3^0(u_{\qb},u_{q},j_q)\big.\big]$} 
			\\\cline{2-4}
			& $\Jh{1}(1_g,j_g)$	& $\frac{1}{2}\XIFint{G}{3}{0}{g}(s_{1i})$ & $\frac{1}{4}\left[G_{3,g}^0(1_g,u_{q},u_{\qb})+G_{3,g}^0(1_g,u_{\qb},u_{q})\right]$ \\\cline{2-4}
			\hline
		\end{tabular}
		\caption{Replacement rules to convert integrated antenna functions to their unintegrated counterparts when the final-state gluon splits into an unresolved quark-antiquark pair, denoted with $(u_q,u_\qb)$. Symmetrization over the inserted quark-antiquark pair is always considered.}
		\label{tab:insqqb}
	\end{table}
	
	After the insertion, we need to relabel the partonic indices. First of all, the real kinematics has one gluon less, so the gluons are re-assigned to indices $(i,\dots,i+\nfsg-2)$. The actual way this relabelling happens is not relevant, since, once again, we will sum over all possible colour orderings. Then, the newly added quark $u_q$ is assigned to the free index $i+\nfsg-1$, while the antiquark is indexed as $n+3$. This strategy ensures that a quark-antiquark pair is consistently inserted in any possible ordering of the original set of gluons. As a result, the subtraction term for each limit is generated exactly $\nfsg$ times, and we must normalize by  $1/\nfsg$.
		
	If the LO process already has hard quark-antiquark pairs, for the insertion procedure we assume the newly added pair has a flavour different from any other pair already present in the process. Indeed, collinear $q\parallel \qb$ singularities are only present at the squared amplitude level when $q$ and $\qb$ belong to the same fermionic line. Therefore, the $q\parallel \qb$ singular behaviour of real emissions corrections with two or more identical-flavour fermionic lines are described by the same structures required for the different-flavour case. 

	For an example, we can consider once again the $q\qb\to gggg$ process in the context of the NLO correction to $pp\to 4j$. In this case $\nfsg=4$ and the final-state gluons are labelled with indices $3_g$, $4_g$, $5_g$ and $6_g$. After the insertion, we will have three gluons labelled as $3_g$, $4_g$ and $5_g$, a quark $6_q$ and an antiquark $7_\qb$. We take the following term from the leading-$N_f$ contribution to the virtual subtraction term:
	\begin{eqnarray}\label{exqqb1}
		&&-\left(\Jh{1}(1_q,3_g)+\Jh{1}(3_g,4_g)+\Jh{1}(4_g,5_g)+\Jh{1}(5_g,6_g)+\Jh{1}(2_\qb,6_g)\right)\nn\\
		&&\hspace{5cm}\times A^{2q,0}_{6}(1_q,3_g,4_g,5_g,6_g,2_\qb)\jet{4}{4}{p_3,p_4,p_5,p_6}.
	\end{eqnarray}
The insertion yields, after summing over all gluons and multiplying with an overall factor	$1/\nfsg=1/4$:
    \begingroup
    \allowdisplaybreaks
	\begin{eqnarray}
		&&-\dfrac{1}{16}E_{3,q}^0(1_q,6_q,7_\qb)A^{2q,0}_{6}(\wt{(16)}_q,\wt{(67)}_g,3_g,4_g,5_g,2_\qb)\jet{4}{4}{p_3,p_4,p_5,\wt{p_{67}}}\nn\\
		&&-\dfrac{1}{16}E_{3,q}^0(2_\qb,6_q,7_\qb)A^{2q,0}_{6}(1_q,3_g,\wt{(67)}_g,4_g,5_g,\wt{(26)}_\qb)\jet{4}{4}{p_3,p_4,p_5,\wt{p_{67}}}\nn\\
		&&-\dfrac{1}{16}G_{3}^0(3_g,6_q,7_\qb)\Big[\Big.A^{2q,0}_{6}(1_q,\wt{(36)}_g,\wt{(67)}_g,4_g,5_g,2_\qb)\nn\\
			&&\hspace{3.2cm}A^{2q,0}_{6}(1_q,\wt{(67)}_g,\wt{(36)}_g,4_g,5_g,2_\qb)\Big.\Big]\jet{4}{4}{\wt{p_{36}},p_4,p_5,\wt{p_{67}}}\nn\\
		&&-\dfrac{1}{16}G_{3}^0(4_g,6_q,7_\qb)\Big[\Big.A^{2q,0}_{6}(1_q,3_g,\wt{(46)}_g,\wt{(67)}_g,5_g,2_\qb)\nn\\
			&&\hspace{3.2cm}A^{2q,0}_{6}(1_q,3_g,\wt{(67)}_g,\wt{(46)}_g,5_g,2_\qb)\Big.\Big]\jet{4}{4}{p_3,\wt{p_{46}},p_5,\wt{p_{67}}}\nn\\
		&&-\dfrac{1}{16}G_{3}^0(5_g,6_q,7_\qb)\Big[\Big.A^{2q,0}_{6}(1_q,3_g,4_g,\wt{(56)}_g,\wt{(67)}_g,2_\qb)\nn\\
			&&\hspace{3.2cm}A^{2q,0}_{6}(1_q,3_g,4_g,\wt{(67)}_g,\wt{(56)}_g,2_\qb)\Big.\Big]\jet{4}{4}{p_3,p_4\wt{p_{56}},\wt{p_{67}}}\nn\\
		&&+\left(6\leftrightarrow 7\right).	
	\end{eqnarray}
	\endgroup
	The insertion proceeds analogously even beyond the leading-$N_f$ contributions, with squared coherent partial amplitudes possibly replaced by incoherent interferences between different colour orderings. Contrary to the gluon insertion case, no additional sum over permutations of real-kinematics momenta is required here, because the obtained expression completely encapsulates the singular behaviour in the $q\parallel \qb$ limit.
	
	\paragraph{Insertion of an unresolved parton in IC limits}
	
	IC limits at NLO occur when a final-state quark (antiquark), denoted in the following by $u_q$ ($u_\qb$) becomes collinear to either an initial-state gluon ($g\to q$) or initial-state quark (antiquark) of the same flavour ($q\to g$). The appropriate replacement rules for IC insertions are listed in Table~\ref{tab:insIC}, where parton $1$ is the one involved in the identity-changing limit, while the second parton in the integrated dipoles (antenna functions) is just a spectator. Symmetrization over the collinear quark-antiquark pair is considered for G- and $E$-type antenna functions. 
	\renewcommand{\arraystretch}{1.5} 
	\addtolength{\tabcolsep}{-2pt} 
	\begin{table}
		\centering
		\begin{tabular}{c|ccc}
			& $\J{1}$ & $\mathcal{X}_3^0$ & $X_3^0$ 
			\\ \hline
			
			\multirow{2}{*}{$q-\bar{q}$}  
			& $\Jic{1}{g\to q}(1_q,i_{\qb})$	& $-\frac{1}{2}\XIFint{A}{3}{0}{g}(s_{1i})$ & $-A_{3,g}^0(u_q,1_g,i_\qb)$ \\\cline{2-4}
			& $\Jic{1}{g\to q}(1_q,2_{\qb})$	& $-\XIIint{A}{3}{0}{gq}(s_{12})$ & $-A_{3,gq}^0(u_q,1_g,2_{\qb})$ \\\cline{2-4}
			\hline
			
			\multirow{4}{*}{$q-g$} 
			& $\Jic{1}{g\to q}(1_q,i_g)$	& $-\XIFint{D}{3}{0}{g\to q}(s_{1i})$ & $-d_{3,g}^0(u_q,1_g,i_g)$ \\\cline{2-4}
			& $\Jic{1}{q\to g}(1_g,i_q)$	& $-\XIFint{E}{3}{0}{q'}(s_{1i})$ & $-\frac{1}{2}\left[E_{3,q'}^0(i_q,1_q,u_\qb)+E_{3,q'}^0(i_q,u_\qb,1_q)\right]$ \\\cline{2-4}
			& $\Jic{1}{g\to q}(1_q,2_g)$	& $-\XIFint{D}{3}{0}{gg}(s_{1i})$ & $-D_{3,gg}^0(u_q,1_g,2_g)$ \\\cline{2-4}
			& $\Jic{1}{q\to g}(1_g,2_q)$	& $-\XIIint{E}{3}{0}{q'q}(s_{12})$ & $-\frac{1}{2}\left[E_{3,q'q}^0(2_q,1_q,u_\qb)+E_{3,q'q}^0(2_q,u_\qb,1_q)\right]$ \\\cline{2-4}
			\hline
			
			\multirow{2}{*}{$g-g$}  
			& $\Jic{1}{q\to g}(1_q,i_g)$	& $-\XIFint{G}{3}{0}{q}(s_{1i})$ & $-\frac{1}{2}\left[G_{3,q}^0(i_g,1_q,u_\qb)+G_{3,q}^0(i_g,u_\qb,1_q)\right]$ \\\cline{2-4}
			& $\Jic{1}{q\to g}(1_q,2_g)$	& $-\XIIint{G}{3}{0}{qg}(s_{12})$ & $-\frac{1}{2}\left[G_{3,qg}^0(2_g,1_q,u_\qb)+G_{3,qg}^0(2_g,u_\qb,1_q)\right]$ \\\cline{2-4}
			\hline
		\end{tabular}
		\caption{Replacement rules to convert integrated antenna functions to their unintegrated counterparts for identity-changing insertions. The final-state quark (antiquark) causing the change of identity of parton $1$ is denoted with $u_q$ ($u_\qb$). Symmetrization over the collinear quark-antiquark pair is considered for $G$- and $E$-type antenna functions.}
		\label{tab:insIC}
	\end{table}
	
	Analogously to the insertion of a quark-antiquark pair, IC insertions require a suitable relabelling of the parton indices within the affected structures. The quark (antiquark) $u_q$ ($u_\qb$) responsible for the IC limit is labelled with the proper index that it has in the real-emission kinematics, while the indices of the other final-state partons are shifted accordingly. The procedure can be clarified with an example: we consider the LO process $qg\to qggg$ and we address a $g\to q$ IC limits. Namely we aim to construct the subtraction term for the real-emission process $gg\to q\qb ggg$, for the limit when the final state antiquark becomes collinear to the initial state gluon. We have $u_\qb=4$, which means that the additional final-state gluons, in the transition from the LO to the real kinematics, must have their indices shifted from $(4,5,6)$ to $(5,6,7)$.  
	
	We start from the following contribution in the IC component of the virtual subtraction term:
	\begin{equation}
		-\Jic{1}{g\to q}(1_q,2_g)M_6^{0,2q}(1_q,2_g,3_q,4_g,5_g,6_g)\jet{4}{4}{p_3,p_4,p_5,p_6},
	\end{equation}
	where $M_6^{0,2q}$ represents the full LO matrix element. The transition to the unintegrated level, including the appropriate relabelling of momenta gives:
	\begin{equation}
		d_{3,gg}^0(4_\qb,1_g,2_g)M_6^{0,2q}(\wt{(14)}_q,\wt{(24)}_g,3_q,5_g,6_g,7_g)\jet{4}{4}{p_3,p_5,p_6,p_7}.
	\end{equation}
	The same logic applies to contributions beyond the leading-colour approximation and to $g\to q$ collinear limits. Also in this case, the generated expressions completely cover the considered IC collinear limit and no additional sum over permutations of the external momenta is needed.

	\vspace{1cm}
	
\begin{figure}
		\centering
		\begin{tikzpicture}[node distance=1.8cm, 
			box/.style={draw, rectangle, text width=10cm, minimum height=1.2cm, align=center, fill=lightblue}]
			
			\definecolor{lightblue}{RGB}{173,216,230}
			
			\node[box] (box1) {Removal of splitting kernels from $\J{1}$};
			
			\node[box, below of=box1] (box2) {Insertion of a single unresolved parton: $\mathcal{X}_3^0\to X_3^0$};
			
			\node[box, below of=box2] (box3) {Relabelling/mapping of momenta};
			
			\node[box, below of=box3] (box4) {Fix overall factors};
			
			\draw[->, thick] (box1) -- (box2);
			\draw[->, thick] (box2) -- (box3);
			\draw[->, thick] (box3) -- (box4);
			
		\end{tikzpicture}
		\caption{General outline of the generation of the real subtraction term from the virtual subtraction term, via the application of the $\ins{\cdot}$ operator.}
		\label{fig:sigS_generation}
	\end{figure}
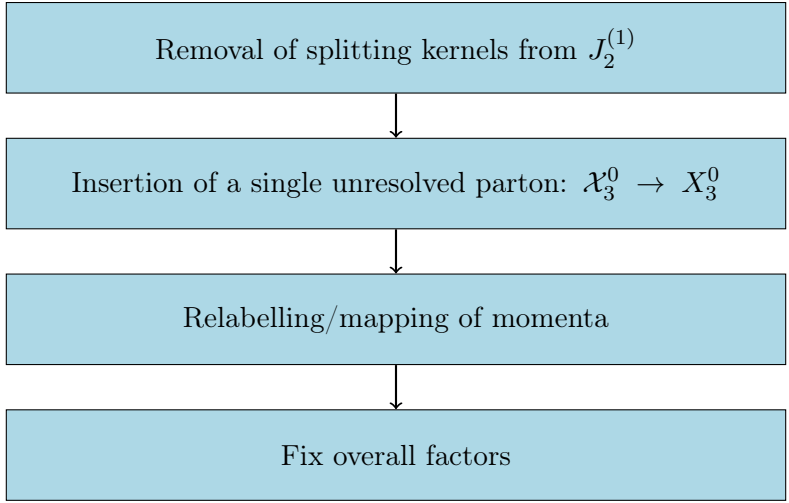

	The procedure described above, taking into account all different types of insertions, is comprehensively summarized by the following notation:
	\begin{equation}\label{unint}
		\dsigSNLO{ab}=-\ins{\dsigTNLO{ab}}.
	\end{equation}
	The $\ins{\cdot}$ operator systematically applies the transition from integrated to unintegrated quantities, as well as the required adjustments, such as momentum mapping, relabelling and appropriate sum over permutations of the external momenta. 
	 We summarize the action of the $\ins{\cdot}$ operator in Figure~\ref{fig:sigS_generation}.
	
	We will now clarify some important aspects related to the procedure we have described. First of all, we observe that the real subtraction term obtained by~\eqref{unint} satisfies~\eqref{sigTNLO} \textit{by construction}. Indeed, the analytical integration of $\dsigSNLO{ab}$ is realized in the context of antenna subtraction by integrating the antenna functions appearing in it. Since the action of the $\ins{\cdot}$ operator is based on the univocal correspondence in~\eqref{insertion}, by converting the unintegrated antenna functions in $\dsigSNLO{ab}$ into their integrated counterparts (analytical integration as in the traditional approach), one would obtain the exact same expression for the virtual subtraction term used as the input of~\eqref{unint}, modulo the mass factorization counterterm which is added later. In this regard, equation~\eqref{unint} should be viewed as the unambiguous inverse operation with respect to~\eqref{sigTNLO}. Namely, up to mass factorization counterterms, one has:
	\begin{equation}
		\int_1\ins{f}=f,
	\end{equation}
	where $f$ is used instead of $\dsigTNLO{ab}$ to emphasize the generality of such property.
	
	The momentum mapping relating the $(n+3)$- and the $(n+2)$-momentum sets is induced in each term of an expression like~\eqref{sigSexample} by the antenna function. The specific form of the mapping does not affect the final integrated result, provided it preserves the exact factorization of the antenna phase space~\cite{Gehrmann-DeRidder:2005btv,Currie:2013vh}. Therefore, when the $\ins{\cdot}$ operator is applied, the unintegrated antenna functions straightforwardly dictate which kind of momentum mapping is required within the associated reduced matrix element, namely what are the hard radiators, the unresolved parton and the kinematical configuration (FF, IF or II). The specific form of the mapping can then be chosen at the level of the numerical implementation. As in the traditional antenna subtraction approach, we always consider the mappings described in~\cite{Kosower:2002su}.
	
	What is in principle left to discuss is whether $\dsigSNLO{ab}$ obtained via the outlined approach correctly subtracts \textit{all} the infrared-divergent behaviour of the real-emission correction. First of all, we remark once again that the relation in~\eqref{sigTNLO} is automatically fulfilled. Although equation~\eqref{sigTNLO} alone is not sufficient to claim that $\dsigSNLO{ab}$ works as intended, it is a necessary condition for the cancellation of the infrared singularities between real and virtual corrections and significantly restricts the form of the real subtraction term. In addition, with~\eqref{unint} we are effectively dressing each pair of hard radiators with an unresolved emission, taking then into account all possible unresolved limits of the real-emission matrix element. The potential double-counting of an unresolved configuration is prevented by the fact that NLO antenna functions have well defined unresolved partons and hard radiators. As discussed in the insertion examples above, the correct infrared behaviour in collinear limits is only recovered when a sum over all potentially unresolved partons is performed, due to the ambiguity in the identification of a hard emitter and an unresolved parton in such limits. We anticipate that at NNLO some of these features are less clearly manifest in the final form of the subtraction terms. This is the reason why we will need to suitably decompose the subtraction framework into several structures with specific roles in order to define a process-independent strategy to assemble the subtraction terms. The knowledge and insights accumulated during previous applications of the traditional antenna subtraction method will be particularly relevant for the formulation of the colourful antenna subtraction approach at NNLO. 
	
	We conclude by mentioning here that angular correlations in collinear limits require particular attention. Such terms average to zero after integration over the phase space of the collinear particles, yielding no explicit poles at the virtual level. Within the antenna subtraction method, angular terms are never subtracted locally. As explained in detail in~\cite{Glover:2010kwr}, a suitable point-by-point angular average is performed to locally enforce the vanishing of the angular terms. For this reason, the absence of explicit poles associated to such divergences at the virtual level does not represent an issue for the colourful antenna subtraction method.   
	
	\subsection{Observations}
	
	We carefully treated the insertion of unresolved partons at NLO because the techniques established at NLO are repeatedly used at NNLO as well. Indeed, the majority of the NNLO subtraction terms can be generated through the iterated application of $\ins{\cdot}$ to integrated quantities. This was recently observed also in the context of the nested soft-collinear subtraction scheme~\cite{Devoto:2023rpv}. The only exception is represented by configurations that require a simultaneous double insertion of unresolved partons. 
		
	The construction of the NLO real subtraction term concludes the description of the colourful antenna method at this perturbative order. 
	As in the traditional subtraction schemes, the key requirement is the analytical integrability of the real subtraction term over the phase space of the unresolved radiation. In addition, here it is crucial that the virtual subtraction term is expressed in a language suitable for the \textit{unintegration} procedure. The correspondence between integrated antenna functions, collected in \eqref{J21}, and their unintegrated versions guarantees this property. 
	
	We remark that the knowledge of the unintegrated antenna functions required for the construction of the real subtraction term is a crucial premise for the application of this method. Indeed, the described unintegration procedure allows for a systematic assembly of such ingredients and not for an actual direct generation of the structures required to remove the infrared divergences of real emission corrections, such as eikonal factors or splitting kernels. This observation, of course, applies to the NNLO case too. 
	
	
	\section{Colourful antenna subtraction at NNLO}\label{sec:subNNLO}
	
	In the following, we document the colourful antenna subtraction method at NNLO. We recall the structure of the NNLO correction to a partonic sub-process initiated by partons $a$ and $b$:
	\begin{eqnarray}\label{NNLOcssub2}
		\dsigNNLO{ab}&=&\int_n\left[\dsigVV{ab}-\dsigUNNLO{ab}\right]\nonumber\\
		&+&\int_{n+1}\left[\dsigRV{ab}-\dsigTNNLO{ab}\right]\nonumber\\
		&+&\int_{n+2}\left[\dsigRR{ab}-\dsigSNNLO{ab}\right],
	\end{eqnarray}
	with
	\begin{eqnarray}\label{subtermsNNLO2}
		\dsigSNNLO{ab}&=&\dsigSNNLOspe{ab}{1}+\dsigSNNLOspe{ab}{2}\,,\nonumber\\
		\dsigTNNLO{ab}&=&\dsigVSNNLO{ab}-\int_1 \dsigSNNLOspe{ab}{1}-\dsigMFNNLO{ab}{1}\,,\nonumber\\
		\dsigUNNLO{ab}&=&-\int_1 \dsigVSNNLO{ab}-\int_2 \dsigSNNLOspe{ab}{2}-\dsigMFNNLO{ab}{2}\,.
	\end{eqnarray}
    
    As for the NLO case, typically one would study the infrared-divergent behaviour of the real-virtual and double-real matrix elements to construct $\dsigVSNNLO{ab}$, $\dsigSNNLOspe{ab}{1}$ and $\dsigSNNLOspe{ab}{2}$ first, which are then integrated and combined with the mass factorization counterterms to complete the subtraction infrastructure. In contrast, the application of the colourful antenna subtraction method at NNLO begins by addressing the double-virtual correction in colour space. Once the double-virtual subtraction term is constructed, the real-virtual and double-real subtraction terms are generated via the insertion of unresolved partons, exploiting the relations between integrated and unintegrated structures in the subtraction terms in~\eqref{subtermsNNLO}.
	
\begin{figure}[t]
		\centering
		\includegraphics[width=\linewidth,keepaspectratio]{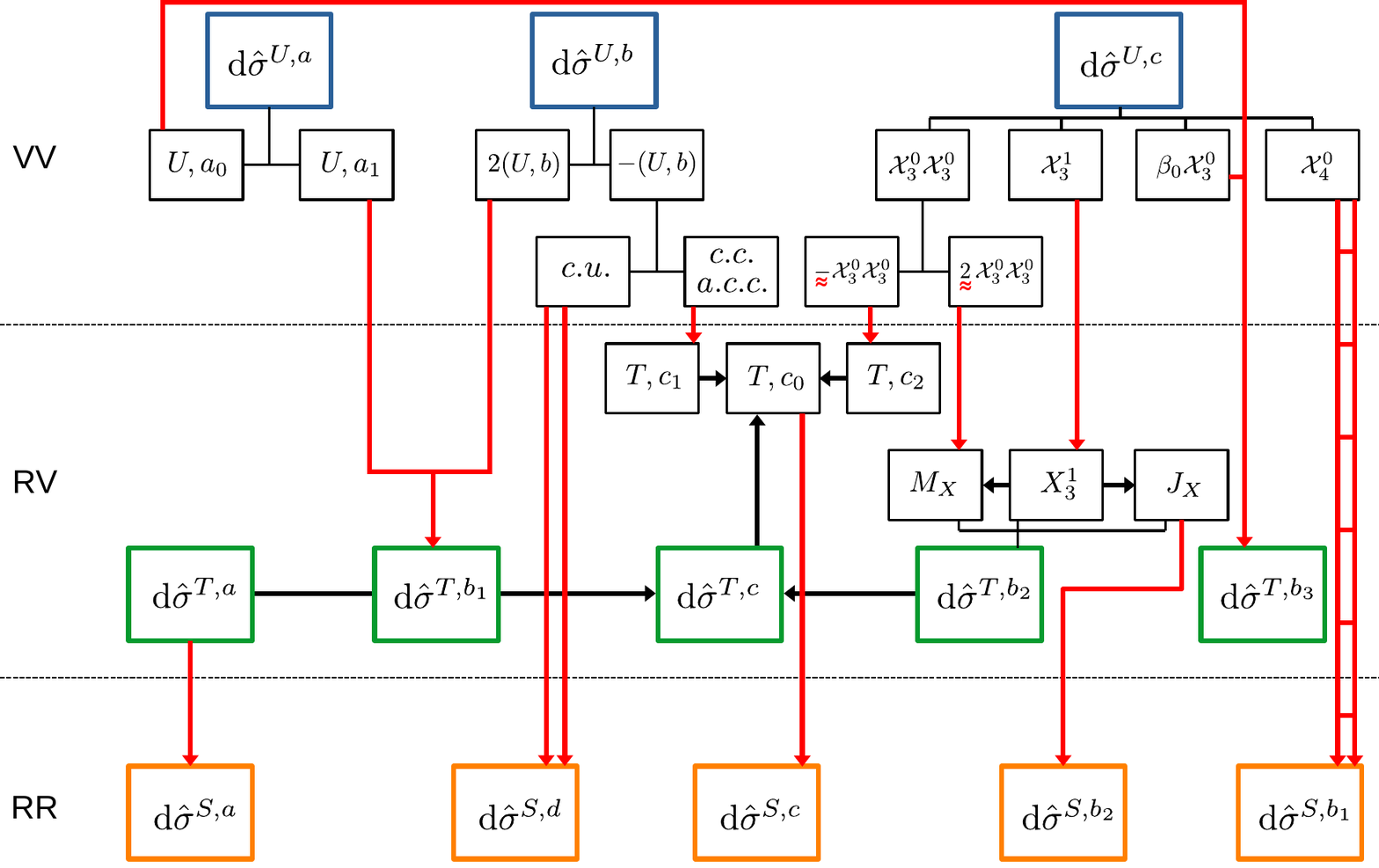}
		\caption{Structure of the colourful antenna subtraction at NNLO. Descendant red arrows represent the transition from an integrated quantity to its unintegrated counterpart via single insertion (single arrow), two iterated single insertions (two disjoint arrows) or double simultaneous insertion (two connected arrows) of unresolved partons. The definitions of each component are listed in Table~\ref{tab:equations}.}\label{fig:colant_scheme}
	\end{figure}

	To outline the colourful antenna subtraction method at NNLO, we summarize its procedure in Figure~\ref{fig:colant_scheme}. 	Single descendant red arrows represent the transition from an integrated quantity to its unintegrated counterpart by means of the insertion of an unresolved parton. Two disjoint red arrows indicate the iterated insertion of two unresolved partons, while two connected red arrows indicate the simultaneous insertion of two unresolved partons.

	\begin{table}[t]
		\begin{minipage}{.4\linewidth}
			\centering
			\setlength{\tabcolsep}{4pt}
			\begin{tabular}{cc|cc}
				\hline
				\multirow{2}{*}{$U,a$} & \multirow{2}{*}{eq. \eqref{sigUa}}
				& $U,a_0$ & eq. \eqref{sigUa0} \\\cline{3-4}
				&& $U,a_1$ & eq. \eqref{sigUa1} \\\cline{3-4}
				\hline
				\multirow{2}{*}{$U,b$}  & \multirow{2}{*}{eq. \eqref{sigUb}}
				& $U,b,c.u.$ & eq. \eqref{sigUbcu} \\\cline{3-4}
				&& $U,b,a.c.c.$ & eq. \eqref{sigUbacc} \\\cline{3-4}
				\hline	
				$U,c$ & eq. \eqref{sigUc} & & \\\cline{3-4}
				\hline
				\hline
				$S,a$ & eq. \eqref{sigSa} &&\\
				$S,b_1$ & eq. \eqref{sigSb1} && \\
				$S,b_2$ & eq. \eqref{sigSb2} && \\
				$S,c$ & eq. \eqref{sigSc} && \\
				$S,d$ & eq. \eqref{sigSd} && \\
				\hline
			\end{tabular}
		\end{minipage}%
		\hspace{0.7cm}
		\begin{minipage}{.4\linewidth}
			\vspace{-0.86cm}
			\centering
			\setlength{\tabcolsep}{4pt}
			\begin{tabular}{cc|cc}
				\hline
				$T,a$ & eq. \eqref{sigTaIP} & & \\
				$T,b_1$ & eq. \eqref{sigTb1} & & \\\cline{3-4}
				\multirow{3}{*}{$T,b_2$} & \multirow{3}{*}{eq. \eqref{sigTb2}}
				& $T,b_2,X_3^1$ & eq. \eqref{sigTb2X31} \\\cline{3-4}
				&& $T,b_2,J_X$ & Section \ref{sec:dsigTb} \\\cline{3-4}
				&& $T,b_2,M_X$ & eq. \eqref{sigTb2MXrel} \\\cline{3-4}
				$T,b_3$ & eq. \eqref{sigTb3} & & \\\cline{3-4}
				\multirow{3}{*}{$T,c$} & \multirow{3}{*}{eq. \eqref{finalTc}}
				& $T,c_1$ & eq. \eqref{sigTc1} \\\cline{3-4}
				&& $T,c_2$ & eq. \eqref{sigTc2} \\\cline{3-4}
				&& $T,c_0$ & eq. \eqref{sigTc0}\\\cline{3-4}
				\hline
			\end{tabular}
		\end{minipage} 
		\caption{Definitions of each term appearing in Figure \ref{fig:colant_scheme}.}\label{tab:equations}
	\end{table}
	To facilitate the navigation through this section, we list in Table \ref{tab:equations} the location of the definition of each term appearing in Figure \ref{fig:colant_scheme}.
	
	\subsection{NNLO mass factorization in colour space}\label{sec:MFNNLOcol}
	
	We start by casting the NNLO mass factorization counterterm in a form which suits the colourful approach. At NNLO, as indicated in~\eqref{NNLOcssub} and~\eqref{subtermsNNLO}, we have two different contributions: the double-virtual and the real-virtual mass factorization terms.
	
	\subsubsection{Double virtual mass factorization term}
	
	The double-virtual mass factorization term at NNLO reads:
	\begin{eqnarray}\label{MFVVbis}
		\dd\sigpart{MF,2}{ab,\mathrm{NNLO}}&=&-\int\dr{x_1}\dr{x_2}\sum_{c,d}\Bigg\lbrace\Bigg.\coeff\left[\Gammaone{ab;cd}{x_1,x_2}\left(\dd\sigpart{V}{cd,\mathrm{NLO}}-\dd\sigpart{T}{cd,\mathrm{NLO}}\right)\right]\nonumber\\
		&&\hspace{0.5cm}+\coeff^2\Big[\Big.\Gammatwo{ab;cd}{x_1,x_2}-\dfrac{\beta_0}{\e}\Gammaone{ab;cd}{x_1,x_2}\nonumber\\
		&&\hspace{3cm}+\dfrac{1}{2}\sum_{\alpha,\beta}\left[\Gamma^{(1)}_{ab;\alpha\beta}\otimes\Gamma^{(1)}_{\alpha\beta;cd}\right](x_1,x_2)\Big.\Big]\dd\sigpart{}{cd,\mathrm{LO}}\Bigg.\Bigg\rbrace,
	\end{eqnarray}
	and collects contributions factorizing onto an $n$-particle phase space, where $n$ is the final-state multiplicity of the LO process. As usual we separate the IP and IC components:
	\begin{equation}
		\dd\sigpart{MF,2}{ab,\mathrm{NNLO}}=\dd\sigpart{MF,2,\mathrm{IP}}{ab,\mathrm{NNLO}}+\dd\sigpart{MF,2,\mathrm{IC}}{ab,\mathrm{NNLO}}.
	\end{equation}
	
	As in Section~\ref{sec:MFNLOcol} at NLO, we express the two-loop identity-preserving mass factorization kernels in colour space as:
	\begin{equation}
		\BGammatwo{ab;ab}{x_1,x_2}=\BGammatwo{aa,\text{full}}{x_1}\delta(1-x_2)+\BGammatwo{bb,\text{full}}{x_2}\delta(1-x_1),
	\end{equation}
	where 
	\begin{equation}
		\BGammatwo{aa,\text{full}}{x_i}=-\Gammatwo{aa,\text{full}}{x_i}\dfrac{1}{C_a}\sum_{j\ne i}\T_i\cdot\T_j,\quad i=1,2.
	\end{equation}
	Hence, we can write the IP component of the mass factorization counterterm as:
	\begin{eqnarray}\label{MFVVIP}
		\dd\sigpart{MF,2,\mathrm{IP}}{ab,\mathrm{NNLO}}&=&-\coeffVVNNLO\int\dr{x_1}\dr{x_2}\dphi{n}(p_3,\dots,p_{n+2};x_1p_1,x_2p_2)\jet{n}{n}{\pset}\nonumber\\
		&&\Big\lbrace\Big.\braket{\ampnum{n+2}{0}|\BGammaone{ab;ab}{x_1,x_2}|\ampnum{n+2}{1}}+\braket{\ampnum{n+2}{1}|\BGammaone{ab;ab}{x_1,x_2}|\ampnum{n+2}{0}}\nonumber\\
		&&\hspace{2mm}-2\braket{\ampnum{n+2}{0}|\left[\BGammaonenodep{ab;ab}\otimes\Jcol{1}(\e)\right](x_1,x_2)|\ampnum{n+2}{0}}
		\nonumber\\
		&&\hspace{2mm}
		+\dfrac{1}{2}\braket{\ampnum{n+2}{0}|\left[\BGammaonenodep{ab;ab}\otimes\BGammaonenodep{ab;ab}\right](x_1,x_2)|\ampnum{n+2}{0}}\nonumber\\
		&&\hspace{2mm}-\dfrac{\beta_0}{\e}\braket{\ampnum{n+2}{0}|\BGammaone{ab;ab}{x_1,x_2}|\ampnum{n+2}{0}}
		\nonumber\\
		&&\hspace{2mm}+\braket{\ampnum{n+2}{0}|\BGammatwo{ab;ab}{x_1,x_2}|\ampnum{n+2}{0}}\Big.\Big\rbrace,
	\end{eqnarray}
	where we inserted the expression of the identity-preserving virtual subtraction term $\dd\sigpart{T,\mathrm{IP}}{cd,\mathrm{NLO}}$ given in~\eqref{sigTIP}.
	
	The IC part of the two-loop mass factorization counterterms is not expressed in colour space. It reads:
	\begingroup
	\allowdisplaybreaks
	\begin{eqnarray}\label{MFVVIC}
		\dd\sigpart{MF,2,\mathrm{IC}}{ab,\mathrm{NNLO}}&=&-\int\dr{x_1}\dr{x_2}\sum_{(cd)\neq (ab)}\Bigg\lbrace\Bigg.\coeff\left[\Gammaone{ab;cd}{x_1,x_2}\left(\dd\sigpart{V}{cd,\mathrm{NLO}}-\dd\sigpart{T}{cd,\mathrm{NLO}}\right)\right]\nonumber\\
		&&\hspace{-1.5cm}+\coeff^2\Bigg[\Bigg.\sum_{(ef)\neq(ab),(cd)}\dfrac{1}{2}\left[\Gamma^{(1)}_{ab;ef}\otimes\Gamma^{(1)}_{ef;cd}\right](x_1,x_2)\nonumber\\	&&\hspace{1cm}+\overline{\Gamma}^{(2)}_{ab;cd}(x_1,x_2)-\dfrac{\beta_0}{\e}\Gammaone{ab;cd}{x_1,x_2}\nonumber\\
		&&\hspace{1cm}+\dfrac{1}{2}\left[\Gamma^{(1)}_{ab;ab}\otimes\Gamma^{(1)}_{ab;cd}\right](x_1,x_2)+\dfrac{1}{2}\left[\Gamma^{(1)}_{ab;cd}\otimes\Gamma^{(1)}_{cd;cd}\right](x_1,x_2)\Bigg.\Bigg]\dd\sigpart{}{cd,\mathrm{LO}}\Bigg.\Bigg\rbrace\nonumber \\
		&&\hspace{-1.5cm}+\coeff^2\sum_{(ef)\neq(ab)}\Bigg[\Bigg.\dfrac{1}{2}\left[\Gamma^{(1)}_{ab;ef}\otimes\Gamma^{(1)}_{ef;ab}\right](x_1,x_2)+\overline{\Gamma}^{(2)}_{ab;ef;ab}(x_1,x_2)\Bigg.\Bigg]\dd\sigpart{}{ab,\mathrm{LO}},
	\end{eqnarray}
	\endgroup
	where the last line includes the flip-flopping contributions, with the flip-flopping kernel defined as
	\begin{equation}
		\Gammatwo{ab;ef;ab}{x_1,x_2}=\Gammatwo{a\to e\to a}{1}\delta(1-x_2)\delta_{ae}+\Gammatwo{b\to f\to b}{2}\delta(1-x_1)\delta_{bf}.
	\end{equation}
	
	\subsubsection{Real virtual mass factorization term}\label{sec:MFRV}
	
	The real-virtual mass factorization term is given by~\ref{MFRV}:
	\begin{displaymath} 
		\dd\sigpart{MF,1}{ab,\mathrm{NNLO}}=-\coeff\sum_{c,d}\int\dr{x_1}\dr{x_2}\Gammaone{ab;cd}{x_1,x_2}\left(\dd\sigpart{R}{cd,\mathrm{NLO}}-\dd\sigpart{S}{cd,\mathrm{NLO}}\right).
	\end{displaymath}
	and can be split into two contributions:
	\begin{equation}
		\dd\sigpart{MF,1}{ab,\mathrm{NNLO}}=\dd\sigpart{MF,1,a}{ab,\mathrm{NNLO}}+\dd\sigpart{MF,1,b}{ab,\mathrm{NNLO}},
	\end{equation}
	with 
	\begin{eqnarray}
		\label{MF1a}\dd\sigpart{MF,1,a}{ab,\mathrm{NNLO}}&=&-\coeff\sum_{c,d}\int\dr{x_1}\dr{x_2}\Gammaone{ab;cd}{x_1,x_2}\dd\sigpart{R}{cd,\mathrm{NLO}},\\
		\dd\sigpart{MF,1,b}{ab,\mathrm{NNLO}}&=&\coeff\sum_{c,d}\int\dr{x_1}\dr{x_2}\Gammaone{ab;cd}{x_1,x_2}\dd\sigpart{S}{cd,\mathrm{NLO}}.
	\end{eqnarray}
	Both $\dd\sigpart{MF,1,a}{ab,\mathrm{NNLO}}$ and $\dd\sigpart{MF,1,b}{ab,\mathrm{NNLO}}$ can be split into their IP and IC components. In analogy with~\eqref{MFIP}, the IP component of $\dd\sigpart{MF,1,a}{ab,\mathrm{NNLO}}$ can be rewritten in colour space as:
	\begin{eqnarray}\label{MF1aIP}
		\dd\sigpart{MF,1,a,\mathrm{IP}}{ab,\mathrm{NNLO}}&=&-\coeffRVNNLO\int\dr{x_1}\dr{x_2}\int\dphi{n}(p_3,\dots,p_{n+2};x_1 p_1,x_2 p_2)\,\jet{n}{n+1}{\lb p \rb_{n+1}}\nonumber\\
		&&\times\braket{\ampnum{n+3}{0}|\BGammaone{ab;ab}{x_1,x_2}|\ampnum{n+3}{0}},
	\end{eqnarray}
	where $\ket{\ampnum{n+3}{0}}$ indicates the single-real correction amplitude and $\coeffRVNNLO$ is the appropriate overall coefficient at the real-virtual level:
	\begin{equation}\label{coeffRV}
		\coeffRVNNLO=s_{RV}(4\pi\alpha_s)\coeff\coeffLO,\quad\text{with}\quad s_{RV}=s_{R}.
	\end{equation}
	
	Terms in $\dd\sigpart{MF,1,b}{ab,\mathrm{NNLO}}$ have a structure like $\Gamma^{(1)}_{ij}X_{3}^{0}\anum{n+2}{0}$. As we show in Section \ref{sec:dsigTb} below, this term is used to reconstruct one-loop integrated dipoles that are needed in the real-virtual subtraction term to remove the explicit poles of one-loop reduced matrix elements.
	
	\subsection{NNLO double-virtual subtraction term}\label{sec:VV}
	
	The double-virtual subtraction term at NNLO $\dd\sigpart{U}{ab,\mathrm{NNLO}}$, reproduces the explicit poles of the two-loop matrix element and contains the double-virtual mass factorization counterterm. In the following we see how to construct $\dd\sigpart{U}{ab,\mathrm{NNLO}}$ in a general way relying on the results of the previous sections.
	
	We focus first on the identity-preserving component. Using the dipole operators defined in Sections~\ref{sec:int_dip_1} and~\ref{sec:int_dip_2}, we can construct:
	\begin{eqnarray}\label{sigUIP}
		\dd\sigpart{U,\mathrm{IP}}{ab,\mathrm{NNLO}}&=&\coeffVVNNLO\int\dr{x_1}\dr{x_2}\dphi{n}(p_3,\dots,p_{n+2};x_1p_1,x_2p_2)\jet{n}{n}{\pset}\nonumber\\
		&&\times2\Big\lbrace\Big.\braket{\ampnum{n+2}{0}|\Jcol{1}(\e)|\ampnum{n+2}{1}}+\braket{\ampnum{n+2}{1}|\Jcol{1}(\e)|\ampnum{n+2}{0}}\nonumber\\
		&&\hspace{2mm}-\braket{\ampnum{n+2}{0}|\Jcol{1}(\e)\otimes\Jcol{1}(\e)|\ampnum{n+2}{0}}\nonumber\\
		&&\hspace{2mm}-\dfrac{\beta_0}{\e}\braket{\ampnum{n+2}{0}|\Jcol{1}(\e)|\ampnum{n+2}{0}}\nonumber\\
		&&\hspace{2mm}+\braket{\ampnum{n+2}{0}|\Jcol{2}(\e)|\ampnum{n+2}{0}}-\braket{\ampnum{n+2}{0}|\Jcolb{2}(\e)|\ampnum{n+2}{0}}\Big.\Big\rbrace.
	\end{eqnarray}
	One can verify that the expression above reproduces the same infrared singularity structure described by~\eqref{VVpoles}, thanks to the identities relating the $\e$-poles of colour-stripped integrated dipoles and infrared insertion operators, given in equations~\eqref{dipids1}, \eqref{dipids2} and~\eqref{dipids3}. We remark the presence of $\Jcolb{2}$ in the last line, to compensate for the unphysical poles present in quark-gluon integrated dipoles. Equation~\eqref{sigUIP} provides a fully general result for the removal of explicit double-virtual singularities in terms of integrated antenna functions. 
	
    According to the usual decomposition of the double-virtual subtraction term~\cite{Currie:2013vh}, we split~\eqref{sigUIP} into the following contributions:
	\begin{eqnarray}\label{sigUa}
		\dd\sigpart{U,a,\mathrm{IP}}{ab,\mathrm{NNLO}}&=&\coeffVVNNLO\int\dr{x_1}\dr{x_2}\dphi{n}(p_3,\dots,p_{n+2};x_1p_1,x_2p_2)\jet{n}{n}{\pset}\nonumber\\
		&&\times2\,\Big\lbrace\Big.\braket{\ampnum{n+2}{0}|\Jcol{1}(\e)|\ampnum{n+2}{1}}+\braket{\ampnum{n+2}{1}|\Jcol{1}(\e)|\ampnum{n+2}{0}}\nonumber\\
		&&\hspace{1cm}-\dfrac{\beta_0}{\e}\braket{\ampnum{n+2}{0}|\Jcol{1}(\e)|\ampnum{n+2}{0}}\Big.\Big\rbrace,
	\end{eqnarray}
	which collects dipole insertions within the one-loop correction,
	\begin{eqnarray}\label{sigUb}
		\dd\sigpart{U,b,\mathrm{IP}}{ab,\mathrm{NNLO}}&=&\coeffVVNNLO\int\dr{x_1}\dr{x_2}\dphi{n}(p_3,\dots,p_{n+2};x_1p_1,x_2p_2)\jet{n}{n}{\pset}\nonumber\\
		&&\times2\,\Big\lbrace\Big.-\braket{\ampnum{n+2}{0}|\Jcol{1}(\e)\otimes\Jcol{1}(\e)|\ampnum{n+2}{0}}\Big.\Big\rbrace,
	\end{eqnarray}
	which addresses double dipole insertions at tree-level and 
	\begin{eqnarray}\label{sigUc}
		\dd\sigpart{U,c,\mathrm{IP}}{ab,\mathrm{NNLO}}&=&\coeffVVNNLO\int\dr{x_1}\dr{x_2}\dphi{n}(p_3,\dots,p_{n+2};x_1p_1,x_2p_2)\jet{n}{n}{\pset}\nonumber\\
		&&\times2\,\Big\lbrace\Big.\braket{\ampnum{n+2}{0}|\Jcol{2}(\e)|\ampnum{n+2}{0}}-{\ampnum{n+2}{0}|\Jcolb{2}(\e)|\ampnum{n+2}{0}}\Big.\Big\rbrace,
	\end{eqnarray}
	which contains two-loop integrated dipoles. We further decompose $\dsigUNNLOshort{a}$ into	
	\begin{eqnarray}\label{sigUa0}
		\dd\sigpart{U,a_0,\mathrm{IP}}{ab,\mathrm{NNLO}}&=&\coeffVVNNLO\int\dr{x_1}\dr{x_2}\dphi{n}(p_3,\dots,p_{n+2};x_1p_1,x_2p_2)\jet{n}{n}{\pset}\nonumber\\
		&&\times2\,\Big\lbrace\Big.-\dfrac{\beta_0}{\e}\braket{\ampnum{n+2}{0}|\Jcol{1}(\e)|\ampnum{n+2}{0}}\Big.\Big\rbrace,
	\end{eqnarray}
	\begin{eqnarray}\label{sigUa1}
		\dd\sigpart{U,a_1,\mathrm{IP}}{ab,\mathrm{NNLO}}&=&\coeffVVNNLO\int\dr{x_1}\dr{x_2}\dphi{n}(p_3,\dots,p_{n+2};x_1p_1,x_2p_2)\jet{n}{n}{\pset}\nonumber\\
		&&\times2\,\Big\lbrace\Big.\braket{\ampnum{n+2}{0}|\Jcol{1}(\e)|\ampnum{n+2}{1}}+\braket{\ampnum{n+2}{1}|\Jcol{1}(\e)|\ampnum{n+2}{0}}\Big.\Big\rbrace,
	\end{eqnarray}
	where we separated the contribution of the one-loop amplitude from the $\beta_0$ term, which only contains tree-level amplitudes. 
	
	We also introduce a decomposition of $\dd\sigpart{U,b,\mathrm{IP}}{ab,\mathrm{NNLO}}$ in~\eqref{sigUb} according to the colour connections among the hard radiators in the two integrated dipoles. After the evaluation of the colour algebra, $\dd\sigpart{U,b,\mathrm{IP}}{ab,\mathrm{NNLO}}$ has the form:
	\begin{equation}
		\dd\sigpart{U,b,\mathrm{IP}}{ab,\mathrm{NNLO}}\sim\sum_{c,c'}\sum_{ij,kl}\Jfull{1}(i,j)\Jfull{1}(k,l)a_{n+2}^0(c,c';.,i,.,j,.,k,.,l,.), 
	\end{equation}
	which we split into the following contributions~\cite{Gehrmann-DeRidder:2005btv,Currie:2013vh}:
	\begin{itemize}
		\item colour-connected ($c.c.$) contributions, where the pair of hard radiators coincides in the two integrated dipoles:
		\begin{equation}\label{sigUbcc}
			\dd\sigpart{U,b,\mathrm{IP},c.c.}{ab,\mathrm{NNLO}}\sim\sum_{c,c'}\sum_{ij}\Jfull{1}(i,j)\Jfull{1}(i,j)a_{n+2}^0(c,c';.,i,.,j,.), 
		\end{equation}
		\item almost colour-connected ($a.c.c.$) contributions, where the two integrated dipoles share one hard radiator $i$:
		\begin{equation}\label{sigUbacc}
		\dd\sigpart{U,b,\mathrm{IP},a.c.c.}{ab,\mathrm{NNLO}}\sim\sum_{c,c'}\sum_{ijk}\Jfull{1}(i,j)\Jfull{1}(i,k)a_{n+2}^0(c,c';.,i,.,j,.,k,.), 
		\end{equation}
		\item colour-unconnected ($c.u.$) contributions, where there is no overlap between the hard radiators in the two integrated dipoles:
		\begin{equation}\label{sigUbcu}
		\dd\sigpart{U,b,\mathrm{IP},u.c.}{ab,\mathrm{NNLO}}\sim\sum_{c,c'}\sum_{ij}\sum_{kl\ne ij}\Jfull{1}(i,j)\Jfull{1}(k,l)a_{n+2}^0(c,c';.,i,.,j,.,k,.,l,.). 
		\end{equation}
	\end{itemize}
	
	Finally,
	 we label the contributions in $\dd\sigpart{U,c}{ab,\mathrm{NNLO}}$ according to the different types of integrated antenna functions which appear in them:
	\begin{itemize}
		\item $\dd\sigpart{U,c,\mathcal{X}_{4}^{0}}{ab,\mathrm{NNLO}}$: integrated four-parton tree-level antenna functions;
		\item $\dd\sigpart{U,c,\mathcal{X}_{3}^{1}}{ab,\mathrm{NNLO}}$: integrated three-parton one-loop antenna functions;
		\item $\dd\sigpart{U,c,{\mathcal{X}_{3}^{0}\otimes\mathcal{X}_{3}^{0}}}{ab,\mathrm{NNLO}}$: convolution of two integrated three-parton tree-level antenna functions;
		\item $\dd\sigpart{U,c,\beta_0}{ab,\mathrm{NNLO}}$: $\beta_0/\e$ terms, with a single integrated three-parton tree-level antenna function.
	\end{itemize}
	Other structures appearing in the two-loop integrated dipoles are not listed here since they originate from mass factorization and do not propagate beyond the double-virtual level. Such structures are two-loop mass factorization kernels, convolutions of two one-loop mass factorization kernels and convolutions of a one-loop mass factorization kernel with an integrated three-parton tree-level antenna function. 
	
	The identity-changing part of the double-virtual subtraction term reflects the structure of the IC mass factorization counterterm:
	\begingroup
	\allowdisplaybreaks
	\begin{eqnarray}\label{sigUIC}
		\dd\sigpart{U,\mathrm{IC}}{ab,\mathrm{NNLO}}&=&-\int\dr{x_1}\dr{x_2}\sum_{(cd)\neq (ab)}\Bigg\lbrace\Bigg.\coeff\left[\Jfulltotic{1}{ab;cd}(x_1,x_2)\left(\dd\sigpart{V}{cd,\mathrm{NLO}}-\dd\sigpart{T}{cd,\mathrm{NLO}}\right)\right]\nonumber\\
		&&\hspace{-2cm}+\coeff^2\Bigg[\Bigg.\sum_{(ef)\neq(ab),(cd)}\dfrac{1}{2}\left[\Jfulltotic{1}{ab;ef}\otimes\Jfulltotic{1}{ef;cd}\right](x_1,x_2)\nonumber\\	&&\hspace{2.7cm}+\,\Jfulltotic{2}{ab;cd}(x_1,x_2)-\dfrac{\beta_0}{\e}\Jfulltotic{1}{ab;cd}(x_1,x_2)\Bigg.\Bigg]\dd\sigpart{}{cd,\mathrm{LO}}\Bigg.\Bigg\rbrace\nonumber \\
		&&\hspace{-2cm}+\coeff^2\sum_{(ef)\neq(ab)}\Bigg[\Bigg.\dfrac{1}{2}\left[\Jfulltotic{1}{ab;ef}\otimes\Jfulltotic{1}{ef;ab}\right](x_1,x_2)+\Jfulltotic{2}{ab;ef;ab}(x_1,x_2)\Bigg.\Bigg]\dd\sigpart{}{ab,\mathrm{LO}}.
	\end{eqnarray}
	\endgroup
	The expression above collects all the structures in the double-virtual IC mass factorization counterterm and combines them with IC integrated NNLO antenna functions. It is overall free from infrared singularities. We notice that in the first line, the difference between the virtual correction and its corresponding subtraction term has no $\e$-poles. Therefore, even if we were to express the virtual subtraction term in colour space to expose its singularity structure, it would 
	not be necessary to do so in this context.
	
	Despite not being expressed in colour space, the IC double-virtual subtraction term exhibits analogous structures to the ones present in its IP counterpart. We can indeed decompose it in the following components:
	\begin{eqnarray}\label{sigUICa0}
		\dd\sigpart{U,\mathrm{IC},a_0}{ab,\mathrm{NNLO}}&=&\coeff^2\sum_{(cd)\neq (ab)}\int\dr{x_1}\dr{x_2}\left(\dfrac{\beta_0}{\e}\Jfulltotic{1}{ab;cd}(x_1,x_2)\right)\dd\sigpart{}{cd,\mathrm{LO}},
	\end{eqnarray}
	\begin{eqnarray}\label{sigUICa1}
		\dd\sigpart{U,\mathrm{IC},a_1}{ab,\mathrm{NNLO}}&=&-\coeff\sum_{(cd)\neq (ab)}\int\dr{x_1}\dr{x_2}\Jfulltotic{1}{ab;cd}(x_1,x_2)\nn\\
		&&\hspace{3cm}\times\left(\dd\sigpart{V}{cd,\mathrm{NLO}}-\dd\sigpart{T}{cd,\mathrm{NLO}}\right)\dd\sigpart{}{cd,\mathrm{LO}},
	\end{eqnarray}
	\begin{eqnarray}\label{sigUICb}
		\dd\sigpart{U,\mathrm{IC},b}{ab,\mathrm{NNLO}}&=&-\dfrac{1}{2}\coeff^2\int\dr{x_1}\dr{x_2}\Bigg\lbrace\Bigg.\nn\\
		&&\sum_{(cd)\neq (ab)}\sum_{(ef)\neq(ab),(cd)}\left[\Jfulltotic{1}{ab;ef}\otimes\Jfulltotic{1}{ef;cd}\right](x_1,x_2)\dd\sigpart{}{cd,\mathrm{LO}}\nn\\
		&&+\sum_{(ef)\neq(ab)}\left[\Jfulltotic{1}{ab;ef}\otimes\Jfulltotic{1}{ef;ab}\right](x_1,x_2)\dd\sigpart{}{ab,\mathrm{LO}}\Bigg.\Bigg\rbrace,
	\end{eqnarray}
	with a subsequent decomposition into colour-connected, almost colour-connected and colour-unconnected contributions as for the IP counterparts, and
	\begin{eqnarray}\label{sigUICc}
		\dd\sigpart{U,\mathrm{IC},c}{ab,\mathrm{NNLO}}&=&-\coeff^2\int\dr{x_1}\dr{x_2}\Bigg\lbrace\Bigg.\sum_{(cd)\neq (ab)}\Jfulltotic{2}{ab;cd}(x_1,x_2)\dd\sigpart{}{cd,\mathrm{LO}}\nonumber \\
		&&\hspace{3cm}+\sum_{(ef)\neq(ab)}\Jfulltotic{2}{ab;ef;ab}(x_1,x_2)\dd\sigpart{}{ab,\mathrm{LO}}\Bigg.\Bigg\rbrace,
	\end{eqnarray}
	which can be further decomposed according to the specific type of integrated structures appearing in each term.
	
	Since the IP and IC components of the double-virtual subtraction term exhibit the same structures in terms of integrated dipoles, in the derivation of the real-virtual and double-real subtraction terms we can always consider the full combination of IP and IC terms, given that the required manipulations to translate integrated quantities to unintegrated ones are analogous. For this reason, in the following sections we will often drop the superscripts IP and IC, indicating that we are working with the complete subtraction terms. We will restore the labels only when needed.
	
	At NLO, once the virtual subtraction term is obtained, it is straightforward to systematically construct the real subtraction term. At NNLO, the structure of the subtraction is significantly more involved, due to the presence of two additional layers besides the double-virtual correction: real-virtual and double-real.
	
	\subsection{NNLO real-virtual subtraction term}\label{sec:RV}
	
    The real-virtual subtraction term $\dd\sigpart{T}{ab,\mathrm{NNLO}}$ cancels the explicit $\e$-poles in the real-virtual matrix element, contains the real-virtual mass factorization counterterm and removes the divergent behaviour in single-unresolved infrared limits. In the following, we illustrate how the real-virtual subtraction term can be generated in the context of the colourful antenna subtraction method. 
    
    We anticipate that the current section is the densest one, with the construction of the real-virtual subtraction term really being the core step of the colourful antenna subtraction method. This may sound surprising, since we are dealing with just a single unresolved emission. However, as we show in Section~\ref{sec:RR}, the price to pay for a particularly straightforward generation of the double-real subtraction term for a generic process, is that we must carefully arrange all the structures in the real-virtual subtraction term in order to achieve this. The reader may notice that some components of the real-virtual subtraction term seem redundant, and that significant cancellations are possible when one considers the full expressions. This is indeed true, but, once again, to properly prepare the derivation of the double-real subtraction term, we are forced to isolate individual structures, even if they yield a simpler result when summed. Of course, such separation is necessary only for the illustration purposes. In any practical implementation of the subtraction terms, one can consider the sum over all components, likely resulting in more compact expressions.
    
    The construction of $\dd\sigpart{T}{ab,\mathrm{NNLO}}$ is performed in two main steps: 
	
	\begin{itemize}
		\item integrated terms are translated from $\dd\sigpart{U}{ab,\mathrm{NNLO}}$ to $\dd\sigpart{T}{ab,\mathrm{NNLO}}$ inserting an unresolved parton, via the application of $\ins{\cdot}$, in complete analogy to what is done at NLO;
		\item additional terms are systematically generated to remove leftover explicit and implicit infrared singularities.
	\end{itemize}
	Any additional contribution which is added at the real-virtual level and does not have a direct correspondence to terms in $\dd\sigpart{U}{ab,\mathrm{NNLO}}$ will eventually generate corresponding terms at the double-real level after the insertion of a second unresolved parton. We note that not the entirety of $\dd\sigpart{U}{ab,\mathrm{NNLO}}$ will be translated to $\dd\sigpart{T}{ab,\mathrm{NNLO}}$, some terms will undergo a double insertion and directly move from $\dd\sigpart{U}{ab,\mathrm{NNLO}}$ to $\dd\sigpart{S}{ab,\mathrm{NNLO}}$.
	
\begin{figure}[t]
		\centering
		\includegraphics[width=\linewidth,keepaspectratio]{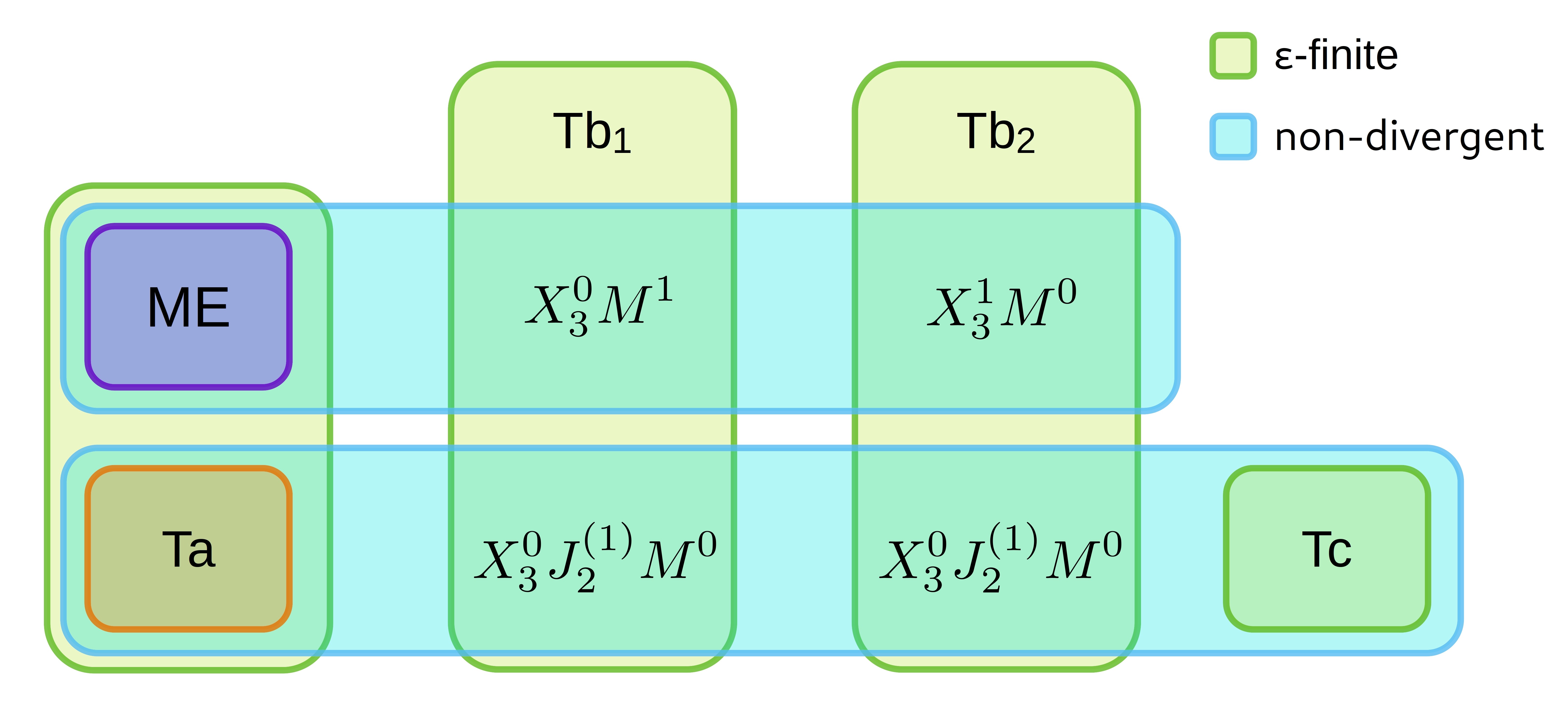}
		\caption{Structure of the infrared-subtracted real-virtual correction. Green boxes enclose $\e$-finite quantities, while inside light-blue ones the infrared-divergent behaviour has been completely subtracted. The combination of the matrix element (ME) and the subtraction terms is free from any explicit and implicit infrared singularity.}\label{fig:RV_struct}
	\end{figure}

	We recall the usual decomposition of $\dd\sigpart{T}{ab,\mathrm{NNLO}}$ in the context of antenna subtraction~\cite{Currie:2013vh}:
	\begin{equation}
		\dd\sigpart{T}{ab,\mathrm{NNLO}}=\dd\sigpart{T,a}{ab,\mathrm{NNLO}}+\dd\sigpart{T,b}{ab,\mathrm{NNLO}}+\dd\sigpart{T,c}{ab,\mathrm{NNLO}}.
	\end{equation}
	The meaning of this decomposition is the following:
	\begin{itemize}
		\item $\dd\sigpart{T,a}{ab,\mathrm{NNLO}}$ removes the explicit poles of the real-virtual matrix element;
		\item $\dd\sigpart{T,b}{ab,\mathrm{NNLO}}$ reproduces the divergent behaviour of the real-virtual matrix element in single unresolved limits, but requires additional contributions to ensure $\e$-finiteness;
		\item $\dd\sigpart{T,c}{ab,\mathrm{NNLO}}$  removes the overlap in the
		single unresolved behaviour between the two terms above.
	\end{itemize}
	We summarize in Figure~\ref{fig:RV_struct} the interplay between different contributions in the real-virtual subtraction term and the matrix element, denoted with ME. Green boxes are free from explicit $\e$-poles, namely the quantities they contain are constructed to be $\e$-finite. Light-blue boxes denote combinations of terms which are not divergent in single-unresolved limits. One can observe that the sum of the matrix element and the subtraction terms is overall free from any infrared singularity. Each term appearing in Figure~\ref{fig:RV_struct} is discussed in detail in the following.
	
	\subsubsection{\text{d}\boldmath{$\sigma^{T,a}_{ab,\mathrm{NNLO}}$}}\label{sec:dsigTa}
	
	This component of the real-virtual subtraction term removes the explicit poles of the $(n+3)$-particle one-loop matrix element. Moreover, it contains the mass factorization contribution $\dd\sigpart{MF,1,a}{ab,\mathrm{NNLO}}$ given in~\eqref{MF1a}. It is indicated as the orange box in Figure~\ref{fig:RV_struct}. The construction of $\sigma^{T,a}_{ab,\mathrm{NNLO}}$ is completely analogous to the one adopted for the NLO virtual subtraction term in Section~\ref{sec:NLOV}, with the only difference given by the presence of an additional particle. Hence, we have:
	\begin{equation}
		\dd\sigpart{T,a,\mathrm{IP}}{}=\dd\sigpart{T,a,\mathrm{IP}}{ab,\mathrm{NNLO}}+\dd\sigpart{T,a,\mathrm{IC}}{ab,\mathrm{NNLO}},
	\end{equation}
	with
	\begin{eqnarray}\label{sigTaIP}
		\dd\sigpart{T,a,\mathrm{IP}}{ab,\mathrm{NNLO}}&=&\coeffRVNNLO\int\dr{x_1}\dr{x_2}\dphi(p_3,\dots,p_{n+3};x_1 p_1,x_2 p_2)\jet{n}{n+1}{\lb p \rb_{n+1}}\nonumber\\
		&&\times 2\braket{\ampnum{n+3}{0}|\Jcol{1}(\e)|\ampnum{n+3}{0}},
	\end{eqnarray}
	and
	\begin{equation}\label{sigTaIC}
		\dd\sigpart{T,a,\mathrm{IC}}{ab,\mathrm{NNLO}}=\coeff\sum_{(c,d)\neq(a,b)}\int\dr{x_1}\dr{x_2}\Jfulltotic{1}{ab;cd}(x_1,x_2)\,\dd\sigpart{R}{cd,\mathrm{NLO}}.
	\end{equation}
	This term is not directly related to any contribution at the double-virtual level. Indeed, its unintegrated counterpart removes the single-unresolved behaviour of the double-real matrix element and will be subtracted there.
	
	\subsubsection{\text{d}\boldmath{$\sigma^{T,b}_{ab,\mathrm{NNLO}}$}}\label{sec:dsigTb}
	
	This term reproduces the divergent behaviour of the real-virtual matrix element in single unresolved infrared limits. The subtraction of infrared divergences from a one loop matrix element is more involved than the one required at tree-level. In particular, along with tree-level antenna functions and reduced one-loop matrix elements (tree $\times$ loop), suitable combinations of three-particle one-loop antenna functions and tree-level reduced matrix elements (loop $\times$ tree) have to be used. Both these structures are present, in their integrated form, in the double-virtual subtraction term, since they describe the emission of soft and collinear particles from the one-loop amplitude. It is indeed possible to systematically generate these terms at the real-virtual level starting from $\dsigUNNLO{ab}$, as we show in the following. According to~\cite{Currie:2013vh}, we introduce a suitable decomposition of $\dsigTNNLOshort{b}$:
	\begin{equation}\label{Tb_decomposition}
		\dsigTNNLOshort{b}=\dsigTNNLOshort{b_1}+\dsigTNNLOshort{b_2}+\dsigTNNLOshort{b_3}, 
	\end{equation}
	where the elements on the right-hand-side respectively contain (tree $\times$ loop) contributions, (loop $\times$ tree) contributions and suitable terms needed to ensure the correct renormalization of loop quantities in the real-virtual subtraction term. Moreover, $\dsigTNNLOshort{b_1}$ and $\dsigTNNLOshort{b_2}$ need to be made $\e$-finite including additional terms, as shown in Figure~\ref{fig:RV_struct}. $\dsigTNNLOshort{b_1}$ and $\dsigTNNLOshort{b_2}$ are indicated in Figure~\ref{fig:RV_struct} by the two central green boxes.
	
	\paragraph{\text{d}\boldmath{$\sigma^{T,b_1}_{ab,\mathrm{NNLO}}$}}
	
	We first focus on $\dsigTNNLOshort{b_1}$, which addresses configurations where the infrared divergences can be captured by the same tree-level unresolved factors used at NLO. The respective integrated counterpart appears in the double-virtual subtraction term as a combination of a one-loop integrated dipole (integrated tree-level three-parton antenna function) and a one-loop reduced matrix element in $\dsigUNNLOshort{a_1}$. The procedure of inserting an unresolved parton at this level is the same as the one depicted in Section~\ref{sec:NLOR}, with the tree-level amplitudes replaced by the one-loop ones. The result has the (tree $\times$ loop) structure:
	\begin{equation}\label{treexloop_1}
		\ins{\dsigUNNLOshort{a_1}}\sim \sum_{ij,u}\sum_{c,c'} \,X_{3}^{0}(i,u,j)\,\anum{n+2}{1}(c,c';\{.,\widetilde{iu},.,\widetilde{uj},.\}),
	\end{equation}
	where $u$ denotes the inserted unresolved parton. Even if the used notation is technically only appropriate in the context of unresolved-gluon insertions, we rely on it here and in the following examples for the sake of simplicity. 
	
	Terms in $\ins{\dsigUNNLOshort{a_1}}$ partially take care of the divergent behaviour of the real-virtual matrix element. However, the one-loop colour interferences $\anum{n+2}{1}$ contain explicit $\e$-poles which must be removed to ensure the finiteness of the real-virtual subtraction term. This only affects the IP sector of the subtraction terms, with the IC one being free of any explicit $\e$-poles.
	
	 Omitting the phase space integration and other factors which are not relevant for the current discussion, these poles read:
{	\allowdisplaybreaks	\begin{eqnarray}\label{UaUbrel1}
		\lefteqn{\poles\lb\ins{\dsigUNNLOshort{a_1}}\rb } \nonumber \\
&	\sim& \poles\lb\mathcal{I}ns\Bigg[\Bigg.2\left(\braket{\ampnum{n+2}{0}|\Jcol{1}|\ampnum{n+2}{1}}+\braket{\ampnum{n+2}{1}|\Jcol{1}|\ampnum{n+2}{0}}\right)\Bigg.\Bigg]\rb \nonumber\\
	    &=&\lefteqn{\poles\lb\mathcal{I}ns\Bigg[\Bigg.2\sum_{ij}\Jfull{1}(i,j)\left(\braket{\ampnum{n+2}{0}|\left(\T_{i}\cdot\T_{j}\right)|\ampnum{n+2}{1}}+\braket{\ampnum{n+2}{1}|\left(\T_{i}\cdot\T_{j}\right)|\ampnum{n+2}{0}}\right)\Bigg.\Bigg]\rb}\nonumber\\
		&=&\poles\lb4\,\mathcal{I}ns\Bigg[\Bigg.\sum_{ij}\Jfull{1}(i,j)\text{Re}\left(\braket{\ampnum{n+2}{0}|\left(\T_{i}\cdot\T_{j}\right)|\ampnum{n+2}{1}}\right)\Bigg.\Bigg]\rb\nonumber\\
		&=&4\sum_{ij,u}X_3^0(i,u,j)\,\poles\lb\text{Re}\left(\braket{\ampnum{\wt{n+2}}{0}|\left(\T_{i}\cdot\T_{j}\right)|\ampnum{\wt{n+2}}{1}}\right)\rb\nonumber\\
		&=&4\sum_{ij,u}X_3^0(i,u,j)\,\poles\lb\text{Re}\left(\braket{\ampnum{\wt{n+2}}{0}|\left(\T_{i}\cdot\T_{j}\right)\left(\sum_{kl}\left(\T_{k}\cdot\T_{l}\right)\ourIop{1}{kl}{\e,\mu_r^2}\right)|\ampnum{\wt{n+2}}{0}}\right)\rb\nonumber\\
		&=&4\sum_{ij,u}X_3^0(i,u,j)\sum_{kl}\,\poles\lb\text{Re}\left[\ourIop{1}{kl}{\e,\mu_r^2}\right]\rb\braket{\ampnum{\wt{n+2}}{0}|\left(\T_{i}\cdot\T_{j}\right)\left(\T_{k}\cdot\T_{l}\right)|\ampnum{\wt{n+2}}{0}},
	\end{eqnarray}}
	where $(\wt{n+2})$ is a shorthand notation to indicate that the momenta of the external partons in the colour-ordered amplitudes underwent a suitable relabelling according to the action of the $\ins{\cdot}$ operator. In particular, the momenta associated to partons $k$ and $l$, which enter in the one-loop infrared operator, belong to the relabelled set of momenta.

 These poles can be reproduced systematically by
 applying the unresolved parton insertion to the 
 $\dd\sigpart{U,b}{ab,\mathrm{NNLO}}$, namely within the double dipole structure:
	\begin{equation}
		\ins{\dsigUNNLOshort{b}}\sim\mathcal{I}ns\left[-2\braket{\ampnum{n+2}{0}|\Jcol{1}(\e)\otimes\Jcol{1}(\e)|\ampnum{n+2}{0}}\right].
	\end{equation}

	We need to clarify how the unresolved parton insertion works in the presence of more than one integrated dipole. The $\ins{\cdot}$ operator discards the splitting kernels from both the integrated dipoles $\Jcol{1}$. The transition to unintegrated antenna functions is then performed in only one of the integrated dipoles (integrated antenna functions). The choice of integrated dipole should be done in such a way that any pair of hard radiators is addressed once and only once. In practice, this can be easily achieved by symmetrizing:
	\begin{equation}
		\Jcol{1}_a\otimes\Jcol{1}_b=\dfrac{1}{2}\Jcol{1}_a\otimes\Jcol{1}_b+\dfrac{1}{2}\Jcol{1}_b\otimes\Jcol{1}_a,
	\end{equation}
	where the subscripts $a$ and $b$ have been introduced to distinguish the two dipoles, and then fixing the first dipole in each term of the symmetrized expression to be the one subjected to the insertion of an unresolved parton. Finally, the momenta relabelling has to be performed, not only within the colour interferences and the jet function, but also within the surviving integrated antenna functions. This applies in general at NNLO, namely the transition to higher multiplicities via momenta relabelling occurs within any function of the external momenta accompanying the integrated dipole (antenna function) which is converted into an unintegrated antenna function. With this, we can write:
	\begin{eqnarray}\label{UaUbrel2}
		\lefteqn{\poles\lb\ins{-2\,\dsigUNNLOshort{b}}\rb\sim4\,\poles\lb\mathcal{I}ns\left[\braket{\ampnum{n+2}{0}|\Jcol{1}(\e)\otimes\Jcol{1}(\e)|\ampnum{n+2}{0}}\right]\rb}\nn \\
		&=&4\,\poles\lb\mathcal{I}ns\left[\sum_{ij}\sum_{kl}\Jfull{1}(i,j)\Jfull{1}(k,l)\braket{\ampnum{n+2}{0}|\left(\T_i\cdot\T_j\right)\left(\T_k\cdot\T_l\right)|\ampnum{n+2}{0}}\right]\rb\nn\\
		&=&4\sum_{ij,u}X_3^0(i,u,j)\sum_{k,l}\poles\lb s_{\mathcal{X}}\mathcal{X}_3^0(s_{kl})\rb\braket{\ampnum{\wt{n+2}}{0}|\left(\T_i\cdot\T_j\right)\left(\T_k\cdot\T_l\right)|\ampnum{\wt{n+2}}{0}},
	\end{eqnarray}
	where $s_{\mathcal{X}}$ indicates the specific symmetry factor and sign relating integrated dipoles and antenna functions. The obtained results is already quite similar to~\eqref{UaUbrel1}, but one final manipulation is needed to compare the two expressions. The goal is to convert the integrated antenna function in~\eqref{UaUbrel2} back to an integrated one-loop dipole. One can indeed verify that the mass factorization kernels which restore the integrated dipoles are precisely provided by $\dd\sigpart{MF,1,b}{ab,\mathrm{NNLO}}$. Hence, one has:
	\begin{eqnarray}\label{Uaubrel3}
		\poles\lb\ins{-2\dsigUNNLOshort{b}}-\dd\sigpart{MF,1,b}{ab,\mathrm{NNLO}}\rb&&\nn \\
		&&\hspace{-6.5cm}\sim4\sum_{ij,u}X_3^0(i,u,j)\sum_{k,l}\poles\lb \Jfull{1}(k,l)\rb\braket{\ampnum{\wt{n+2}}{0}|\left(\T_i\cdot\T_j\right)\left(\T_k\cdot\T_l\right)|\ampnum{\wt{n+2}}{0}},
	\end{eqnarray}
	where we note that the momenta $k$ and $l$ belong to the relabelled set according to the action of the insertion operator. We can finally use~\eqref{J21relation} to conclude that~\eqref{UaUbrel1} and~\eqref{Uaubrel3} are equivalent, so:
	\begin{equation}\label{UaUbrel4}
		\poles\lb\ins{\dsigUNNLOshort{a_1}}\rb=\poles\lb-2\,\ins{\dsigUNNLOshort{b}}-\dd\sigpart{MF,1,b}{ab,\mathrm{NNLO}}\rb. 
	\end{equation}

	This relation provides a natural and general way to produce terms to remove the residual explicit poles in $\poles\lb\ins{\dsigUNNLOshort{a_1}}\rb$. Schematically, these terms look like:
	\begin{equation}\label{Ubstructure}
		-2\,\ins{\dsigUNNLOshort{b}}-\dd\sigpart{MF,1,b}{}\sim\sum_{ij,u}\sum_{kl}\sum_{c,c'}X_{3}^{0}(i,u,j)\Jfull{1}(k,l)\anum{n+2}{0}(c,c';\{.,\widetilde{iu},.,\widetilde{uj},.\}).
	\end{equation}
		It is particularly convenient that this can be achieved by the action of the $\ins{\cdot}$ operator on pre-existent structures at the double-virtual level and fully absorbing $\dd\sigpart{MF,1,b}{ab,\mathrm{NNLO}}$. 
	
	The final result for $\dsigTNNLOshort{b_1}$ then reads:
	\begin{equation}\label{sigTb1}
		\dsigTNNLOshort{b_1}=-\ins{\dsigUNNLOshort{a_1}}-2\,\ins{\dsigUNNLOshort{b}}-\dd\sigpart{MF,1,b}{ab,\mathrm{NNLO}},
	\end{equation}
	which, thanks to~\eqref{UaUbrel4}, is free of explicit $\e$-singularities. To compensate $\dsigUNNLOshort{b}$ in the double-virtual subtraction term, we should add back a factor $+\ins{\dsigUNNLOshort{b}}$, which will appear below in the generation of the real-virtual and double-real subtraction terms. In Figure~\ref{fig:RV_struct}, the (tree $\times$ loop) component of $\dsigTNNLOshort{b_1}$ is indicated by the label $X_3^0M^1$, while additional contributions to remove the $\e$-poles are indicated by $X_3^0\J{1}M^1$.

	As a cross-check for the strategy described above, we can extract the explicit singularity structure of the one-loop colour-ordered partial amplitudes, following closely the procedures used 
	for the real-virtual corrections to $pp\to jj$ in~\cite{Chen:2022clm}. 
	For a general process, one can extract the amplitude-level pole structure via suitable projectors in colour space. A generic renormalized $n$-parton one-loop amplitude in colour space is given by:
    \begin{equation}
    	\ket{\ampnum{n}{1}(\lb p\rb_n)}=\sum_{c\in I^{1}} \mathbfcal{C}_{n,c}^{1} \, A^{1}_{n,c}(\lb p\rb_n),
    \end{equation}
    with the coefficients $A^{1}_{n,c}$ being one-loop colour-ordered amplitudes. Each vector $\mathbfcal{C}_{n,c}^{1}$ defines a direction in colour space. By linearly combining the generating vectors $\lb \mathbfcal{C}_{n,c}^{\ell} \rb_c$ one can construct a projector $\boldsymbol{\mathcal{P}}_c$ to single out the component of the full amplitude along a specific direction:
    \begin{equation}
    	\boldsymbol{\mathcal{P}}_c \ket{\ampnum{n}{1}(\lb p\rb_n)} = A^{1}_{n,c}(\lb p\rb_n).
    \end{equation}
    One can then exploit the relation above to isolate the infrared poles of the colour-ordered amplitude $A^{1}_{n,c}(\lb p\rb_n)$:
    \begin{eqnarray}\label{proj1loop}
    	\poles\left(A^{1}_{n,c}(\lb p\rb_n)\right)&=&\poles\left(\boldsymbol{\mathcal{P}}_c \ket{\ampnum{n}{1}(\lb p\rb_n)}\right)\nonumber \\
    	&=&\boldsymbol{\mathcal{P}}_c\,\,\poles\left(\ket{\ampnum{n}{1}(\lb p\rb_n)}\right)\nonumber \\
    	&=&\boldsymbol{\mathcal{P}}_c\,\,\Iop{1}{\e,\mu_r^2}\ket{\ampnum{n}{0}(\lb p\rb_n)},
    \end{eqnarray}
    where we used that the projector commutes with the extraction of the poles. The obtained pole structure can then be inserted in~\eqref{treexloop_1} to 
 obtain   $\poles\lb\ins{\dsigUNNLOshort{a_1}}\rb$, in order to validate~\eqref{UaUbrel4}.

	\paragraph{\text{d}\boldmath{$\sigma^{T,b_2}_{ab,\mathrm{NNLO}}$}}
	
	We consider now $\dsigTNNLOshort{b_2}$, namely (loop $\times$ tree) contributions, addressing configurations which require novel one-loop unresolved factors to described the divergent behaviour due to the unresolved emission. The core part of this term is given by unintegrated three-parton one-loop antenna functions $X_{3}^{1}$ combined with tree-level reduced matrix elements. The integrated counterpart of these terms is contained in $\dsigUNNLOshort{c,\mathcal{X}_{3}^{1}}$. Once again, we can insert an unresolved parton and obtain from $\dsigUNNLOshort{c,\mathcal{X}_{3}^{1}}$ the contribution needed at the real-virtual level. The replacement rules for integrated three-parton one-loop antenna functions are completely analogous to the ones followed by tree-level antenna functions and are listed in Appendix~\ref{app:X31}. We label the resulting contribution $\dsigTNNLOshort{b_2,X_3^1}$. It has the following form:
	\begin{equation}\label{sigTb2X31}
		\dsigTNNLOshort{b_2,X_3^1}=-\ins{\dsigUNNLOshort{c,\mathcal{X}_{3}^{1}}}\sim\sum_{ij,u}\sum_{c,c'} X_{3}^{1}(i,u,j)\anum{n+2}{0}(c,c';\{.,\widetilde{iu},.,\widetilde{uj},.\}).
	\end{equation}
	As it happened for $\dsigTNNLOshort{b_1}$, the expression above contains explicit $\e$-poles coming from the one-loop antenna functions. These singularities need to be removed to obtain a finite subtraction term. This can be done systematically since the singularity structure of the unintegrated three-parton one-loop antenna functions can be expressed by means of one-loop integrated dipoles and three-parton tree-level antenna functions~\mbox{\cite{Gehrmann-DeRidder:2005btv,Currie:2013vh}}. The poles of a generic one-loop three-parton antenna function can be absorbed by the following replacement:
	\begin{equation}\label{X31decomp}
		X_{3}^{1}(i,u,j)\to X_{3}^{1}(i,u,j) + \sum_{(l,m)=1}^{N_X}\J{1}(l,m)X_{3}^{0}(i,u,j) - M_X\J{1}(\widetilde{iu},\widetilde{uj})X_{3}^{0}(i,u,j),
	\end{equation}
	where the sum in the second term runs over the pairs of colour-connected partons in the antenna configuration. The replacement is valid in any kinematics (FF, IF or II). The number of colour-connected pairs $N_X$, as well as the coefficient $M_X$ depend on the specific antenna function. Values for $N_X$ and $M_X$ and the explicit replacements required for all the one-loop antenna functions are listed in Appendix~\ref{app:X31}. The expression obtained inserting~\eqref{X31decomp} in~\eqref{sigTb2X31} is free of poles and becomes part of the real-virtual subtraction term. 
	
	To use a similar notation to~\cite{Currie:2013vh}, we label $\dsigTNNLOshort{b_2,J_X}$ and $\dsigTNNLOshort{b_2,M_X}$ the two additional blocks coming from the newly added terms in~\eqref{X31decomp}. As we will explain in detail in the following,  $\dsigTNNLOshort{b_2,M_X}$ can be related to $\dsigUNNLOshort{c,\mathcal{X}_{3}^{0}\otimes\mathcal{X}_{3}^{0}}$ after the insertion of an unresolved parton, while $\dsigTNNLOshort{b_2,J_X}$ is a genuinely new contribution added at the real-virtual level, which needs to be compensated by its unintegrated counterpart at the double-real level. This can be noticed by looking at the arguments of the integrated dipoles appearing in the two blocks. $\dsigTNNLOshort{b_2,M_X}$ depends on mapped momenta, which come from the insertion of an extra unresolved parton in the $n$-particle phase space, where the double-virtual subtraction term lives. On the other hand, the integrated dipoles in $\dsigTNNLOshort{b_2,J_X}$ depend on $(n+3)$-particle phase space momenta, which are not accessible at the double-virtual level. It is trivial to verify that the mass factorization kernels between the integrated dipoles in $\dsigTNNLOshort{b_2,J_X}$ and $\dsigTNNLOshort{b_2,M_X}$ exactly cancel for any configuration of partons $i$, $j$ and $u$. For bookkeeping purposes we label this mass factorization $\dd\sigpart{MF,1,b_2}{ab,\mathrm{NNLO}}$.
	
	The term  $\dsigTNNLOshort{b_2,M_X}$ can be constructed as follows: 
	\begin{equation}\label{sigTb2MXrel}
		\dsigTNNLOshort{b_2,M_X}=- m_X \ins{\dsigUNNLOshort{c,\mathcal{X}_{3}^{0}\otimes\mathcal{X}_{3}^{0}}}-\dd\sigpart{MF,1,b_2}{ab,\mathrm{NNLO}},
	\end{equation}
	where $m_X$ is an integer. For purely gluonic processes, it was found previously~\cite{Chen:2022ktf} that $m_X=2$. Any multiple of $\ins{\dsigUNNLOshort{c,\mathcal{X}_{3}^{0}\otimes\mathcal{X}_{3}^{0}}}$ that is not compensated by the double-virtual subtraction term will be re-introduced in the construction of $\sigma^{T,c}_{ab,\mathrm{NNLO}}$ below. 
	
	In conclusion, the (loop $\times$ tree) block is given by:
	\begin{equation}\label{sigTb2}
		\dsigTNNLOshort{b_2}=-\ins{\dsigUNNLOshort{c,\mathcal{X}_{3}^{1}}}+\dsigTNNLOshort{b_2,J_X}+\dsigTNNLOshort{b_2,M_X},
	\end{equation}
	which is free of $\e$-poles. In Figure~\ref{fig:RV_struct}, the (loop $\times$ tree) component of $\dsigTNNLOshort{b_2}$ is indicated by the label $X_3^1M^0$, while contributions in $\dsigTNNLOshort{b_2,J_X}$ and $\dsigTNNLOshort{b_2,M_X}$ are generically indicated by $X_3^0\J{1}M^0$.
	
	\paragraph{\text{d}\boldmath{$\sigma^{T,b_3}_{ab,\mathrm{NNLO}}$}}
	
	The last contribution to $\dsigTNNLOshort{b}$ is required to fix the renormalization of one-loop antenna functions in $\dsigTNNLOshort{b_2}$. To fix the correct renormalization of the one-loop antenna functions it is sufficient to perform the following replacements~\cite{Gehrmann-DeRidder:2005btv}:
	\begin{eqnarray}
		X_{3}^{1}(i,u,j)&\to& X_{3}^{1}(i,u,j) +\dfrac{b_0}{\e}X_{3}^{0}(i,u,j)\left(\left(\dfrac{\left|s_{iuj}\right|}{\mu_r^2}\right)^{-\e}-1\right),\\
		\hat{X}_{3}^{1}(i,u,j)&\to& X_{3}^{1}(i,u,j) +\dfrac{b_{0,F}}{\e}X_{3}^{0}(i,u,j)\left(\left(\dfrac{\left|s_{iuj}\right|}{\mu_r^2}\right)^{-\e}-1\right),
	\end{eqnarray}
	for the leading-colour and the fermionic-loop one-loop antenna functions. The subleading-colour one-loop antenna functions do not require any renormalization. $\dsigTNNLOshort{b_3}$ is entirely constructed with terms coming from the double-virtual subtraction term:
	\begin{equation}\label{sigTb3}
		\dsigTNNLOshort{b_3}=-\ins{\dsigUNNLOshort{a_0}}-\ins{\dsigUNNLOshort{c,\beta_0}},
	\end{equation}
	as can be easily checked keeping track of the coefficient $\beta_0$.
	
	The contribution $\dsigTNNLOshort{b_3}$ is not depicted explicitly in Figure~\ref{fig:RV_struct}, since we assume it is absorbed in the (loop $\times$ tree) term of $\dsigTNNLOshort{b_2}$, when the three-parton one-loop antenna functions are evaluated at the correct renormalization scale.
	
	\subsubsection{\text{d}\boldmath{$\sigma^{T,c}_{ab,\mathrm{NNLO}}$}}\label{sec:dsigTc}
	
	The last block of the real-virtual subtraction term is $\dsigTNNLOshort{c}$. This contribution is significantly less straightforward to derive than $\sigma^{T,a}_{ab,\mathrm{NNLO}}$ and $\sigma^{T,b}_{ab,\mathrm{NNLO}}$, discussed in the previous sections. The reason for this lies in less stringent constraints for the generation of new structures, which propagate to the double-real subtraction term. Typically, the correctness of the $\sigma^{T,c}_{ab,\mathrm{NNLO}}$ term can only be fully assessed by cross-checking its influence at the real-virtual and double-real level. Moreover, the $\sigma^{T,c}_{ab,\mathrm{NNLO}}$ contribution in general has an exiguous numerical impact, which makes it hard to clearly detect mistakes during the validation process. We also notice that the remainder of this section differs significantly from Section $4.4.3$ of~\cite{Chen:2022ktf}, because the extension beyond gluonic processes required a reinterpretation of the derivation strategy. Indeed, the completely symmetric structure of the subtraction terms in the gluons-only case allowed for a straightforward generation of the $\sigma^{T,c}_{ab,\mathrm{NNLO}}$ term, which is not directly applicable to other partonic configurations.
	
	We begin by observing Figure~\ref{fig:RV_struct}: $\dsigTNNLOshort{c}$ has to remove residual divergent behaviour present in the combination of $\dsigTNNLOshort{a}$ and the $X_3^0\J{1}M^0$ terms of  $\dsigTNNLOshort{b}$ and has to be individually free from $\e$-poles. We address these two requirements in this order. 
	
	\paragraph{Removal of single-unresolved behaviour}
	
	The first requirement is fulfilled designing a preliminary candidate for $\dsigTNNLOshort{c}$, denoted with $\dsigTNNLOshort{c,\text{prel.}}$, given by:
	\begin{equation}\label{sigTcprel}
		\dsigTNNLOshort{c,\text{prel.}}=-\left[\dsigTNNLOshort{a}\Bigg|_{\text{singular}}+\dsigTNNLOshort{b_1}\Bigg|_{X_3^0\J{1}M^0}+\dsigTNNLOshort{b_2}\Bigg|_{X_3^0\J{1}M^0}\right],
	\end{equation}
	where the first term represents the singular behaviour of $\dsigTNNLOshort{a}$ in single-unresolved limits, while the remaining contributions are simply identical to the terms with a $X_3^0\J{1}M^0$ structure in $\dsigTNNLOshort{b_1}$ and $\dsigTNNLOshort{b_2}$. By construction, $\dsigTNNLOshort{c,\text{prel.}}$ trivially fulfils its purpose, but a systematic procedure to construct $\dsigTNNLOshort{a}\big|_{\text{singular}}$ must be defined. The caveat is to have $\dsigTNNLOshort{a}\big|_{\text{singular}}$ extracted by means of three-parton tree-level antenna functions, to produce a $\dsigTNNLOshort{c,\text{prel.}}$ which coherently fits within the subtraction terms.
	
	The structure of $\dsigTNNLOshort{a}$ discussed in Section~\ref{sec:dsigTa} is the one for an NLO virtual subtraction term for a process with $(n+3)$ partons:
	\begin{equation}\label{exTa}
		\dsigTNNLOshort{a}\sim\sum_{c,c'}\sum_{ij} \J{1}(i,j)a_{n+3}^0(c,c';\{.,i,.,j,.\}).
	\end{equation}
	The colour correlations $a_{n+3}^0(c,c';\{.\})$ exhibit a divergent behaviour in single-unresolved limits. The expression in~\eqref{exTa} is clearly not a physical matrix element, therefore we cannot reproduce its divergent behaviour with the same techniques we employed for the construction of NLO subtraction terms. Nevertheless, the full $\dsigTNNLOshort{a}\big|_{\text{singular}}$ can be obtained if in each term of~\eqref{exTa} we replace the colour interference $a_{n+3}^0(c,c';\{.,i,.,j,.\})$ with the description of its divergent behaviour in terms of antenna functions. In other words, we need a systematic procedure to construct a real subtraction term for an individual colour interference $a_{n+3}^0(c,c';\{.\})$. Unfortunately, the paradigm of the colourful antenna subtraction: infer real emission subtraction terms from virtual structures, cannot be applied in this case. The reason lies in the fact that it is not possible to identify a one-to-one correspondence between single terms in the colour decomposition of the matrix elements for virtual and real corrections. Hence, for this specific task, we need to rely on the factorization properties of tree-level amplitudes in single-unresolved limits.
	
	The strategy we adopt is similar to the one described in Section~\ref{sec:dsigTb} for the extraction of the explicit singularities of one-loop colour-ordered amplitudes. We first exploit universal factorization properties of QCD to describe the singular behaviour of tree-level amplitudes and then rely on suitable projection operators to single out a specific direction in colour space, effectively extracting the singular behaviour of single colour-ordered amplitudes. We recall the decomposition of a $n$-parton tree-level amplitude:
	\begin{equation}
		\ket{\ampnum{n}{0}(\lb p\rb_n)}=\sum_{c\in I^{0}} \mathbfcal{C}_{n,c}^{0} \, A^{0}_{n,c}(\lb p\rb_n),
	\end{equation}
	where the vectors $\lb \mathbfcal{C}_{n,c}^{0} \rb_n$ span the entire $n$-parton tree-level colour space. We can construct projectors $\proj{c}$ which satisfy:
	\begin{equation}
		\proj{c}\ket{\ampnum{n}{0}(\lb p\rb_n)}=A^{0}_{n,c}(\lb p\rb_n).
	\end{equation}
	
	We now consider the emission of a soft gluon, labelled as usual with $u$. The factorization reads:
	\begin{equation}
		\ket{\ampnum{n}{0}(\lb p\rb_{n})}\xrightarrow{u\to 0}\sum_{i=1}^{n-1}\T_i\dfrac{p_i^{\mu}}{p_u\cdot p_i}\ket{\ampnum{n-1}{0}(\lb p\rb_{n-1})},
	\end{equation}
	where $\mu$ is the Lorentz index associated to the soft gluon and $\T_i$ are the colour charges of the hard partons. The $(n-1)$-parton amplitude on the RHS is obtained removing the soft gluon $u$ from the partonic content of the original amplitude. If we are interested in the soft behaviour of a specific colour-ordered amplitude $A^{0}_{n,c}(\lb p\rb_{n})$, we can make use of the projectors defined above:
	\begin{eqnarray}
		A^{0}_{n,c}(\lb p\rb_{n})\xrightarrow{u\to 0}\proj{c}\left(\sum_{i=1}^{n-1}\T_i\dfrac{p_i^{\mu}}{p_u\cdot p_i}\ket{\ampnum{n-1}{0}(\lb p\rb_{n-1})}\right).
	\end{eqnarray}
	We can apply this result to a generic  interference between amplitudes with colour-ordering $c$ and $c'$:
	\begin{eqnarray}\label{Tasing0}
		&&\hspace{-0.5cm}a^{0}_{n}(c,c';\lb p\rb_{n})=A^{0}_{n,c}(\lb p\rb_{n})^{\dagger}A^{0}_{n,c'}(\lb p\rb_{n})\xrightarrow{u\to 0}\nn\\
		&&\hspace{-0.5cm}\left[\proj{c}\left(\sum_{i=1}^{n-1}\T_i\dfrac{p_i^{\mu}}{p_u\cdot p_i}\ket{\ampnum{n-1}{0}(\lb p\rb_{n-1})}\right)\right]^{\dagger}\left[\proj{c'}\left(\sum_{j=1}^{n-1}\T_j\dfrac{p_j^{\mu}}{p_u\cdot p_j}\ket{\ampnum{n-1}{0}(\lb p\rb_{n-1})}\right)\right].
	\end{eqnarray}
	The expression above can be evaluated contracting colour and Lorentz indices to obtain a real result describing the soft behaviour of a generic colour interference. The results, up to numerical factors, has the following form:
	\begin{equation}\label{Tasing1}
		a^{0}_{n}(c,c';\lb p\rb_{n})\xrightarrow{u\to 0}\sum_{\tilde{c},\tilde{c}'}\sum_{ij}S^0_{iuj}a^{0}_{n-1}(\tilde{c},\tilde{c}';.,\wt{iu},.,\wt{uj},.),
	\end{equation}
	where the first sum runs over colour structures $\tilde{c}$ and $\tilde{c}'$ in the $(n-1)$-parton tree-level colour space and $S^0_{iuj}$ is the soft eikonal factor given:
	\begin{equation}\label{eikonal}
		S^0_{ijk}=\dfrac{2s_{ik}}{s_{ij}s_{jk}}.
	\end{equation} 
	
	We can build on the result in~\eqref{Tasing1} exchanging the eikonal factors with suitable three-parton tree level antenna functions. Indeed, $X_3^0(i,u,j)$ antenna functions, exactly reproduce the eikonal factor $S^0_{iuj}$ when a soft gluon $u$ is emitted between the pair of hard radiators $(i,j)$. The specific choice of antenna function depends on the species and the kinematical configurations of partons $i$ and $j$. Additionally, antenna functions also contain the collinear behaviour in the limits $i\parallel u$ and $j\parallel u$, with the collinear limit $g\parallel g$ is not captured by a single $X_3^0$ antenna function, but is distributed between pairs of antenna functions where the role of the hard and unresolved gluons are interchanged. We can therefore take the result in~\eqref{Tasing1}, with eikonal factors suitably replaced by antenna functions, and perform a sum over all possible choices of soft gluon $u$. The final result has the following structure:
	\begin{equation}\label{Tasing2}
		a^{0}_{n}(c,c';\lb p\rb_{n})\xrightarrow{u\to 0}\sum_{\tilde{c},\tilde{c}'}\sum_{ij,u}X_3^0(i,u,j)a^{0}_{n-1}(\tilde{c},\tilde{c}';.,\wt{iu},.,\wt{uj},.),
	\end{equation}
	which is capturing not only the soft behaviour of the considered colour correlation in all possible soft limits, but also the correct collinear behaviour in any collinear $g\parallel g$ and $g\parallel q$ configuration. In particular, we notice that the described procedure also retains IC collinear limits for a hard initial-state gluon and a final-state quark. This is straightforwardly achieved by letting the initial-state gluon play the role of the unresolved one, and then by replacing the eikonal factor with suitable IC IF or II antenna functions, which do not contain any soft limit.
	
 Finally the  treatment $q'\parallel \qb'$ collinear limits is quite simple, since these limits are only present in colour-interferences where $q'$ and $\qb'$ are colour connected in both ordering $c$ and $c'$. In this case, the factorization reads:
	\begin{equation}\label{qqbcoll}
		a^{0}_{n}(c,c';.,q',\qb',.)\xrightarrow{u\to 0}\dfrac{1}{s_{q'\qb'}}P^0_{q\qb\to g}(x)\,a^{0}_{n-1}(\tilde{c},\tilde{c}';.,\wt{q'\qb'},.),
	\end{equation}
	where $\wt{q'\qb'}$ represent a gluon, which, in the colour orderings $\tilde{c}$ and $\tilde{c}'$ of the reduced matrix element, occupies the same positions of the original $q'\qb'$ pair. As far as the other hard partons are concerned, $\tilde{c}$ and $\tilde{c}'$ coincide with $c$ and $c'$. The splitting function $P^0_{q\qb\to g}/s_{q'\qb'}$ can then be replaced by an appropriate antenna function choosing a spectator parton for the collinear limit. If the spectator is chosen to be a gluon $g$, $G_3^0(g,q',\qb')$ is used, otherwise, if it is a quark $q$,  $E_3^0(q,q',\qb')$ is considered. Any choice yields the same result in the collinear limit, given that the spectator parton in these antenna functions is not associated with any singular behaviour. 
	
	We notice that the extraction of the single-unresolved behaviour of colour-ordered interferences discussed here is completely analogous to the description of infrared limits done in the context of the dipole subtraction formalism~\cite{Catani:1996jh}. We focused here on specific colour orderings, or equivalently specific directions in colour space, rather than on the full matrix element, in order to derive the correct factorization behaviour of a more generic object, namely $\dsigTNNLOshort{a}$.
	
	With the described procedure, we can systematically write down the singular behaviour of $\dsigTNNLOshort{a}$ in single-unresolved limits and consequently assemble $\dsigTNNLOshort{c,\text{prel.}}$.

	\paragraph{Pole cancellation}
	
	The outcome of the first part of the current derivation, $\dsigTNNLOshort{c,\text{prel.}}$, has explicit $\e$-poles, which need to be removed without spoiling the infrared behaviour of the whole expression.
	
	We consider the generic structure of $\dsigTNNLOshort{c,\text{prel.}}$, which is given by~\cite{Currie:2013dwa,Chen:2022ktf}:
	\begin{eqnarray}\label{structTc1}
		&&\dsigTNNLOshort{c,\text{prel.}}\sim \sum_{c,c'}\sum_{ij,u}\sum_{a,b}X_3^0(i,u,j)\Big[\Big.\nn\\
		&&\hspace{2cm}\phantom{-}\left(\Jfull{1}(\wt{iu},\wt{uj})-\Jfull{1}(i,j)\right)\nn\\
		&&\hspace{2cm}-\left(\Jfull{1}(\wt{iu},a)-\Jfull{1}(i,a)\right)\nn\\
		&&\hspace{2cm}-\left(\Jfull{1}(b,\wt{uj})-\Jfull{1}(b,j)\right)\Big.\Big]a^0_{n+2}(c,c';.,a,.,\wt{iu},.,\wt{ju},.,b,.),
	\end{eqnarray} 
	where we chose to use integrated colour-stripped one-loop dipoles, but we could have equivalently used integrated three-parton tree-level antenna functions $\mathcal{X}_3^0$, since the mass factorization kernels cancel exactly in the three differences of integrated dipoles. The combination above has residual $\e^{-1}$ poles, whose origin can be related to soft gluons emitted at large angle in the double-real emission correction. As discussed in~\cite{Weinzierl:2008iv,Gehrmann-DeRidder:2007foh}, such poles cancel against a suitable combination of integrated soft eikonal factors. One can verify that the poles structure given by the arrangement of integrated dipoles in~\eqref{structTc1} is:
	\begin{eqnarray}\label{structTc2}
		&&\poles\Big\lbrace\Big.\left(\Jfull{1}(\wt{iu},\wt{uj})-\Jfull{1}(i,j)\right)-\left(\Jfull{1}(\wt{iu},a)-\Jfull{1}(i,a)\right)\nn\\
		&&\hspace{7cm}-\left(\Jfull{1}(b,\wt{uj})-\Jfull{1}(b,j)\right)\Big.\Big\rbrace\nn\\
		&&\hspace{2cm}=\dfrac{1}{\e}\left\lbrace-\ln\left(\dfrac{s_{(iu)(uj)}}{s_{ij}}\right)+\ln\left(\dfrac{s_{(iu)a}}{s_{ia}}\right)+\ln\left(\dfrac{s_{b(uj)}}{s_{bj}}\right)\right\rbrace.
	\end{eqnarray}
	We note that there are no residual poles in the $N_f$ colour factor~\cite{Gehrmann-DeRidder:2007foh}. The poles structure above is removed introducing the so-called \textit{integrated large angle soft terms}~\cite{Weinzierl:2008iv,Gehrmann-DeRidder:2007foh,Glover:2010kwr,Gehrmann-DeRidder:2011jwo,Currie:2013vh}. We consider the integral of an eikonal factor over the unresolved antenna phase space:
	\begin{equation}\label{LASTint}
		\mathcal{S}(s_{ij},s_{i'j'},x_{ij,i'j'})=\dfrac{8\pi^2}{\overline{C}(\e)}\int\dd\Phi_{X_{iuj}}S^0_{iuj}.
	\end{equation}
	In the expression above, $i$ and $j$ represent any pair of hard partons, while partons $i'$ and $j'$ can be chosen arbitrarily since there is a priori no singular behaviour associated with them. They appear as reference momenta in a phase space mapping for soft gluon radiation between the hard radiators $i$ and $j$, which can, but do not have to, be different from $i'$ and $j'$. All specific arrangements of $i, j$ and $i',j'$ are discussed in Section~\ref{sec:dsigSc} below, when these structures are propagated at the double-real level. In particular, for process involving four or more partons at LO, $i'$ and $j'$ can be chosen to be two final-state partons~\cite{Gehrmann-DeRidder:2011jwo}. This is the reason why in~\eqref{LASTint} we used the FF antenna phase space. Explicit expressions for~\eqref{LASTint} are given in~\cite{Gehrmann-DeRidder:2011jwo}. The key point for the current discussion is that:
	\begin{eqnarray}\label{structTc3}
		\poles\Bigg\lbrace\Bigg.&-&\left(\mathcal{S}(s_{(iu)(uj)},s_{i'j'},x_{(iu)(uj),i'j'})-\mathcal{S}(s_{ij},s_{i'j'},1)\right)\nn\\
		&+&\left(\mathcal{S}(s_{(iu)a},s_{i'j'},x_{(iu)a,i'j'})-\mathcal{S}(s_{ia},s_{i'j'},x_{ia,i'j'})\right)\nn\\
		&+&\left(\mathcal{S}(s_{b(uj)},s_{i'j'},x_{b(uj),i'j'})-\mathcal{S}(s_{bj},s_{i'j'},x_{bj,i'j'})\right)\Bigg.\Bigg\rbrace\nn\\
		&&\hspace{1cm}=-\dfrac{1}{\e}\left\lbrace-\ln\left(\dfrac{s_{(iu)(uj)}}{s_{ij}}\right)+\ln\left(\dfrac{s_{(iu)a}}{s_{ia}}\right)+\ln\left(\dfrac{s_{b(uj)}}{s_{bj}}\right)\right\rbrace,
	\end{eqnarray}
	namely the pole structure of the arrangement of integrated eikonal factors above precisely cancels against the one in~\eqref{structTc2}. The combination:
	\begingroup
	\allowdisplaybreaks
	\begin{eqnarray}\label{structTc4}
		&&\Bigg[\Bigg.\left(\Jfull{1}(\wt{iu},\wt{uj})-\Jfull{1}(i,j)\right)-\left(\Jfull{1}(\wt{iu},a)-\Jfull{1}(i,a)\right)-\left(\Jfull{1}(b,\wt{uj})-\Jfull{1}(b,j)\right)\Bigg.\Bigg]\nn\\
		&&-\Bigg[\Bigg.\left(\mathcal{S}(s_{(iu)(uj)},s_{i'j'},x_{(iu)(uj),i'j'})-\mathcal{S}(s_{ij},s_{i'j'},1)\right)\nn\\
		&&-\left(\mathcal{S}(s_{(iu)a},s_{i'j'},x_{(iu)a,i'j'})-\mathcal{S}(s_{ia},s_{i'j'},x_{ia,i'j'})\right)\nn\\
		&&-\left(\mathcal{S}(s_{b(uj)},s_{i'j'},x_{b(uj),i'j'})-\mathcal{S}(s_{bj},s_{i'j'},x_{bj,i'j'})\right)\Bigg.\Bigg]
	\end{eqnarray}
	\endgroup
	is indeed free from explicit infrared singularities, for any partonic configuration. Therefore, the expression for $\dsigTNNLOshort{c,\text{prel.}}$ in~\eqref{structTc1} can be made $\e$-finite by suitably inserting integrated eikonal factors according to the following replacements:
	\begin{eqnarray}\label{replaceTc}
		\Jfull{1}(\widetilde{iu},\widetilde{uj})&\to& \Jfull{1}(\widetilde{iu},\widetilde{uj})+\mathcal{S}(s_{ij},s_{i'j'},1)-\mathcal{S}(s_{(iu)(uj)},s_{i'j'},x_{(iu)(uj),i'j'}),\\
		\Jfull{1}(\widetilde{iu},a)&\to& \Jfull{1}(\widetilde{iu},a)+\mathcal{S}(s_{ia},s_{i'j'},x_{ia,i'j'})-\mathcal{S}(s_{(iu)a},s_{i'j'},x_{(iu)a,i'j'}),\\
		\Jfull{1}(b,\widetilde{uj})&\to& \Jfull{1}(b,\widetilde{uj})+\mathcal{S}(s_{bj},s_{i'j'},x_{bj,i'j'})-\mathcal{S}(s_{b(uj)},s_{i'j'},x_{b(uj),i'j'}).
	\end{eqnarray}
	We notice that the considered arrangement of integrated eikonal factors vanishes in single unresolved limits, therefore no additional unresolved behaviour is introduced. The complete $\dsigTNNLOshort{c}$ is then obtained performing the substitutions above in $\dsigTNNLOshort{c,\text{prel.}}$. For bookkeeping purposes, we denote as $\dsigTNNLOshort{c,\mathcal{S}}$ the collection of integrated eikonal factors we have included in the subtraction term. It has the following structure:
	\begin{eqnarray}\label{structTc5}
		&&\dsigTNNLOshort{c,\mathcal{S}}\sim -\sum_{c,c'}\sum_{ij,u}\sum_{a,b}X_3^0(i,u,j)\Big[\Big.\nn\\
		&&\hspace{2cm}\phantom{-}\left(\mathcal{S}(s_{ij},s_{i'j'},1)-\mathcal{S}(s_{(iu)(uj)},s_{i'j'},x_{(iu)(uj),i'j'})\right)\nn\\
		&&\hspace{2cm}-\left(\mathcal{S}(s_{ia},s_{i'j'},x_{ia,i'j'})-\mathcal{S}(s_{(iu)a},s_{i'j'},x_{(iu)a,i'j'})\right)\nn\\
		&&\hspace{2cm}-\left(\mathcal{S}(s_{bj},s_{i'j'},x_{bj,i'j'})-\mathcal{S}(s_{b(uj)},s_{i'j'},x_{b(uj),i'j'})\right)\nn\\
		&&\hspace{5cm}\Big.\Big]a^0_{n+2}(c,c';.,a,.,\wt{iu},.,\wt{uj},.,b,.),
	\end{eqnarray}

	The procedure described up to here is sufficient to ensure finiteness in the soft gluon as well as in the $g\parallel g$ and $q\parallel \qb$ collinear limits. In fact, it turns out that in such limits, the contribution of  $\dsigTNNLOshort{c,\text{prel.}}$  is even suppressed by several orders of magnitude with respect to other components of the real-virtual subtraction term, due to a basically complete cancellation of the divergent behaviour in the RHS of~\eqref{sigTcprel}. However, the strategy above requires further extension to handle also $q\parallel g$ collinear limits. 
	 Indeed, in $\dsigTNNLOshort{c,\text{prel.}}$, constructed according to~\eqref{sigTcprel}, contributions proportional to $A$- and $D$-type unintegrated antenna functions do not actually have the structure given in~\eqref{structTc1}. In particular, they do not come with the complete set of three pairs of integrated dipoles, but rather with one or two pairs only. This is explained by the interplay of two phenomena. First of all, we have seen in Section~\ref{sec:int_dip_2} that it is necessary to introduce quark-antiquark integrated dipoles by means of the colour operator $\Jcolb{2}$ to remove unphysical limits present in $D$-type antenna functions. Hence, a propagation of $A$-type antenna functions occurs from $\dsigUNNLOshort{c}$ to the real-virtual subtraction term, which are not matched by corresponding structures coming from $\dsigUNNLOshort{b}$. Additionally, the colour-charge operators for a quark-gluon dipole $\T_q\cdot\T_g$ and quark-antiquark dipole $\T_q\cdot\T_\qb$ contribute to different colour factors when evaluated on a given tree-level or one-loop amplitude. Specifically, $\T_q\cdot\T_\qb$ does not contribute at leading-colour. This yields incomplete sets of integrated dipoles in~\eqref{structTc1} for some choices of hard partons $i$, $j$, $a$ and $b$. The incompleteness of~\eqref{structTc1} prevents the removal of the leftover $\e$-poles by the inclusion of the combination of integrated eikonal factors described above. However, according to our understanding,~\eqref{structTc4} is the only fundamental $\e$-finite structure we can construct for arbitrary partonic configurations, which correctly propagates at the double-real level. The solution to this issue consists of restoring the complete structure in~\eqref{structTc1} by introducing suitable combinations of integrated dipoles, to then proceed with the inclusion of integrated eikonal factors as before. We illustrate how this is done in the following.
	
	We consider a process with a colour-connected quark-antiquark pair denoted with $(q,\qb)$. We focus first on contributions proportional to $D$-type antenna functions, generically indicated in the following by $D_3^0(q,u,g)$, with $g$ a hard gluon serving as the second hard radiator for the antenna, and $u$, as usual an unresolved gluon. The incomplete structure they come with has the following form:
	\begin{eqnarray}\label{structTc6}
		&&\sum_{c,c'}D_3^0(q,u,g)\Bigg[\Bigg.\left(\Jfull{1}(\wt{qu},\wt{ug})-\Jfull{1}(q,g)\right)\nn\\
		&&\hspace{2.3cm}-\left(\Jfull{1}(b,\wt{ug})-\Jfull{1}(b,g)\right)\Bigg.\Bigg]a^0_{n+2}(c,c';.,\wt{qu},.,\wt{ug},.,b,.),
	\end{eqnarray}
	where $b$ indicates a generic hard spectator different from $\qb$. One can see that a pair of integrated dipoles is missing. To fix this, we can add the following term:
	\begin{equation}\label{structTc7}
		-\sum_{c,c'}D_3^0(q,u,g)\left(\Jfull{1}(\wt{qu},\qb)-\Jfull{1}(q,\qb)\right)a^0_{n+2}(c,c';.,\qb.,\wt{qu},.,\wt{ug},.,b,.),
	\end{equation}
	with $\qb$ acting as a spectator along with $q$ in quark-antiquark integrated dipoles. The resulting structure then reads
	\begin{eqnarray}\label{structTc8}
		&&\sum_{c,c'}D_3^0(q,u,g)\Bigg[\Bigg.\left(\Jfull{1}(\wt{qu},\wt{ug})-\Jfull{1}(q,g)\right)\nn\\
		&&\hspace{2.3cm}-\left(\Jfull{1}(b,\wt{ug})-\Jfull{1}(b,g)\right)\nn\\
		&&\hspace{2.3cm}-\left(\Jfull{1}(\wt{qu},\qb)-\Jfull{1}(q,\qb)\right)\Bigg.\Bigg]a^0_{n+2}(c,c';.,\qb.,\wt{qu},.,\wt{ug},.,b,.),
	\end{eqnarray}
	which is analogous to~\eqref{structTc1}.
	
	We address now the case of an generic unintegrated $A$-type antenna function $A_3^0(q,u,\qb)$. In $\dsigTNNLOshort{c,\text{prel.}}$ it comes with a single pair of integrated dipoles:
	\begin{eqnarray}\label{structTc9}
		&&\sum_{c,c'}A_3^0(q,u,\qb)\left(\Jfull{1}(\wt{qu},\wt{u\qb})-\Jfull{1}(q,\qb)\right)a^0_{n+2}(c,c';.,\wt{qu},.,\wt{u\qb},.),
	\end{eqnarray}
	hence we need to add two pairs of integrated dipoles:
	\begin{eqnarray}\label{structTc10}
		&&-\sum_{c,c'}A_3^0(q,u,\qb)\Big[\Big.\left(\Jfull{1}(\wt{qu},q)-\Jfull{1}(q,a)\right)\nn\\
		&&\hspace{2.5cm}-\left(\Jfull{1}(b,\wt{u\qb})-\Jfull{1}(b,\qb)\right)\Big.\Big]a^0_{n+2}(c,c';.,a,.,\wt{qu},.,\wt{u\qb},.,b,.),
	\end{eqnarray}
	with $a$ and $b$ two hard spectators. Differently from the $D$-type antenna case, here there is no natural choice for spectators $a$ and $b$ and they can in principle be represented by any hard parton in the process different from $q$ and $\qb$. However, inconsistent choices of $a$ and $b$ across multiple contributions may produce subtraction terms which perform in a suboptimal way in some unresolved limits, either at the real-virtual or double-real level after insertion of a second unresolved parton. For specific partonic configurations or colour factors it is possible to define guidelines for the choice of $a$ and $b$. For example, at leading-colour the choice is dictated by the adjacency patterns in the colour ordering of the reduced matrix elements. In general, the numerical validation of the subtraction terms, discussed in Section~\ref{sec:validation}, typically indicates which choice of spectators is preferable. 
	
	We collect all the inserted dipoles in~\eqref{structTc7} and~\eqref{structTc10} in a contribution that we denote as $\dsigTNNLOshort{c,\text{extra}}$:
	\begin{eqnarray}\label{structTc11}
		&&\dsigTNNLOshort{c,\text{extra}}=\nn\\ 
		&&\hspace{0.5cm}-\sum_{q,\qb,g,u}\sum_{c,c'}D_3^0(q,u,g)\left(\Jfull{1}(\wt{qu},\qb)-\Jfull{1}(q,\qb)\right)a^0_{n+2}(c,c';.,\qb.,\wt{qu},.,\wt{ug},.,b,.)\nn\\
		&&\hspace{0.5cm}-\sum_{q,\qb,u,a,b}\sum_{c,c'}A_3^0(q,u,\qb)\Big[\Big.\left(\Jfull{1}(\wt{qu},q)-\Jfull{1}(q,a)\right)\nn\\
		&&\hspace{0.5cm}\hspace{2.3cm}-\left(\Jfull{1}(b,\wt{u\qb})-\Jfull{1}(b,\qb)\right)\Big.\Big]a^0_{n+2}(c,c';.,a,.,\wt{qu},.,\wt{u\qb},.,b,.).
	\end{eqnarray} 
	
	Adding $\dsigTNNLOshort{c,\text{extra}}$ to $\dsigTNNLOshort{c,\text{prel.}}$ and removing of the residual poles with $\dsigTNNLOshort{c,\mathcal{S}}$ yields a finite and complete result. However, $\dsigTNNLOshort{c,\text{extra}}$ inevitably introduces additional singularities in $g\parallel q$ collinear limits, deviating from the original design of $\dsigTNNLOshort{c,\text{prel.}}$. One finds that the divergent behaviour ends up being over-subtracted exactly twice, and simply adding an overall factor $1/2$ to the final expression (after the removal of the $\e$-poles) compensates for the additional terms. As we explained before, other single-unresolved limits are not affected by this, since there $\dsigTNNLOshort{c}$ is vanishing by construction. The characteristic factor $1/2$ derived here can be also explicitly found for example in~\cite{Gehrmann-DeRidder:2011jwo,Currie:2013dwa}. There, its presence was inferred at leading-colour proceeding with the traditional logic: from the double-real correction to the real-virtual one. The recovery of the same structures in the general colourful approach is a solid cross-check.
	
	To summarize, the final expression for $\dsigTNNLOshort{c}$ is:
	\begin{equation}\label{finalTc}
		\dsigTNNLOshort{c}=\dfrac{1}{2}\left[\dsigTNNLOshort{c,\text{prel.}}+\dsigTNNLOshort{c,\mathcal{S}}+\dsigTNNLOshort{c,\text{extra}}\right],
	\end{equation}
	with the different components given in equations~\eqref{sigTcprel},~\eqref{structTc5} and~\eqref{structTc11} respectively. We conclude the treatment of $\dsigTNNLOshort{c}$ in the following section, illustrating which part of the structures discussed above is actually derived from the double-virtual subtraction terms via the insertion of an unresolved partons and which terms, on the contrary, constitute a new addition at the real-virtual level, which will be inherited by the double-real subtraction term.
	
	\paragraph{Extraction of \text{d}\boldmath{$\sigma^{T,c_0}_{ab,\mathrm{NNLO}}$}}
	
	The $\sigma^{T,c}_{ab,\mathrm{NNLO}}$ derived above contains contributions which can be related to structures in the double-virtual subtraction terms that underwent the insertion of an unresolved parton. These contributions are denoted as $\sigma^{T,c_1}_{ab,\mathrm{NNLO}}$ and $\sigma^{T,c_2}_{ab,\mathrm{NNLO}}$~\cite{Currie:2013vh,Chen:2022ktf}. 
	
	The first block $\dsigTNNLOshort{c_1}$ comes from part of the leftover $+\ins{\dsigUNNLOshort{b}}$ that we did not use in Section~\ref{sec:dsigTb} during the construction of $\dsigTNNLOshort{b_1}$. In particular, it comes from the insertion of an unresolved parton in the colour-connected and almost colour-connected components of $\dsigUNNLOshort{b}$ defined in equations~\eqref{sigUbcc} and~\eqref{sigUbacc}:
	\begin{equation}\label{sigTc1}
		\dsigTNNLOshort{c_1}= +\ins{\dsigUNNLOshort{b,c.c.}}+\ins{\dsigUNNLOshort{b,a.c.c.}}.
	\end{equation}
	The remaining colour-unconnected component $+\ins{\dsigUNNLOshort{b,u.c.}}$ produces structures in the double-real subtraction term, as we will explain in Section~\ref{sec:dsigSd}. 
	
	To generate $\dsigTNNLOshort{c_2}$, we directly use the residual term from~\ref{sigTb2MXrel}:
	\begin{equation}\label{sigTc2}
		\dsigTNNLOshort{c_2}= -\ins{\dsigUNNLOshort{c,\mathcal{X}_{3}^{0}\otimes\mathcal{X}_{3}^{0}}}-\left[\dsigTNNLOshort{b_2,M_X}+\dd\sigpart{MF,1,b_2}{ab,\mathrm{NNLO}}\right]\sim +\ins{\dsigUNNLOshort{c,\mathcal{X}_{3}^{0}\otimes\mathcal{X}_{3}^{0}}},
	\end{equation}
	where the proportionality factor depends on the  $m_X$ required in~\ref{sigTb2MXrel} for the respective term. 
	We observe that, to ensure that the insertion is properly performed in both integrated antenna functions, the same symmetrization procedure employed for $\dsigUNNLOshort{b}$ can be used here as well.
	
	We can now define the following term:
	\begin{equation}\label{sigTc0}
		\dsigTNNLOshort{c_0}=\dsigTNNLOshort{c}-\dsigTNNLOshort{c_1}-\dsigTNNLOshort{c_2},
	\end{equation} 
	which collects all contributions in $\dsigTNNLOshort{c}$ that are not derived from the unintegration of structures in the double-virtual subtraction term and, hence, are introduced at the real-virtual level for the first time. As we see in Section~\ref{sec:dsigSc}, the insertion of an unresolved parton within $\dsigTNNLOshort{c_0}$ produces contributions in the double-real subtraction term. 
	
	\subsubsection{Summary and observations}\label{sec:sumRV}
	
	For ease of reference, we collect here the main results of this section. The real-virtual subtraction term has been decomposed according to:
	\begin{equation}
		\dd\sigpart{T}{ab,\mathrm{NNLO}}=\dd\sigpart{T,a}{ab,\mathrm{NNLO}}+\dd\sigpart{T,b}{ab,\mathrm{NNLO}}+\dd\sigpart{T,c}{ab,\mathrm{NNLO}}.
	\end{equation}
	The first term $\dd\sigpart{T,a}{ab,\mathrm{NNLO}}$ has an NLO-like structure and is presented in section~\ref{sec:dsigTa}. The other terms are given by
	\begin{equation}
		\dsigTNNLOshort{b}=\dsigTNNLOshort{b_1}+\dsigTNNLOshort{b_2}+\dsigTNNLOshort{b_3}, 
	\end{equation}
	with
	\begin{eqnarray}
		\dsigTNNLOshort{b_1}&=&-\ins{\dsigUNNLOshort{a_1}}-2\,\ins{\dsigUNNLOshort{b}}-\dd\sigpart{MF,1,b}{ab,\mathrm{NNLO}},\\
		\dsigTNNLOshort{b_2}&=&-\ins{\dsigUNNLOshort{c,\mathcal{X}_{3}^{1}}}+\dsigTNNLOshort{b_2,J_X}+\dsigTNNLOshort{b_2,M_X},\\
		\dsigTNNLOshort{b_3}&=&-\ins{\dsigUNNLOshort{a_0}}-\ins{\dsigUNNLOshort{c,\beta_0}},
	\end{eqnarray}
	and
	\begin{equation}
		\dsigTNNLOshort{c}=\dsigTNNLOshort{c_1}+\dsigTNNLOshort{c_2}+\dsigTNNLOshort{c_0}, 
	\end{equation}
	with
	\begin{eqnarray}
		\dsigTNNLOshort{c_1}&=& +\ins{\dsigUNNLOshort{b,c.c.}}+\ins{\dsigUNNLOshort{b,a.c.c.}},\\
		\dsigTNNLOshort{c_2}&=& -\ins{\dsigUNNLOshort{c,\mathcal{X}_{3}^{0}\otimes\mathcal{X}_{3}^{0}}}-\left[\dsigTNNLOshort{b_2,M_X}+\dd\sigpart{MF,1,b_2}{ab,\mathrm{NNLO}}\right].
	\end{eqnarray}
	All the contributions to $\dd\sigpart{T,b}{ab,\mathrm{NNLO}}$ and $\dd\sigpart{T,c}{ab,\mathrm{NNLO}}$ are discussed in detail respectively in sections~\ref{sec:dsigTb} and~\ref{sec:dsigTc}.
	
	To illustrate the origin of the different contributions, 
	 we also recall  the full decomposition of the double virtual subtraction terms, described in detail in section~\ref{sec:VV}:
	\begin{eqnarray}
		\dd\sigpart{U}{ab,\mathrm{NNLO}}&=& \dsigUNNLOshort{a_0} + \dsigUNNLOshort{a_1}\nonumber \\
		&+&\dsigUNNLOshort{b,c.c.} + \dsigUNNLOshort{b,a.c.c.}+ \dsigUNNLOshort{b,c.u.}\nonumber \\
		&+&\dsigUNNLOshort{c,\mathcal{X}_{4}^{0}}+\dsigUNNLOshort{c,\mathcal{X}_{3}^{1}}+\dsigUNNLOshort{c,\mathcal{X}_{3}^{0}\otimes\mathcal{X}_{3}^{0}}+\dsigUNNLOshort{c,\beta_0}.
	\end{eqnarray}
	One can notice that almost the entire double-virtual subtraction term has been converted into its unintegrated counterpart and has been used at the real-virtual level, with the only exceptions being $\dsigUNNLOshort{b,c.u.}$ and $\dsigUNNLOshort{c,\mathcal{X}_{4}^{0}}$, which are directly converted to the double-real subtraction term. Moreover, new components had to be added at the real-virtual level and require a corresponding counterpart in the double-real subtraction term. These components are $\dsigTNNLOshort{a}$, $\dsigTNNLOshort{b_2,J_X}$ and $\dsigTNNLOshort{c_0}$.
	
	\subsection{NNLO double-real subtraction term}\label{sec:RR}
	
	The last ingredient for an NNLO calculation is the double-real subtraction term $\dsigSNNLO{ab}$, which removes the divergent behaviour of the double-real matrix element in single and double unresolved limits. In the colourful antenna subtraction approach, the generation of $\dsigSNNLO{ab}$ is the last step of the procedure. It is a feature of the colourful formalism that 
	 once the double-virtual and the real-virtual subtraction terms are available, it is quite straightforward to complete the subtraction procedure with the missing blocks needed to cancel the unmatched contributions in those two layers.
	 	
	The double-real subtraction term is constructed by inserting a second unresolved parton in contributions coming from $\dsigTNNLO{ab}$ and two unresolved partons in terms coming from $\dsigUNNLO{ab}$. As we discuss below, the only genuinely new procedure at this level is represented by the simultaneous insertion of two colour-connected unresolved partons within the integrated four-parton antennae $\mathcal{X}_{4}^{0}$. 
	
	We recall the usual decomposition of $\dsigSNNLO{ab}$~\cite{Currie:2013vh}:
	\begin{equation}
		\dsigSNNLO{ab}=\dsigSNNLOshort{a}+\dsigSNNLOshort{b}+\dsigSNNLOshort{c}+\dsigSNNLOshort{d}.
	\end{equation}
	The first term $\dsigSNNLOshort{a}$ removes single unresolved limits and it is analogous to an NLO real subtraction term for an underlying $(n+1)$-particle final state LO process. The remaining terms respectively reproduce the divergent behaviour of the double-real correction in colour-connected, almost colour-connected and colour-unconnected configurations~\cite{Glover:2010kwr,Currie:2013vh}. $\dsigSNNLOshort{c}$ also contains the large angle soft terms~\cite{Weinzierl:2008iv,Gehrmann-DeRidder:2007foh}. In the following, we describe how to systematically generate each contribution.
	
	\subsubsection{\text{d}\boldmath{$\sigma^{S,a}_{ab,\mathrm{NNLO}}$}}\label{sec:dsigSa}
	
	This contribution to the subtraction term is directly generated from $\dsigTNNLOshort{a}$, since it can be seen as its corresponding real NLO subtraction term:
	\begin{equation}
		\dsigTNNLOshort{a}=-\int_1\dsigSNNLOshort{a}-\dd\hat{\sigma}^{MF,1,a}_{ab,\mathrm{NNLO}},
	\end{equation}
	which reflects equation~\eqref{sigTNLO} and therefore, following what is done in Section~\ref{sec:NLOR} for the NLO real subtraction term, we can simply write:
	\begin{equation}\label{sigSa}
		\dsigSNNLOshort{a}=-\ins{\dsigTNNLOshort{a}}.
	\end{equation}
	
	\subsubsection{\text{d}\boldmath{$\sigma^{S,b}_{ab,\mathrm{NNLO}}$}}\label{sec:dsigSb}
	
	In the colour-connected configuration, the two unresolved partons are emitted between the same pair of hard radiators. According to~\cite{Currie:2013vh}, we further decompose $\dsigSNNLOshort{b}$ into two contributions:
	\begin{equation}
		\dsigSNNLOshort{b}=\dsigSNNLOshort{b_1}+\dsigSNNLOshort{b_2}.
	\end{equation}
	The $\dsigSNNLOshort{b_1}$ terms contains four-parton antenna functions $X_{4}^{0}$, while the $\dsigSNNLOshort{b_2}$ term contains convolutions of two three-parton antenna functions $X_{3}^{0}\otimes X_{3}^{0}$, which are needed to remove the singular behaviour of $X_{4}^{0}$ antenna functions in single-unresolved limits. 
	
	The generation of $\dsigSNNLOshort{b_2}$ is straightforward, since its integrated counterpart is exactly $\dsigTNNLOshort{b_2,J_X}$, so:
	\begin{equation}\label{sigSb2}
		\dsigSNNLOshort{b_2}=-\ins{\dsigTNNLOshort{b_2,J_X}}, 
	\end{equation}
	where the momenta relabelling due to the insertion of an unresolved parton must also occur within the unintegrated antenna functions which appear in $\dsigTNNLOshort{b_2,J_X}$. The general structure of the resulting term is the following:
	\begin{eqnarray}
		&&\dsigSNNLOshort{b_2}\sim \sum_{ij,u_1,u_2}\sum_{c,c'}\Bigg[\Bigg.X_3^0(i,u_2,j)X_3^0(\wt{iu_2},u_1,\wt{ju_2})a^0_{n+2}(c,c';.,\wt{(iu_2)u_1},.,\wt{(u_2j)u_1},.)\nn\\
		&&\hspace{1cm}+X_3^0(i,u_2,u_1)X_3^0(\wt{iu_2},\wt{u_1u_2},j)a^0_{n+2}(c,c';.,\wt{(iu_2)(u_1u_2)},.,\wt{j(u_1u_2)},.)\nn\\
		&&\hspace{1cm}+X_3^0(u_1,u_2,j)X_3^0(i,\wt{u_1u_2},\wt{ju_2})a^0_{n+2}(c,c';.,\wt{i(u_1u_2)},.,\wt{(u_2j)(u_1u_2)},.)\Bigg.\Bigg],
	\end{eqnarray}
	where $u_1$ and $u_2$ represent the two unresolved partons.
	
	$\dsigSNNLOshort{b_1}$ comes from the insertion of two unresolved partons in $\dsigUNNLOshort{c,\mathcal{X}_{4}^{0}}$, namely from the transition of integrated tree-level four-parton antenna functions to the unintegrated level. In Figure \ref{fig:colant_scheme}, such insertion was depicted with two connected descendant red arrows to emphasize the fact that it cannot be performed relying on the $\ins{\cdot}$ operator defined in Section~\ref{sec:subNLO}. Nevertheless, we can still identify a one-to-one correspondence between four-parton integrated and unintegrated antenna functions:
	\begin{equation}\label{double_insertion}
		\mathcal{X}_{4}^{0}(s_{ij})\anum{n+2}{0}(c,c',\{.,i,.,j,.\})\leftrightarrow X_{4}^{0}(i,u_1,u_2,j)\anum{n+2}{0}(c,c',\{.,\widetilde{iu_1u_2},.,\widetilde{u_1u_2j},.\}).
	\end{equation}
    The momentum mapping relating the two different phase-space multiplicities is the one appropriate at NNLO~\cite{Kosower:2002su,Gehrmann-DeRidder:2005btv,Currie:2013vh}. The simultaneous insertion of two unresolved partons cannot be achieved by the single-insertion operator $\ins{\cdot}$.  We need to define a new procedure, which is however analogous to the one described in Section~\ref{sec:subNLO}. The insertion of two unresolved colour-connected partons occurs as follows:
	\begin{enumerate}
		\item[\textbf{1)}] remove the mass factorization kernels if present;
		
		\item[\textbf{2)}] replace each four-parton integrated antenna function $\mathcal{X}_{4}^{0}(s_{ij})$ with its unintegrated counterpart $X_{4}^{0}(i,u_1,u_2,j)$ (see below);
		
		\item[\textbf{3)}] suitably replace the momenta in the colour interferences, according to the accompanying integrated antenna, following~\eqref{double_insertion};
		
		\item[\textbf{4)}] apply the same momenta relabelling to the jet function;
		
		\item[\textbf{5)}] promote the phase-space measure to the appropriate one for $(n+2)$ final-state momenta;
		
		\item[\textbf{6)}] replace the overall factor with appropriate one for the real correction:
		\begin{equation}
			\coeffVVNNLO\to\coeffRRNNLO=s_{RR}\left(4\pi\alpha_s\right)^2\coeffLO,
		\end{equation}
		where $s_{RR}$ compensates the potentially different final state symmetry factors in the presence two real emissions;
	\end{enumerate}
	As for the single-insertion case, we need to clarify how exactly point $2$ is performed. At NNLO we can identify the following possibilities for the insertion of two unresolved partons:
	\begin{itemize}
		\item insertion of two unresolved gluons;
		\item insertion of an unresolved quark-antiquark pair, which becomes soft or collinear to hard partons;
		\item insertion of two collinear quark-antiquark pairs;
		\item insertion of unresolved partons in IC limits.
	\end{itemize}
	We comment in the following about the different types of insertions. The appropriate replacement rules to convert integrated antenna functions to their unintegrated counterparts are presented in Appendix~\ref{app:X40}.
	
	\paragraph{Insertion of two unresolved gluons}
	
	The insertion of two unresolved gluons $u_{g_1}$ and $u_{g_2}$ was presented in~\cite{Chen:2022ktf}. It is the natural extension of the insertion of a single unresolved gluon. If the underlying LO process has $(n+2)$ external partons, the additional gluonic emissions are labelled with indices $u_{g_1}=n+3$ and $u_{g_2}=n+4$. The expressions obtained performing steps 1 to 4 above provides the appropriate structure to remove the double-unresolved limits of partons $u_1=n+3$ and $u_2=n+4$, which we denote with:
	\begin{equation}\label{xxx}
		f(p_1,\dots,p_{n+2};p_{n+3},p_{n+4}),
	\end{equation}
	with the last two entries representing the two momenta which can become unresolved. The ordering of the hard momenta $p_1,\dots,p_{n+2}$ is irrelevant, since a sum over all possible colour orderings is considered. A suitable sum of structures like the one in~\eqref{xxx} is required to cover all possible two-gluon unresolved configurations of the double-real matrix elements:\\
	\begin{equation}
		\sum_{g_1,g_2\,\in\,\text{f.s. gluons}}f(\dots;p_{g_1},p_{g_2}),
	\end{equation}
	where the sum runs over the momenta of ordered pairs of final-state gluons and the dots indicate all the other external momenta. The final result reproduces the full singular behaviour of the double-real matrix element when any pair of colour-connected gluons becomes unresolved. The replacement rules for the insertion of two unresolved gluons are given in Appendix~\ref{app:X40} in Table~\ref{tab:insggX40}.
	
	\paragraph{Insertion of an unresolved quark-antiquark pair}
	
	This type of insertion differs from the corresponding one at NLO because in the infrared limits it addresses, the extra quark-antiquark pair $(u_q,u_{\qb})$ can become soft or collinear to another hard parton. 
	
	On the practical side, the insertion at NNLO is actually simpler to perform than the one at NLO. Indeed, at NLO, the splitting of a gluon into a quark-antiquark pair required a proper shift of all the partonic indices and a dedicated procedure to cover all possible splittings. At NNLO, given that this is a double-unresolved limit, we can proceed as for the insertion of two unresolved gluons and simply append the indices of the newly added quark and antiquark to the original set of momenta, namely $u_q=n+3$ and $u_\qb=n+4$ for a process with $(n+2)$ partons at LO. Moreover, no additional sum over partonic permutations is needed. The associated replacement rules are given in Appendix~\ref{app:X40} in Table~\ref{tab:insqqbX40}.
	
	\paragraph{Insertion of two unresolved quark-antiquark pairs}
	
	This insertion targets the situation in which two hard gluons individually split into two collinear quark-antiquark pairs $(u_q,u_\qb)$ and $(u_{q'},u_{\qb'})$. Since the two pairs belong to distinct fermionic lines, their flavours can be considered different, with the same-flavour case not adding any contribution to this specific unresolved limit. 
	
	The only IP antenna function related to such configuration is the gluon-gluon final-final antenna function $H_4^0(q,\qb,q',\qb')$, which only contains the double-collinear limit $q\parallel \qb$, $q'\parallel \qb'$. The transition from the integrated to the unintegrated antenna function is given in Appendix~\ref{app:X40} in Table~\ref{tab:insqqbqqbX40}. Symmetrization within each pair and over the exchange of the two pairs is considered. As it happens for the quark-antiquark insertion at NLO, besides adding two extra partons, two gluons which are present in the original set of partons need to be converted into quarks. We consider an $(n+2)$-parton process with $n_g\ge 2$ final-state gluons labelled as $i,\dots,i+n_g-1$. After the insertions, the final-state gluons can be relabelled as $i,\dots,i+n_g-3$, while the added quarks are labelled, for example, as:
	\begin{equation}
		u_q=i+n_g-2,\quad\quad u_\qb=i+n_g-1, \quad\quad u_{q'}=n+3,\quad\quad u_{\qb'}=n+4.
	\end{equation}
	The insertion is done by letting all possible pairs of external gluons split into the quark-antiquark pairs, so no additional sums are required to reproduce the full singular behaviour of the matrix element.
	
	\paragraph{Insertion of two unresolved partons in IC limits}
	
	At NNLO, significantly more IC configurations are present with respect to the NLO case. The possibilities parallel the two-loop mass factorization kernels listed in Section~\ref{sec:subNNLO_pre}. The replacement rules to convert IC four-parton antenna functions to their unintegrated counterparts are given in Appendix~\ref{app:X40} in Table~\ref{tab:insICX401} for quark-antiquark antenna functions, in Table~\ref{tab:insICX402} for quark-gluon antenna functions and in Table~\ref{tab:insICX401} for gluon-gluon antenna functions. 
	
	\vspace{1cm}
	
\begin{figure}
		\centering
		\begin{tikzpicture}[node distance=1.8cm, 
			box/.style={draw, rectangle, text width=10cm, minimum height=1.2cm, align=center, fill=lightblue}]
			
			\definecolor{lightblue}{RGB}{173,216,230}
			
			\node[box] (box1) {Removal of splitting kernels};
			
			\node[box, below of=box1] (box2) {Insertion of two unresolved partons: $\mathcal{X}_4^0\to X_4^0$};
			
			\node[box, below of=box2] (box3) {Relabelling/mapping of momenta};
			
			\node[box, below of=box3] (box4) {Fix overall factors};
			
			\draw[->, thick] (box1) -- (box2);
			\draw[->, thick] (box2) -- (box3);
			\draw[->, thick] (box3) -- (box4);
			
		\end{tikzpicture}
		\caption{Application of the $\insdouble{\cdot}$ operator for the simultaneous insertion of two colour-connected unresolved partons.}
		\label{fig:ins2operator}
	\end{figure}
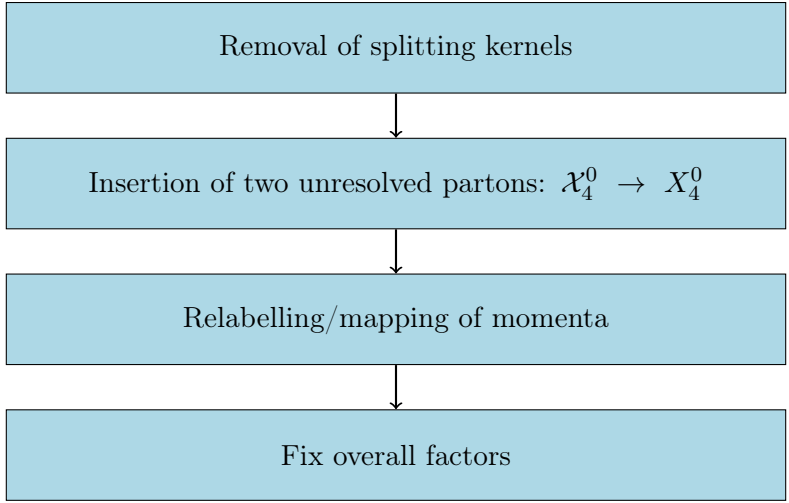

	We introduce a new operator denoted with $\insdouble{\cdot}$ for the simultaneous double insertion of two unresolved partons, applying the procedure outlined above addressing each insertion possibility. The application of $\insdouble{\cdot}$ is summarized in Figure~\ref{fig:ins2operator}. 
	
	With the new operator, we can finally write:
	\begin{equation}\label{sigSb1}
		\dsigSNNLOshort{b_1}=-\insdouble{\dsigUNNLOshort{c,\mathcal{X}_{4}^{0}}},
	\end{equation}
	which has the following structure:
	\begin{equation}\label{sigSb1struct}
		\dsigSNNLOshort{b_1}\sim\sum_{c,c'}\sum_{ij,u_1,u_2}X_4^0(i,u_1,u_2,j)a_{n+2}^0(c,c';.,\wt{iu_1u_2},.,\wt{ju_1u_2},.).
	\end{equation}
	
	With this, we completed the generation of the $\dsigSNNLOshort{b}$ component of the double-real subtraction term.
	
	We observe that, in the outlined procedure, $\dsigSNNLOshort{b_1}$ and $\dsigSNNLOshort{b_2}$ are inferred from apparently independent contributions. However, a precise relation between these two terms should hold to ensure the removal of single unresolved limits from the four-particle antenna functions at the double-real level. In fact, the structure of $\dsigSNNLOshort{b_2}$ directly descends from the one-loop three-parton antennae appearing in the two-loop integrated dipoles, which are in turn related to the four-parton antenna functions. Indeed, the relation among $\dsigSNNLOshort{b_1}$ and $\dsigSNNLOshort{b_2}$ is actually mirrored by the inner structure of the two-loop integrated dipoles and the interplay between the $\mathcal{X}_{4}^{0}$ and the $\mathcal{X}_{3}^{1}$, which gives the correct $\e$-poles at the double-virtual level, manifest here in the form of the correct arrangement of $X_{4}^{0}$ and $X_{3}^{0}\otimes X_{3}^{0}$ contributions. The removal of the single-unresolved behaviour of $\dsigSNNLOshort{b_1}$ by $\dsigSNNLOshort{b_2}$ can be used as a strong check of the correctness of the derivation we illustrated.
	
	\subsubsection{\text{d}\boldmath{$\sigma^{S,c}_{ab,\mathrm{NNLO}}$}}\label{sec:dsigSc}
	
	In the almost colour-connected configuration, the unresolved partons are emitted between two pairs of hard radiators which share one common hard radiator. The structure of the blocks needed to remove the divergences associated to these configurations is shared by the large angle soft terms too, which are thus naturally incorporated in $\dsigSNNLOshort{c}$~\cite{Glover:2010kwr,Currie:2013vh}. The integrated counterpart of $\dsigSNNLOshort{c}$ is $\dsigTNNLOshort{c_0}$ generated at the real-virtual level, as described in Section~\ref{sec:dsigTc}. Hence we have:
	\begin{equation}\label{sigSc}
		\dsigSNNLOshort{c}=-\ins{\dsigTNNLOshort{c_0}}.
	\end{equation}
	The action of the $\ins{\cdot}$ operator on $\dsigTNNLOshort{c_0}$ requires the insertion of an unresolved parton within integrated large angle soft terms. With $u_1$ and $u_2$ respectively the previously inserted unresolved parton and the newly inserted unresolved parton at the double-real level, the required replacements are:
	\begin{equation}\label{Softreplace}
		\begin{aligned}
			\mathcal{S}(s_{IJ},s_{I'J'},x)\quad&\to& S(I,u_2,J)&\quad\text{for}\quad(I,J)\ne(I',J'),\\
			\mathcal{S}(s_{I'J},s_{I'J'},x)\quad&\to& S(\wt{I'u_2},u_2,J)&\quad\text{for}\quad J\ne J',\\
			\mathcal{S}(s_{IJ'},s_{I'J'},x)\quad&\to& S(I,u_2,\wt{u_2J'})&\quad\text{for}\quad I\ne I',\\
			\mathcal{S}(s_{I'J'},s_{I'J'},x)\quad&\to& S(\wt{I'u_2},u_2,\wt{u_2J'})&,\\
		\end{aligned}
	\end{equation}
	where $S(a,b,c)=S^0_{abc}$ is the usual soft factor. $I$ and $J$ represent any unmapped or mapped parton and $I'$, $J'$ indicate either $i'$, $j'$ or $\wt{i'u_1}$, $\wt{u_1j'}$. In this latter case, namely when parton $i'$ or $j'$ acts as a hard radiator for both partons $u_1$ and $u_2$, the correct order of the momentum mapping is $i'\to\wt{i'u_2}\to\wt{(i'u_2)u_1}$ or $j'\to\wt{j'u_2}\to\wt{(j'u_2)u_1}$, since the first parton which is integrated over at the double-real level is $u_2$. The eikonal factors generated via \eqref{Softreplace} remove the remnant soft gluon divergent behaviour associated to colour-connected and almost colour-connected contributions at the double-real level \cite{Gehrmann-DeRidder:2007foh,Weinzierl:2008iv,Weinzierl:2009nz,Glover:2010kwr,Currie:2013vh}.
		
	\subsubsection{\text{d}\boldmath{$\sigma^{S,d}_{ab,\mathrm{NNLO}}$}}\label{sec:dsigSd}
	
	In the colour-unconnected configuration, the two unresolved partons are emitted between two distinct pairs of hard radiators. These terms do not appear at the real-virtual level but can be inherited directly from the double-virtual subtraction term to the double-real one. This is achieved inserting two unresolved partons in $\dsigUNNLOshort{b,c.u.}$, one in each of the two one-loop integrated dipoles. Since the two pairs of hard radiators are distinct, the two insertions can be performed independently. Therefore, $\dsigSNNLOshort{d}$ is generated through the iterated application of the $\ins{\cdot}$ operator:
	\begin{equation}\label{sigSd}
		\dsigSNNLOshort{d}=+\ins{\ins{\dsigUNNLOshort{b,c.u.}}}.
	\end{equation}
	In Figure~\ref{fig:colant_scheme}, we indicated the iterated insertion of two unresolved partons as two disjoint descendant red arrows, to differentiate it from the simultaneous double insertion discussed in Section \ref{sec:dsigSb}.
	
	\subsubsection{Summary and observations}\label{sec:sumRR}
	For ease of reference, we collect here the main results of this section. The double-real subtraction term has been decomposed according to:
	\begin{equation}
		\dsigSNNLO{ab}=\dsigSNNLOshort{a}+\dsigSNNLOshort{b}+\dsigSNNLOshort{c}+\dsigSNNLOshort{d},
	\end{equation}
	with
	\begin{eqnarray}
		\dsigSNNLOshort{a}&=&-\ins{\dsigTNNLOshort{a}},\\
		\dsigSNNLOshort{b}&=&\dsigSNNLOshort{b_1}+\dsigSNNLOshort{b_2},\\
		\dsigSNNLOshort{b_1}&=&-\insdouble{\dsigUNNLOshort{c,\mathcal{X}_{4}^{0}}},\\
		\dsigSNNLOshort{b_2}&=&-\ins{\dsigTNNLOshort{b_2,J_X}},\\
		\dsigSNNLOshort{c}&=&-\ins{\dsigTNNLOshort{c_0}},\\
		\dsigSNNLOshort{d}&=&+\ins{\ins{\dsigUNNLOshort{b,c.u.}}}.
	\end{eqnarray}
	
	Comparing the summary above with the one in section~\ref{sec:sumRV} and with the visual support of Figure~\ref{fig:colant_scheme}, one can easily track down the relations among all the components of the three layers of the NNLO subtraction infrastructure. In particular, since, up to mass factorization counterterms which are absorbed in the redefinition of the PDFs, the following relations hold:
	\begin{eqnarray}
		\int_1 \ins{f} &=& f,\\
		\int_2 \insdouble{f} &=& f,
	\end{eqnarray}
	where $f$ indicates a generic contribution to the subtraction terms, it is possible to verify that the relations in~\eqref{subtermsNNLO} are fulfilled and therefore the total sum of the subtraction terms vanishes after integration over the unresolved partons.
	
	\section{Subtraction terms for three-jet production at hadron colliders}\label{sec:3jet}
	
	As a first application, we re-derived the NNLO subtraction terms for $pp\to 2$~jets and $pp\to H+j$ in the colourful antenna subtraction method and compared 
	them to their original implementations~\cite{Chen:2016zka,Chen:2022clm} in \textsc{NNLOjet}. 
	Exact pointwise agreement has been found for most of the structures, with few discrepancies showing up away from the infrared limits. This is explained by the fact that in some cases different choices can be made for the construction of subtraction terms. In general, the systematic procedure described in this paper to construct the virtual subtraction terms by means of the integrated dipoles defined in sections~\ref{sec:int_dip_1} and~\ref{sec:int_dip_2} may yield different structures at the integrated level with respect to the \textit{ad-hoc} construction done in the past for specific processes. However, the two implementations clearly provide the same virtual poles and only differ by $\e$-finite contributions.
	
	Differences in the virtual subtraction terms obtained from the two approaches imply that also the real subtraction terms should not coincide exactly, to ensure that after integration the net contribution of the subtraction terms vanishes. In this case, the discrepancies only show up in the resolved regions and do not affect the infrared limits. If consistent choices are made at the integrated and unintegrated level, namely equations~\eqref{sigTNLO} and~\eqref{subtermsNNLO} are satisfied, any implementation provides the same numerical result for the cross section when all the layers of a fixed-order calculation are combined. 
	
	A typical example is the choice of the spectator parton in an initial-state collinear limit: different options reproduce the same singular behaviour, but do not give the same numerical outcome in the hard regions. Such discrepancies in the real-emission layers are compensated by different $\e$-finite terms in the virtual subtraction terms, in such a way the final result is unaffected by the specific choice. 
	
	We used the colourful antenna subtraction method to construct the subtraction terms relevant to 
	$pp\to 3$~jets. Three-jet production and in general multi-jet events at hadron colliders provide an optimal environment for investigating the properties of QCD. The reduced sensitivity to electroweak effects allows for precision studies of parton dynamics in QCD and the extraction of the strong coupling constant $\alpha_s$~\cite{ATLAS:2011qvj,ATLAS:2020vup,CMS:2013vbb,CMS:2014tkl,CMS:2018svp}.
	
	Three-jet production, with five external partons at LO, presents an unprecedented complexity in the context of antenna subtraction. In general, the computation of the NNLO correction to $pp\to3j$ stands as the most challenging fully massless NNLO calculation among the ones which are in principle feasible at present, given the available matrix elements. Such a calculation lies beyond the practical reach of most of the current subtraction schemes at NNLO. It has been computed in~\cite{Czakon:2021mjy,Alvarez:2023fhi}, with the sector-improved residue subtraction technique~\cite{Czakon:2010td,Czakon:2014oma}. The only approximation present in these calculations is the truncation to the leading-colour component of the infrared-finite remainder of the two-loop 
	amplitudes~\cite{Abreu:2021oya}, since the complete subleading-colour contributions to these amplitudes  are not yet available.  
	
	Along with the complicated infrared structure, due to the large number of partons and, hence, possible infrared limits, this calculation has an extremely high computational cost~\cite{Alvarez:2023fhi}. The stability over the whole phase space of the high-multiplicity one- and two-loop matrix elements must be ensured exploiting quadruple-precision arithmetic when the standard evaluation fails. Moreover, the numerical integration of real-virtual and double-real corrections and their respective subtraction terms requires a very substantial number of evaluations to reach a satisfactory precision. 
	
	We exploited the colourful antenna subtraction method to systematically construct NNLO subtraction terms for $pp\to3j$. In the following we discuss their validation. The first proof-of-principle application of the colourful antenna subtraction approach was the calculation of the NNLO correction to the subchannel $gg\to ggg$ in the gluons-only scenario, presented in~\cite{Chen:2022ktf}. The construction and validation of the subtraction terms discussed below anticipates the complete calculation of the NNLO correction for three-jet production at hadron colliders.
	
	The computation is performed within the \textsc{NNLOjet} framework~\cite{Gehrmann-DeRidder:2015wbt}. \textsc{NNLOjet} is a parton-level 
	Monte Carlo event generator which implements the antenna subtraction method to compute NNLO QCD corrections to a series of processes. 
	
	For three-jet production at NNLO, high multiplicity tree and loop amplitudes are needed. The computation relies on a mixture of analytical results and numerical implementations for the amplitudes. The helicity amplitudes the LO and NLO processes, namely five- and six-parton scattering at tree-level~\cite{Berends:1987cv,Kunszt:1985mg,Gunion:1985bp,Gunion:1986cb,Gunion:1986zh,Dixon:2010ik} and five-parton scattering at one-loop~\cite{Bern:1993mq,Bern:1994fz,Kunszt:1994nq,Signer:1995a} are incorporated in a fully analytical form in \textsc{NNLOjet}. The planar five-parton two-loop amplitudes have been computed in~\cite{Abreu:2019odu} using a basis of pentagon functions~\cite{Chicherin:2017dob,Gehrmann:2018yef,Chicherin:2020oor}. We rely on a public C\texttt{++} code~\cite{Abreu:2021oya} which implements the aforementioned amplitudes and computes the renormalized infrared-finite remainder of the five-parton two-loop matrix elements. The implementation in~\cite{Abreu:2021oya} provides the leading-colour part of the finite remainder, including leading-$N_f$ and leading-$N_f^2$ contributions. However, for the infrared poles of the two-loop matrix element we employ the complete full-colour result, derived from its infrared factorization properties~\cite{Catani:1998bh,Becher:2009qa,Gardi:2009qi}. The six-parton one-loop matrix elements are computed with the \textsc{OpenLoops} generator~\cite{Cascioli:2011va,Buccioni:2017yxi,Buccioni:2019sur}. We rely on the built-in quadruple-precision rescue system to increase the accuracy in infrared limits and deal with occasional unstable points. We rely on \textsc{OpenLoops} also for the tree-level seven-parton matrix elements, with the only exception of the seven-gluon matrix element. For the tests we performed, quadruple-precision is not needed to have a satisfactory stability of tree-level matrix elements in single- and double-unresolved limits.
	
	\subsection{Validation of the subtraction terms}\label{sec:validation}
	
	In this section we discuss how the process of validation of the subtraction terms occurs. Three-jet production at NNLO has a large number of partonic sub-channels, for an overall number of possible single- and double-unresolved limits of $\mathcal{O}(1000)$. The computational costs of high-multiplicity matrix elements is such that even the validation procedure requires a non-negligible amount of CPU-hours. For this reasons, we limit the following presentation to some exemplary unresolved configurations. 
   
   \subsubsection{Double-virtual subtraction terms}\label{sec:VVvalid}
   
   The validation of the double-virtual subtraction term consists of checking the cancellation of all the explicit $\e$-poles against the ones of the double-virtual matrix element, as predicted by~\eqref{VVpoles}, as well as in the recovery of the complete structure of the mass factorization kernels at two loops, including IC contributions. Both requirements are validated at the symbolic level. As explained above, even if the finite-remainder of the two-loop correction is only available at leading colour, these checks are performed considering the complete full-colour results. Retaining the full-colour double-virtual subtraction terms is actually mandatory to be consistent with the full-colour calculation of the real-virtual and double-real corrections.
   
   In the colourful antenna subtraction approach, the double-virtual subtraction term is constructed mirroring the infrared structure of the two-loop matrix element. Hence, one may expect this part of the validation procedure to be trivial. However, several crucial aspects of the colourful approach are assessed during this phase, such as:
   \begin{itemize}
   		\item the definition of IP and IC two-loop colour-stripped integrated dipoles, in particular the correct removal of spurious limits from IP ones;
   		\item the colour decomposition of tree-level and one-loop amplitudes and the insertion of operators in colour space;
   		\item the assembling of real quantities at the squared level.
   \end{itemize}
   We also note that the extraction of the singularity structure of one-loop colour-ordered amplitudes discussed in Section~\ref{sec:dsigTb} becomes useful in this context, since one-loop squared amplitudes and colour interferences are part of the infrared structure at two loops. 
   
   We successfully generated and validated the double-virtual subtraction terms for all the sub-process of three-jet production at hadron colliders in full-colour.
   
   \subsubsection{Real-virtual subtraction terms}
	
    The assessment of the real-virtual subtraction terms occurs in two phases: cancellation of explicit infrared poles and correct subtraction of the singular behaviour in single-unresolved limits. 
    
    First of all, during the automated generation, we check that, aside from the $\dsigTNNLOshort{a}$ term, the real-virtual subtraction terms is $\e$-finite. In particular, we check that the contributions $\dsigTNNLOshort{b_1}$, $\dsigTNNLOshort{b_2}$ (with the one-loop antenna functions properly renormalized) and the $\dsigTNNLOshort{c}$ are individually free from explicit infrared singularities, since, as explained in Section~\ref{sec:RV}, they are constructed to be $\e$-finite. Secondly, the explicit poles of the $\dsigTNNLOshort{a}$ component are compared at the symbolic level against the singularity structure of one-loop matrix elements, assessing their complete correspondence. Finally, the coefficients of the $\e$-poles in the full subtraction terms at the level of the numerical implementation are validated against the ones of the real-virtual matrix element. The cancellation of the poles has to occur up to machine-level precision. Once the validation is successful, only the finite part of both the matrix element and the subtraction term is computed.
    
    The assessment of the subtraction of implicit divergent behaviour occurs with pointwise tests against the real-virtual matrix element, which we call \textit{spike-tests}. The procedure is outlined in~\cite{Glover:2010kwr}: we generate a sample of $10000$ phase space points at $\sqrt{s}=13$ TeV close to a given infrared limit and we compute:
	\begin{equation}
		R_{\text{RV}}=\dfrac{\dsigRV{ab}}{\dsigTNNLO{ab}}.
	\end{equation} 
	We then bin the events according to the following quantity:
	\begin{equation}\label{tvar}
		t_{RV}=\log_{10} \left(\left|1-R_{RV}\right|\right),
	\end{equation}
	which provides an estimate of the number of correct digits reproduced by the subtraction terms. We probe each unresolved limit through the variable $x$, which parametrizes the depth at which each infrared limit is tested. The definition of $x$ varies according to the considered unresolved configuration and is given in Table~\ref{tab:smallx}. The squared centre-of-mass energy is indicated as $s$, while the other invariants are defined as:
	\begin{equation}
		\begin{aligned}
			s_{i_1\dots i_m}&=(p_{i_1}+\dots+p_{i_m})^2\quad&\text{small when }i_1,\dots,i_m\text{ are collinear},\\
			s_{-i_1\dots i_m}&=\left(\sum_{j\geq 3,\,j\ne i_1\dots i_m}p_j\right)^2\quad&\text{close to }s\text{ when }i_1,\dots,i_m\text{ are soft}.
		\end{aligned}
	\end{equation}
	The smaller $x$ becomes, the more enhanced is the divergent behaviour of matrix elements. 
	\begin{table}
		\centering
		\begin{tabular}{c|c|c|c|c}
			\hline
			Configuration & Soft & Collinear & $x$ & $y$\\
			\hline
			Single soft & $i$ & - & $(s-s_{-i})/s$ & - \\
			Single collinear & - & $i\parallel j$ & $s_{ij}/s$ & - \\
		\end{tabular}
		\caption{Variable $x$ used to probe single-unresolved infrared limits.}\label{tab:smallx}
	\end{table}
    To remove angular correlations and achieve a proper subtraction in infrared limits with collinear partons, a point-by-point angular average is considered, as described in detail in~\cite{Glover:2010kwr}. 
    
    We present here the spike-tests for the assessment of the real-virtual subtraction terms across a sample of partonic subprocesses of $pp\to 3j$. The validation is carried out in full-colour. The titles of the plots below state the infrared configuration, the unresolved parton, counted from $1$ to $6$ with (1,2) in the initial state, 
    and the considered subprocess. The pairs of numbers reported under the label `outside' respectively indicate how many events fell on the left and on the right of the displayed range in $t_{RV}$. 
    
\begin{figure}
    	\begin{center}
    		\includegraphics[width=0.30\columnwidth]{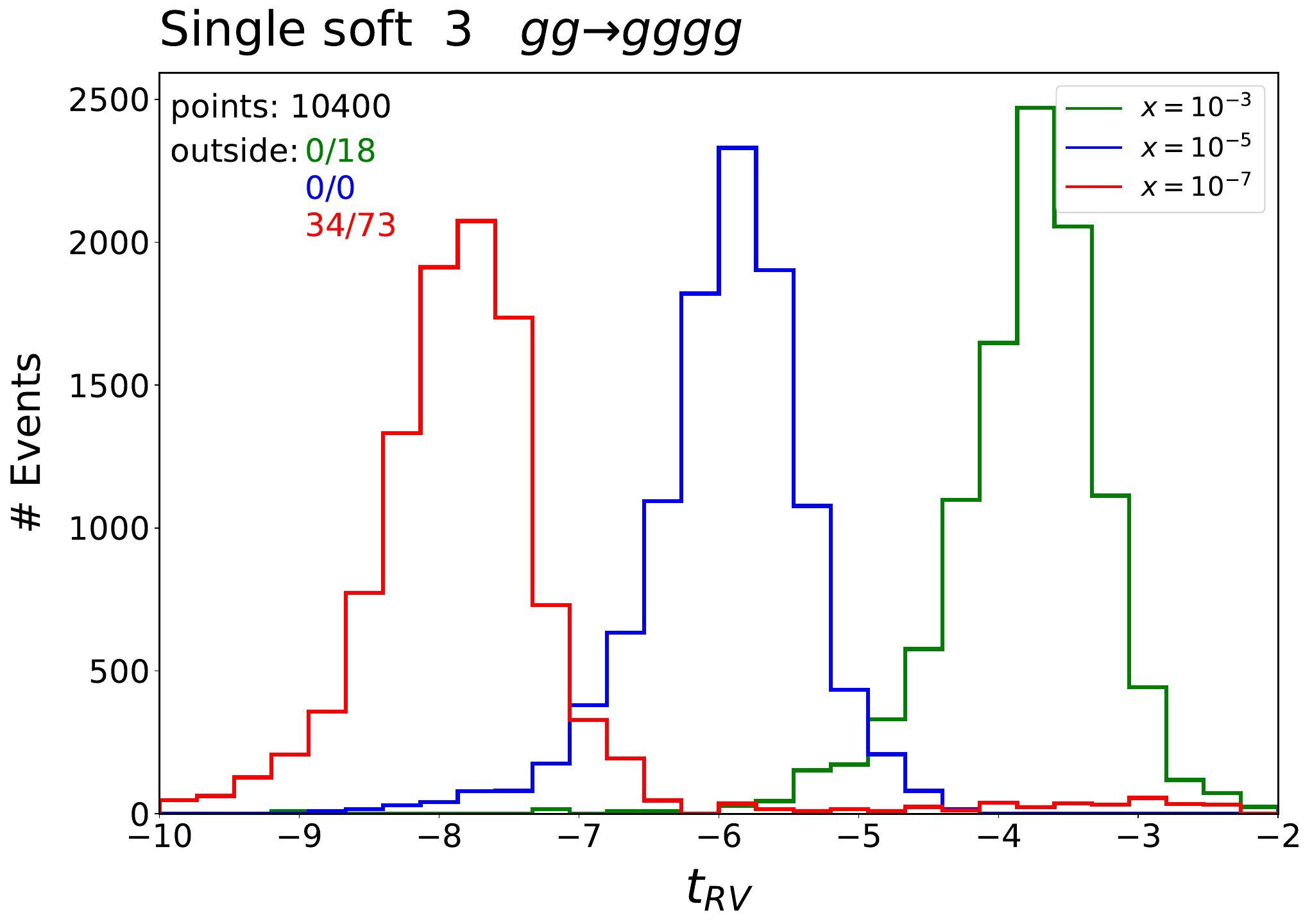}
    		\hspace{0.3cm}
    		\includegraphics[width=0.30\columnwidth]{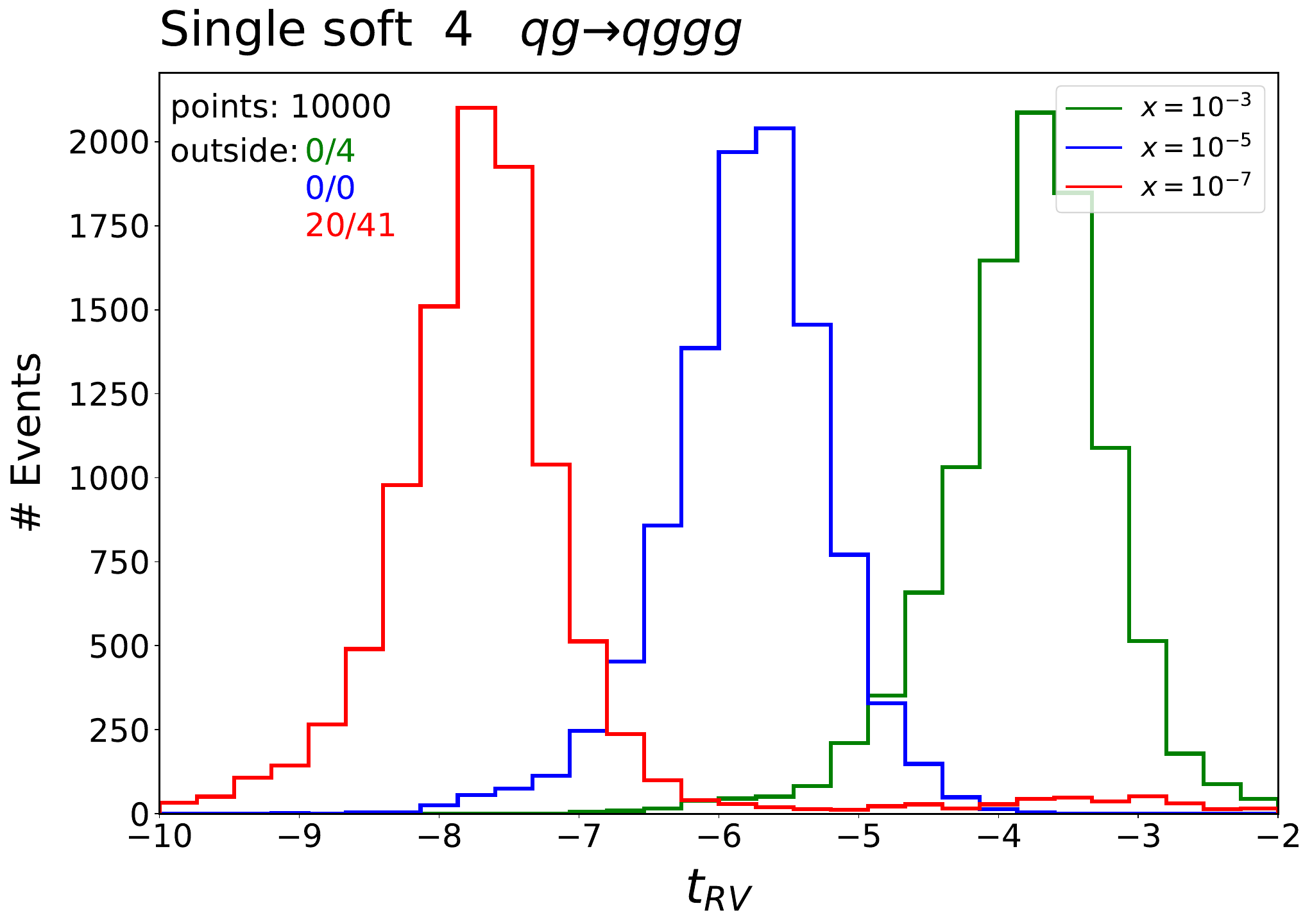}
    		\hspace{0.3cm}
    		\includegraphics[width=0.30\columnwidth]{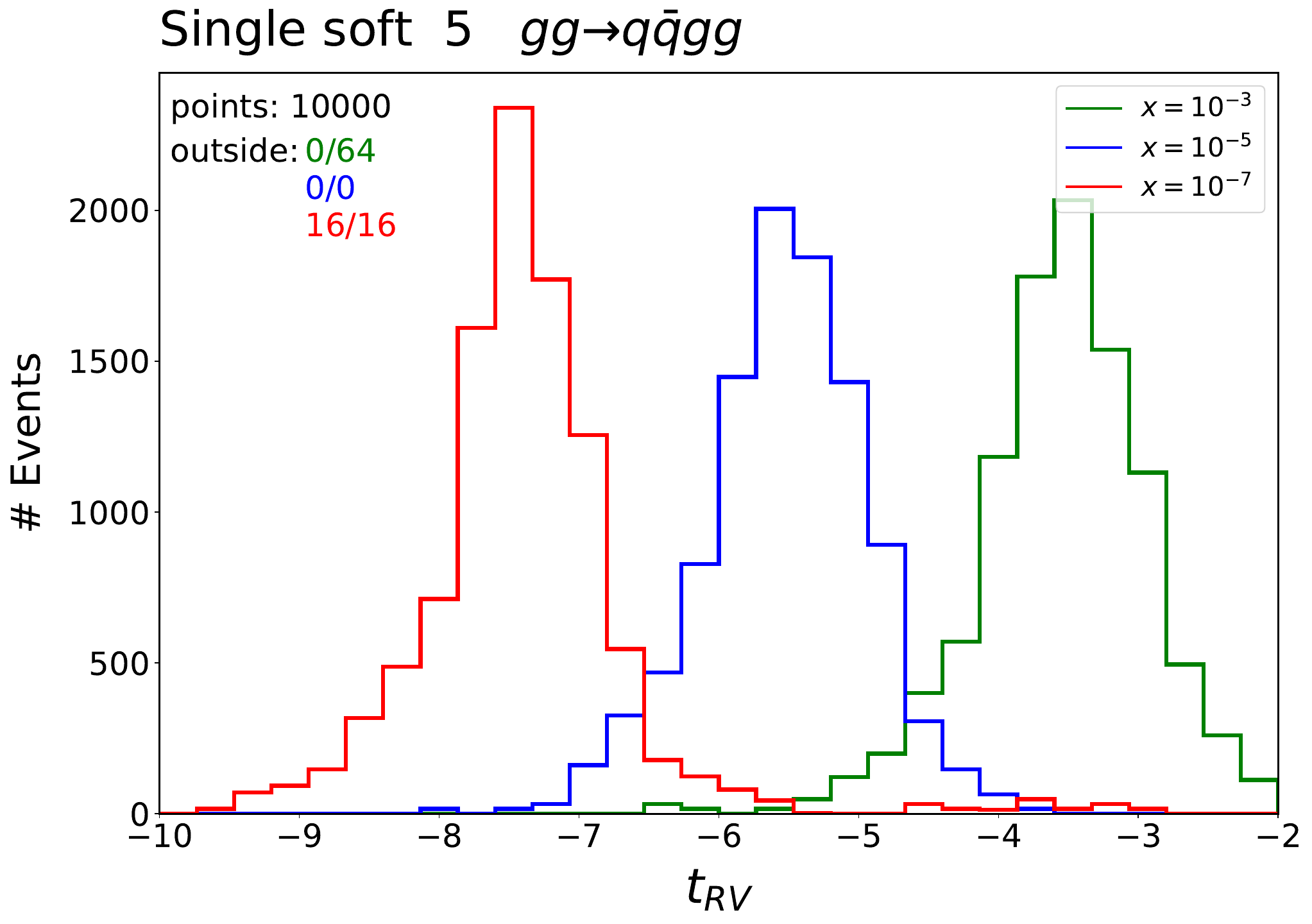}\\
    		\includegraphics[width=0.30\columnwidth]{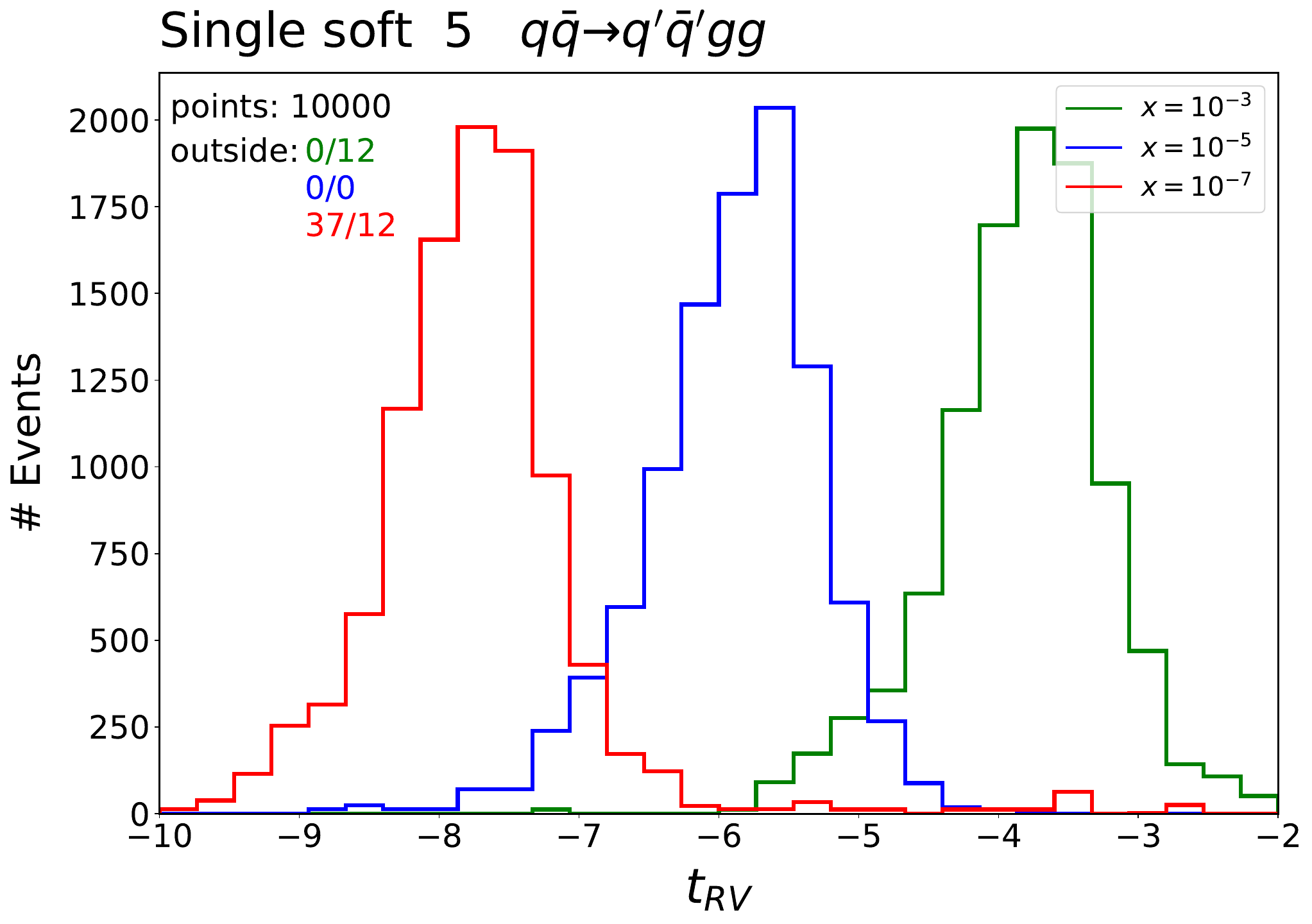}
    		\hspace{0.3cm}
    		\includegraphics[width=0.30\columnwidth]{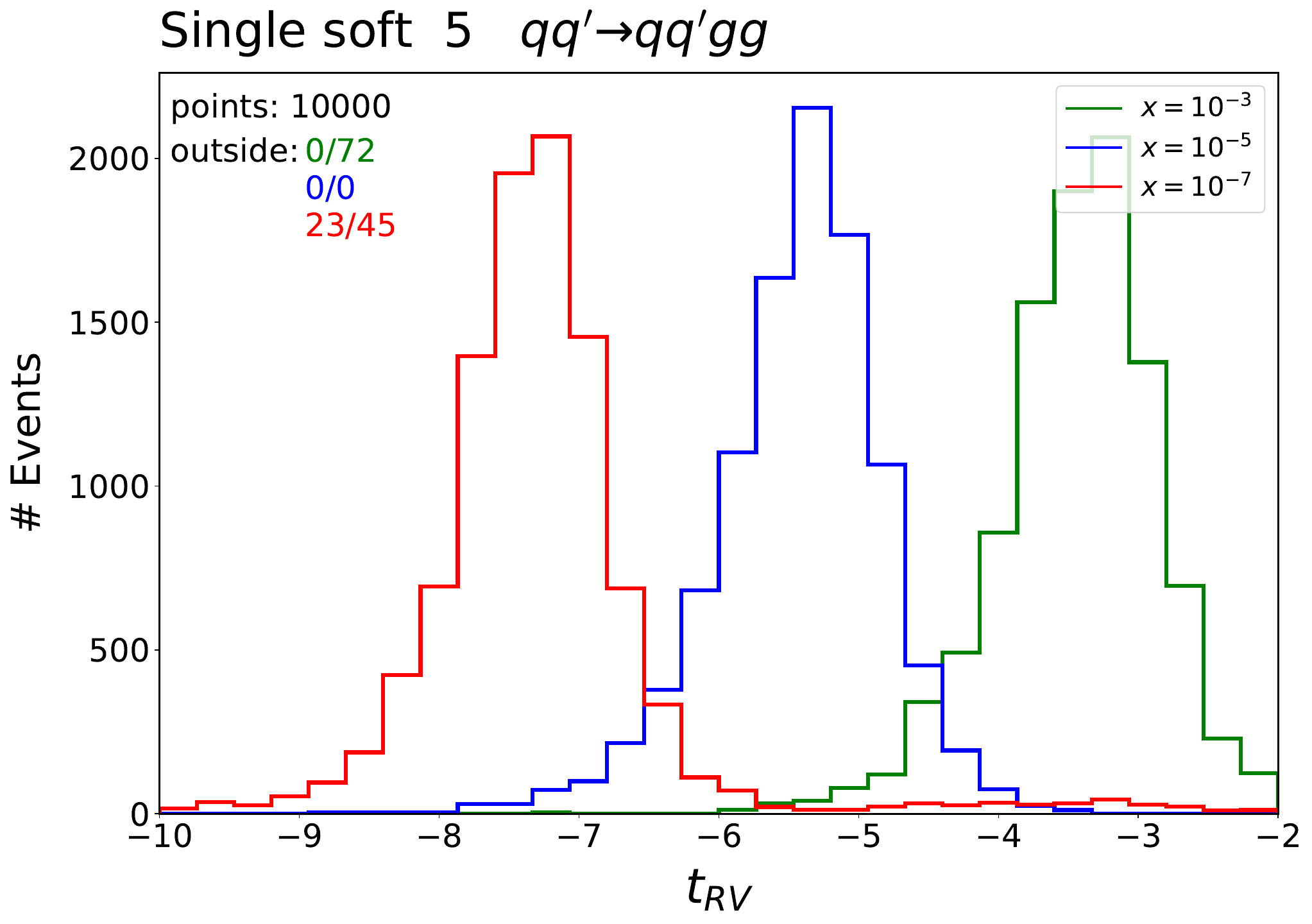}
    		\hspace{0.3cm}
    		\includegraphics[width=0.30\columnwidth]{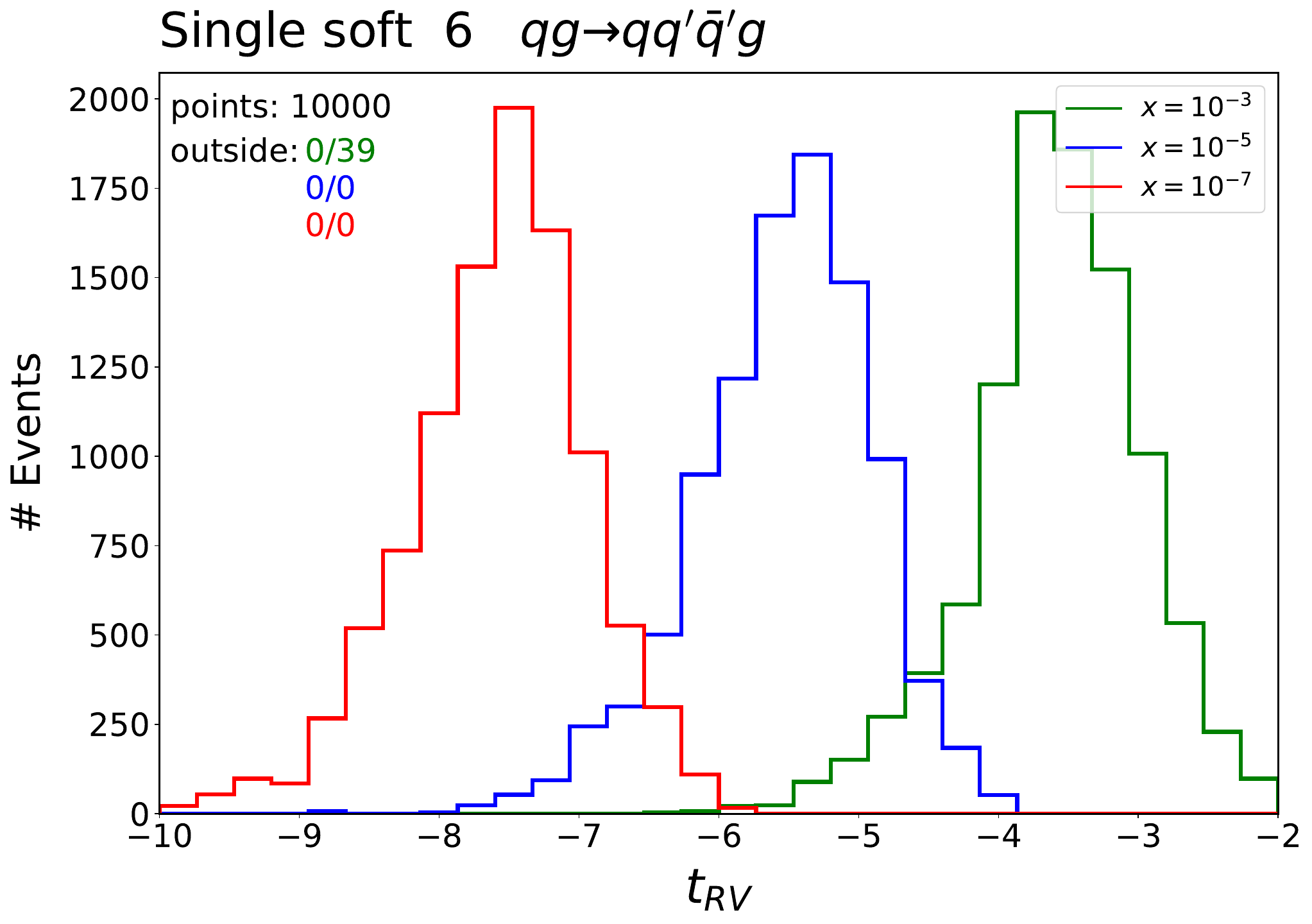}
    	\end{center}
    	\caption{Validation of the real-virtual subtraction terms for single-soft limits.}
    	\label{fig:spikeRVsoft}
    \end{figure}

    We first consider the soft limit. In Figure~\ref{fig:spikeRVsoft} we present the validation of the subtraction terms for a selection of subprocesses. The first plot, which refers to the $gg\to gggg$ channel, differs from the results discussed in the gluons-only case in~\cite{Chen:2022ktf} by the inclusion of fermionic-loop contributions ($N_f\neq0$).
    In general, we observe a very good convergence pattern: the deeper the infrared limit is probed, the better is the agreement between the matrix element and the subtraction terms. The distributions appear particularly smooth, indicating that potential numerical noise from the multitude of terms in subleading-colour factors is not spoiling the overall convergence. 
    
    We notice however a small fraction of points ($\lesssim 0.5\%$) leaking to the right side of the plots for the deepest choice of the $x$ variable (red curve) form some of the considered subprocesses. Given the good quality of the three spikes, it is unlikely that this is due to a mistake in the construction of the subtraction terms, which would show up with much more dramatic effects. We can assume that for $x=10^{-7}$, it is inevitable to hit a few unstable points, where either the full-colour matrix element returned by $\textsc{OpenLoops}$ or the large expressions in our subtraction terms have percent-level deviations from the true result. During the full calculation, such extreme regions of the phase space are highly suppressed. A small number of unstable points should not significantly affect the importance sampling techniques employed to optimize a Monte Carlo integration. 
        
\begin{figure}
    	\begin{center}
    		\includegraphics[width=0.3\columnwidth]{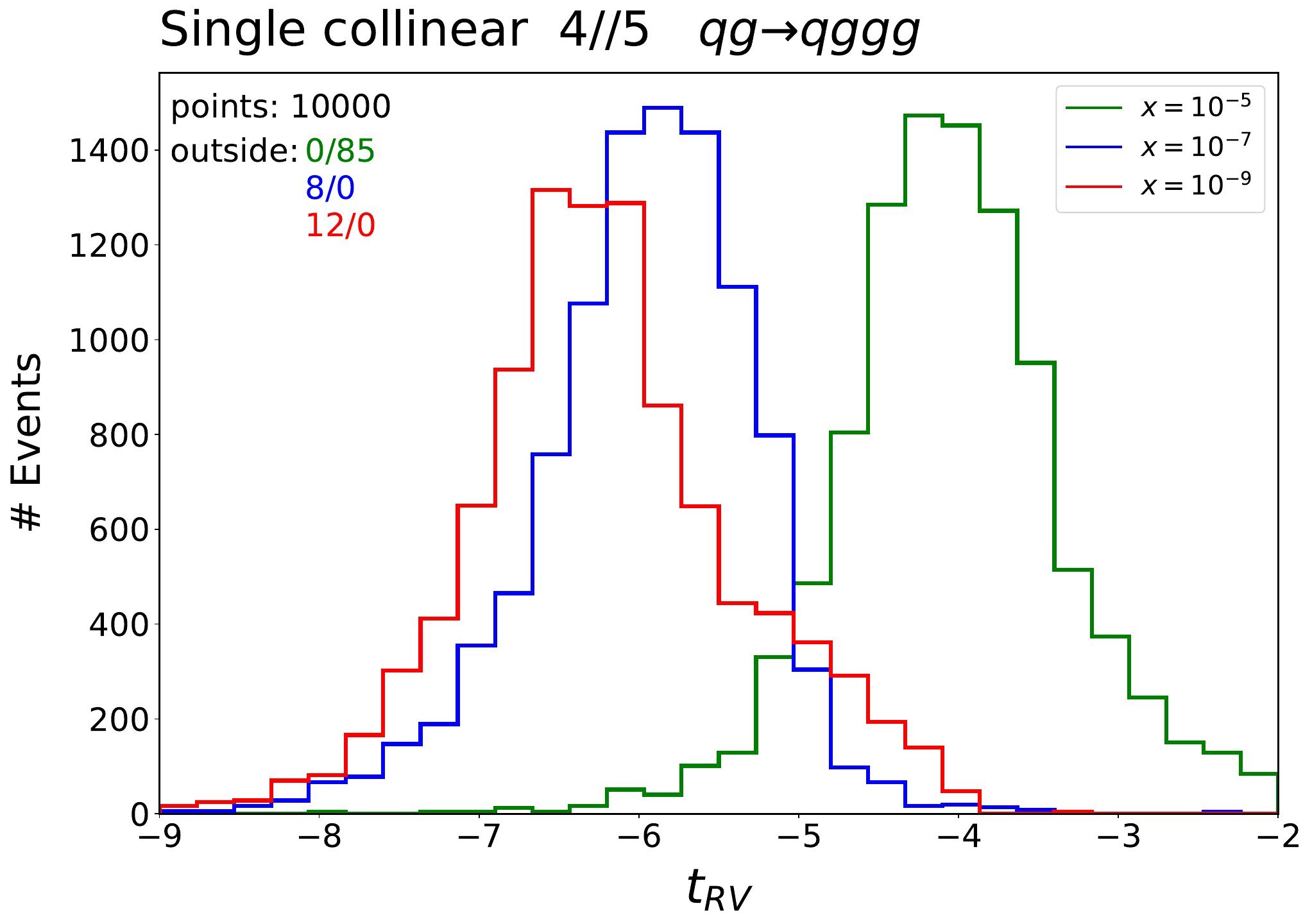}
    		\hspace{0.3cm}
    		\includegraphics[width=0.3\columnwidth]{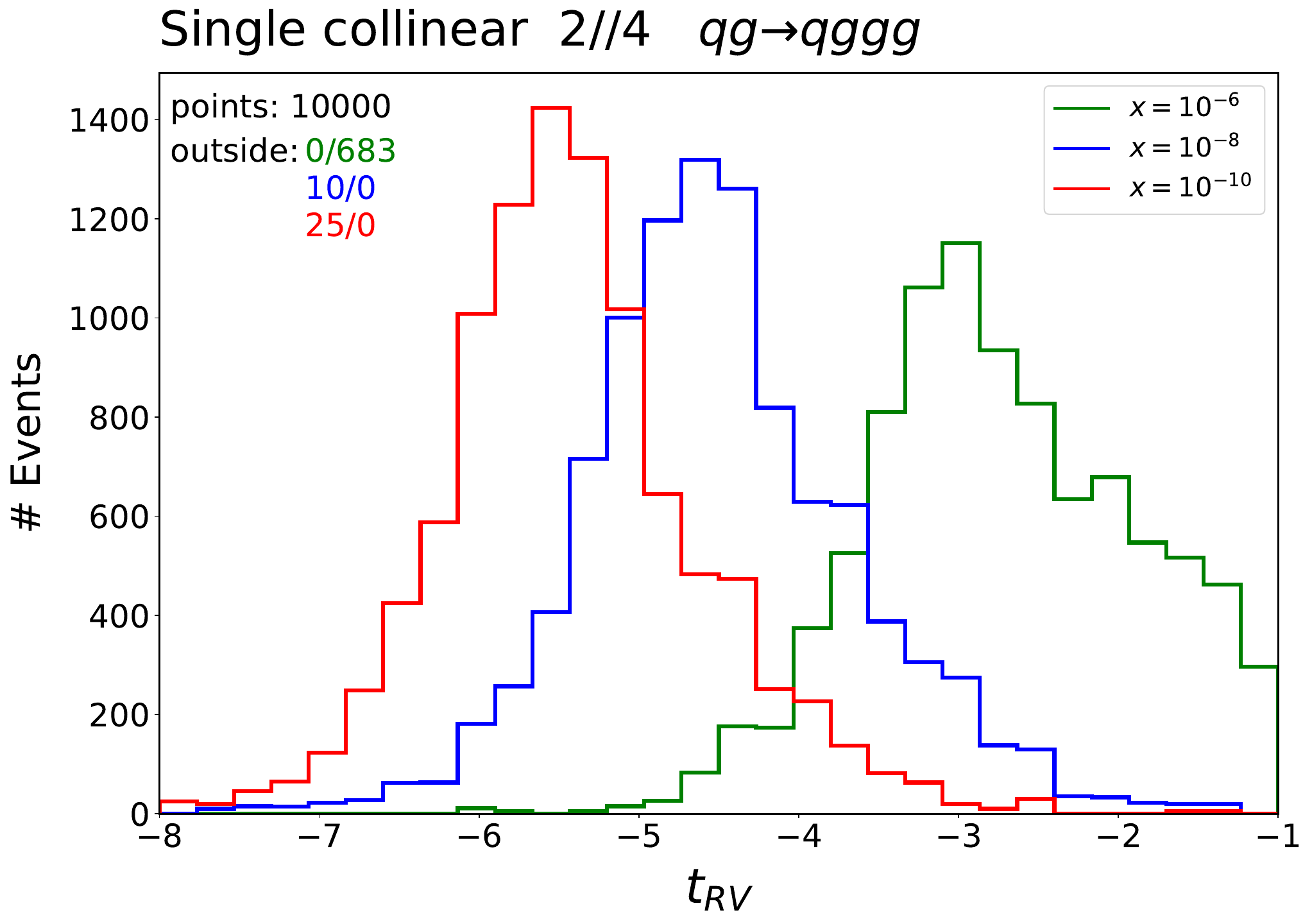}
    		\hspace{0.3cm}
    		\includegraphics[width=0.3\columnwidth]{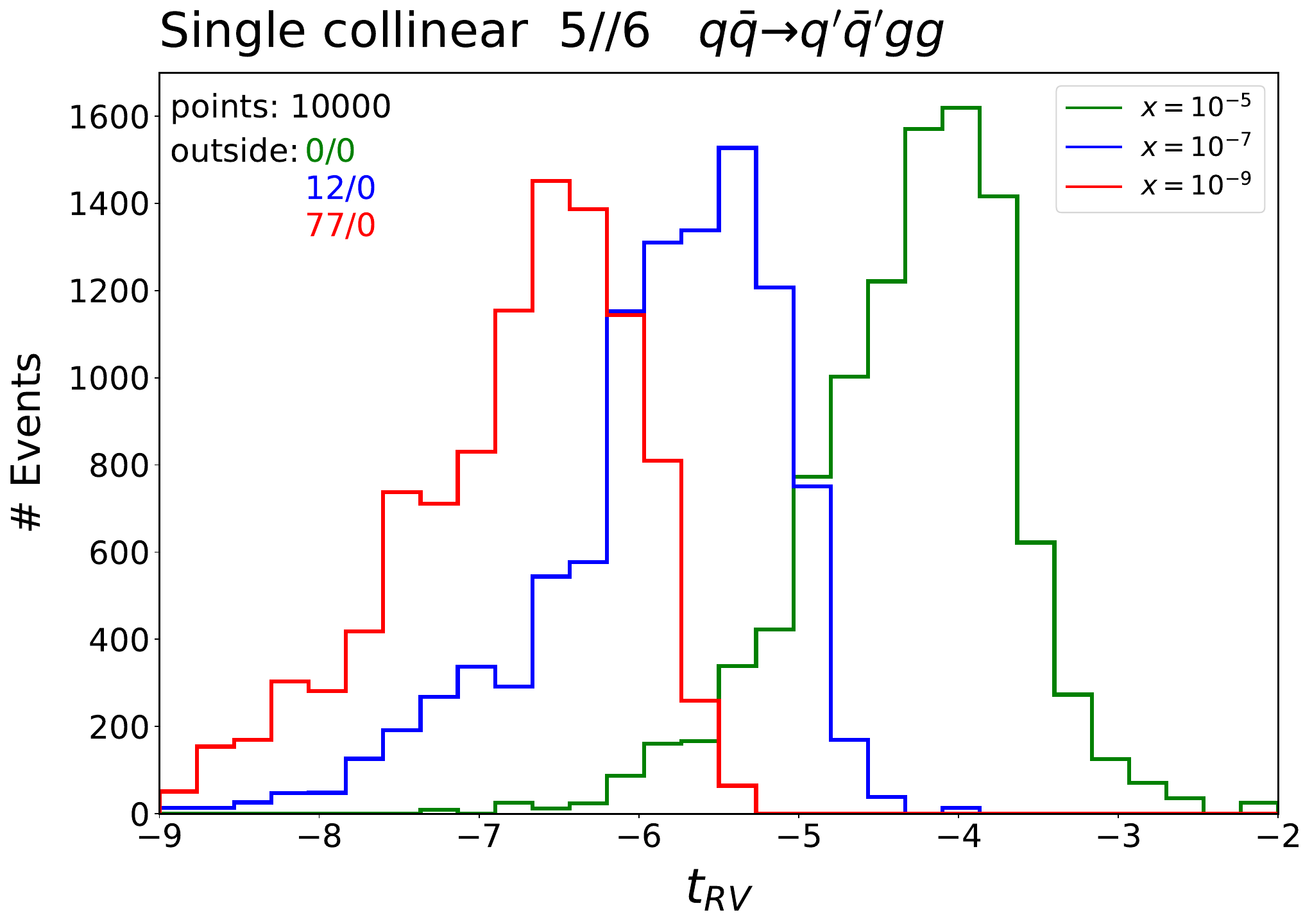}\\
    		\includegraphics[width=0.3\columnwidth]{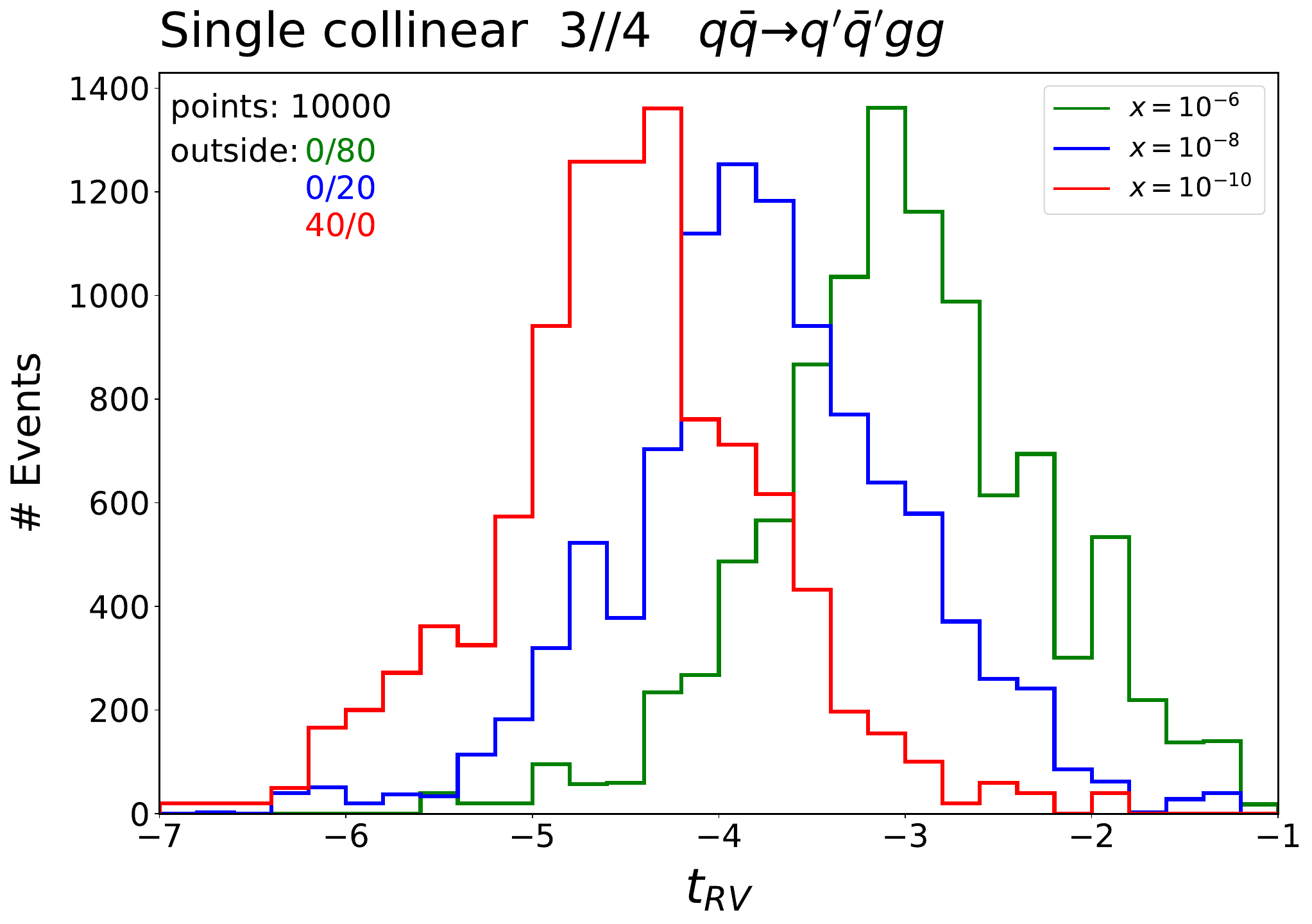}
    		\hspace{0.3cm}
    		\includegraphics[width=0.3\columnwidth]{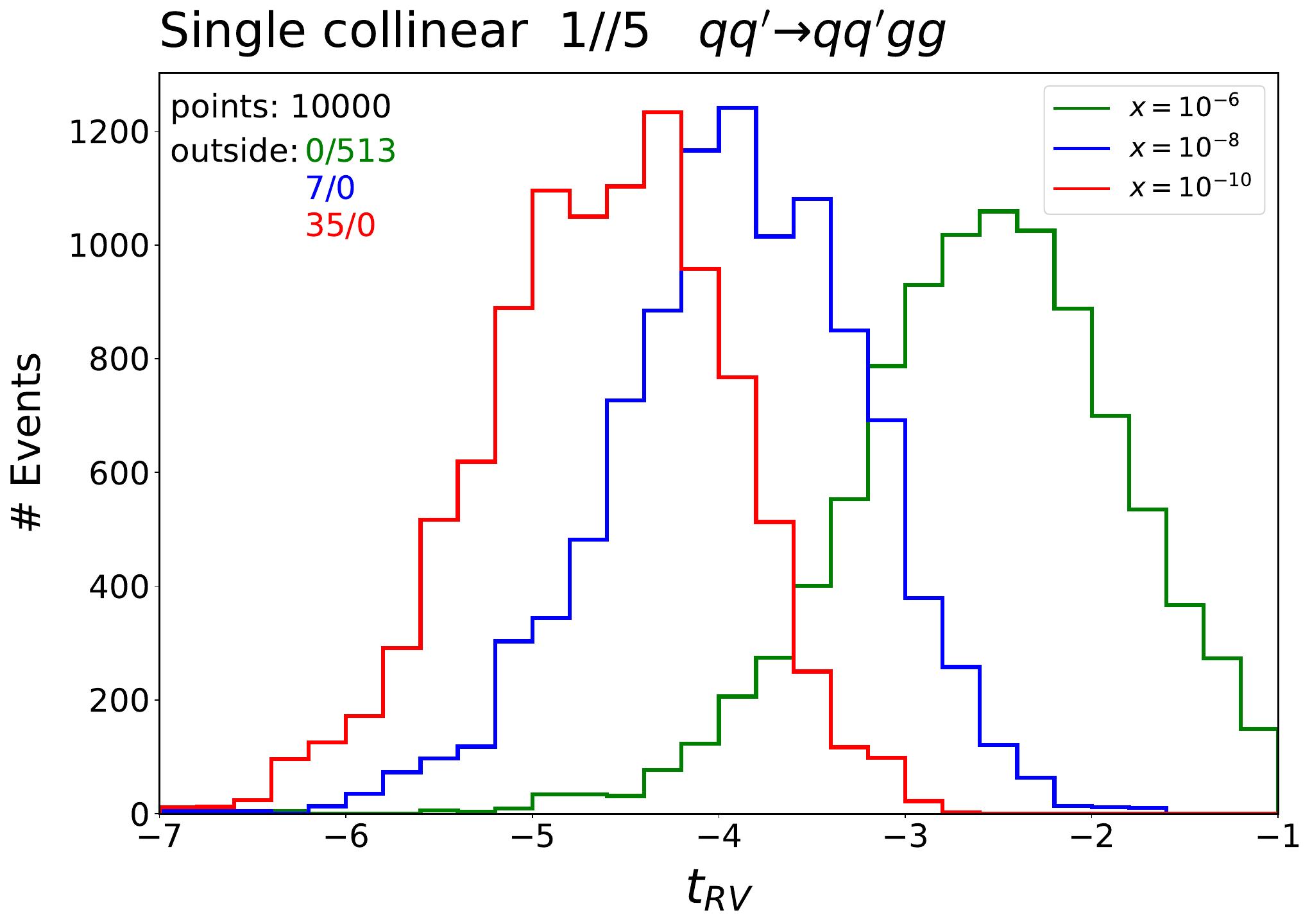}
    		\hspace{0.3cm}
    		\includegraphics[width=0.3\columnwidth]{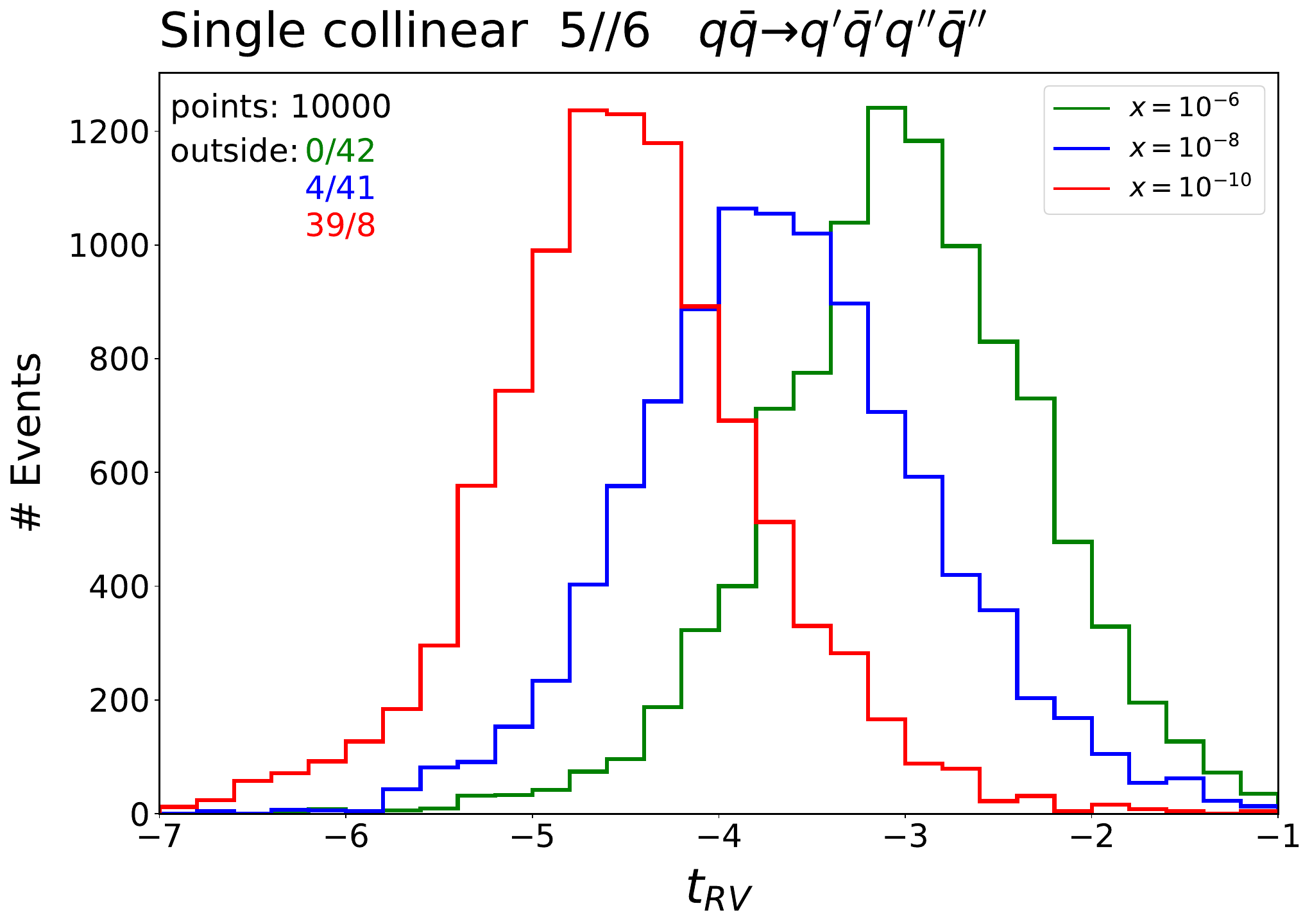}
    	\end{center}
    	\caption{Validation of the real-virtual subtraction terms for single-collinear limits.}
    	\label{fig:spikeRVcoll}
    \end{figure}

    In Figure~\ref{fig:spikeRVcoll} we show examples of the validation for single-collinear limits. We note that the spikes look less peaked with respect to the soft limits, for the same reasons mentioned in the gluons-only case~\cite{Chen:2022ktf}. Nevertheless, we observe a satisfactory agreement between the matrix elements and the subtraction terms. For collinear limits, we do not see the appearance of numerical instabilities. 
	
	\subsubsection{Double-real subtraction terms}
	
	The assessment of the subtraction at the double-real level is performed with the same strategy employed for the real-virtual correction. We generate $10000$ phase space points at $\sqrt{s}=13$ TeV in single- and double-unresolved limits and study the ratio
	\begin{equation}
		\quad R_{\text{RR}}=\dfrac{\dsigRR{ab}}{\dsigSNNLO{ab}},
	\end{equation}
	binned according to 
	\begin{equation}\label{tvarRR}
		t_{RR}=\log_{10} \left(\left|1-R_{RR}\right|\right).
	\end{equation}
	To parametrize the depth we reach in unresolved limits we use the variables $x$ and $y$. For single-unresolved limits, the definition of $x$ is given above in Table~\ref{tab:smallx}, while $x$ and $y$ for double-unresolved ones are given in Table~\ref{tab:smallx2}. For configurations that require both $x$ and $y$ (soft-collinear and double-collinear), we choose to fix $x=y$. 
	\begin{table}
		\centering
		\begin{tabular}{c|c|c|c|c}
			\hline
			Configuration & Soft & Collinear & $x$ & $y$\\
			\hline
			Double soft & $i,j$ & - & $(s-s_{-ij})/s$ & - \\
			Triple collinear & - & $i\parallel j\parallel k$ & $s_{ijk}/s$ & - \\
			Soft and collinear & $i$ & $j\parallel k$ & $(s-s_{-i})/s$ & $s_{jk}/s$ \\
			Double collinear & - & $i\parallel j,\,k\parallel l$ & $s_{jk}/s$ & $s_{kl}/s$ \\
		\end{tabular}
		\caption{Variables $x$ and $y$ used to probe double-unresolved infrared limits.}\label{tab:smallx2}
	\end{table}
	Also for the double-real case, a point-by-point angular average is considered~\cite{Glover:2010kwr}, to enforce a local subtraction of the singular behaviour.
	
	We present the spike-tests for the assessment of the double-real subtraction terms across a sample of partonic subprocesses of $pp\to 3j$. The validation is carried out in full-colour. The titles of the plots below state the infrared configuration, the unresolved parton involved counting from $1$ to $7$ with (1,2) in the initial state, and the considered subprocess. As before, the pairs of numbers reported under the label `outside' respectively indicate how many events fell on the left and on the right of the displayed range in $t_{RR}$. 
	
\begin{figure}
		\begin{center}
			\includegraphics[width=0.3\columnwidth]{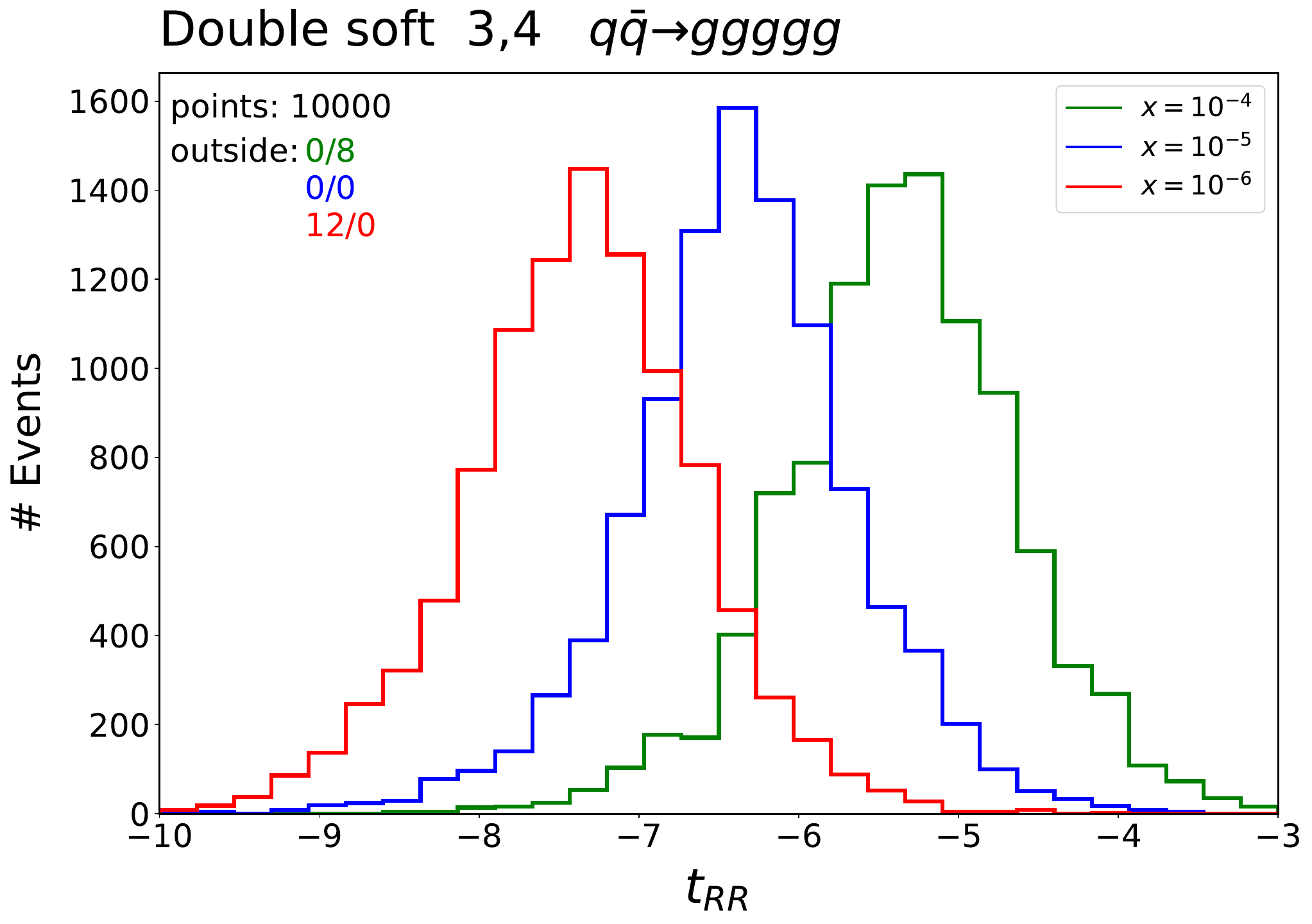}
			\hspace{0.3cm}
			\includegraphics[width=0.3\columnwidth]{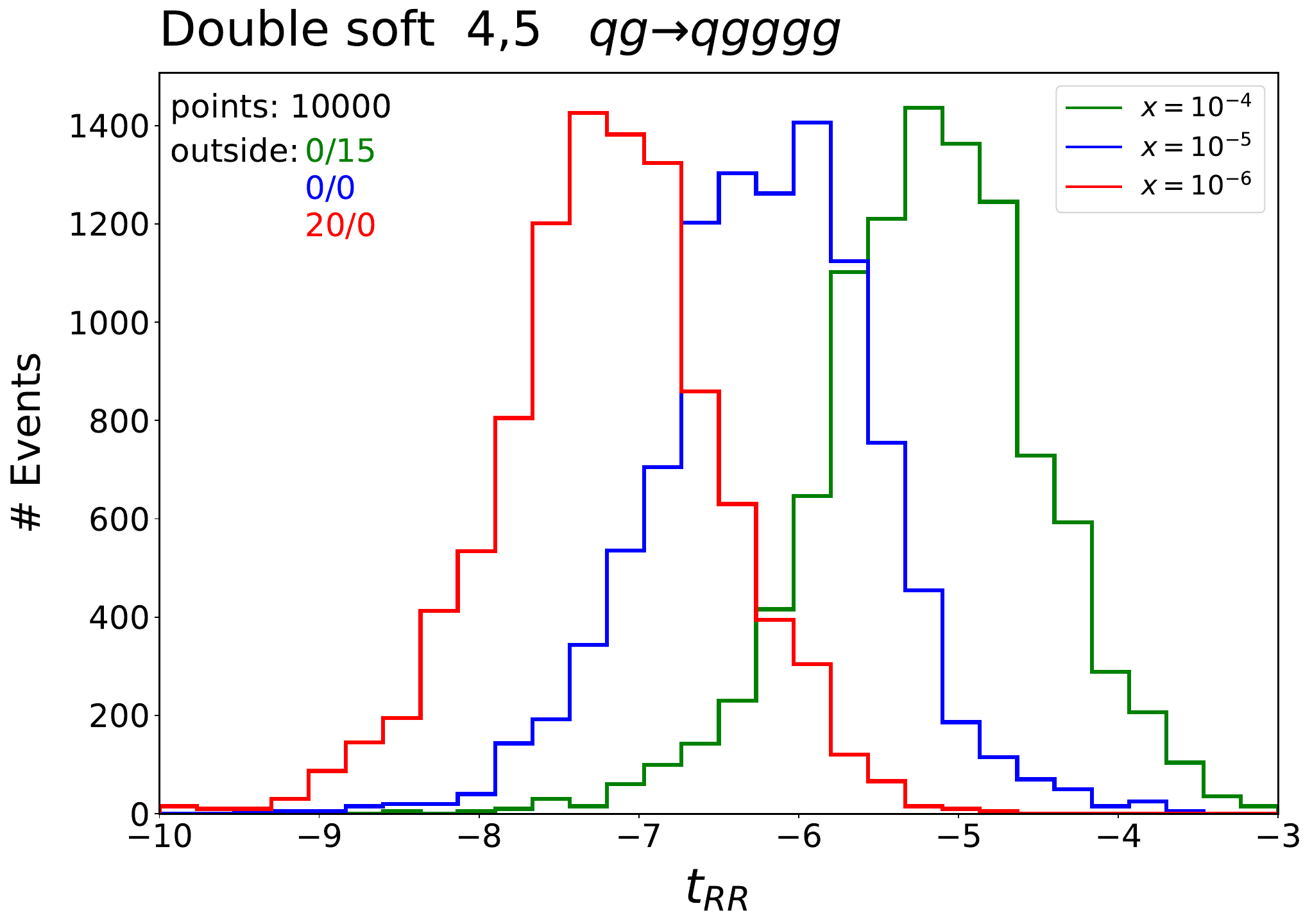}
			\hspace{0.3cm}
			\includegraphics[width=0.3\columnwidth]{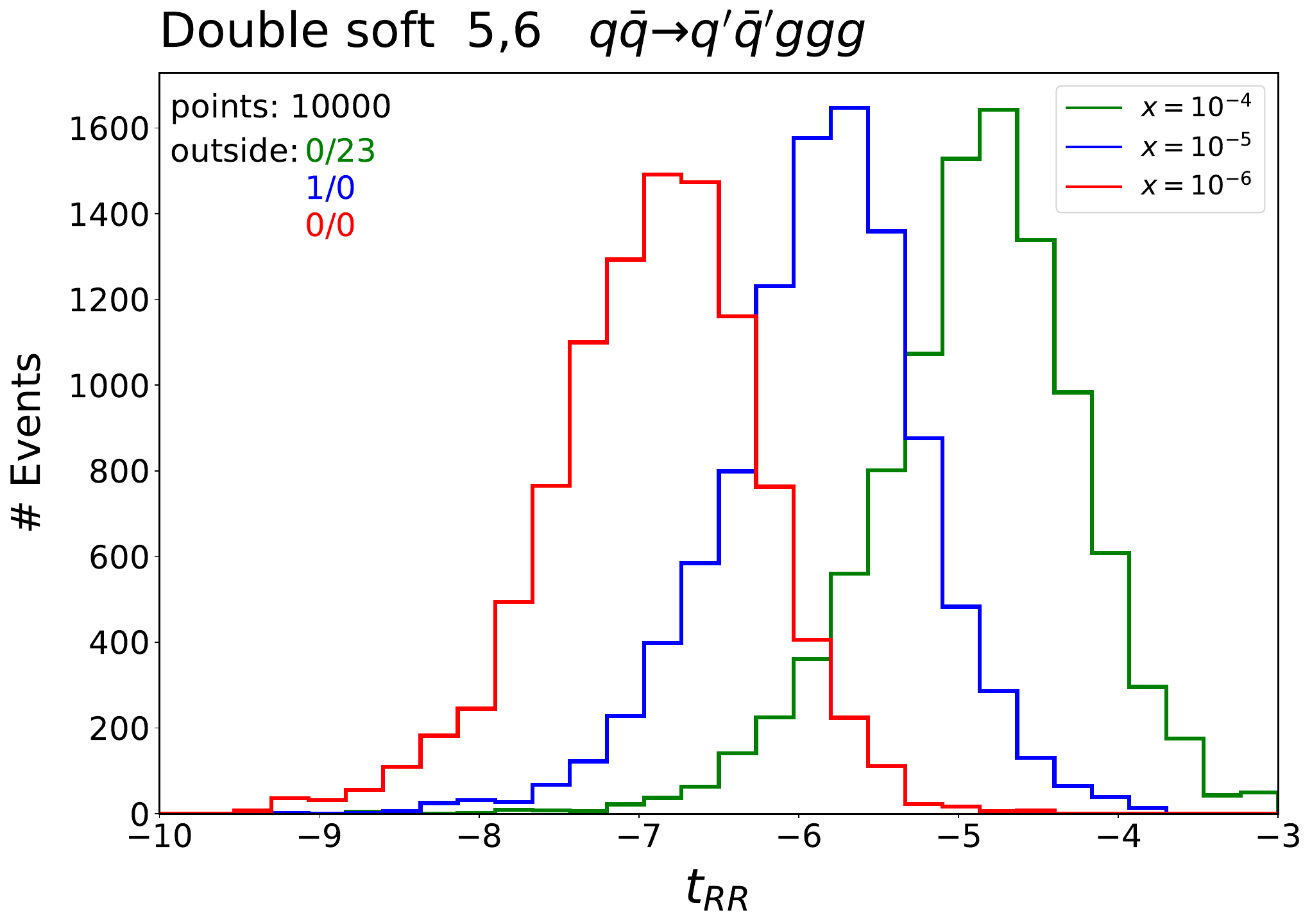}\\
			\includegraphics[width=0.3\columnwidth]{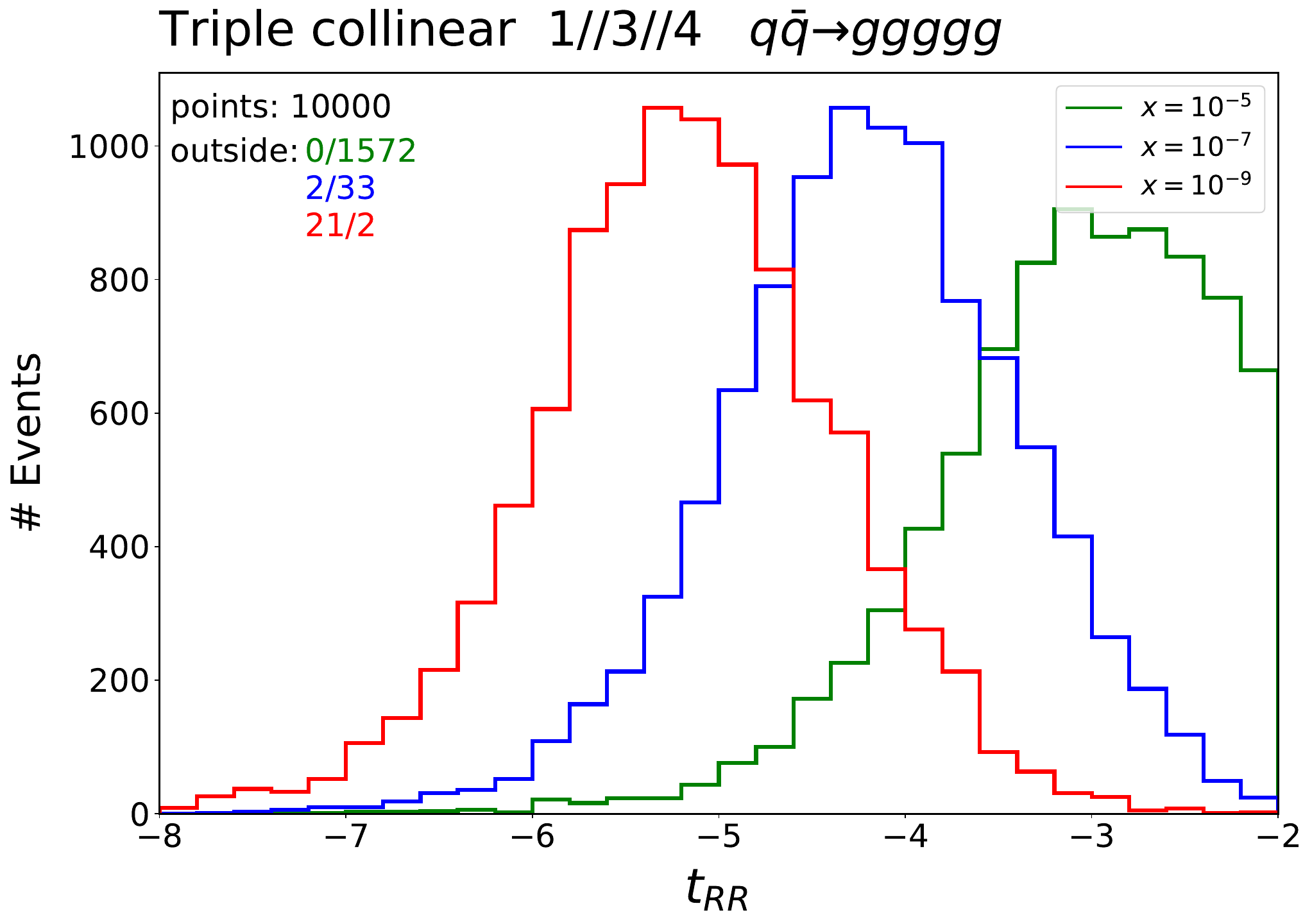}
			\hspace{0.3cm}
			\includegraphics[width=0.3\columnwidth]{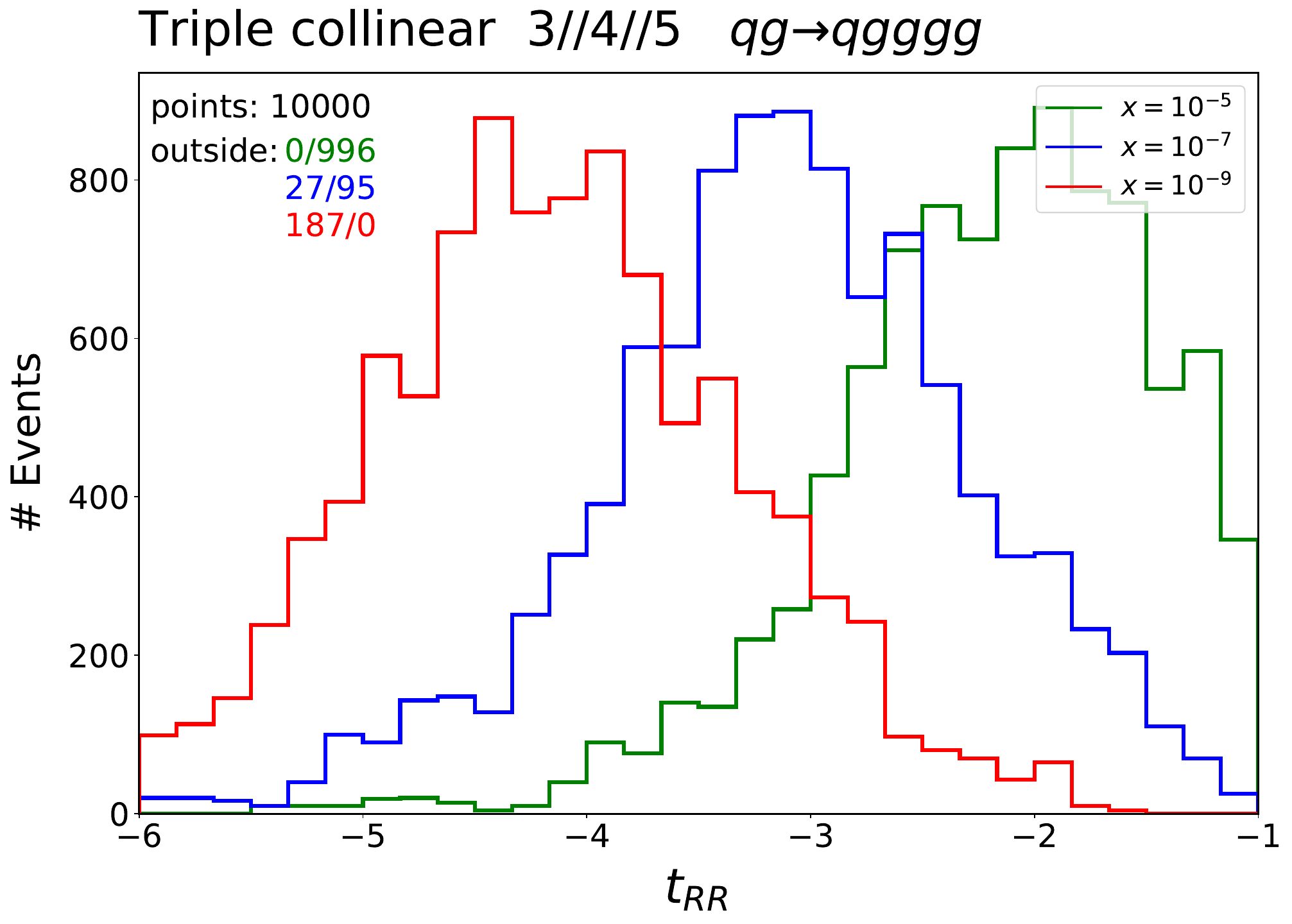}
			\hspace{0.3cm}
			\includegraphics[width=0.3\columnwidth]{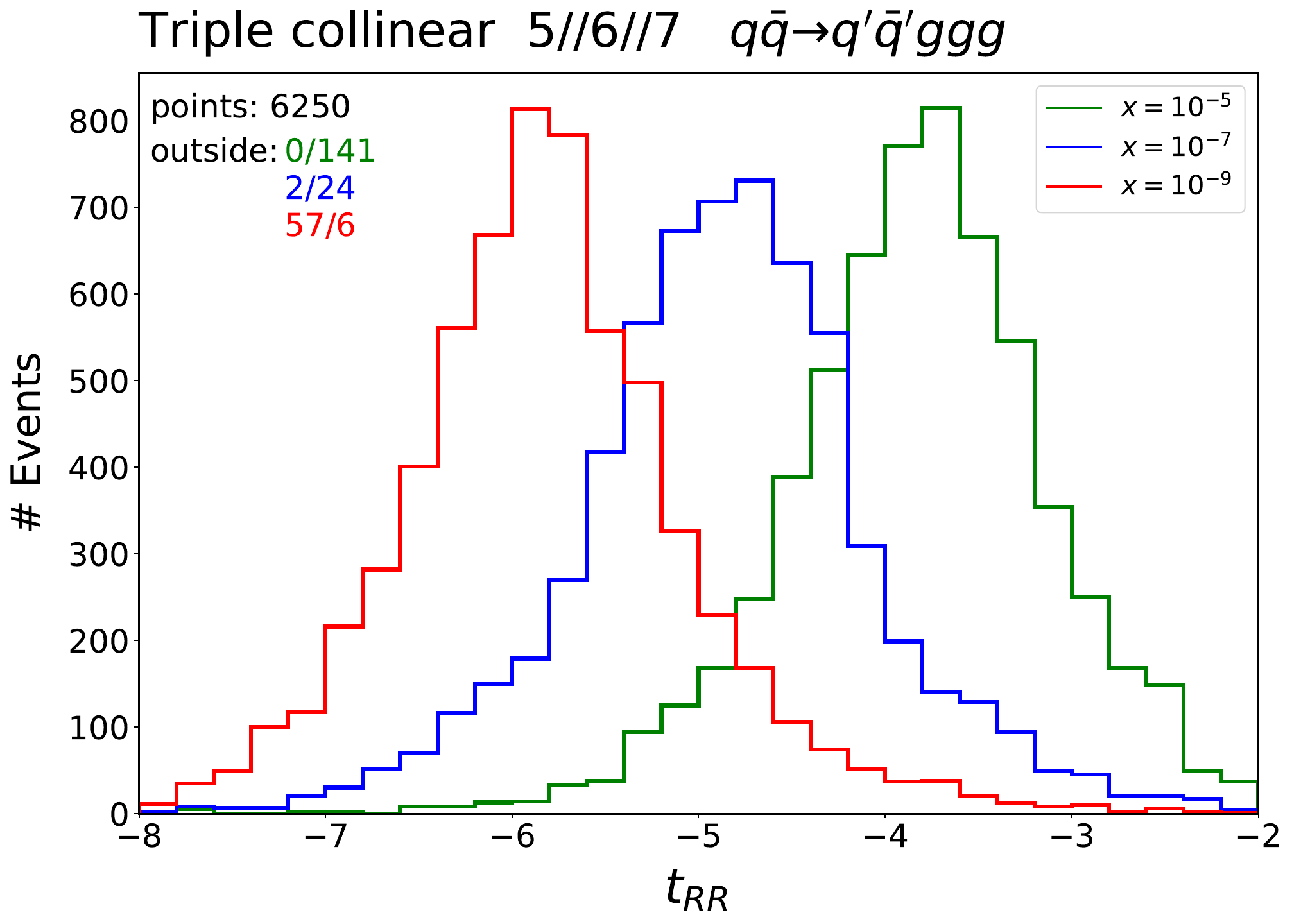}
		\end{center}
		\caption{Validation of the double-real subtraction terms in double-soft (upper row) and triple-collinear (lower row) limits for a selection of subprocesses.}
		\label{fig:spikeRR_DS_TC}
	\end{figure}

	In Figure~\ref{fig:spikeRR_DS_TC}, we display some examples of double-unresolved configurations related to colour-connected unresolved partons: double-soft and triple-collinear limits. We observe a very good convergence. The double-real matrix elements and subtraction terms, not containing any loop amplitude, are in general much more stable in double-unresolved limits than the real-virtual ones in single-unresolved configurations. For this reason, we do not notice any numerical instability in the point-by-point tests and we can exclude the necessity of employing quadruple-precision arithmetic for the computation of the double-real correction.
	
\begin{figure}
		\begin{center}
			\includegraphics[width=0.3\columnwidth]{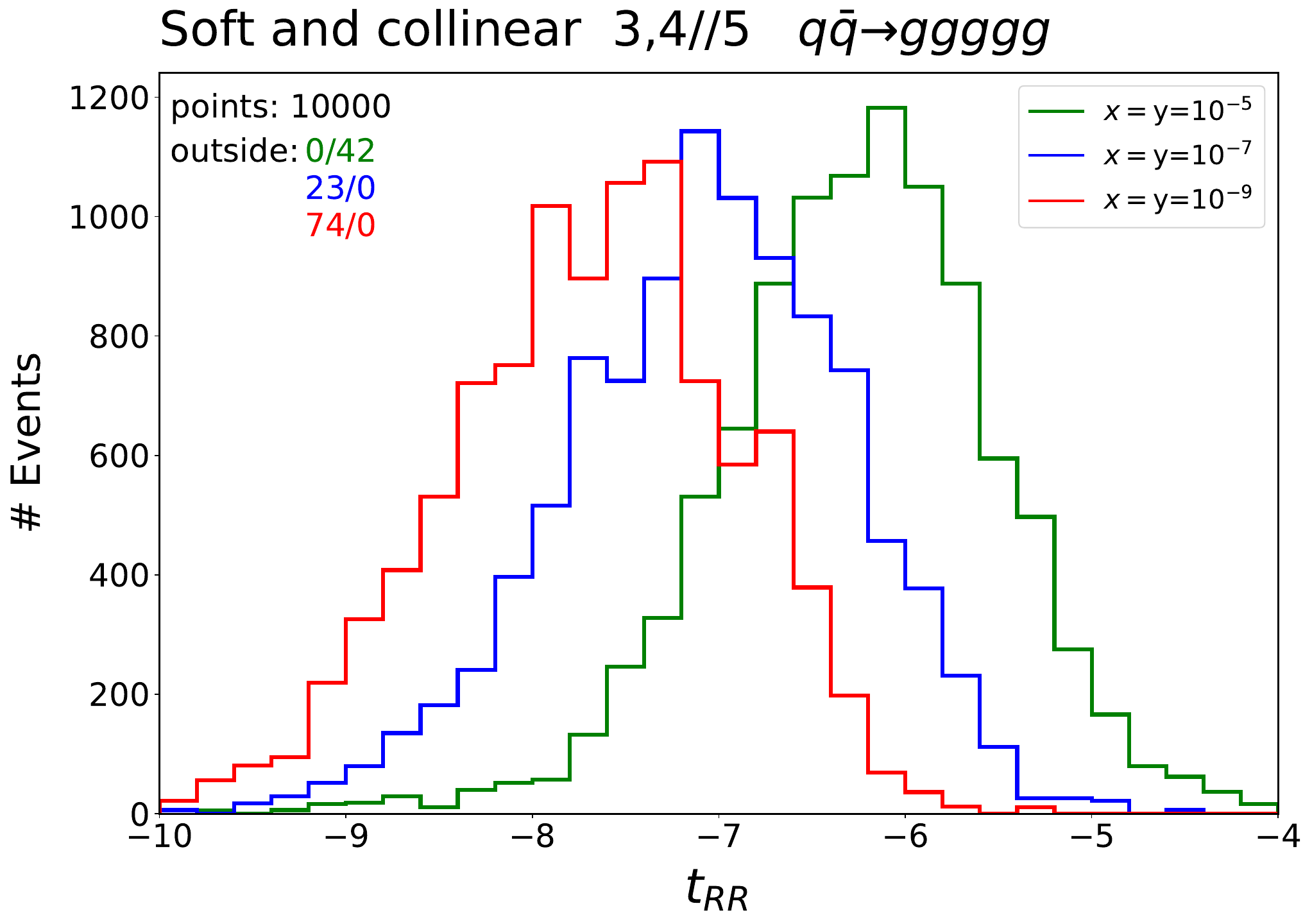}
			\hspace{0.3cm}
			\includegraphics[width=0.3\columnwidth]{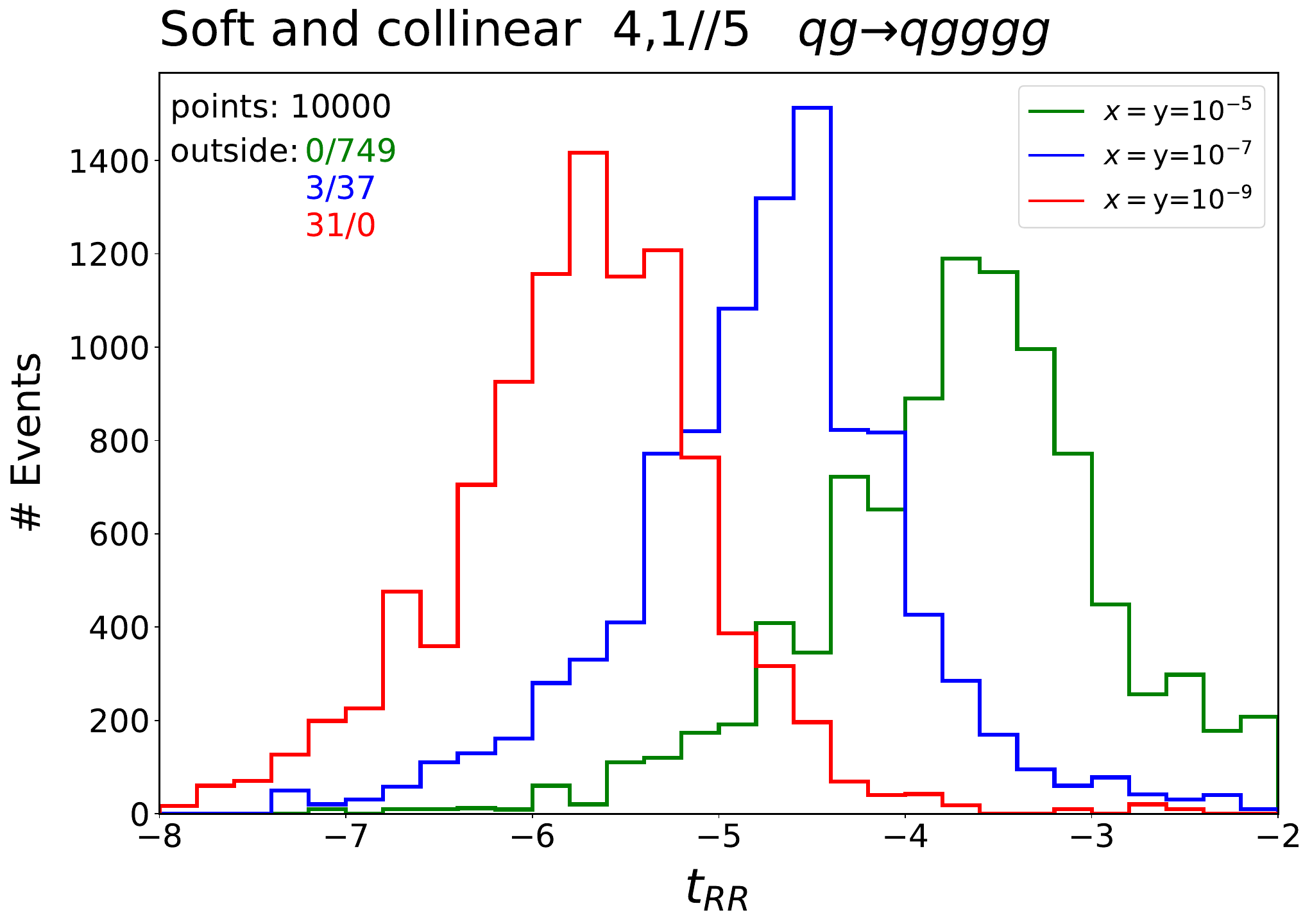}
			\hspace{0.3cm}
			\includegraphics[width=0.3\columnwidth]{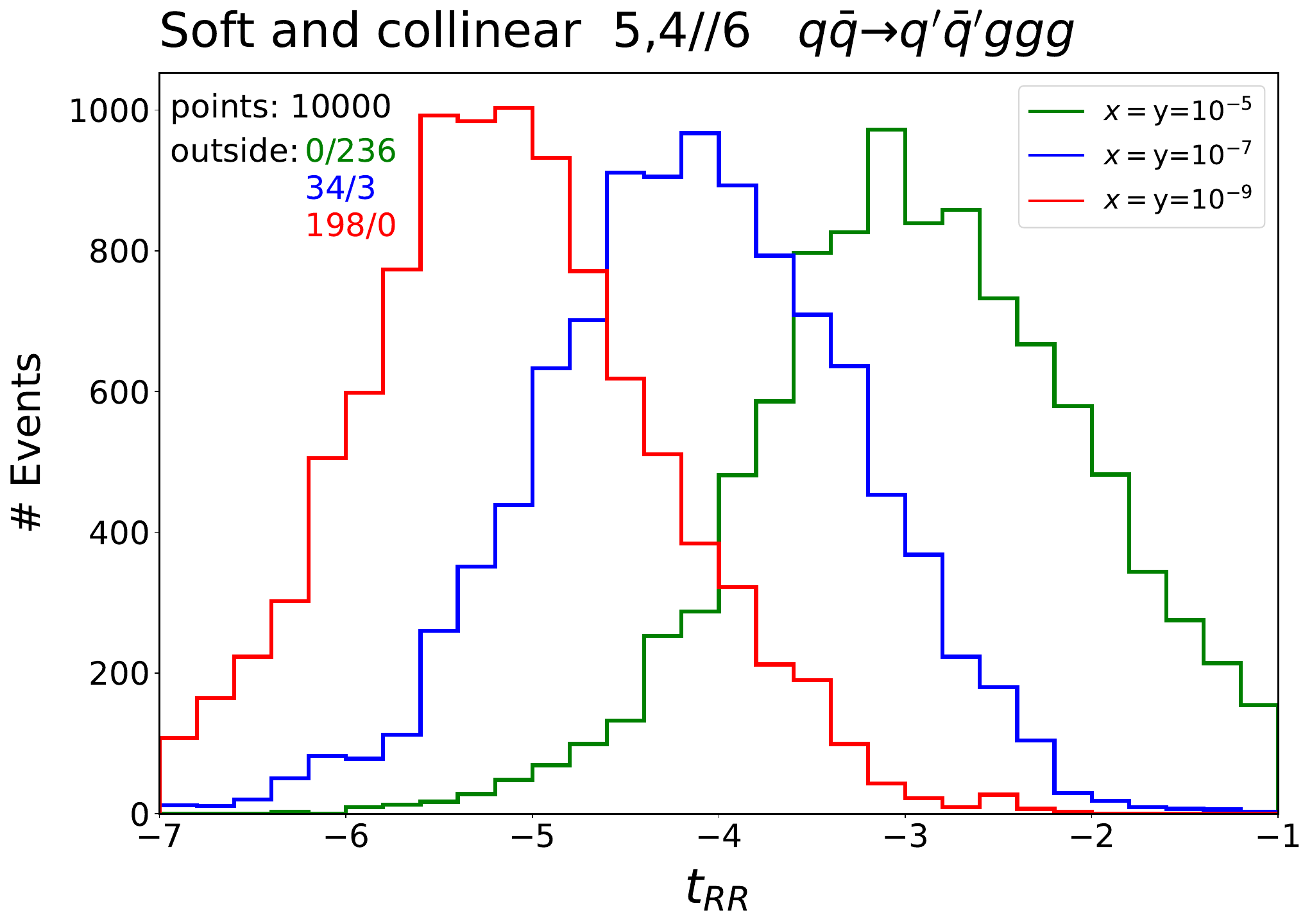}\\
			\includegraphics[width=0.3\columnwidth]{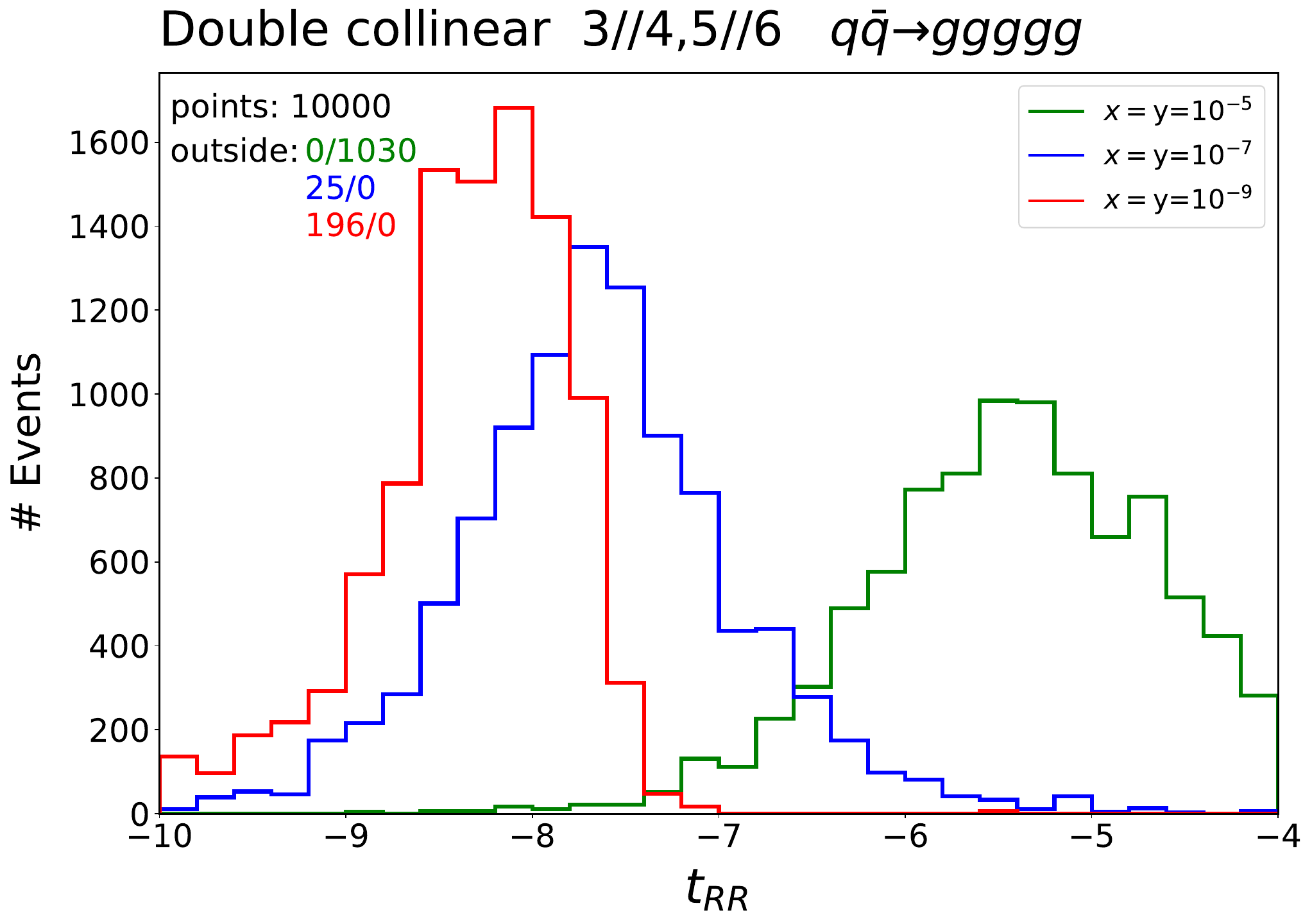}
			\hspace{0.3cm}
			\includegraphics[width=0.3\columnwidth]{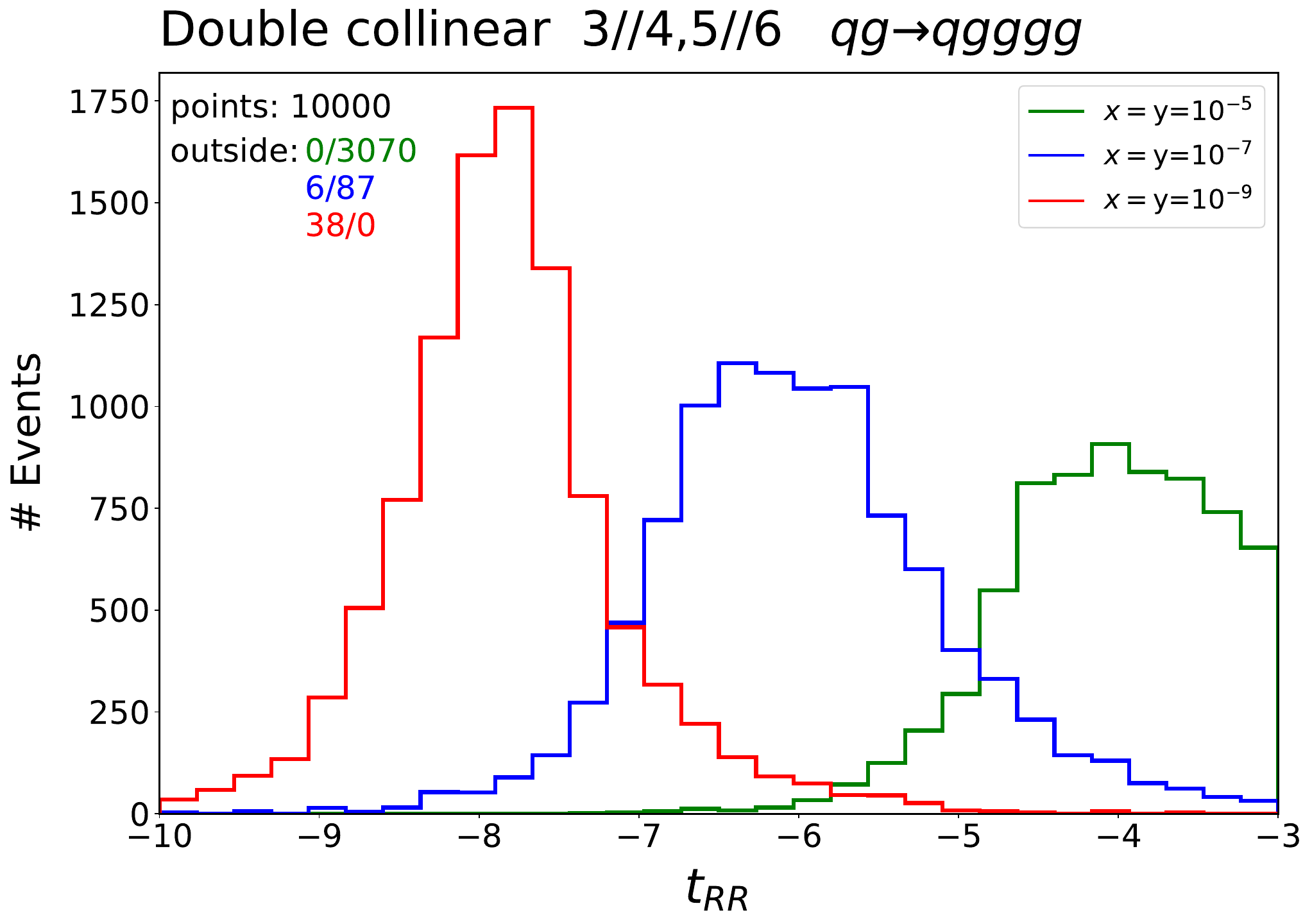}
			\hspace{0.3cm}
			\includegraphics[width=0.3\columnwidth]{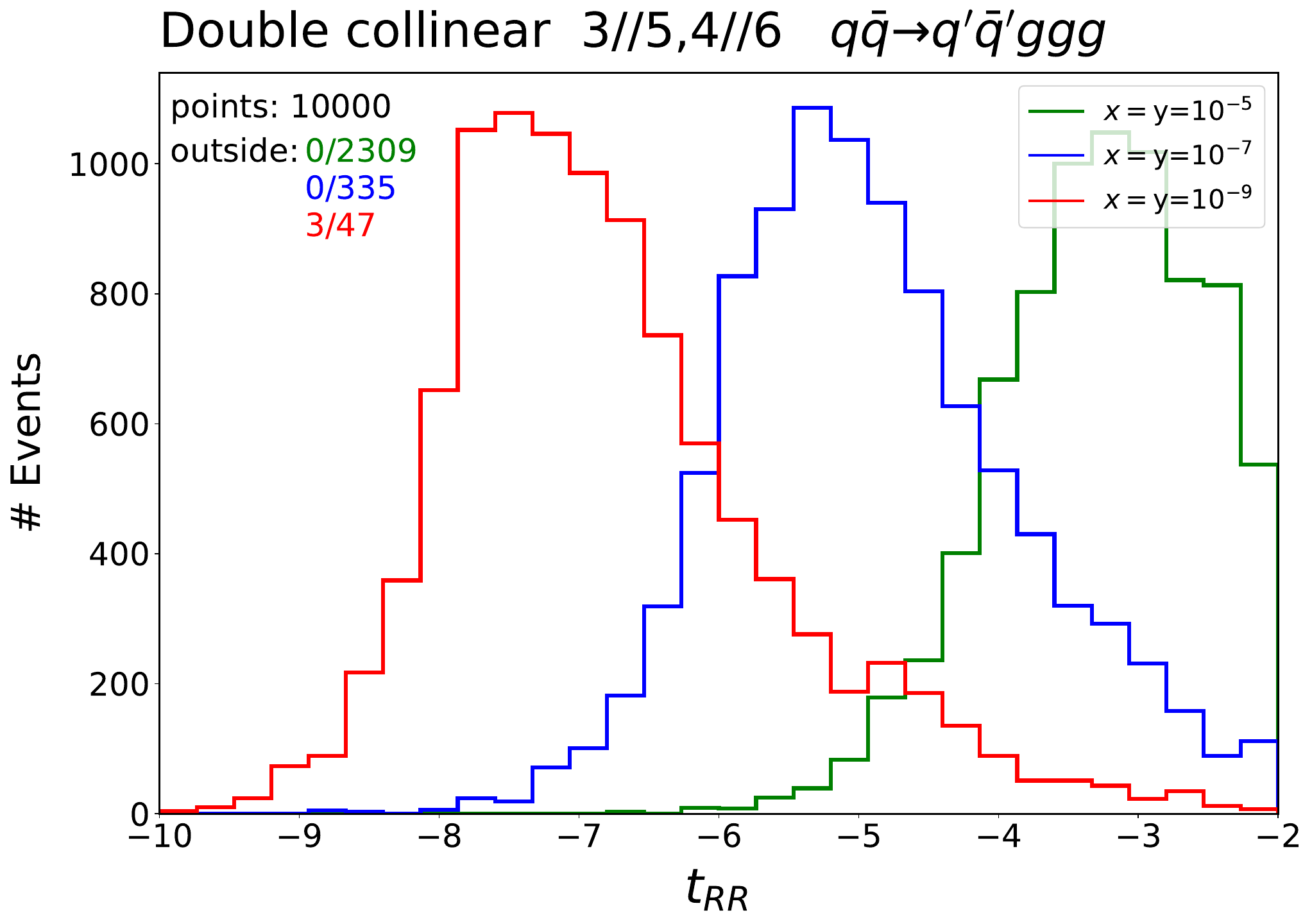}
		\end{center}
		\caption{Validation of the double-real subtraction terms in soft-collinear (upper row) and double-collinear (lower row) limits for a selection of subprocesses.}
		\label{fig:spikeRR_SC_DC}
	\end{figure}

	In Figure~\ref{fig:spikeRR_SC_DC} we show a few examples of double-unresolved limits given by the overlap of two single-unresolved configurations: soft-collinear and double-collinear limits. For the $q\qb\to q'\qb'ggg$ subprocess we have an example of suboptimal behaviour, with unusually large tails towards the right edge of the $t_{RR}$ range. Nevertheless, the behaviour of the distribution for decreasing $x$ clearly indicates the proper functioning of the subtraction. 
	
\begin{figure}
		\begin{center}
			\includegraphics[width=0.3\columnwidth]{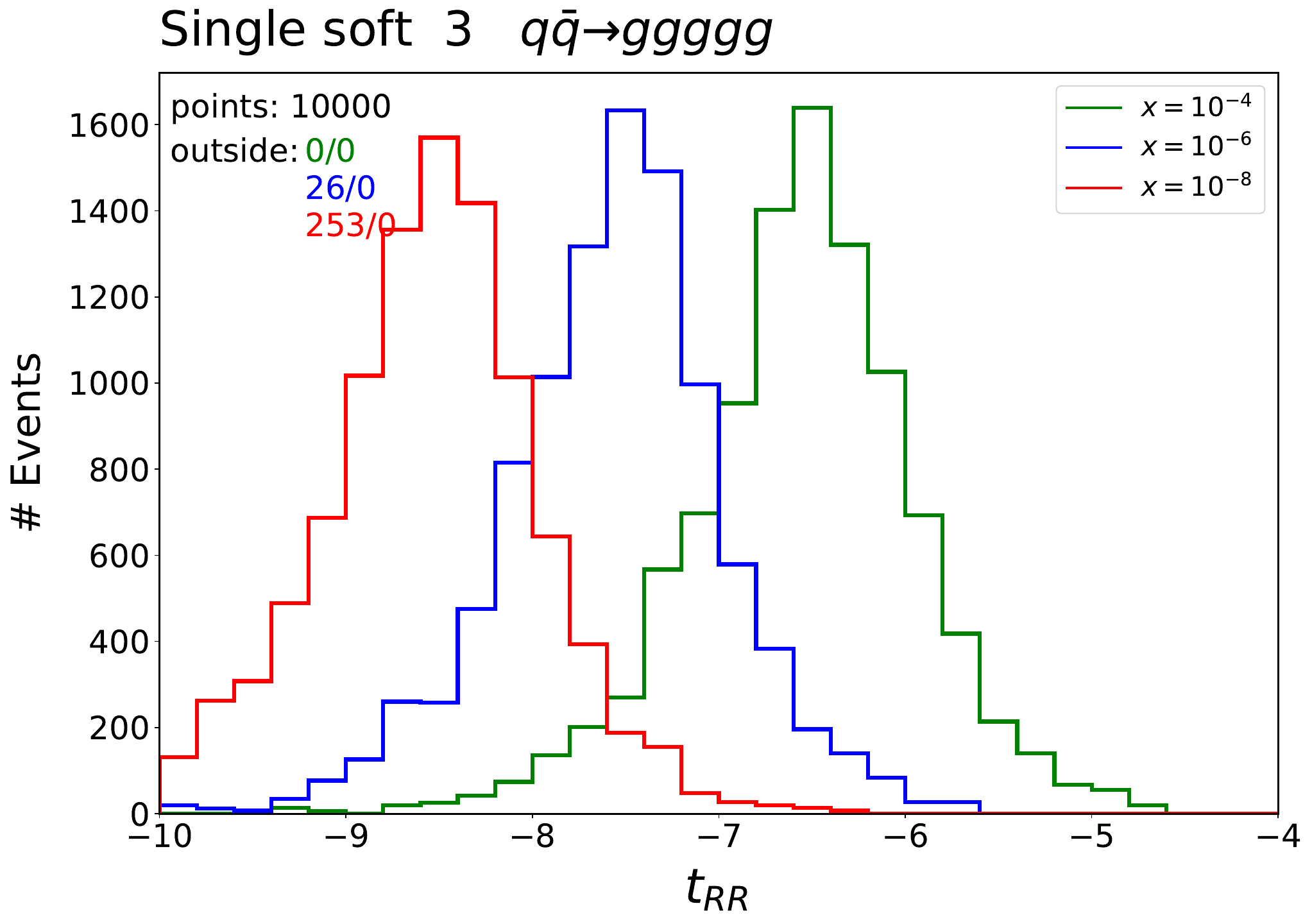}
			\hspace{0.3cm}
			\includegraphics[width=0.3\columnwidth]{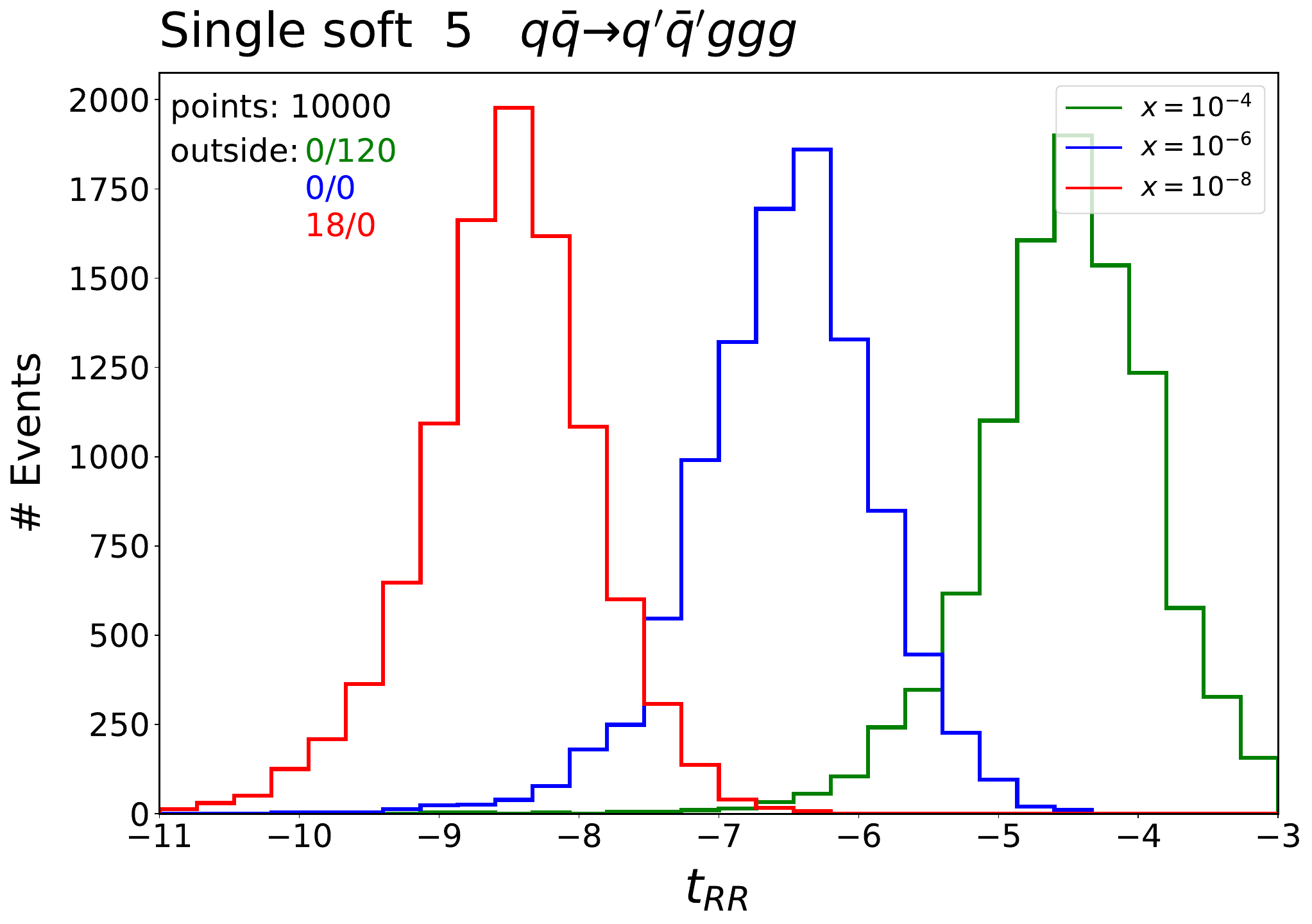}
			\hspace{0.3cm}
			\includegraphics[width=0.3\columnwidth]{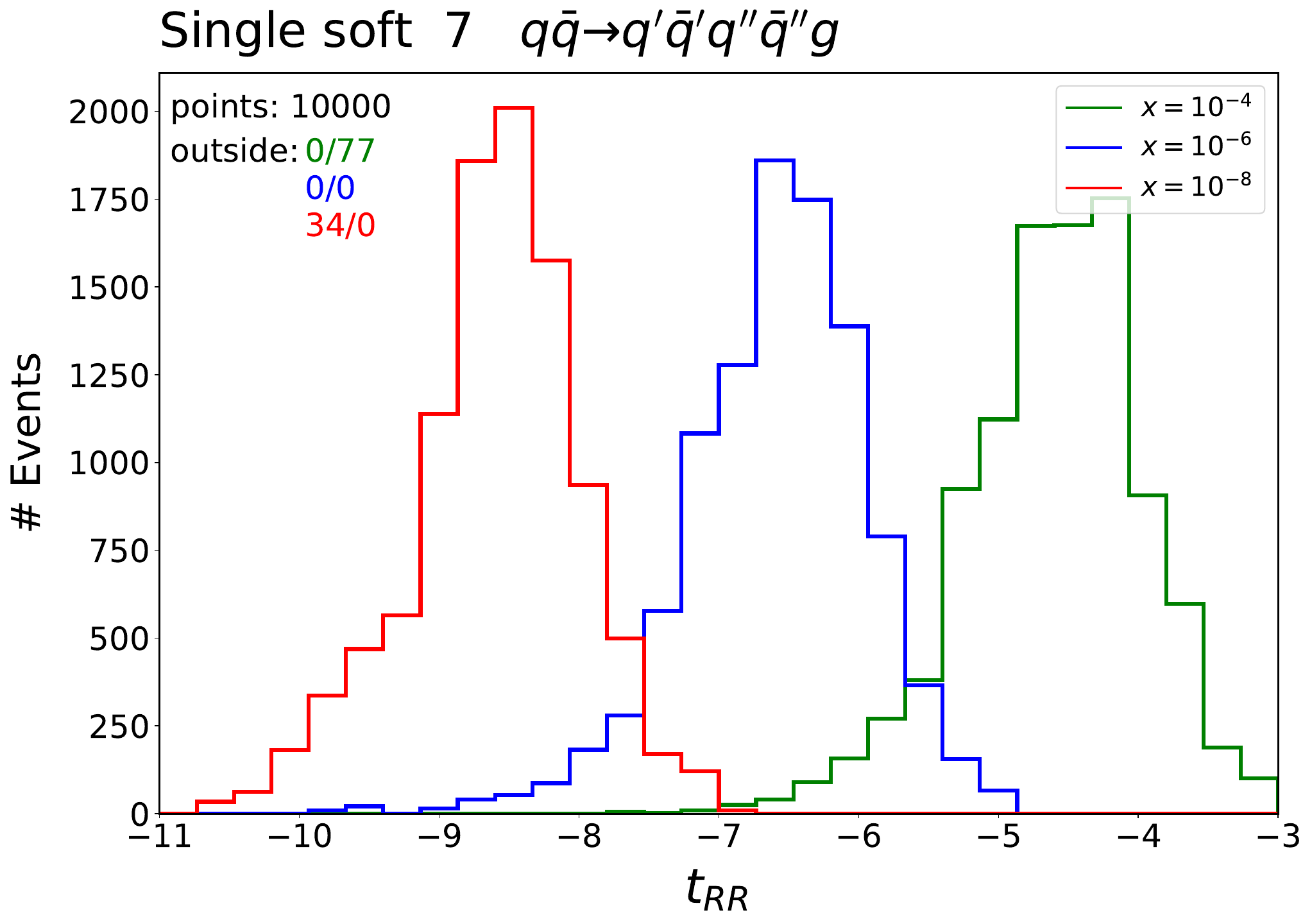}\\
			\includegraphics[width=0.3\columnwidth]{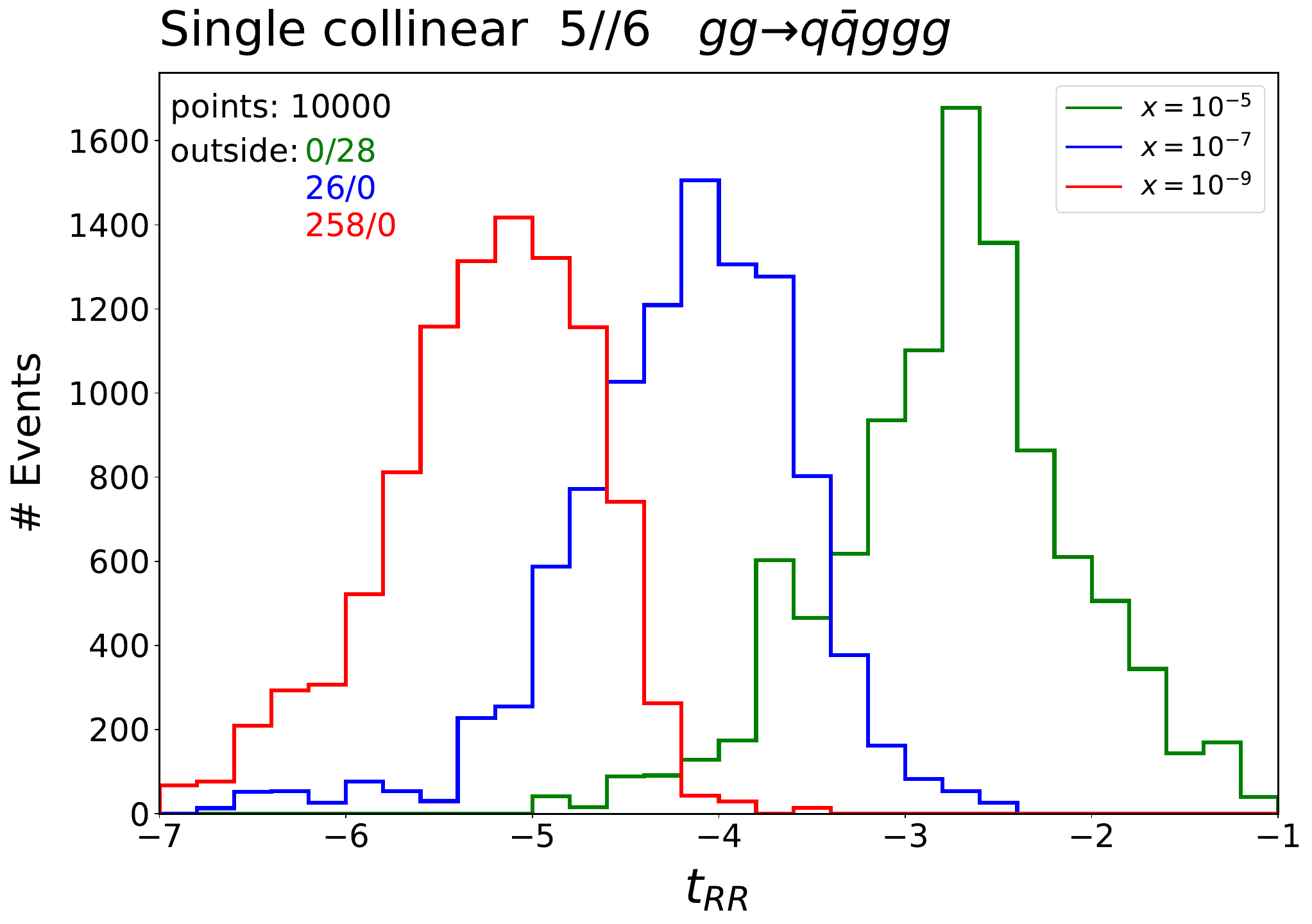}
			\hspace{0.3cm}
			\includegraphics[width=0.3\columnwidth]{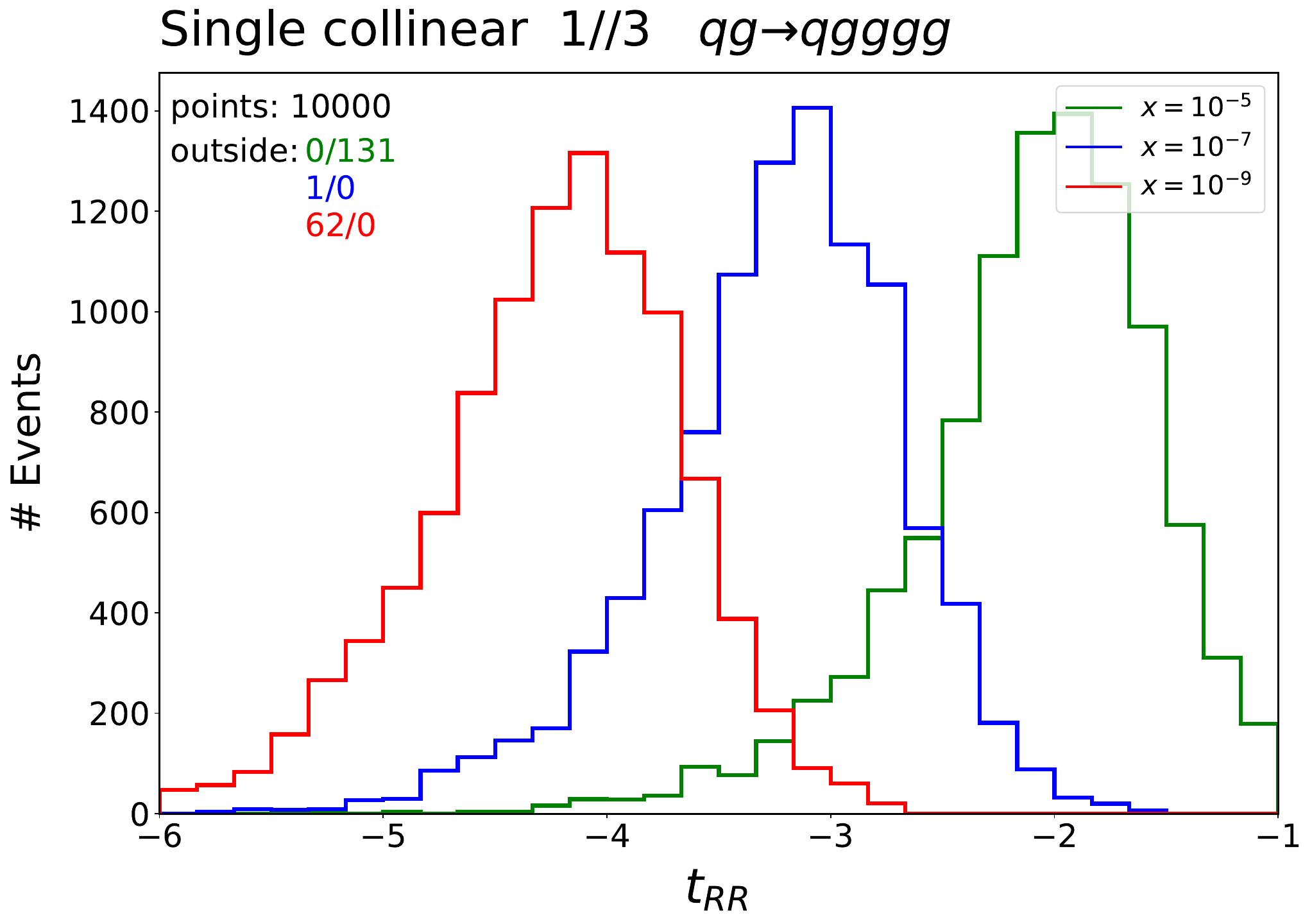}
			\hspace{0.3cm}
			\includegraphics[width=0.3\columnwidth]{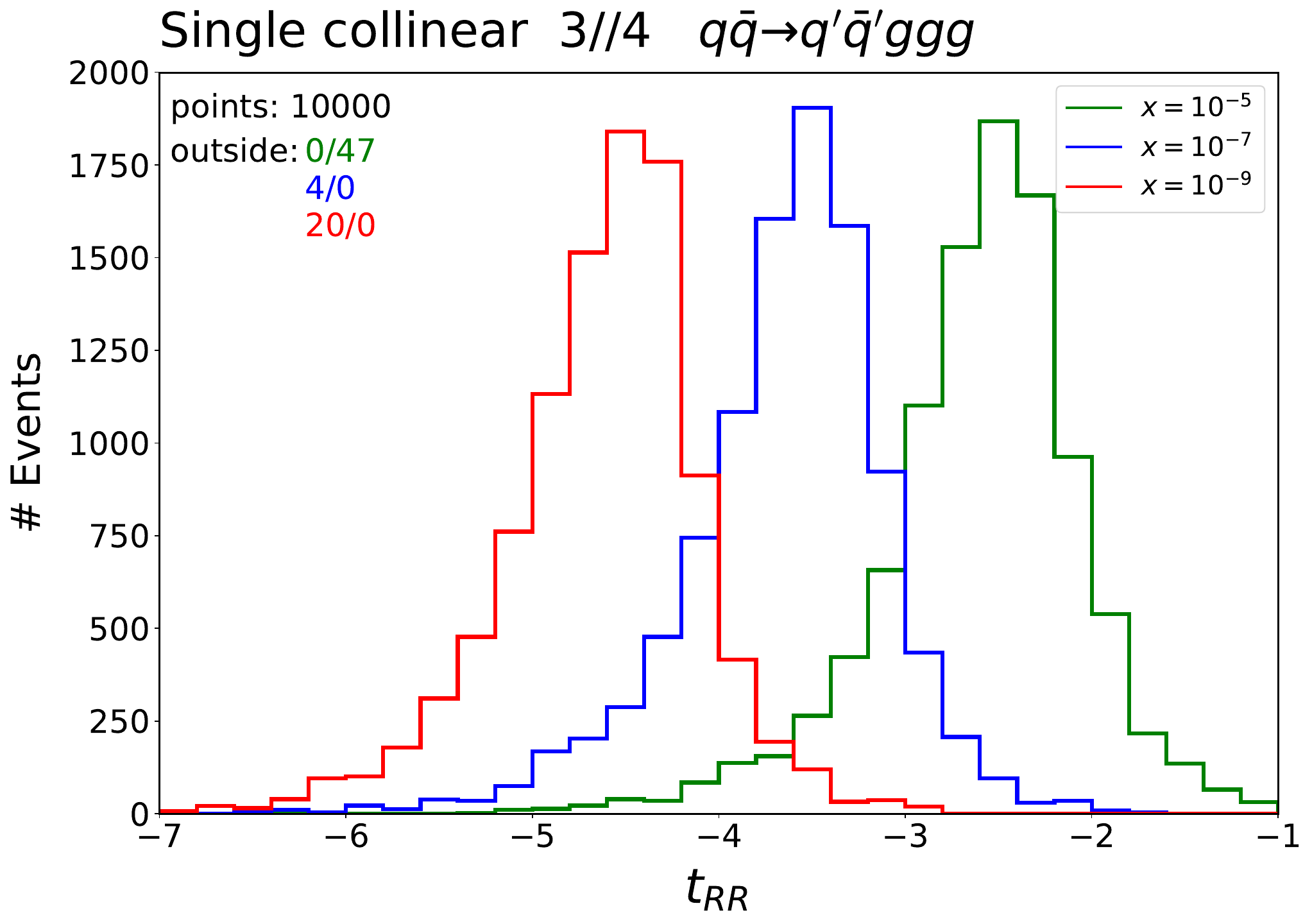}
		\end{center}
		\caption{Validation of the double-real subtraction terms in single-soft (upper row) and single-collinear (lower row) limits for a selection of subprocesses.}
		\label{fig:spikeRR_single}
	\end{figure}

	Double-real subtraction terms also need to properly remove the single-unresolved behaviour of the matrix elements. A non-trivial interplay occurs between all the components to yield the correct result. In particular, the singular behaviour of four-parton tree-level antenna functions needs to be properly subtracted by combinations of three-parton tree-level antenna functions. In Figure~\ref{fig:spikeRR_single} we demonstrate the correct behaviour of a sample of subtraction terms in single-soft and single-collinear limits.

	\section{Conclusions and outlook}\label{sec:conc}
	
	We presented the colourful antenna subtraction method, a general and fully automatable framework for NNLO calculations in massless QCD. We outlined in detail how the new formalism is applied for the generation of the infrared subtraction terms for NLO and NNLO calculations, respectively in Sections~\ref{sec:subNLO} and~\ref{sec:subNNLO}. The derivation is carried out in a completely process-independent fashion. The key aspect of the colourful antenna subtraction method is the derivation of the subtraction terms for real emissions from the subtraction terms for virtual corrections by exploiting the one-to-one correspondence between integrated and unintegrated antenna functions. We described in detail how the transition from integrated to unintegrated quantities occurs, providing a list of replacement rules and algorithmic steps to perform it. We delineated methodical procedures for the construction of each component and discussed the interplay among the different layers of the subtraction, demonstrating the coherence of the approach and the complete cancellation of infrared singularities at NLO and NNLO. We show that the large majority of the contributions to the NNLO subtraction infrastructure can be obtained by suitably iterating the operations used at NLO, with the only exception being the emission of two colour-connected unresolved partons.
	
	The new formalism overcomes some of the intrinsic limitations present in the original formulation of the antenna subtraction method, addressing in particular leading- and subleading-colour contributions in a unified and systematic framework. One of its main advantages lies in the significant simplifications it yields in constructing the double-real subtraction term at NNLO, which is essentially derived as a by-product of the double-virtual and real-virtual subtraction terms. For these reasons, in comparison to the traditional approach, the colourful antenna subtraction method is better suited for the application to high-multiplicity processes. Furthermore, it is naturally suited to be fully automated. We anticipate that in the future the colourful antenna subtraction method can be readily modified to accommodate different types of antenna functions, such as designer antenna functions~\cite{Braun-White:2023sgd,Braun-White:2023zwd,Fox:2023bma}, antenna functions for identified final-state particles~\cite{Gehrmann:2022cih,Gehrmann:2022pzd} or antenna functions with massive radiators~\cite{Gehrmann-DeRidder:2009lyc}. 
	
	As a first application of the colourful antenna subtraction approach, we generated the subtraction terms for the computation of the NNLO corrections to hadronic three-jet production. In Section~\ref{sec:3jet} we discuss the validation procedure for the large variety of subtraction terms, showing how we can assess their correct behaviour in single- and double-unresolved limits. The natural next step is the calculation of the complete NNLO correction to three-jet production at hadron colliders, which will represent the ultimate validation of the colourful antenna subtraction formalism. Its heavy computational cost has to be necessarily taken into account while designing the phenomenological study.

	\section{Acknowledgements}
	We thank Xuan Chen, James Currie, Aude Gehrmann-De Ridder, Alexander Huss, Jonathan Mo and Joao Pires for their contributions to the jet production processes in \textsc{NNLOjet}. We are very grateful to Oscar Braun-White and Xuan Chen for helpful discussions and suggestions on the manuscript. This work was supported by the Swiss National Science Foundation (SNF) under contract 200020-204200, by the European Research Council (ERC) under the European Union's Horizon 2020 research and innovation programme grant agreement 101019620 (ERC Advanced Grant TOPUP) and by the UK Science and Technology Facilities Council (STFC) under grant ST/X000745/1.

	\appendix
	
	\section{Treatment of NNLO antenna functions}\label{app:NNLOant}
	
	In this appendix we collect results needed for the conversion of integrated NNLO antenna functions into their unintegrated counterparts.
		
	\subsection{One-loop three-parton antenna functions}\label{app:X31}
	
	The insertion of unresolved partons within one-loop antenna functions is completely analogous to the insertion at tree-level, presented in Table~\ref{tab:insg}, Table~\ref{tab:insqqb} and Table~\ref{tab:insIC}. The insertion of an unresolved gluon is presented in Table~\ref{tab:insg1loop}, the insertion of an unresolved quark-antiquark pair is presented in Table~\ref{tab:insqqb1loop} and the insertion a parton within IC limits is presented in Table~\ref{tab:insIC1loop}.
	
	\renewcommand{\arraystretch}{2} 
	\addtolength{\tabcolsep}{-7pt} 
	\begin{table}[h]
		\centering
		\begin{tabular}{c|cc|cc|cc}
			 & $\mathcal{X}_3^1$ & $X_3^1$ & $\wh{\mathcal{X}}_3^1$ & $\wh{X}_3^1$ & $\wt{\mathcal{X}}_3^1$ & $\wt{X}_3^1$
			\\ \hline
			
			\multirow{3}{*}{$q-\bar{q}$} & 
			$\XFFint{A}{3}{1}(s_{ij})$ & $A_3^1(i_q,u_g,j_{\qb})$ & $\XhFFint{A}{3}{1}(s_{ij})$ & $\hat{A}_3^1(i_q,u_g,j_{\qb})$ & $\XtFFint{A}{3}{1}(s_{ij})$ & $\tilde{A}_3^1(i_q,u_g,j_{\qb})$ \\\cline{2-7}
			& $\XIFint{A}{3}{1}{q}(s_{1i})$ & $A_{3,q}^1(1_q,u_g,i_{\qb})$ & $\XhIFint{A}{3}{1}{q}(s_{1i})$ & $\hat{A}_{3,q}^1(1_q,u_g,i_{\qb})$ & $\XtIFint{A}{3}{1}{q}(s_{1i})$ & $\tilde{A}_{3,q}^1(1_q,u_g,i_{\qb})$ \\\cline{2-7}
			& $\XIIint{A}{3}{1}{q\qb}(s_{12})$ & $A_{3,q\qb}^1(1_q,u_g,2_{\qb})$ & $\XhIIint{A}{3}{1}{q\qb}(s_{12})$ & $\hat{A}_{3,q\qb}^1(1_q,u_g,2_{\qb})$ & $\XtIIint{A}{3}{1}{q\qb}(s_{12})$ & $\tilde{A}_{3,q\qb}^1(1_q,u_g,2_{\qb})$ \\\cline{2-7}
			\hline
			
			\multirow{4}{*}{$q-g$} & 
			$\XFFint{D}{3}{1}(s_{ij})$ & $2d_3^1(i_q,u_g,j_g)$ & $\XhFFint{D}{3}{1}(s_{ij})$ & $2\hat{d}_3^1(i_q,u_g,j_g)$ & & \\\cline{2-7}
			& $\XIFint{D}{3}{1}{q}(s_{1i})$ & $2d_{3,q}^1(1_q,u_g,i_g)$ & $\XhIFint{D}{3}{1}{q}(s_{1i})$ & $2\hat{d}_{3,q}^1(1_q,u_g,i_g)$ & & \\\cline{2-7}
			& $\XIFint{D}{3}{1}{g}(s_{1i})$ & $D_{3,g}^1(i_q,u_g,1_g)$ & $\XhIFint{D}{3}{1}{g}(s_{1i})$ & $\hat{D}_{3,g}^1(i_q,u_g,1_g)$ & & \\\cline{2-7}
			& $\XIIint{D}{3}{1}{qg}(s_{12})$ & $D_{3,qg}^1(1_q,u_g,2_g)$ & $\XhIIint{D}{3}{1}{qg}(s_{12})$ & $\hat{D}_{3,qg}^1(1_q,u_g,2_g)$ & & \\\cline{2-7}
			\hline
			
			\multirow{3}{*}{$g-g$} & 
			$\XFFint{F}{3}{1}(s_{ij})$ & $3f_3^1(i_g,u_g,j_g)$ & $\XhFFint{F}{3}{1}(s_{ij})$ & $3\hat{f}_3^1(i_g,u_g,j_g)$ & & \\\cline{2-7}
			& $\XIFint{F}{3}{1}{g}(s_{1i})$ & $2f_{3,g}^1(1_g,u_g,i_g)$ & $\XhIFint{F}{3}{1}{g}(s_{1i})$ & $2\hat{f}_{3,g}^1(1_g,u_g,i_g)$ & & \\\cline{2-7}
			& $\XIIint{F}{3}{1}{gg}(s_{12})$ & $F_{3,gg}^1(1_g,u_g,2_g)$ & $\XhIIint{F}{3}{1}{gg}(s_{12})$ & $\hat{F}_{3,gg}^1(1_g,u_g,2_g)$ & & \\\cline{2-7}
			\hline
		\end{tabular}
		\caption{Replacement rules to convert integrated one-loop antenna functions to their unintegrated counterparts for the insertion of an unresolved gluon (denoted with $u_g$) between the pair of hard radiators. The three columns address the leading-colour, the fermionic-loop and the subleading-colour antenna functions respectively.}
		\label{tab:insg1loop}
	\end{table}
	
	\renewcommand{\arraystretch}{2} 
	\addtolength{\tabcolsep}{+15pt} 
	\begin{table}
		\centering
		\begin{tabular}{c|cc}
			& $\mathcal{X}_3^1$ & $X_3^1$ 
			\\ \hline
			
			\multirow{2}{*}{$q-g$} & 
			$\XFFint{E}{3}{1}(s_{ij})$ & $\frac{1}{2}\left[E_3^1(i_q,u_{q},u_{\qb})+E_3^1(i_q,u_{\qb},u_{q})\right]$ \\\cline{2-3}
			&$\XIFint{E}{3}{1}{q}(s_{1i})$ & $\frac{1}{2}\left[E_{3,q}^1(1_q,u_{q},u_{\qb})+E_{3,q}^1(1_q,u_{\qb},u_{q})\right]$ \\\cline{2-3}
			\hline
			
			\multirow{2}{*}{$g-g$} & 
			$\XFFint{G}{3}{1}(s_{ij})$ & \thead{$\frac{1}{2}\big[\Big.G_3^1(i_g,u_{q},u_{\qb})+G_3^1(i_g,u_{\qb},u_{q})$ \\ $\hspace{0.5cm}+G_3^1(u_{q},u_{\qb},j_q)+G_3^1(u_{\qb},u_{q},j_q)\big.\big]$} 
			\\\cline{2-3}
			&$\XIFint{G}{3}{1}{g}(s_{1i})$ & $\frac{1}{2}\left[G_{3,g}^1(1_g,u_{q},u_{\qb})+G_{3,g}^1(1_g,u_{\qb},u_{q})\right]$ \\\cline{2-3}
			\hline
			\hline
			
			& $\wh{\mathcal{X}}_3^1$ & $\wh{X}_3^1$ 
			\\ \hline
			
			\multirow{2}{*}{$q-g$} & 
			$\XhFFint{E}{3}{1}(s_{ij})$ & $\frac{1}{2}\left[\wh{E}_3^1(i_q,u_{q},u_{\qb})+\wh{E}_3^1(i_q,u_{\qb},u_{q})\right]$ \\\cline{2-3}
			&$\XhIFint{E}{3}{1}{q}(s_{1i})$ & $\frac{1}{2}\left[\wh{E}_{3,q}^1(1_q,u_{q},u_{\qb})+\wh{E}_{3,q}^1(1_q,u_{\qb},u_{q})\right]$ \\\cline{2-3}
			\hline
			
			\multirow{2}{*}{$g-g$} & 
			$\XhFFint{G}{3}{1}(s_{ij})$ & \thead{$\frac{1}{2}\big[\Big.\wh{G}_3^1(i_g,u_{q},u_{\qb})+\wh{G}_3^1(i_g,u_{\qb},u_{q})$ \\ $\hspace{0.5cm}+\wh{G}_3^1(u_{q},u_{\qb},j_q)+\wh{G}_3^1(u_{\qb},u_{q},j_q)\big.\big]$} 
			\\\cline{2-3}
			&$\XhIFint{G}{3}{1}{g}(s_{1i})$ & $\frac{1}{2}\left[\wh{G}_{3,g}^1(1_g,u_{q},u_{\qb})+\wh{G}_{3,g}^1(1_g,u_{\qb},u_{q})\right]$ \\\cline{2-3}
			\hline
			\hline
			
			& $\wt{\mathcal{X}}_3^1$ & $\wt{X}_3^1$ 
			\\ \hline
			
			\multirow{2}{*}{$q-g$} & 
			$\XtFFint{E}{3}{1}(s_{ij})$ & $\frac{1}{2}\left[\wt{E}_3^1(i_q,u_{q},u_{\qb})+\wt{E}_3^1(i_q,u_{\qb},u_{q})\right]$ \\\cline{2-3}
			&$\XtIFint{E}{3}{1}{q}(s_{1i})$ & $\frac{1}{2}\left[\wt{E}_{3,q}^1(1_q,u_{q},u_{\qb})+\wt{E}_{3,q}^1(1_q,u_{\qb},u_{q})\right]$ \\\cline{2-3}
			\hline
			
			\multirow{2}{*}{$g-g$} & 
			$\XtFFint{G}{3}{1}(s_{ij})$ & \thead{$\frac{1}{2}\big[\Big.\wt{G}_3^1(i_g,u_{q},u_{\qb})+\wt{G}_3^1(i_g,u_{\qb},u_{q})$ \\ $\hspace{0.5cm}+\wt{G}_3^1(u_{q},u_{\qb},j_q)+\wt{G}_3^1(u_{\qb},u_{q},j_q)\big.\big]$} 
			\\\cline{2-3}
			&$\XtIFint{G}{3}{1}{g}(s_{1i})$ & $\frac{1}{2}\left[\wt{G}_{3,g}^1(1_g,u_{q},u_{\qb})+\wt{G}_{3,g}^1(1_g,u_{\qb},u_{q})\right]$ \\\cline{2-3}
			\hline
		\end{tabular}
		\caption{Replacement rules to convert integrated one-loop antenna functions to their unintegrated counterparts when the final-state gluon splits into an unresolved quark-antiquark pair, denoted with $(u_q,u_\qb)$. Symmetrization over the inserted quark-antiquark pair is always considered.}
		\label{tab:insqqb1loop}
	\end{table}
	
	\renewcommand{\arraystretch}{1.6} 
	\begin{table}
		\centering
		\begin{tabular}{c|cc}
			& $\mathcal{X}_3^1$ & $X_3^1$ 
			\\ \hline
			
			\multirow{2}{*}{$q-\bar{q}$}  
				& $\XIFint{A}{3}{1}{g}(s_{1i})$ & $2A_{3,g}^1(i_\qb,1_q,u_q)$ \\\cline{2-3}
				& $\XIIint{A}{3}{1}{gq}(s_{12})$ & $A_{3,gq}^1(u_q,1_g,2_{\qb})$ \\\cline{2-3}
			\hline
			
			\multirow{4}{*}{$q-g$} 
				& $\XIFint{E}{3}{1}{q'}(s_{1i})$ & $\frac{1}{2}\left[E_{3,q'}^1(i_q,1_q,u_\qb)+E_{3,q'}^1(i_q,u_\qb,1_q)\right]$ \\\cline{2-3}
				& $\XIFint{D}{3}{1}{gg}(s_{1i})$ & $D_{3,gg}^1(u_q,1_g,2_g)$ \\\cline{2-3}
				& $\XIIint{E}{3}{1}{q'q}(s_{12})$ & $\frac{1}{2}\left[E_{3,q'q}^1(2_q,1_q,u_\qb)+E_{3,q'q}^1(2_q,u_\qb,1_q)\right]$ \\\cline{2-3}
			\hline
			
			\multirow{2}{*}{$g-g$}  
				& $\XIFint{G}{3}{1}{q}(s_{1i})$ & $\frac{1}{2}\left[G_{3,q}^1(i_g,1_q,u_\qb)+G_{3,q}^1(i_g,u_\qb,1_q)\right]$ \\\cline{2-3}
				& $\XIIint{G}{3}{1}{qg}(s_{12})$ & $\frac{1}{2}\left[G_{3,qg}^1(2_g,1_q,u_\qb)+G_{3,qg}^1(2_g,u_\qb,1_q)\right]$ \\\cline{2-3}
			\hline
			\hline
			
			& $\wh{\mathcal{X}}_3^1$ & $\wh{X}_3^1$ 
			\\ \hline
			
			\multirow{2}{*}{$q-\bar{q}$}  
			& $\frac{1}{2}\XhIFint{A}{3}{1}{g}(s_{1i})$ & $A_{3,g}^1(i_\qb,1_q,u_q)$ \\\cline{2-3}
			& $\XhIIint{A}{3}{1}{gq}(s_{12})$ & $A_{3,gq}^1(u_q,1_g,2_{\qb})$ \\\cline{2-3}
			\hline
			
			\multirow{3}{*}{$q-g$} 
			& $\XhIFint{E}{3}{1}{q'}(s_{1i})$ & $\frac{1}{2}\left[\wh{E}_{3,q'}^1(i_q,1_q,u_\qb)+\wh{E}_{3,q'}^1(i_q,u_\qb,1_q)\right]$ \\\cline{2-3}
			& $\XhIFint{D}{3}{1}{gg}(s_{1i})$ & $D_{3,gg}^1(u_q,1_g,2_g)$ \\\cline{2-3}
			& $\XhIIint{E}{3}{1}{q'q}(s_{12})$ & $\frac{1}{2}\left[\wh{E}_{3,q'q}^1(2_q,1_q,u_\qb)+\wh{E}_{3,q'q}^1(2_q,u_\qb,1_q)\right]$ \\\cline{2-3}
			\hline
			
			\multirow{2}{*}{$g-g$}  
			& $\XhIFint{G}{3}{1}{q}(s_{1i})$ & $\frac{1}{2}\left[\wh{G}_{3,q}^1(i_g,1_q,u_\qb)+\wh{G}_{3,q}^1(i_g,u_\qb,1_q)\right]$ \\\cline{2-3}
			& $\XhIIint{G}{3}{1}{qg}(s_{12})$ & $\frac{1}{2}\left[\wh{G}_{3,qg}^1(2_g,1_q,u_\qb)+\wh{G}_{3,qg}^1(2_g,u_\qb,1_q)\right]$ \\\cline{2-3}
			\hline
			\hline
			
			& $\wt{\mathcal{X}}_3^1$ & $\wt{X}_3^1$ 
			\\ \hline
			
			\multirow{2}{*}{$q-\bar{q}$}  
			& $\XtIFint{A}{3}{1}{g}(s_{1i})$ & $2A_{3,g}^1(i_\qb,1_q,u_q)$ \\\cline{2-3}
			& $\XtIIint{A}{3}{1}{gq}(s_{12})$ & $A_{3,gq}^1(u_q,1_g,2_{\qb})$ \\\cline{2-3}
			\hline
			
			\multirow{2}{*}{$q-g$} 
			& $\XtIFint{E}{3}{1}{q'}(s_{1i})$ & $\frac{1}{2}\left[\wt{E}_{3,q'}^1(i_q,1_q,u_\qb)+\wt{E}_{3,q'}^1(i_q,u_\qb,1_q)\right]$ \\\cline{2-3}
			& $\XtIIint{E}{3}{1}{q'q}(s_{12})$ & $\frac{1}{2}\left[\wt{E}_{3,q'q}^1(2_q,1_q,u_\qb)+\wt{E}_{3,q'q}^1(2_q,u_\qb,1_q)\right]$ \\\cline{2-3}
			\hline
			
			\multirow{2}{*}{$g-g$}  
			& $\XtIFint{G}{3}{1}{q}(s_{1i})$ & $\frac{1}{2}\left[\wt{G}_{3,q}^1(i_g,1_q,u_\qb)+\wt{G}_{3,q}^1(i_g,u_\qb,1_q)\right]$ \\\cline{2-3}
			& $\XtIIint{G}{3}{1}{qg}(s_{12})$ & $\frac{1}{2}\left[\wt{G}_{3,qg}^1(2_g,1_q,u_\qb)+\wt{G}_{3,qg}^1(2_g,u_\qb,1_q)\right]$ \\\cline{2-3}
			\hline
		\end{tabular}
		\caption{Replacement rules to convert integrated one-loop antenna functions to their unintegrated counterparts for identity-changing insertions. The final-state quark (antiquark) causing the change of identity of parton $1$ is denoted with $u_q$ ($u_\qb$). Symmetrization over the collinear quark-antiquark pair is considered for $G$- and $E$-type antenna functions.}
		\label{tab:insIC1loop}
	\end{table}
	
	We also illustrate here the $\e$-poles structure of the renormalized unintegrated three-parton one-loop antenna functions. We recall that, in any kinematic configuration (II, IF or FF), the $\e$-poles of a generic $X_3^1$ antenna function can be absorbed by the following replacement:
	\begin{equation}\label{finX31}
		X_{3}^{1}(i,u,j) \to X_{3}^{1}(i,u,j) + \sum_{(l,m)=1}^{N_X}\J{1}(l,m)X_{3}^{0}(i,u,j) - M_X\J{1}(\widetilde{iu},\widetilde{uj})X_{3}^{0}(i,u,j),
	\end{equation}
	where the sum in the second term runs over the $N_X$ pairs of colour-connected partons within the antenna configuration. In Table~\ref{tab:NxMx} we give the value for the $N_X$ and $M_X$ constants for the different antenna functions. 
	\renewcommand{\arraystretch}{1.2} 
	\addtolength{\tabcolsep}{-10pt} 
	\begin{table}[h]
		\centering
		\begin{tabular}{c|ccccccccccccc}
			$X_3^1$&$A_3^1$&$\wh{A}_3^1$&$\wt{A}_3^1$&$D_3^1$&$\wh{D}_3^1$&$E_3^1$&$\wh{E}_3^1$&$\wt{E}_3^1$&$F_3^1$&$\wh{F}_3^1$&$G_3^1$&$\wh{G}_3^1$&$\wt{G}_3^1$ \\
			\hline
			$N_X$&2&2&1&3&3&2&0&1&3&3&2&2&1 \\
			$M_X$&1&0&1&2&2&2&2&0&2&2&2&2&0
		\end{tabular}
		\caption{Values of $N_X$ (colour-connected pairs) and $M_X$ to be used in~\eqref{finX31} to remove the $\e$-poles of unintegrated one-loop antenna functions.}
		\label{tab:NxMx}
	\end{table}
	
	We give here the explicit form of~\eqref{finX31} for FF kinematics. Completely analogous formulae hold for II and IF configurations.
	\begingroup
	\allowdisplaybreaks
	\begin{eqnarray}
		     A_3^1(q,g,\qb)&\to&A_3^1(q,g,\qb)     +A_3^0(q,g,\qb)\left(\J{1}(q,g)+\J{1}(g,\qb)-2\J{1}(\wt{qg},\wt{g\qb})\right),\\
		\wh{A}_3^1(q,g,\qb)&\to&\wh{A}_3^1(q,g,\qb)+A_3^0(q,g,\qb)\left(\Jh{1}(q,g)+\Jh{1}(g,\qb)\right),\\
		\wt{A}_3^1(q,g,\qb)&\to&\wh{A}_3^1(q,g,\qb)+A_3^0(q,g,\qb)\left(\J{1}(q,\qb)-\J{1}(\wt{qg},\wt{g\qb})\right),\\
		     d_3^1(q,g_1,g_2)&\to&d_3^1(q,g_1,g_2)     +d_3^0(q,g_1,g_2)\Big(\Big.\J{1}(q,g_1)+\J{1}(q,g_2)+\J{1}(g_1,g_2)\nn\\
		     &&\hspace{7.6cm}-2\J{1}(\wt{qg_1},\wt{g_1g_2})\Big.\Big),\\
		\wh{d}_3^1(q,g_1,g_2)&\to&\wh{d}_3^1(q,g_1,g_2)     +d_3^0(q,g_1,g_2)\Big(\Big.\Jh{1}(q,g_1)+\Jh{1}(q,g_2)+\Jh{1}(g_1,g_2)\nn\\
		&&\hspace{7.6cm}-2\Jh{1}(\wt{qg_1},\wt{g_1g_2})\Big.\Big),\\
			 E_3^1(q,q',\qb')&\to&E_3^1(q,q',\qb')     +E_3^0(q,q',\qb')\left(\J{1}(q,q')+\J{1}(q,\qb')-\J{1}(\wt{qq'},\wt{q\qb'})\right),\\
		\wh{E}_3^1(q,q',\qb')&\to&\wh{E}_3^1(q,q',\qb')+E_3^0(q,q',\qb')\left(-\Jh{1}(\wt{qq'},\wt{q\qb'})\right),\\
		\wt{E}_3^1(q,q',\qb')&\to&\wt{E}_3^1(q,q',\qb')+E_3^0(q,q',\qb')\left(\J{1}(q',\qb')\right),\\
			 f_3^1(g_1,g_2,g_3)&\to&f_3^1(g_1,g_2,g_3)     +f_3^0(g_1,g_2,g_3)\Big(\Big.\J{1}(g_1,g_2)+\J{1}(g_1,g_3)+\J{1}(g_2,g_3)\nn\\
			 &&\hspace{7.4cm}-2\J{1}(\wt{g_1g_2},\wt{g_2g_3})\Big.\Big),\\
		\wh{f}_3^1(g_1,g_2,g_3)&\to&\wh{f}_3^1(g_1,g_2,g_3)+f_3^0(g_1,g_2,g_3)\Big(\Big.\Jh{1}(g_1,g_2)+\Jh{1}(g_1,g_3)+\Jh{1}(g_2,g_3)\nn\\
		&&\hspace{7.4cm}-2\Jh{1}(\wt{g_1g_2},\wt{g_2g_3})\Big.\Big),\\
			 G_3^1(g,q,\qb)&\to&G_3^1(g,q,\qb)     +G_3^0(g,q,\qb)\left(\J{1}(g,q)+\J{1}(g,\qb)-2\J{1}(\wt{gq},\wt{q\qb})\right),\\
		\wh{G}_3^1(g,q,\qb)&\to&\wh{G}_3^1(g,q,\qb)+G_3^0(g,q,\qb)\left(\Jh{1}(g,q)+\Jh{1}(g,\qb)-2\Jh{1}(\wt{gq},\wt{q\qb})\right),\\
		\wt{G}_3^1(g,q,\qb)&\to&\wt{G}_3^1(g,q,\qb)+G_3^0(g,q,\qb)\left(\J{1}(q,\qb)\right).
	\end{eqnarray}
	\endgroup
	
	\subsection{Tree-level four-parton antenna functions}\label{app:X40}
	
	The replacement rules are presented in Table~\ref{tab:insggX40} for the insertion of two unresolved gluons, in Table~\ref{tab:insqqbX40} for the insertion of an unresolved quark-antiquark pair, in Table~\ref{tab:insqqbqqbX40} for the insertion of two collinear quark-antiquark pairs, while IC limits are split into Tables~\ref{tab:insICX401},~\ref{tab:insICX402} and~\ref{tab:insICX403}. For IC antenna functions with more than two quarks (antiquarks) we make use of the subscripts $q'$ and $\qb'$, in addition to $q$ and $\qb$ to better differentiate between the primary and secondary quarks. For IP antenna functions, the secondary quark pair is uniquely identified as the unresolved pair, denoted by the label $u$. The only exception is given by the $H_4^0$ antenna function, where two unresolved quark-antiquark pairs are present, so we distinguish them.
	
	\renewcommand{\arraystretch}{1.8} 
	\begin{table}[h]
		\centering
		\begin{tabular}{c|cc}
			& $\mathcal{X}_4^0$ & $X_4^0$ 
			\\ \hline
			
			\multirow{6}{*}{$q-\bar{q}$} & 
			$\XFFint{A}{4}{0}(s_{ij})$ & $A_4^0(i_q,\ugo,\ugt,j_{\qb})$  \\\cline{2-3}
			&$\XtFFint{A}{4}{0}(s_{ij})$ & $\wt{A}_{4,a}^0(i_q,\ugo,\ugt,j_{\qb})+\wt{A}_{4,a}^0(i_q,\ugt,\ugo,j_{\qb})$  \\\cline{2-3}
			& $\XIFint{A}{4}{0}{q}(s_{1i})$ & $A_{4,q}^0(1_q,\ugo,\ugt,i_{\qb})$  \\\cline{2-3}
			& $\XtIFint{A}{4}{0}{q}(s_{1i})$ & $\wt{A}_{4,q}^0(1_q,\ugo,\ugt,i_{\qb})$  \\\cline{2-3}
			& $\XIIint{A}{4}{0}{q\qb}(s_{12})$ & $A_{4,q\qb}^0(1_q,\ugo,\ugt,2_{\qb})$ \\\cline{2-3}
			& $\XtIIint{A}{4}{0}{q\qb}(s_{12})$ & $\wt{A}_{4,q\qb}^0(1_q,\ugo,\ugt,2_{\qb})$ \\\cline{2-3}
			\hline
			
			\multirow{5}{*}{$q-g$} & 
			$\XFFint{D}{4}{0}(s_{ij})$ &  \thead{$D_{4,a}^0(i_q,j_g,\ugt,\ugo)+D_{4,a}^0(i_q,\ugo,\ugt,j_g)$ \\$D_{4,c}^0(i_q,j_g,\ugt,\ugo)+D_{4,c}^0(i_q,\ugo,\ugt,j_g)$ } \\\cline{2-3}
			& $\XIFint{D}{4}{0}{q}(s_{1i})$ & $D_{4,q}^0(1_q,i_g,\ugo,\ugt)$ \\\cline{2-3}
			& $\XIFint{D}{4}{0}{g}(s_{1i})$ & $D_{4,g}^0(i_q,1_g,\ugo,\ugt)$ \\\cline{2-3}
			& $\XIFint{D}{4}{0}{g'}(s_{1i})$ & $D_{4,g'}^0(i_q,\ugo,1_g,,\ugt)$ \\\cline{2-3}
			& $\XIIint{D}{4}{0}{qg}(s_{12})$ & $D_{4,qg}^0(1_q,2_g,\ugo,\ugt)$ \\\cline{2-3}
			& $\XIIint{D}{4}{0}{qg'}(s_{12})$ & $D_{4,qg}^0(1_q,\ugo,2_g,\ugt)$ \\\cline{2-3}
			\hline
			
			\multirow{4}{*}{$g-g$} & 
			$\XFFint{F}{4}{0}(s_{ij})$ & $4\big(F_{4,a}^0(i_g,\ugo,\ugt,j_g)+F_{4,b}^0(i_g,\ugt,\ugo,j_g)\big)$ \\\cline{2-3}
			& $\XIFint{F}{4}{0}{g}(s_{1i})$ & $F_{4,g}^0(1_g,\ugo,\ugt,i_g)$ \\\cline{2-3}
			& $\XIIint{F}{4}{0}{gg}(s_{12})$ & $F_{4,gg}^0(1_g,\ugo,\ugt,2_g)$ \\\cline{2-3}
			& $\XIIint{F}{4}{0}{gg'}(s_{12})$ & $F_{4,gg'}^0(1_g,\ugo,2_g,\ugt)$ \\\cline{2-3}
			\hline
		\end{tabular}
		\caption{Replacement rules to convert integrated four-parton antenna functions to their unintegrated counterparts for the insertion of two colour-connected unresolved gluons (denoted with $u_{g_1}$ and $u_{g_2}$) between the pair of hard radiators.}
		\label{tab:insggX40}
	\end{table}
	
	\newcommand{\uq}{{u_q}}
	\newcommand{\uqb}{{u_{\qb}}}
	\newcommand{\uu}{\uq\leftrightarrow\uqb}
	\newcommand{\uqp}{{u_{q'}}}
	\newcommand{\uqbp}{{u_{\qb'}}}
	\newcommand{\uup}{\uqp\leftrightarrow\uqbp}
	
	\newcommand{\lra}{\leftrightarrow}
	
	\renewcommand{\arraystretch}{1.7} 
	\begin{table}[h]
		\centering
		\begin{tabular}{c|cc}
			& $\mathcal{X}_4^0$ & $X_4^0$ 
			\\ \hline
			
			\multirow{5}{*}{$q-\bar{q}$} & 
			$\XFFint{B}{4}{0}(s_{ij})$ &  $\frac{1}{2}B_{4}^0(i_q,u_q,u_\qb,j_\qb)+(\uu)$ \\\cline{2-3}
			& $\XFFint{C}{4}{0}(s_{ij})$ &   $\frac{1}{2}C_{4}^0(i_q,u_q,u_\qb,j_\qb)+(\uu)$\\\cline{2-3}
			& $\XIFint{B}{4}{0}{q}(s_{1i})$ &  $\frac{1}{2}B_{4,q}^0(1_q,u_q,u_\qb,i_\qb)+(\uu)$ \\\cline{2-3}
			& $\XIFint{C}{4}{0}{\qb,\qb q'\qb'}(s_{1i})$ &  $\frac{1}{2}C^0_{4,\qb,\qb q'\qb'}(1_q,u_{q},u_{\qb},i_{\qb})+(u_{q}\lra u_{\qb})$ \\\cline{2-3} 
			& $\XIIint{B}{4}{0}{q\qb}(s_{12})$ &  $\frac{1}{2}B_{4,q\qb}^0(1_q,u_q,u_\qb,2_\qb)+(\uu)$\\\cline{2-3}
			\hline
			
			\multirow{8}{*}{$q-g$} & 
			$\XFFint{E}{4}{0}(s_{ij})$ &  $\frac{1}{2}\left(E_{4,a}^0(i_q,u_q,u_\qb,j_g)+E_{4,b}^0(i_q,u_q,u_\qb,j_g)\right)+(\uu)$ \\\cline{2-3}
			&$\XtFFint{E}{4}{0}(s_{ij})$ &  $\frac{1}{2}\wt{E}_{4}^0(i_q,u_q,u_\qb,j_g)+(\uu)$ \\\cline{2-3}
			& $\XIFint{E}{4}{0}{q}(s_{1i})$ &  $\frac{1}{2}E_{4,q}^0(1_q,u_q,u_\qb,i_g)+(\uu)$ \\\cline{2-3}
			& $\XtIFint{E}{4}{0}{q}(s_{1i})$ &  $\frac{1}{2}\wt{E}_{4,q}^0(1_q,u_q,u_\qb,i_g)+(\uu)$ \\\cline{2-3}
			& $\XIFint{E}{4}{0}{g}(s_{1i})$ &  $\frac{1}{2}E_{4,g}^0(i_q,u_q,u_\qb,1_g)+(\uu)$\\\cline{2-3}
			& $\XtIFint{E}{4}{0}{g}(s_{1i})$ & $\frac{1}{2}\wt{E}_{4,g}^0(i_q,u_q,u_\qb,1_g)+(\uu)$ \\\cline{2-3}
			& $\XIIint{E}{4}{0}{qg}(s_{12})$ &  $\frac{1}{2}E_{4,qg}^0(1_q,u_q,u_\qb,2_g)+(\uu)$\\\cline{2-3}
			& $\XtIIint{E}{4}{0}{qg}(s_{12})$ &  $\frac{1}{2}\wt{E}_{4,qg}^0(1_q,u_q,u_\qb,2_g)+(\uu)$\\\cline{2-3}
			\hline
			
			\multirow{7}{*}{$g-g$} & 
			$\XFFint{G}{4}{0}(s_{ij})$ &  \thead{$\frac{1}{4}\big(\big.G_{4,a}^0(i_g,u_q,u_\qb,j_g)+G_{4,b}^0(i_g,u_q,u_\qb,j_g)$ \\ $+G_{4,c}^0(i_g,u_q,u_\qb,j_g)+(i\leftrightarrow j)\big.\big)+(\uu)$}\\\cline{2-3}
			& $\XtFFint{G}{4}{0}(s_{ij})$ &  $\frac{1}{2}\wt{G}_{4}^0(i_g,u_q,u_\qb,j_g)+(\uu)$\\\cline{2-3}
			& $\XIFint{G}{4}{0}{g}(s_{1i})$ &  $\frac{1}{2}G_{4,g}^0(1_g,u_q,u_\qb,i_g)+(\uu)$\\\cline{2-3}
			& $\XtIFint{G}{4}{0}{g}(s_{1i})$ &  $\frac{1}{2}\wt{G}_{4,g}^0(1_g,u_q,u_\qb,i_g)+(\uu)$\\\cline{2-3}
			& $\XIIint{G}{4}{0}{gg}(s_{12})$ &  $\frac{1}{2}G_{4,gg}^0(1_g,u_q,u_\qb,2_g)+(\uu)$\\\cline{2-3}
			& $\XtIIint{G}{4}{0}{gg}(s_{12})$ &  $\frac{1}{2}G_{4,gg}^0(1_g,u_q,u_\qb,2_g)+(\uu)$\\\cline{2-3}
			\hline
		\end{tabular}
		\caption{Replacement rules to convert integrated four-parton antenna functions to their unintegrated counterparts for the insertion of an unresolved quark-antiquark pair $(u_q,u_\qb)$ between the pair of hard radiators.}
		\label{tab:insqqbX40}
	\end{table}
	
	\renewcommand{\arraystretch}{1.7} 
	\begin{table}[h]
		\centering
		\begin{tabular}{c|cc}
			& $\mathcal{X}_4^0$ & $X_4^0$ 
			\\ \hline
			
			\multirow{1}{*}{$g-g$} & 
			$\XFFint{H}{4}{0}(s_{ij})$ &  $\frac{1}{8}\left[\left(H_4^0\left(u_q,u_\qb,u_{q'},u_{\qb'}\right)+H_4^0\left(u_{q'},u_{\qb'},u_q,u_\qb\right)+(\uu)\right)+(u_{q'}\leftrightarrow u_{\qb'})\right]$\\\cline{2-3} 
			\hline
		\end{tabular}
		\caption{Replacement rules to convert integrated four-parton antenna functions to their unintegrated counterparts for the insertion of two unresolved quark-antiquark pairs $(u_q,u_\qb)$ and $(u'_q,u'_\qb)$ between the pair of hard radiators.}
		\label{tab:insqqbqqbX40}
	\end{table}
	
	\renewcommand{\arraystretch}{1.7} 
	\begin{table}[h]
		\centering
		\begin{tabular}{c|cc}
			& $\mathcal{X}_4^0$ & $X_4^0$ 
			\\ \hline
			
			\multirow{16}{*}{$q-\bar{q}$} & 
			 $\XIFint{A}{4}{0}{g}(s_{1i})$ &  $A^0_{4,g}(u_q,1_g,u_g,i_{\qb})$ \\\cline{2-3}
			& $\XtIFint{A}{4}{0}{g}(s_{1i})$ &  $\wt{A}^0_{4,g}(u_q,1_g,u_g,i_{\qb})$ \\\cline{2-3}
			& $\XIFint{B}{4}{0}{q'}(s_{1i})$ &  $\frac{1}{2}B^0_{4,q'}(u_q,1_{q'},u_{\qb'},i_{\qb})+(1_{q'}\lra u_{\qb'})$ \\\cline{2-3}
			& $\XIFint{C}{4}{0}{q}(s_{1i})$ &   $\frac{1}{2}C_{4,q}^0(i_q,u_q,u_\qb,1_\qb)+(\uu)$\\\cline{2-3} 
			& $\XIFint{C}{4}{0}{\qb',q\qb\qb'}(s_{1i})$ &   $\frac{1}{2}C^0_{4,\qb',q\qb\qb'}(i_q,1_{q'},u_{\qb'},u_{\qb})+(1_{q'}\lra u_{\qb'})$\\\cline{2-3} 
			& $\XIIint{A}{4}{0}{qg}(s_{12})$ &  $A^0_{4,qg}(1_q,2_g,u_g,u_{\qb})$ \\\cline{2-3}
			& $\XIIint{A}{4}{0}{qg'}(s_{12})$ &  $A^0_{4,qg}(1_q,u_g,2_g,u_{\qb})$ \\\cline{2-3}
			& $\XtIIint{A}{4}{0}{qg}(s_{12})$ &  $\wt{A}^0_{4,qg}(1_q,2_g,u_g,u_{\qb})$ \\\cline{2-3}
			& $\XIIint{A}{4}{0}{gg}(s_{1i})$ &  $A^0_{4,gg}(u_q,1_g,2_g,u_{\qb})$  \\\cline{2-3}
			& $\XtIIint{A}{4}{0}{gg}(s_{12})$ &  $\wt{A}^0_{4,gg}(u_q,1_g,2_g,u_{\qb})$ \\\cline{2-3}
			& $\XIIint{B}{4}{0}{qq'}(s_{12})$ &  $\frac{1}{2}B^0_{4,qq'}(1_q,2_{q'},u_{\qb'},u_{\qb})+(2_{q'}\lra u_{\qb'})$\\\cline{2-3}
			& $\XIIint{B}{4}{0}{q'\qb'}(s_{12})$ &  $B^0_{4,q'\qb'}(u_q,1_{q'},2_{\qb'},u_{\qb})$\\\cline{2-3}
			& $\XIIint{C}{4}{0}{q\qb}(s_{12})$ &  $\frac{1}{2}C_{4,q\qb}^0(2_q,u_{q'},u_{\qb'},1_\qb)+(\uup)$\\\cline{2-3} 
			& $\XIIint{C}{4}{0}{qq'}(s_{12})$ &  $\frac{1}{2}C_{4,q\qb}^0(u_q,u_q',2_\qb',1_\qb)+(u_{q'}\lra 2_{\qb'})$\\\cline{2-3} 
			& $\XIIint{C}{4}{0}{q'\qb'}(s_{12})$ & $C_{4,q\qb}^0(u_q,2_q',1_\qb',u_\qb)$ \\\cline{2-3} 
			& $\XIIint{C}{4}{0}{\qb\qb'}(s_{12})$ &  $\frac{1}{2}C_{4,q\qb}^0(1_q,2_q',u_\qb',u_\qb)+(2_{q'}\lra u_{\qb'})$\\\cline{2-3} 
			\hline
		\end{tabular}
		\caption{Replacement rules to convert integrated quark-antiquark four-parton antenna functions to their unintegrated counterparts in the presence of identity-changing limits. The label $u_i$, with $i=g,q,\qb,q',\qb'$ denotes partons (even potentially hard ones) participating to the identity-changing limits.}
		\label{tab:insICX401}
	\end{table}
	
	\renewcommand{\arraystretch}{1.7} 
	\begin{table}[h]
		\centering
		\begin{tabular}{c|cc}
			& $\mathcal{X}_4^0$ & $X_4^0$ 
			\\ \hline
			
			\multirow{10}{*}{$q-g$} & 
			$\XIFint{E}{4}{0}{q'}(s_{1i})$ &  $\frac{1}{2}E^0_{4,q'}(u_q,1_q',u_{\qb'},i_g)+(1_{q'}\lra u_{\qb'})$ \\\cline{2-3}
			& $\XtIFint{E}{4}{0}{q'}(s_{1i})$ &  $\frac{1}{2}\wt{E}^0_{4,q'}(u_q,1_q',u_{\qb'},i_g)+(1_{q'}\lra u_{\qb'})$ \\\cline{2-3}
			& $\XIIint{D}{4}{0}{gg}(s_{12})$ &  $D^0_{4,gg}(u_q,1_g,2_g,u_g)$\\\cline{2-3}
			& $\XIIint{D}{4}{0}{gg'}(s_{12})$ &  $D^0_{4,gg'}(u_q,1_g,u_g,2_g)$\\\cline{2-3}
			& $\XIIint{E}{4}{0}{qq'}(s_{12})$ &  $\frac{1}{2}E^0_{4,qq'}(1_q,2_q',u_{\qb'},u_g)+(2_{q'}\lra u_{\qb'})$\\\cline{2-3}
			& $\XtIIint{E}{4}{0}{qq'}(s_{12})$ &  $\frac{1}{2}\wt{E}^0_{4,qq'}(1_q,2_q',u_{\qb'},i_g)+(2_{q'}\lra u_{\qb'})$\\\cline{2-3}
			& $\XIIint{E}{4}{0}{q\qb'}(s_{12})$ &  $\frac{1}{2}E^0_{4,q\qb'}(1_q,u_q',2_{\qb'},u_g)+(u_{q'}\lra 2_{\qb'})$\\\cline{2-3}
			& $\XtIIint{E}{4}{0}{q'\qb'}(s_{12})$ &  $\frac{1}{2}\wt{E}^0_{4,q\qb'}(1_q,u_q',2_{\qb'},u_g)+(u_{q'}\lra 2_{\qb'})$\\\cline{2-3}
			& $\XIIint{E}{4}{0}{q'g}(s_{12})$ &  $\frac{1}{2}E^0_{4,q'g}(u_q,1_q',u_{\qb'},2_g)+(1_{q'}\lra u_{\qb'})$\\\cline{2-3}
			& $\XtIIint{E}{4}{0}{q'g}(s_{12})$ &  $\frac{1}{2}\wt{E}^0_{4,q'g}(u_q,1_q',u_{\qb'},2_g)+(1_{q'}\lra u_{\qb'})$\\\cline{2-3}
			\hline
		\end{tabular}
		\caption{Replacement rules to convert integrated quark-gluon four-parton antenna functions to their unintegrated counterparts in the presence of identity-changing limits. The label $u_i$, with $i=g,q,\qb,q',\qb'$ denotes partons (even potentially hard ones) participating to the identity-changing limits.}
		\label{tab:insICX402}
	\end{table}
	
	\renewcommand{\arraystretch}{1.7} 
	\begin{table}[h]
		\centering
		\begin{tabular}{c|cc}
			& $\mathcal{X}_4^0$ & $X_4^0$ 
			\\ \hline
			
			\multirow{8}{*}{$g-g$} & 
			$\XIFint{G}{4}{0}{q}(s_{1i})$ &  $\frac{1}{2}G^0_{4,q}(i_g,1_q,u_{\qb},u_g)+(1_q\lra u_{\qb})$\\\cline{2-3}
			& $\XtIFint{G}{4}{0}{q}(s_{1i})$ &  $\frac{1}{2}\wt{G}^0_{4,q}(i_g,1_q,u_{\qb},u_g)+(1_q\lra u_{\qb})$\\\cline{2-3}
			& $\XIFint{H}{4}{0}{q}(s_{1i})$ &  $\frac{1}{2}\left(H^0_{4,q}(1_q,u_{q},\uqp,\uqbp)+(\uup)\right)$\\\cline{2-3}
			& $\XIIint{G}{4}{0}{q\qb}(s_{12})$ &  $G^0_{4,q\qb}(u_{g_1},1_q,2_{\qb},u_{g_2})$\\\cline{2-3}
			& $\XtIIint{G}{4}{0}{q\qb}(s_{12})$ &  $\wt{G}^0_{4,qq'}(u_{g_1},1_q,2_{\qb},u_{g_2})$\\\cline{2-3}
			& $\XIIint{G}{4}{0}{qg}(s_{12})$ &  $\frac{1}{2}G^0_{4,qg}(2_g,1_q,u_{\qb},u_g)+(1_q\lra u_{\qb})$\\\cline{2-3}
			& $\XtIIint{G}{4}{0}{qg}(s_{12})$ &  $\frac{1}{2}\wt{G}^0_{4,qg}(2_g,1_q,u_{\qb},u_g)+(1_q\lra u_{\qb})$\\\cline{2-3}
			& $\XIIint{H}{4}{0}{qq'}(s_{12})$ &  $\frac{1}{2}H^0_{4,qq'}(1_q,u_{\qb},2_q',u_{\qb'})+(2_q'\lra u_{\qb'})$\\\cline{2-3}
			\hline
		\end{tabular}
		\caption{Replacement rules to convert integrated gluon-gluon four-parton antenna functions to their unintegrated counterparts in the presence of identity-changing limits. The label $u_i$, with $i=g,q,\qb,q',\qb'$ denotes partons (even potentially hard ones) participating to the identity-changing limits.}
		\label{tab:insICX403}
	\end{table}


	\clearpage
	
	\bibliographystyle{JHEP}
	\bibliography{bib_col_ant}

\end{document}